\documentclass[fleqn,usenatbib]{mnras}

\usepackage{newtxtext,newtxmath}

\usepackage[T1]{fontenc}

\DeclareRobustCommand{\VAN}[3]{#2}
\let\VANthebibliography\thebibliography
\def\thebibliography{\DeclareRobustCommand{\VAN}[3]{##3}\VANthebibliography}

\usepackage{graphicx}	
\usepackage{amsmath}	

\title[Physical properties of NEA (2102) Tantalus]{Physical properties of near-Earth asteroid (2102) Tantalus from multi-wavelength observations \thanks{Based in part on observations collected at the European Organisation for Astronomical Research in the Southern Hemisphere under ESO programme 185.C-1033(C,D,E,I,V) and the Arecibo Planetary Radar observations collected under programme R3037.}}

\author[A. Rożek et al.]{
Agata Rożek,$^{1,2}$\thanks{E-mail: a.rozek@ed.ac.uk (AR)}
Stephen C. Lowry,$^{2}$
Benjamin Rozitis,$^{3}$  
Lord R. Dover,$^{2}$
Patrick A. Taylor,$^{4,5,6}$
\newauthor
Anne Virkki,$^{7,4}$ 
Simon F. Green,$^{3}$   
Colin Snodgrass,$^{1,3}$
Alan Fitzsimmons,$^{8}$
Justyn Campbell-White,$^{9}$ 
\newauthor
Sedighe Sajadian,$^{10}$  
Valerio Bozza,$^{11,12}$  
Martin J. Burgdorf,$^{13}$
Martin Dominik,$^{14}$ 
R. Figuera Jaimes,$^{15}$ 
\newauthor
Tobias C. Hinse,$^{16,17}$  
Markus Hundertmark,$^{18}$ 
Uffe G. J{\o}rgensen,$^{19}$ 
Pen{\'e}lope Longa-Pe{\~n}a,$^{20}$  
Markus Rabus,$^{21}$ 
\newauthor
Sohrab Rahvar,$^{22}$ 
Jesper Skottfelt,$^{23}$  
and 
John Southworth$^{24}$  
\\
$^{1}$ Institute for Astronomy, University of Edinburgh, Royal Observatory, Edinburgh, EH9 3HJ, UK, 
$^{2}$ School of Physical Sciences, University of Kent, Ingram \\ Building, Canterbury,  CT2 7NH, UK, 
$^{3}$ School of Physical Sciences, The Open  University, Milton Keynes, MK7 6AA, UK, 
$^4$ Arecibo Observatory, \\ Universities Space Research Association, HC 3 Box 53995, Arecibo, PR 00612, Puerto Rico, USA, 
$^5$ National Radio Astronomy Observatory, 1180 Boxwood Estate Rd., \\ Charlottesville, VA 22903, USA,
$^6$ Green Bank Observatory, P.O. Box 2, Green Bank, WV 24494, USA,
$^{7}$  Department of Physics, University of Helsinki,  \\ Gustaf Hällströmin katu 2, 00560, Helsinki, Finland,
$^{8}$ Astrophysics Research Centre, Queens University Belfast, BT7 1NN, UK, \\
$^{9}$ SUPA, School of Science and Engineering,  University  of Dundee, DD1 4HN, Scotland, UK, 
$^{10}$ Department~of~Physics,  Isfahan~University~of~Technology, \\ Isfahan~84156-83111,~Iran, 
$^{11}$ Dipartimento di Fisica "E.R. Caianiello", Universit{\`a} di Salerno,  Via Giovanni Paolo II 132, 84084 Fisciano, Italy, \\
$^{12}$ Istituto Nazionale di Fisica Nucleare,  Sezione di Napoli, Strada Comunale Cinthia, 80126 Napoli, Italy, 
$^{13}$ Universit{\"a}t Hamburg, \\  Faculty of Mathematics, Informatics and Natural Sciences, Department of Earth Sciences,  Meteorological Institute, Bundesstra\ss{}e 55,  20146 Hamburg, Germany,  \\ 
$^{14}$ University of St Andrews, Centre for Exoplanet Science, SUPA School of Physics \& Astronomy, North Haugh, St Andrews, KY16 9SS, UK, \\
$^{15}$ Facultad de Ingeniería y Tecnología, Universidad San Sebastian, General Lagos 1163, Valdivia 5110693, Chile,
$^{16}$ Nicolaus Copernicus University, \\ Institute of Astronomy, ul. Grudziądzka 5, 87-100 Toruń, Poland, 
$^{17}$ Chungnam National University, Department of Astronomy,  Space Science and Geology, \\ 34134 Daejeon, South  Korea, 
$^{18}$ Astronomisches Rechen-Institut,  Zentrum f{\"u}r Astronomie der Universit{\"a}t Heidelberg  (ZAH), 69120 Heidelberg, Germany, \\
$^{19}$ Centre for ExoLife Sciences, Niels Bohr Institute, University of Copenhagen, {\O}ster Voldgade 5, 1350 Copenhagen,  Denmark, 
$^{20}$ Centro de Astronom{\'{\i}}a, \\ Universidad de Antofagasta, Av.\ Angamos 601, Antofagasta, Chile, 
$^{21}$ Departamento de Matem\'atica y F\'isica Aplicadas, Facultad de Ingenier\'ia, \\ Universidad Cat\'olica de la Sant\'isima Concepci\'on, Alonso de Rivera 2850, Concepci\'on, Chile, 
$^{22}$ Department of Physics, Sharif University of Technology, \\ PO Box 11155-9161 Tehran, Iran, 
$^{23}$ Centre for Electronic Imaging,  School of Physical  Sciences, The Open University, Milton Keynes, MK7 6AA, UK, \\
$^{24}$ Astrophysics Group, Keele  University, Staffordshire, ST5 5BG, UK. 
}

\date{Accepted XXX. Received YYY; in original form ZZZ}

\pubyear{2022}

\begin{document}
\label{firstpage}
\pagerange{\pageref{firstpage}--\pageref{lastpage}}
\maketitle

\begin{abstract}

Between 2010 and 2017 we have collected new optical and radar observations of the potentially hazardous asteroid (2102)~Tantalus from the ESO NTT and Danish telescopes at the La Silla Observatory and from the Arecibo planetary radar. The object appears to be nearly spherical, showing a low amplitude light-curve variation and limited large-scale features in the radar images. The spin-state is difficult to constrain with the available data; including a certain light-curve subset significantly changes the spin-state estimates, and the uncertainties on period determination are significant. Constraining any change in rotation rate was not possible, despite decades of observations. 
The convex lightcurve-inversion model, with rotational pole at $\lambda=210\pm41\degr$ and $\beta=-30\pm35\degr$, is more flattened than the two models reconstructed by including radar observations: with prograde ($\lambda=36\pm23\degr$, $\beta=30\pm15\degr$), and with retrograde rotation mode ($\lambda=180\pm24\degr$, $\beta=-30\pm16\degr$). Using data from WISE we were able to determine that the prograde model produces the best agreement in size determination between radar and thermophysical modelling. Radar measurements indicate possible variation in surface properties, suggesting one side might have lower radar albedo and be rougher at centimetre-to-decimetre scale than the other. However, further observations are needed to confirm this. 
Thermophysical analysis indicates a surface covered in fine-grained regolith, consistent with radar albedo and polarisation ratio measurements. Finally, geophysical investigation of the spin-stability of Tantalus shows that it could be exceeding its critical spin-rate via cohesive forces.

\end{abstract}

\begin{keywords}
minor planets, asteroids: individual: (2102) Tantalus -- techniques: photometric -- techniques: radar astronomy
\end{keywords}



\section{Introduction}

The asteroidal Yarkovsky-O'Keefe-Radzievskii-Paddack (YORP) effect is a tiny recoil torque resulting from reflection, absorption and re-radiation of thermal photons illuminating surfaces of small bodies \citep{Rubincam2000}.   
It is considered to be one of the main drivers in the physical and dynamical evolution of small asteroids close to the Sun. The YORP effect changes the rotational momentum of small bodies affecting both the spin-axis orientation and the rate of rotation. The latter can be directly detected through ground-based observations \citep[e.g.][]{Lowry2007, Taylor2007}. The effect is strongest on near-Earth asteroids (NEAs) as it scales inversely with the squares of size and solar distance. Notably, all direct detections to date are in the spin-up sense and the detections are not always in line with theoretical predictions due to high sensitivity of YORP to surface properties and shape details \citep[for example:][]{Statler2009,Rozitis2013,Golubov2016}, and internal properties \citep[e.g.][]{Lowry2014}.

In order to increase the number of detections to aid theoretical advances of {the} YORP effect we are conducting a long-term observing campaign of a relatively large sample of NEA to measure their YORP strengths. Observations of 42 NEAs were carried out, primarily {from a} European Southern Observatory (ESO) Large Programme aimed at optical and infrared photometric monitoring. While this initial phase is now completed, additional data are still being collected with associated programmes at other ground-based facilities.
A detection of a change in the sidereal rotation rate can be achieved in a few ways. One method involves direct measurement {of} the rotation period at different times. This was possible for the very first asteroidal YORP detection \citep{Lowry2007}. Asteroid (54509) YORP's rotation period is only around $730\,\textrm{s}$.
Due to the fast rotation and a small variation in viewing geometry, a very precise determination of the sidereal rotation period was possible. 
For this analysis, observations from each apparition were grouped in year-long batches (mid-2001 to mid-2002, mid-2002 to mid-2003, and so on), and a different period value was obtained for each of the 5 years of the observing campaign. 
A linear change in the rotation period was measured and attributed to YORP as the most likely explanation \citep{Taylor2007}. However, this object was very small (about $57\,\textrm{m}$ radius) making the YORP effect relatively strong. For objects with a weaker YORP acceleration and a slower rotation this technique becomes less practical as it becomes more difficult to make sufficiently accurate period measurements for periods on the order of hours to detect a subtle YORP-induced change. 

A second approach combines a shape model determined using other types of observations with long-term light-curve observations. This can be a radar-derived model, which was the case for asteroids (54509) YORP, (101955) Bennu, and (68346) 2001 KZ66 \citep{Taylor2007,Nolan2019,Zegmott2021}, or a spacecraft model, which helped detect {a} spin-state change for (25143) Itokawa \citep{Lowry2014}. A subset of the light curves may be used for breaking pole degeneracy inherent in radar data interpretation, %
like was done for example when determining YORP detection limit for asteroid (85990) 1999 JV6 \citep{Rozek2019JV6}. All available  light-curve observations are then compared with synthetic light curves, generated using a shape model developed assuming a constant rotation period, and any offset in rotational phase required to align them is measured. A quadratic trend in the measured phase offsets indicates a linear change in the rotation rate.

Finally, {another} approach incorporates a linear change in the rotation rate directly in the shape modelling of the object, normally in the light-curve inversion  \citep{Kaasalainen2007,Durech2008,Durech2012,Durech2018,Durech2021}. The method requires a range of shape models to be developed assuming either a constant or varying sidereal rotation period.  The quality of the fit of model light curves to the data is compared between the constant-period and varying-period solutions to assess the spin-state. We adopted this technique here, although we did not obtain a conclusive YORP measurement.

The subject of this study, (2102) Tantalus, is a potentially hazardous asteroid classified as an Sr spectral type \citep{Thomas2014}.  It was observed photometrically in 1995 and 1996 from Bochum and Od{\v{r}}ejov Observatories by \citet{Pravec1997b} (these light curves are publicly available, {and} we include them in our analysis, labelled 1-3 in Table \ref{tab:obs}) {with the reported mean synodic period $2.391 \pm 0.01\,\textrm{h}$, close to the rubble pile rotationally-induced fission limit}.

{Observations} in 2014 and 2017 from Palmer Divide Station \citep[these light curves are labelled 12-30 in Table~\ref{tab:obs}]{Warner2015,Warner2017} suggest a spin period around $2.383\pm0.001\,\rm h$.  
{Additionally, a secondary period of  $16.38\pm0.02\,\textrm{h}$ was detected in the data that suggested presence of an elongated satellite. Further optical observations from a small telescope in Spain (Isaac Aznar Observatory, $0.36\,\rm{m}$) confirmed the approximately $2.39\,\rm{h}$ rotation period and revealed a possible $8.22\,\rm{h}$ secondary period \cite{Vaduvescu2017}.} 
The asteroid diameter was most recently estimated to be $1.762	\pm 0.603 \,\rm km$ with the NEOWISE survey \citep{Masiero2017}.

We observed Tantalus both photometrically with optical telescopes, and using the Arecibo planetary radar. In this paper we discuss the observing campaign (Sec.~\ref{sec:observations}), how we attempted to constrain YORP using photometric light curves alone (Sec.~\ref{sec:light-curves}),  how we combined optical light-curve and radar observations to develop a spin-state and shape model (Sec.~\ref{sec:modelling}), radar surface properties developed from calibrated radar spectra (Sec.~\ref{sec:radarprop}), thermophysical properties developed by combining our shape models with the WISE infrared data (Sec.~\ref{sec:atpm}), and finally geophysical analysis using one of the shape models (Sec.~\ref{sec:geo}). Issues concerning modelling a nearly spherical object are highlighted here.

\section{Observations} %
\label{sec:observations}

\begin{table*}
		\centering          
	\caption{A chronological list of optical light curves of asteroid (2102) Tantalus  used in this study. %
	For each {light curve} a numerical 	`{ID}', Universal Time (UT) `{Date}'  {at} the beginning of the night, the heliocentric (`$R_h$') and geocentric (`$\Delta$') distances measured in AU, the solar phase angle (`$\alpha$'), the {observer-centred} ecliptic longitude (`$\lambda_o$') and latitude (`$\beta_o$'), 
	`{Total}' length of the {light curve}, the apparent peak-to-peak `{Amplitude}', and the `{Observing facility}' used to obtain the {light curve} are listed. %
	Where applicable a `{Reference}' to the already published work is given. 
	Each line represents a single {light curve} (sometimes a few segments were observed on a single night). The shape modelling using light-curve inversion was done twice, once using the full light-curve data set and one restricted to a subset of {light curves}  marked with black circles in the `{LC-subset model}' column (Sec.~\ref{sec:light-curves}). For the model combining radar data with optical light curves we used a subset of the light curves from our observing campaign selected for SNR and observing geometry coverage (black circles in the `{LC+radar model}' column; Sec.~\ref{sec:modelling}).
		{Observing facility key (with MPC site code):} 
		(Bochum, {809}), European Southern Observatory $0.61\,\rm m$ Bochum telescope in La Silla, Chile;
		(Ond\v{r}ejov, 557), Academy of Sciences of the Czech Republic Ond\v{r}ejov Observatory $0.65\, \rm m$ telescope, Czechia;
		(NTT, {809}), European Southern Observatory $3.58\,\rm m$ New Technology Telescope in La Silla, Chile;
		(PDS, {U82}), Palmer Divide Station (various telescopes with $0.3-0.5\,\rm m$ mirrors), California, USA;
		(Danish, {809}), European Southern Observatory $1.54\,\rm m$ Danish telescope in La Silla, Chile.
	}
	\label{tab:obs}             

\resizebox{\hsize}{!}{
		\begin{tabular}{ ccccc ccccc cccc}
			\hline \hline \noalign{\smallskip}
 {ID} 	&	   {UT Date}    	&	 {$R_h$} 	&	 {$\Delta$} 	&	 {$\alpha$}  	&	 {$\lambda_o$} 	&	  $\beta_o$  	&	 Total 	&	 {Ampl.}  	&	 {Filter} 	&	 {Obs.}  	&	 {LC-subset} 	&	 {LC+radar} &	 {Reference} 	\\	
&	    {[yyyy-mm-dd]}    	&	 {[AU]}  	&	   {[AU]}   	&	 {[$\degr$]} 	&	 {[$\degr$]} 	&	 {[$\degr$]} 	&	 {[h]} 	&	 [mag] 	&	      	&	          facility  	&	 model 	&	 model &	  	\\ \hline	
1	&	 1994-Dec-09	&	1.182	&	0.534	&	55.64	&	119.6	&	-79.3	&	7.4	&	0.13	&	R	&	Bochum	&	 \textbullet	&		&	\cite{Pravec1997b}	\\
2	&	1995-Jun-28 	&	1.327	&	0.343	&	21.93	&	255.9	&	20.9	&	3.9	&	0.20	&	R	&	Ond\v{r}ejov	&	 \textbullet	&		&	\cite{Pravec1997b}	\\
3	&	 1995-Jun-29	&	1.331	&	0.348	&	22.01	&	254.3	&	18.3	&	2.5	&	0.23	&	R	&	Ond\v{r}ejov	&	 \textbullet	&		&	\cite{Pravec1997b}	\\
4	&	 2010-Aug-31	&	1.651	&	1.460	&	37.24	&	162.5	&	-81.6	&	2.4	&	0.11	&	R	&	NTT	&	 \textbullet	&	\textbullet	&		\\
5	&	 2010-Oct-13	&	1.568	&	1.439	&	38.46	&	211.3	&	-77.5	&	1.0	&	0.07	&	R	&	NTT	&	 \textbullet	&	&		\\
6	&	 2010-Oct-14	&	1.566	&	1.436	&	38.53	&	212.3	&	-77.5	&	3.0	&	0.21	&	R	&	NTT	&	 \textbullet	&	\textbullet	&		\\
7	&	 2010-Oct-15	&	1.563	&	1.432	&	38.59	&	213.4	&	-77.6	&	1.7	&	0.12	&	R	&	NTT	&	 \textbullet	&		&		\\
8	&	 2011-Jan-30	&	1.138	&	0.887	&	56.59	&	22.7	&	-26.0	&	1.8	&	0.10	&	R	&	NTT	&	 \textbullet	&	\textbullet	&		\\
9	&	 2011-Sep-01	&	1.416	&	1.282	&	43.59	&	233.5	&	-6.3	&	1.4	&	0.11	&	R	&	NTT	&	 \textbullet	&	\textbullet	&		\\
10	&	 2013-Nov-05	&	1.408	&	1.143	&	44.26	&	174.2	&	-78.1	&	2.1	&	0.14	&	V	&	NTT	&	 \textbullet	&	\textbullet	&		\\
11	&	 2013-Nov-07	&	1.400	&	1.124	&	44.59	&	175.2	&	-78.6	&	2.4	&	0.15	&	V	&	NTT	&	 \textbullet	&	\textbullet	&		\\
12	&	2014-Jun-19 	&	1.204	&	0.455	&	55.15	&	229.9	&	73.1	&	3.9	&	0.15	&	none	&	CS3-PDS	&		&		&	\cite{Warner2015}	\\
13	&	 2014-Jun-19	&	1.205	&	0.454	&	55.05	&	229.7	&	72.8	&	2.4	&	0.20	&	none	&	CS3-PDS	&		&		&	\cite{Warner2015}	\\
14	&	 2014-Jun-19	&	1.206	&	0.453	&	54.99	&	229.6	&	72.6	&	1.2	&	0.11	&	none	&	CS3-PDS	&		&		&	\cite{Warner2015}	\\
15	&	 2014-Jun-20	&	1.209	&	0.449	&	54.39	&	229.0	&	71.0	&	3.3	&	0.20	&	none	&	CS3-PDS	&		&		&	\cite{Warner2015}	\\
16	&	 2014-Jun-20	&	1.210	&	0.448	&	54.29	&	228.9	&	70.7	&	3.2	&	0.14	&	none	&	CS3-PDS	&		&		&	\cite{Warner2015}	\\
17	&	 2014-Jun-20	&	1.210	&	0.447	&	54.23	&	228.8	&	70.4	&	0.5	&	0.06	&	none	&	CS3-PDS	&		&		&	\cite{Warner2015}	\\
18	&	 2014-Jun-21	&	1.214	&	0.443	&	53.63	&	228.4	&	68.8	&	3.4	&	0.16	&	none	&	CS3-PDS	&		&		&	\cite{Warner2015}	\\
19	&	 2014-Jun-21	&	1.215	&	0.442	&	53.51	&	228.3	&	68.5	&	3.2	&	0.23	&	none	&	CS3-PDS	&		&		&	\cite{Warner2015}	\\
20	&	 2014-Jun-22	&	1.219	&	0.438	&	52.81	&	227.9	&	66.5	&	4.0	&	0.21	&	none	&	CS3-PDS	&		&		&	\cite{Warner2015}	\\
21	&	 2014-Jun-22	&	1.219	&	0.437	&	52.71	&	227.8	&	66.1	&	2.5	&	0.10	&	none	&	CS3-PDS	&		&		&	\cite{Warner2015}	\\
22	&	 2014-Jun-23	&	1.223	&	0.434	&	52.06	&	227.4	&	64.2	&	3.4	&	0.13	&	none	&	CS3-PDS	&		&		&	\cite{Warner2015}	\\
23	&	 2014-Jun-23	&	1.224	&	0.433	&	51.97	&	227.4	&	63.9	&	2.6	&	0.13	&	none	&	CS3-PDS	&		&		&	\cite{Warner2015}	\\
24	&	 2014-Jun-23	&	1.225	&	0.433	&	51.91	&	227.3	&	63.6	&	0.9	&	0.06	&	none	&	CS3-PDS	&		&		&	\cite{Warner2015}	\\
25	&	2017-Jan-06 	&	1.016	&	0.183	&	74.7	&	21.1	&	15.2	&	2.3	&	0.16	&	none	&	CS3-PDS	&		&		&	\cite{Warner2017}	\\
26	&	 2017-Jan-06	&	1.016	&	0.185	&	74.81	&	21.0	&	15.6	&	1.4	&	0.13	&	none	&	CS3-PDS	&		&		&	\cite{Warner2017}	\\
27	&	 2017-Jan-06	&	1.015	&	0.185	&	74.87	&	20.9	&	15.9	&	0.6	&	0.13	&	none	&	CS3-PDS	&		&		&	\cite{Warner2017}	\\
28	&	 2017-Jan-16	&	0.978	&	0.336	&	81.15	&	12.2	&	38.8	&	1.4	&	0.11	&	none	&	CS3-PDS	&		&		&	\cite{Warner2017}	\\
29	&	 2017-Jan-17	&	0.974	&	0.351	&	81.16	&	11.6	&	40.0	&	2.2	&	0.14	&	none	&	CS3-PDS	&		&		&	\cite{Warner2017}	\\
30	&	 2017-Jan-18	&	0.971	&	0.368	&	81.12	&	11.1	&	41.0	&	2.1	&	0.11	&	none	&	CS3-PDS	&		&		&	\cite{Warner2017}	\\
31	&	 2017-Jun-11	&	1.287	&	0.384	&	38.52	&	300.6	&	36.9	&	2.6	&	0.25	&	R	&	Danish	&	 \textbullet	&	\textbullet	&		\\
32	&	 2017-Jul-02	&	1.381	&	0.386	&	16.39	&	259.3	&	-4.9	&	1.6	&	0.11	&	R	&	Danish	&	 \textbullet	&		&		\\
		 \hline& 
		\end{tabular}
}
\end{table*}

\begin{table}

\caption{Observations of asteroid  (2102) Tantalus with the Arecibo planetary radar.
`{UT Date}' is the universal-time date on which the observations began,
{and} the timespan of the received data is listed by the UT `{Start}' and `{Stop}' times.
`{RTT}' is the round-trip light time to the target.
`{Baud}' is the {time} resolution of the pseudo-random code used for imaging; baud does not apply to cw data.  The baud length is translated to delay resolution (`Res.') expressed in metres. 
`{Runs}' is the number of completed transmit-receive cycles. 
All but the most crude ranging observations, with resolution of $600\,\textrm{m}$, collected on the 1 and 4 January, were used in the shape modelling. The ranging observations are listed here for completeness.	
} 	
\label{tab:radar}

\resizebox{\hsize{}}{!}{  
	\begin{tabular}{cccccc}
		\hline\hline
		\noalign{\smallskip}
			 {UT Date}  	&	 {Start-Stop}           	&	 {RTT} 	&	 {Baud} & Res.    					&	 {Runs}    	\\
			{[yyyy-mm-dd]}               	&	 {[hh:mm:ss-hh:mm:ss]} 	&	 {[s]} 	&	 {[$\mu\rm s$]} 	&	 {[m]} 			&	 {} 	\\ \hline  \noalign{\smallskip}
	 2017-01-01 & 22:48:15-23:05:02 &  144 &  cw  & --    &  4 \\
	            & 23:13:19-23:25:16 &      &  4   & 600 &  3 \\
	            & 23:33:14-00:28:41 &      &  0.5 &  75 & 12 \\ \hline  \noalign{\smallskip}
	 2017-01-04  & 21:19:51-21:46:35 & 169 & cw &    -- &  5 \\
	             & 21:57:00-22:11:02 & 170 &  4 & 600 &  3 \\
	             & 22:20:44-23:57:05 &     &  1 & 150 & 16 \\ \hline  \noalign{\smallskip}
	 2017-01-05 & 21:01:26-21:22:25 &  181 & cw &    -- &  4 \\
	            & 21:34:24-23:38:35 &      &  1 & 150 & 21 \\ \hline  \noalign{\smallskip}
	 2017-01-06 & 20:48:13-21:23:39 &  193 & cw &    -- &  6 \\
	            & 21:33:07-22:47:29 &  194 &  1 & 150 & 12 \\ \hline  \noalign{\smallskip}
	 2017-01-07 & 20:49:31-21:32:27 &  207 & cw &    -- &  6 \\
	            & 21:42:05-22:20:05 &      &  1 & 150 &  6 \\ \hline
		\end{tabular}
}

\end{table}

We observed Tantalus primarily with the EFOSC2 camera on the $3.6\,\rm m$ NTT telescope at ESO's La Silla Observatory in Chile on 8 nights between August 2010 and November 2013 (labelled with light-curve IDs 4-11 and `NTT' as observing facility in Table~\ref{tab:obs}). These light curves were obtained under ESO programme 185.C-1033. 
We collected further optical light curves in June and July 2017 with the Danish $1.54\,\rm m $ telescope, also located at La Silla (label `Danish' in Table~\ref{tab:obs}, IDs 31 and 32) using the DFOSC instrument. These observations were obtained as part of the MiNDSTEp consortium that operates the Danish telescope for 6 months each year.

The relative optical light curves record the asteroid's brightness variation compared with background stars within the frame and relative to other observations in a series. 
A series of images in Bessel R filter (for light-curve IDs 4-9 from the NTT and 31-32 from the Danish telescope) or Bessel V (light-curve IDs 10 and 11) were acquired on each night. Individual frames were reduced using basic CCD reduction techniques, bias subtraction and flat field correction. Additionally, we removed fringe patterns from the R-frames acquired at the NTT by subtracting an EFOSC2 fringe map (provided by ESO) scaled by the computed fringe amplitude in individual images \citep{Snodgrass2013}. Due to the low signal from the asteroid, images on some nights (light-curve IDs 3 and 9) were summed in batches of 3 to improve the signal-to-noise ratio (SNR). The light-curve data can be accessed in Table~A1, available at the CDS.

The asteroid was observed with the Arecibo planetary radar over 5 nights in January 2017 under NEA monitoring programme R3037 (data summary in Table~\ref{tab:radar}). 
Two types of radar data were collected: continuous-wave (cw) observations, 
which display the Doppler shift of the signal reflected off the asteroid's surface due to its rotation, and 
delay-Doppler imaging, which is a two-dimensional diagram of 
radar echo power as a function of both Doppler shift and delay of the returning signal \citep[e.g.][]{Benner2015}. The imaging frames were collected mainly with a $150\,\rm m$ resolution in delay, except on 1 January 2017 when an imaging resolution of $75\,\rm m$ was achieved. Additionally, ranging observations were taken on 1 and 4 January 2017 with delay resolution of $600\,\rm m$, but those have a resolution that is too low for shape reconstruction and were used to improve orbital parameters only.

\section{Light-curve analysis}
\label{sec:light-curves}

\begin{figure}
	\resizebox{.99\hsize}{!}{
		\includegraphics[width=.48\textwidth, trim=0cm 0cm 0cm 0cm, clip=true]{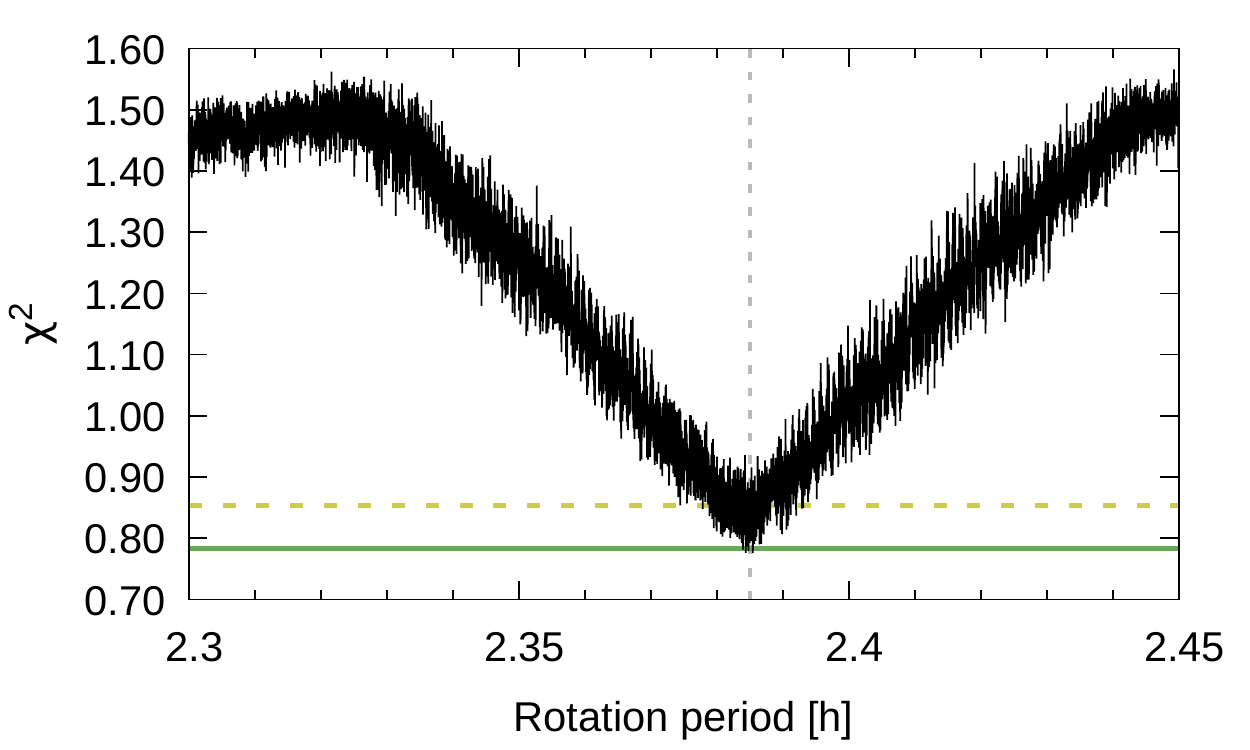} 
	}
	
	\resizebox{.99\hsize}{!}{
		\includegraphics[width=.48\textwidth, trim=0cm 0cm 0cm 0cm, clip=true]{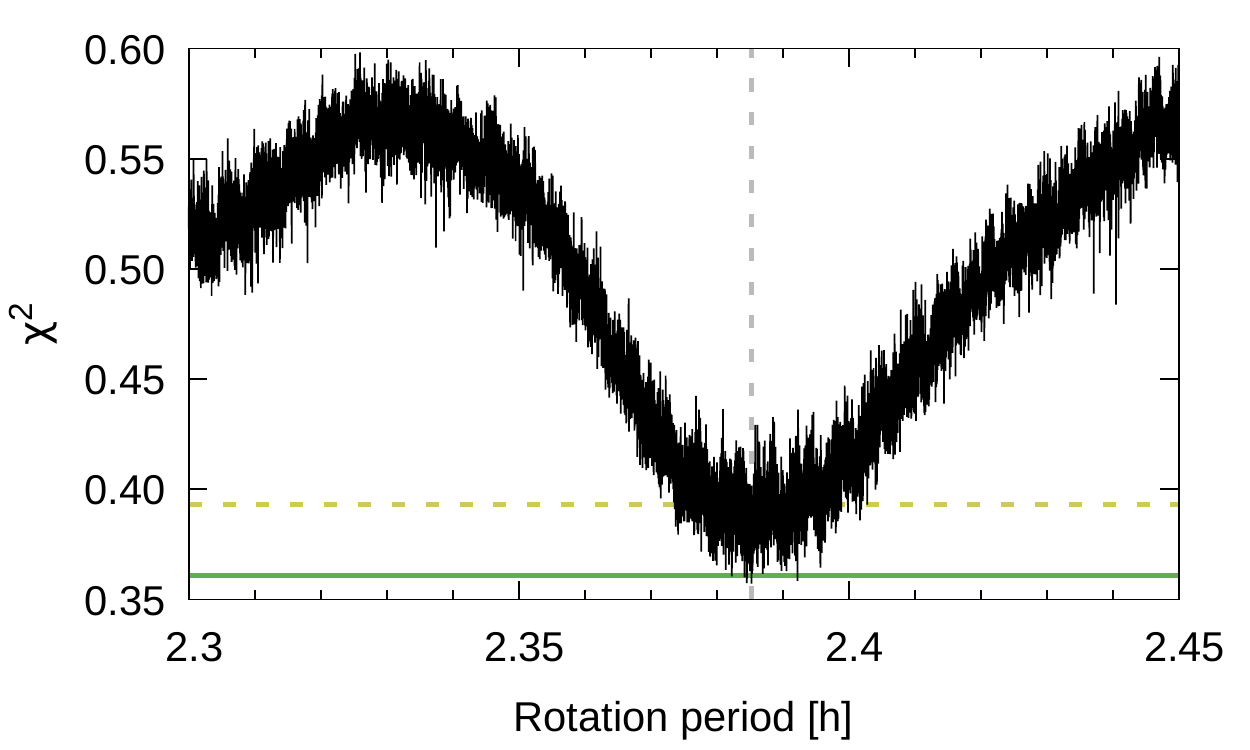}
	}
	
	\caption{ Periodograms obtained using convex light-curve inversion methods searching around the literature value. The top panel shows results of a period search using all available light curves {and} the bottom panel shows the results of the period search excluding some of the light curves, with with the vertical dashed line marking $P = 2.385\, \textrm{h}$, 
	corresponding to the minimum $\chi^2$ value {in the shown interval}. 
	The uncertainties for both period determinations would be $0.003\, \textrm{h}$ for the full period scan and $0.07\, \textrm{h}$ for restricted.  Horizontal continuous lines in both plots mark $1\%$ increase above the minimum $\chi^2$ value for each search, and dashed lines -- $10\%$ increase.
		\label{fig:periods}
	}
\end{figure}

Light-curve data for Tantalus span 23 years and a wide range of observing geometries (see Table~\ref{tab:obs}), which is in many cases a good starting point for shape and spin-state modelling. However, the light-curve peak-to-peak amplitude is not very high, ranging between $0.07$ and $0.25\,\textrm{mag}$ with a few of the light curves having low SNR (for example light curves with IDs 2, 19, and 31). Close inspection shows that in some cases the peak-to-peak amplitude is exaggerated due to the scatter of light-curve points and hence might be actually lower than listed in Table~\ref{tab:obs}. Moreover, some light curves represent only short fragments of rotation (like light curves with IDs 5, 17, and 27). This presents certain difficulties in obtaining spin-state and shape solutions.

Initially, all of the available light curves were used to determine a convex shape model, using established convex inversion procedures described by \citet{Kaasalainen2001a} and \citet{Kaasalainen2001b} and implemented by \cite{Durech2010}. 
For each period tested, a shape and pole were optimised using six different starting pole positions \citep{Kaasalainen2001b}. 
At the end of the optimisation, the quality of the fit,  $\chi^2$ \citep[defined with Eq.13 by][]{Kaasalainen2001a},  for the best of the six models is recorded, but the shape and pole information is discarded. 
{We have run a period search using this method in a wide $2-18\,\rm h$ interval (the periodogram obtained this way,recording the quality of the fit for each tested period, is included in the Appendix  Fig.~\ref{fig:periodogram}). 
There appears to be two significant $\chi^2$ minima, one around the $2.39\,\rm{h}$ literature synodic rotation period \cite{Pravec1997b,Warner2015, Warner2017, Vaduvescu2017}, and another at twice that period. The $16\,\rm{h}$ periodicity reported is not apparent in the full data set. The longer period, around $4.8\,\rm{h}$, is not considered likely as the lighcurve-inversion produces nonviable models, considerably elongated along the z-axis, and initial radar modelling with this period failed to reproduce echo bandwidth at all observed geometries.}
A {close-up of the periodogram around the literature synodic period} (upper panel in Fig.~\ref{fig:periods}),  shows a family of possible solutions around the best-fit solution {within this interval} corresponding to  $\chi^2$ minimum with a few periods having quality of fit within $1\%$ difference. This means there is a certain level of ambiguity to the period solution. 

{Following the same method as \citep{Durech2012} to assess uncertainty,  1-$\sigma$ would correspond to $3.5\%$ increase of $\chi^2$ for the search including all available lightcurves (for around 1600 lightcurve points and 1500 degrees of freedom), and $5\%$ for restricted data set (about 750 lightcurve points and 650 degrees of freedom).  In practice, the fits to data of synthetic light curves generated from models that differ by less than $10\%$  are virtually indistinguishable. Therefore, to assess the uncertainties of parameters such as pole and period we consider standard deviation of models with the $\chi^2$ within $10\%$ of minimal value.}

\begin{figure}
	\includegraphics[width=\columnwidth, trim=3cm 6cm 4cm 6.1cm, clip=true]{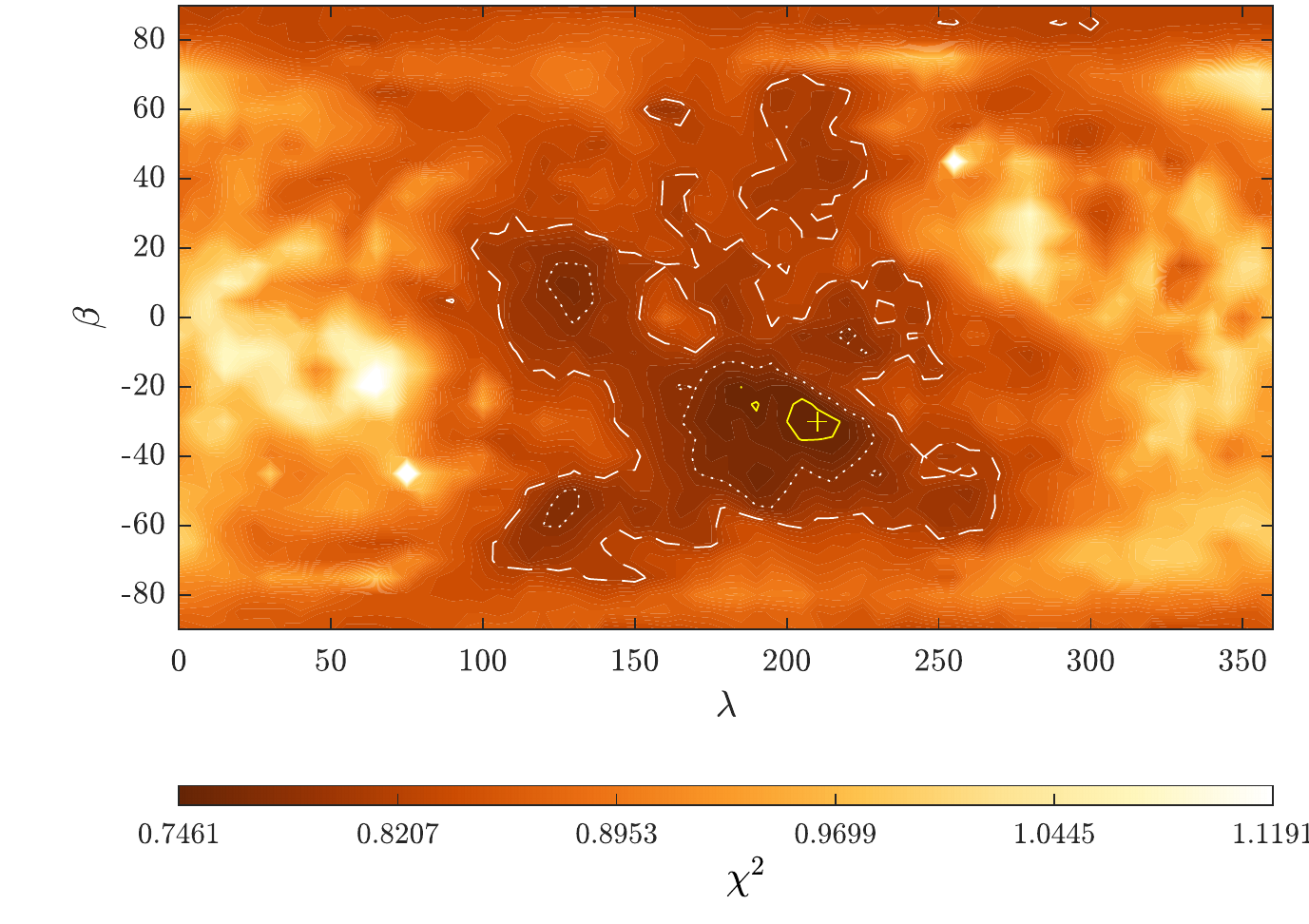} 	
	
	\includegraphics[width=\columnwidth, trim=3cm 6cm 4cm 6.1cm, clip=true]{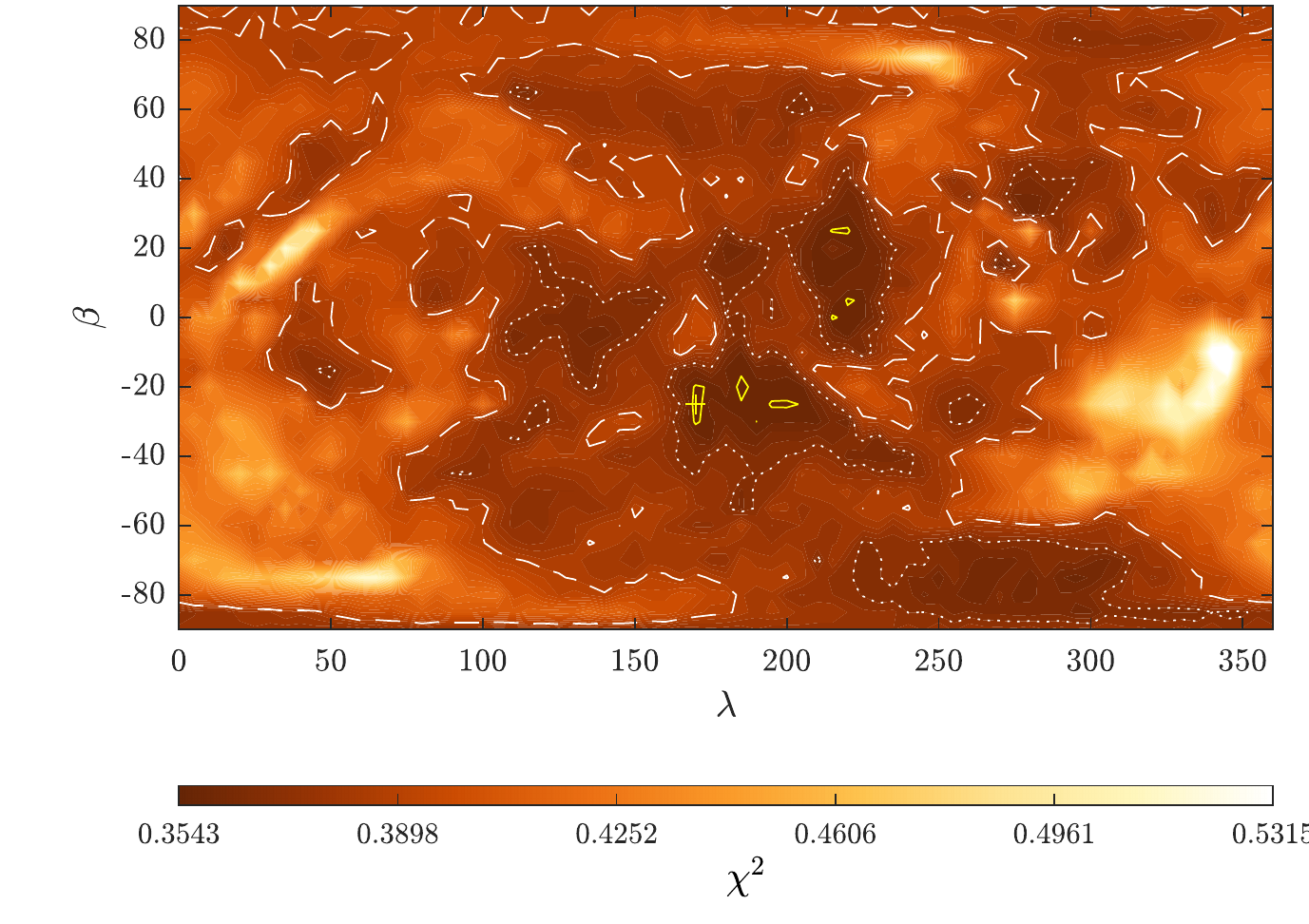}

	\caption{The upper panel shows a $\chi^2$-plane for a constant period pole search using light-curve inversion and the full available light-curve data set, and the lower panel is the same, but for search performed using a selection of the available light curves. 
	Darker colours indicate lower $\chi^2$ {values}, and the lines mark gradual increases over the minimum $\chi^2$ value,  with a solid line for 1\% increase, a dotted line for 5\%, and a dashed line for 10\%. 
	The positions of a rotation pole corresponding to the minimum $\chi^2$ is marked with a cross in each panel, placed at $\lambda=210\degr$ and $\beta=-30\degr$ in the upper panel, and $\lambda=170\degr$ and $\beta=-25\degr$
	in the lower panel. }
	\label{fig:chi:allLC}
\end{figure}

We used the best-fit period, $\textrm{P} = 2.385\, \textrm{h}$,  as a starting point for further convex inversion modelling and searched for the pole solution using a $5\degr\times5\degr$ grid of possible pole positions. At each point of the grid the period and shape were optimised and the quality of the fit recorded. We show the results of such a search, assuming a constant-period solution, in Fig.~\ref{fig:chi:allLC}. The colours in this figure correspond to the quality of the fit, with darker colours marking better solutions with lower $\chi^2$ values and the best-fit pole indicated with a cross. The pole appears not to be very well constrained despite a wide range of observational geometries covered. This is likely due to the symmetry of the body meaning there is not much variation in observed light curves with changing aspect.

\begin{figure}

		\includegraphics[width=.48\textwidth, trim=2.5cm 5cm 3.5cm 5cm, clip=true]{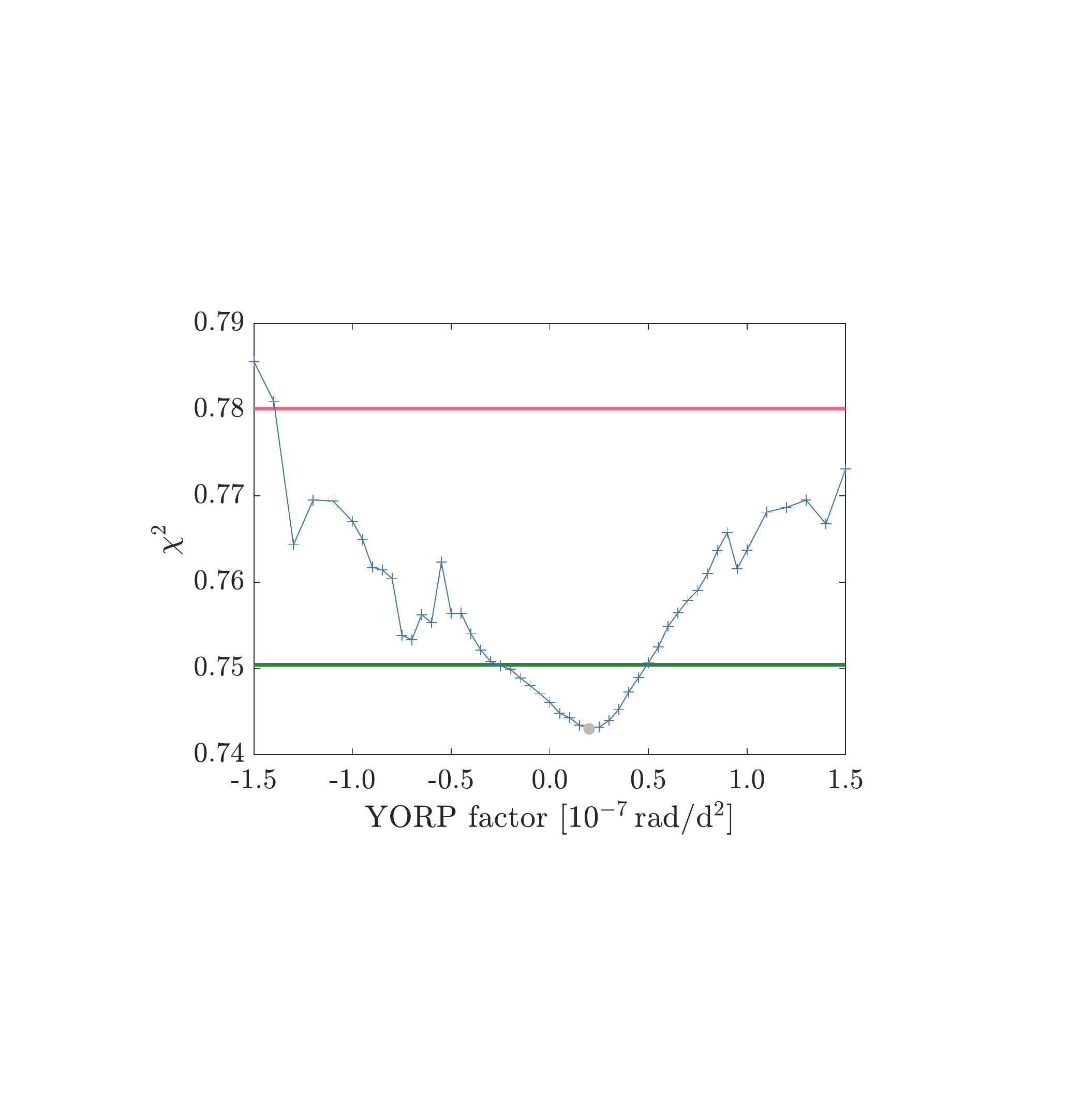}
		 			\includegraphics[width=.48\textwidth, trim=2.5cm 5cm 3.5cm 5cm, clip=true]{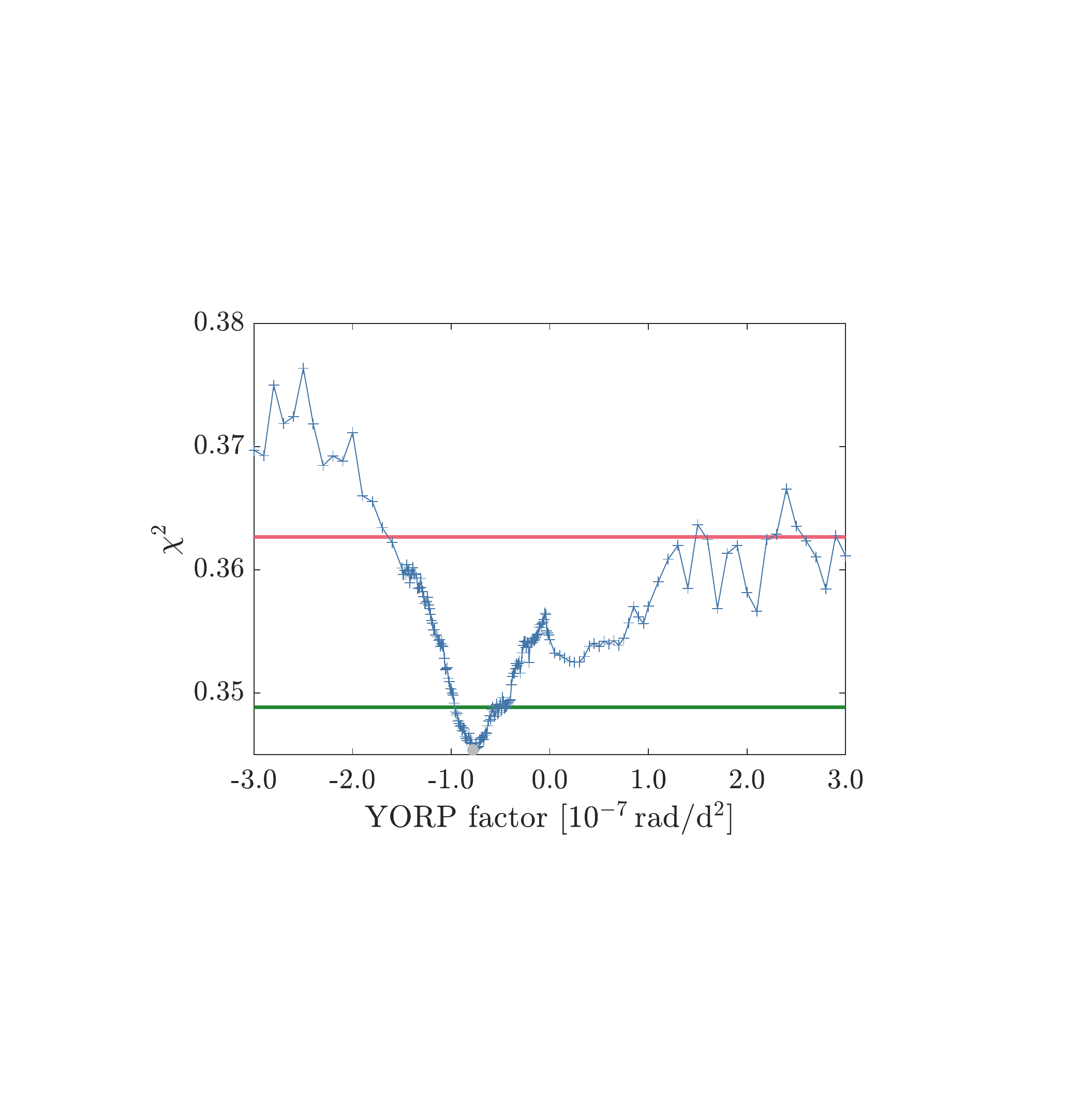}
	\caption{ The results of a combined pole and spin-state change (YORP factor) search using all available light curves are shown in the upper panel, and using a selection of light curves in the lower panel.
	Horizontal axis indicates the rate of change of angular rotation rate expressed in $\textrm{rad}/\textrm{d}^2$ (the YORP factor) and the vertical axis, the quality of the fit for the best shape model developed assuming the given YORP factor.
	The $\chi^2$ differences are very small, with the green horizontal lines marking $1\%$ increase above the minimal values, and  the red {lines} marking $5\%$ increase. 
	{The} grey circle in each panel indicates the minimum $\chi^2$ value.
		\label{fig:chibowl}	}
\end{figure}

Following the procedure outlined by \citet{Rozek2019Cuyo}, the pole search was repeated for different values of possible spin-rate change, which we hereafter call `YORP factor', as the YORP effect is currently the best explanation for gradual changes in rotation rates which can be detected for small NEAs. The YORP factor corresponds to a linear change in a rotation rate (where the rotation rate $\omega\equiv 2\pi/\textrm{P}$) measured in $\textrm{rad}/\textrm{d}^{2}$. The strongest measured YORP factor to date was $3.5\times10^{-6}\, \textrm{rad}/\textrm{d}^{2}$ measured for asteroid (54509) YORP \citep{Lowry2007,Taylor2007}. {Notably, 54509 is the smallest object with YORP detection, relatively close to the Sun,} but this value is an outlier and the other detections are for YORP factors of the order of  $10^{-8}\, \textrm{rad}/\textrm{d}^{2}$, including for objects of similar diameter and orbital semi-major axis to Tantalus \citep[as listed for example in][]{Zegmott2021}. We searched a range of possible YORP factors between $-1.5$ and $1.5\times10^{-7}\, \textrm{rad}/\textrm{d}^{2}$, and the results are illustrated in Fig.~\ref{fig:chibowl}. There appears to be a minimum in the $\chi^2$ distribution corresponding to a YORP factor $2\times10^{-8}\, \textrm{rad}/\textrm{d}^{2}$. However, the quality of the fit for a constant-period solution is within $1\%$ increase above the lowest $\chi^2$ value and the light curve fits are very similar.

\begin{figure*}
	\resizebox{\hsize}{!}{
		\includegraphics[width=.48\textwidth, trim=2cm 4cm 3.8cm 4cm, clip=true]{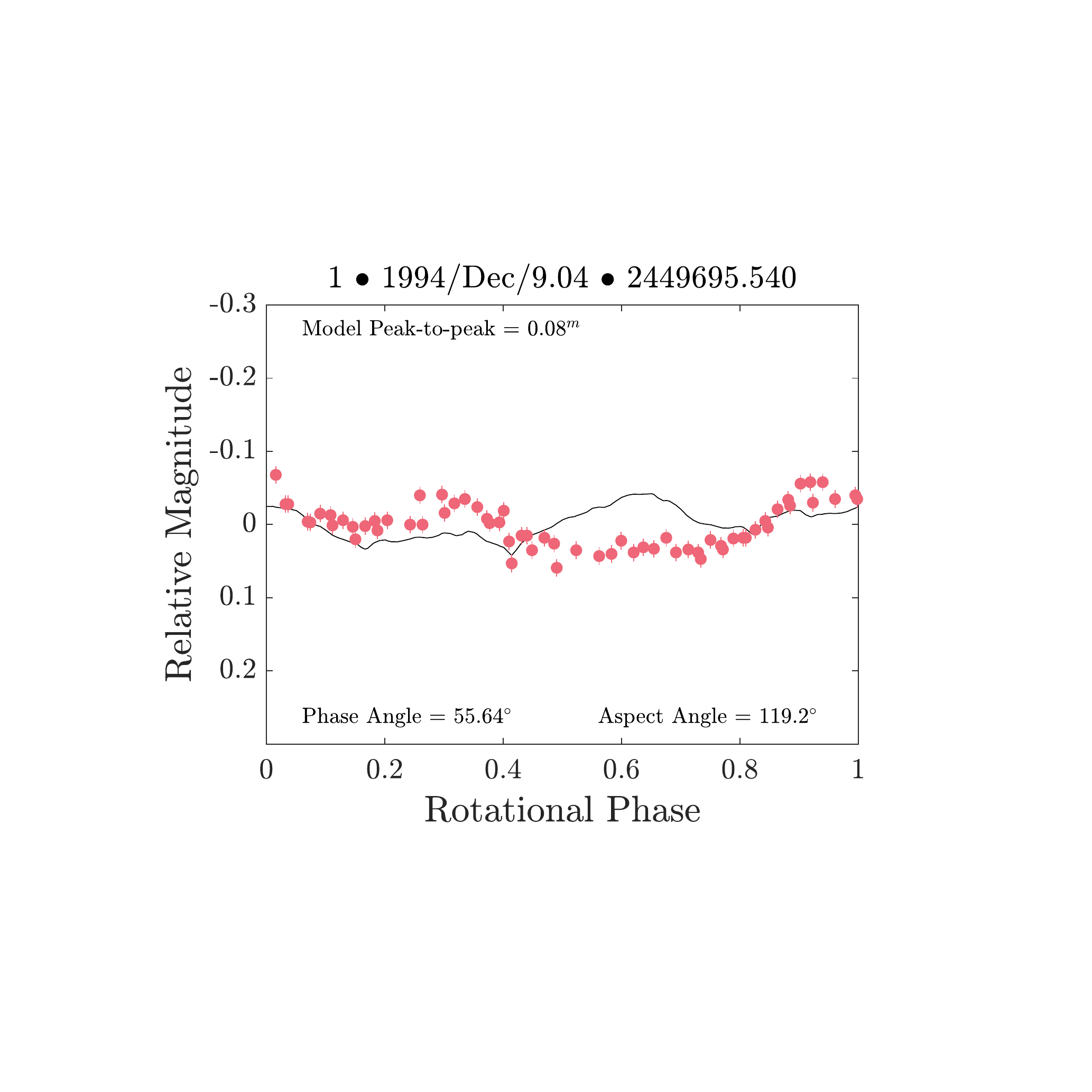} 
		\includegraphics[width=.48\textwidth, trim=2cm 4cm 3.8cm 4cm, clip=true]{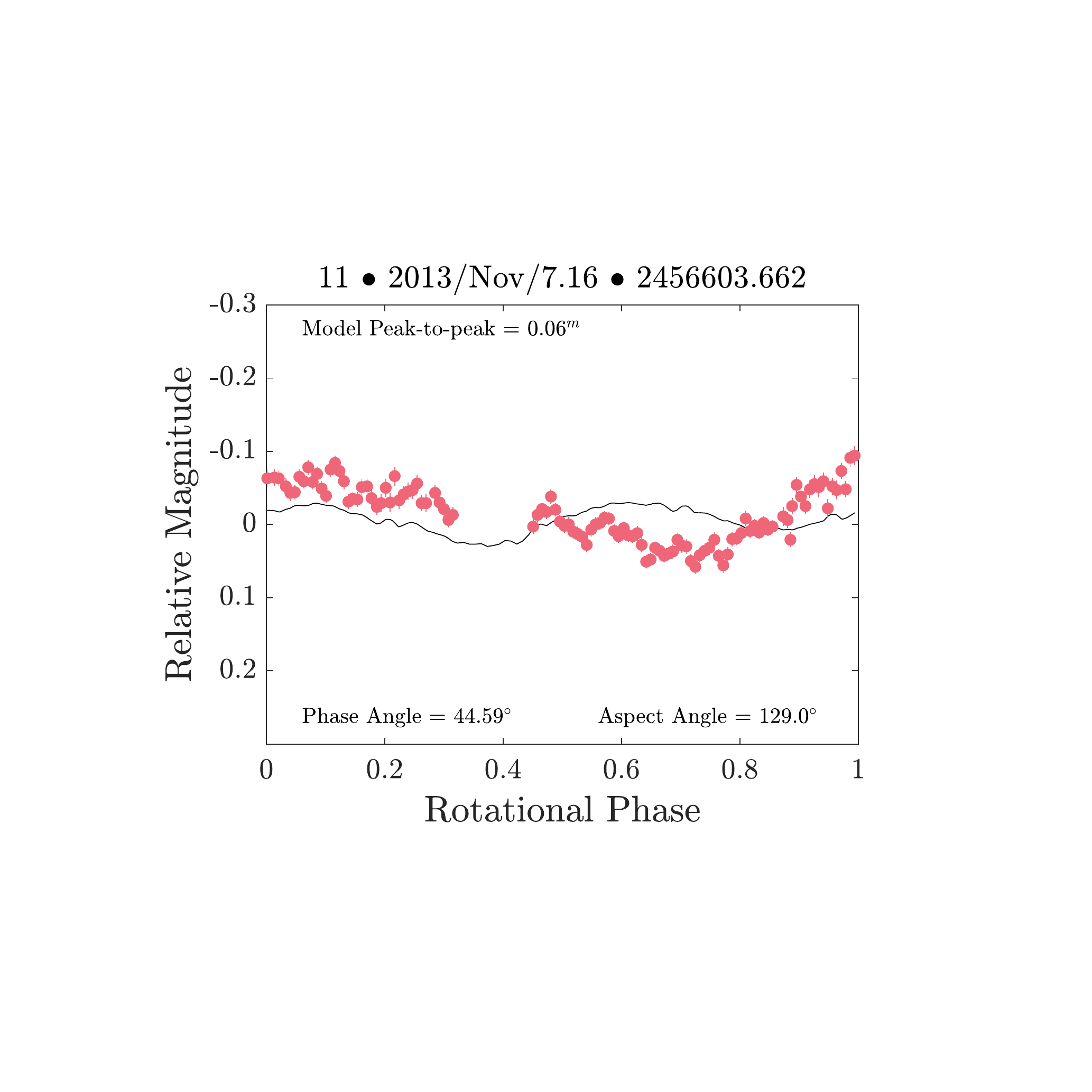} 		
		\includegraphics[width=.48\textwidth, trim=2cm 4cm 3.8cm 4cm, clip=true]{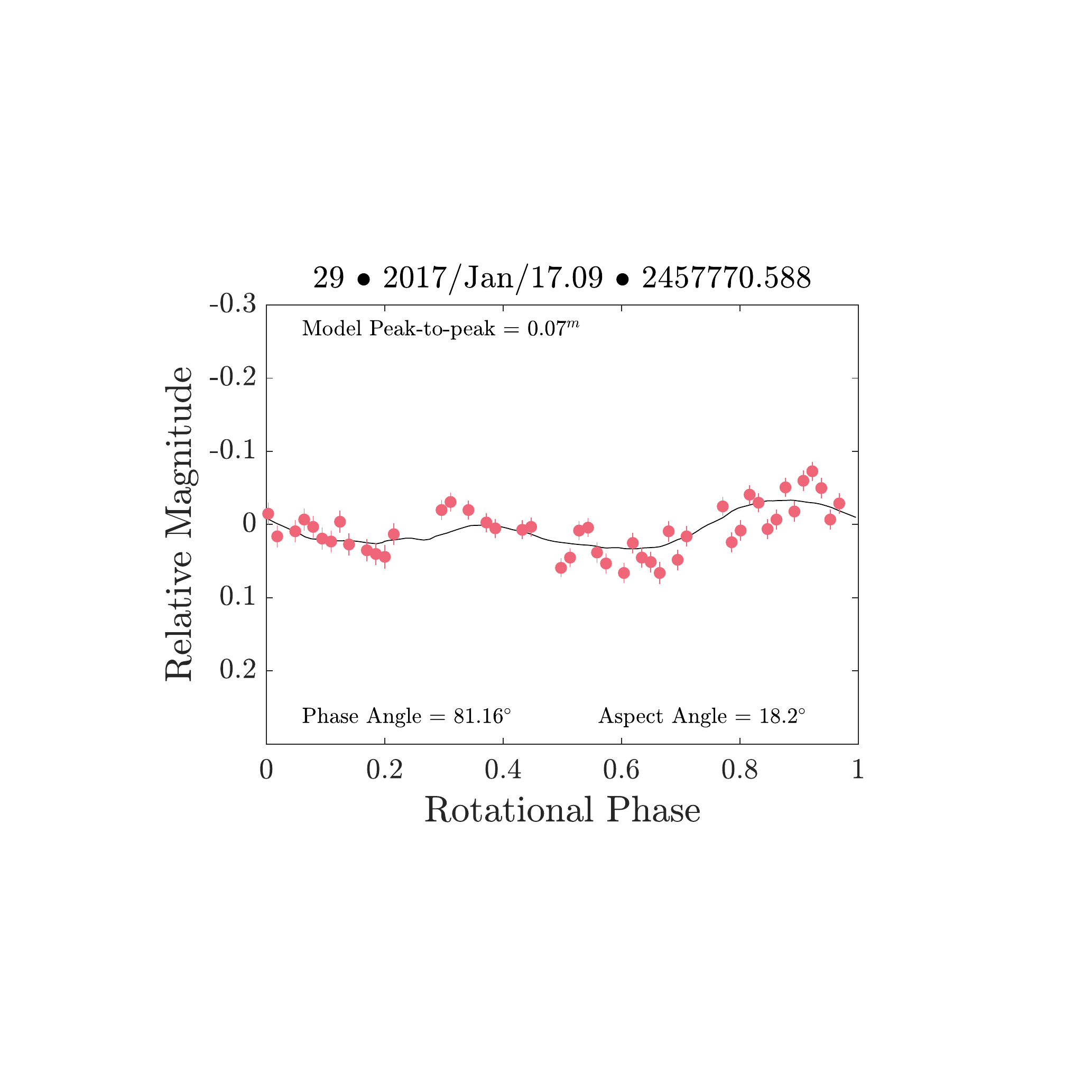} 
	}

		\resizebox{\hsize}{!}{
		\includegraphics[width=.48\textwidth, trim=2cm 4cm 3.8cm 4cm, clip=true]{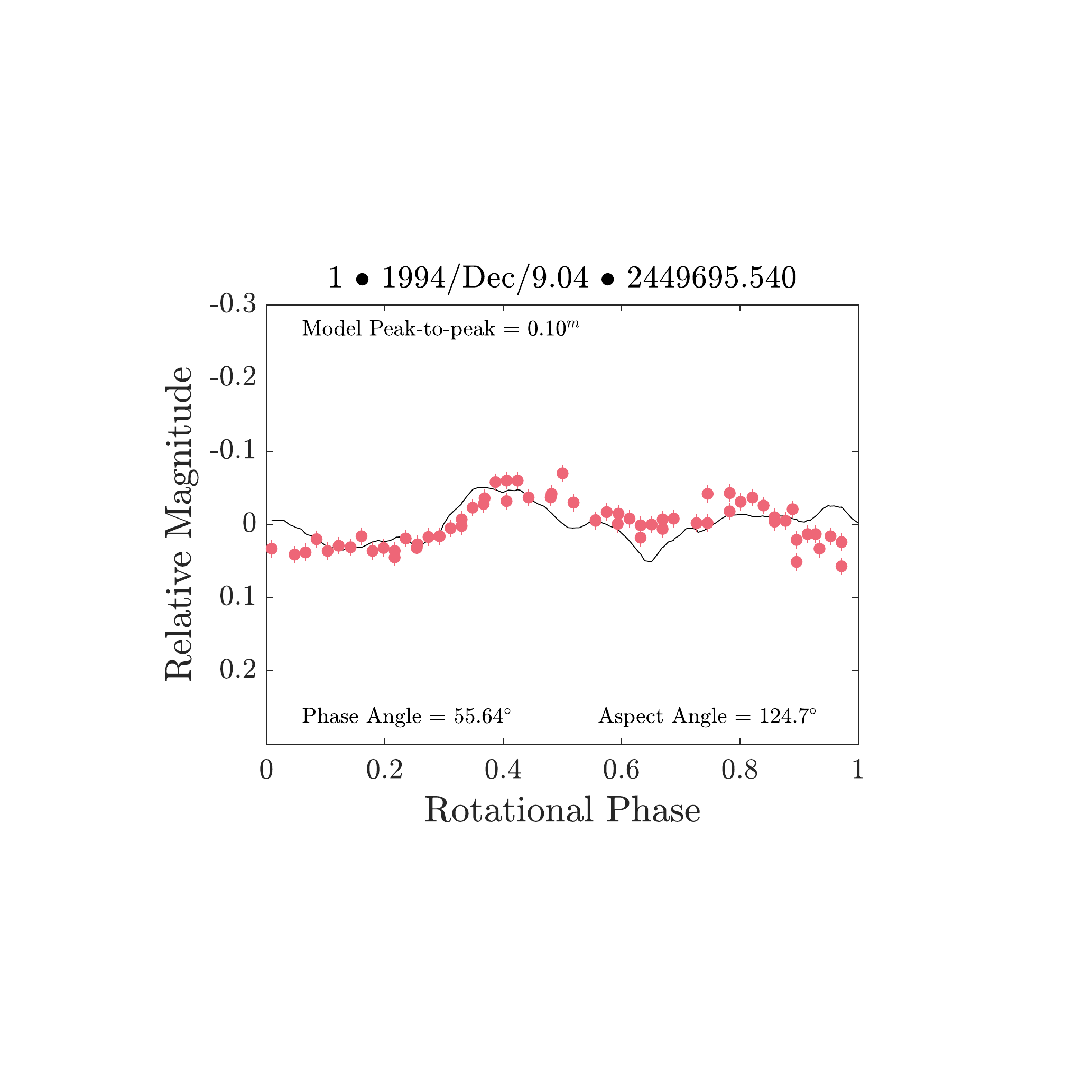} 
		\includegraphics[width=.48\textwidth, trim=2cm 4cm 3.8cm 4cm, clip=true]{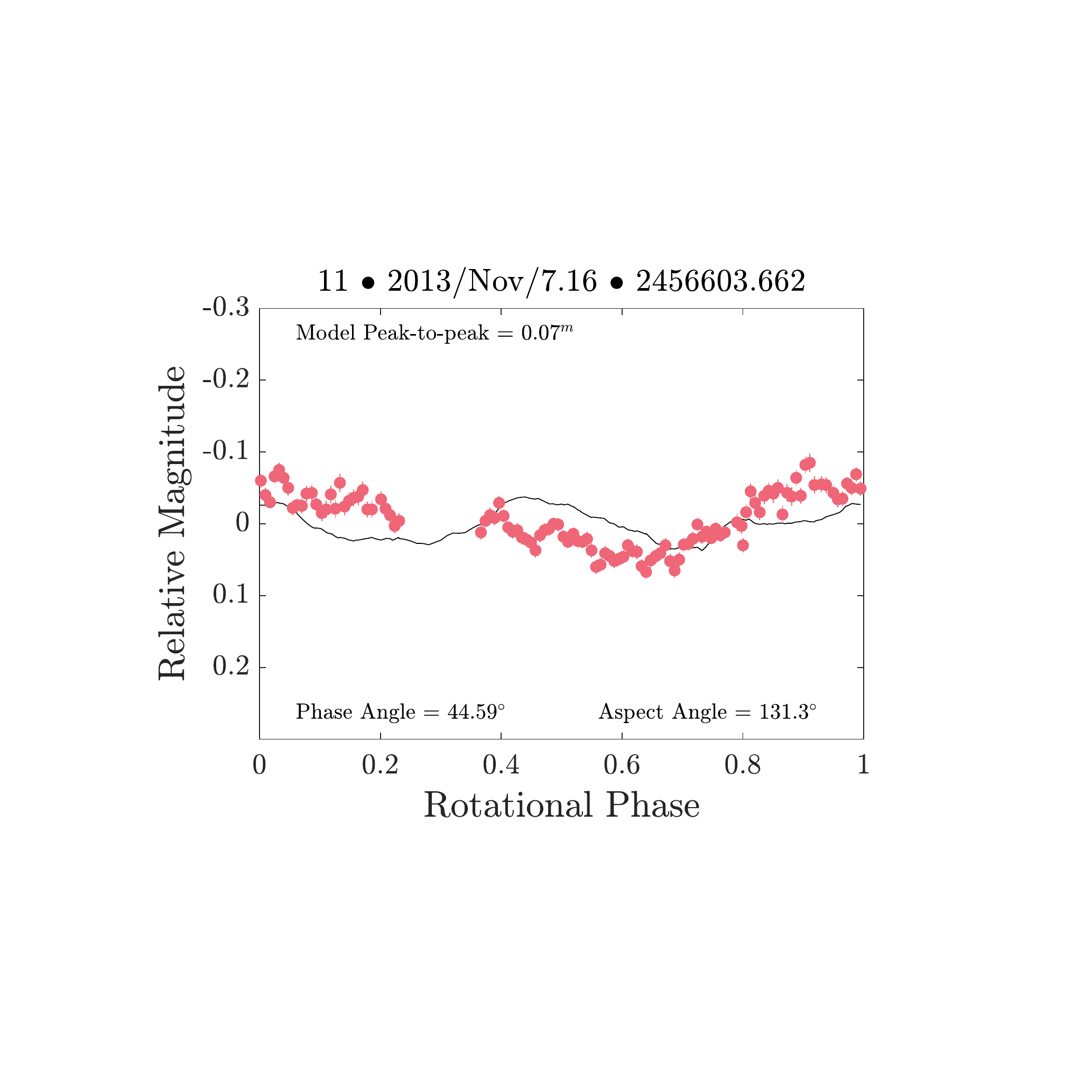} 		
		\includegraphics[width=.48\textwidth, trim=2cm 4cm 3.8cm 4cm, clip=true]{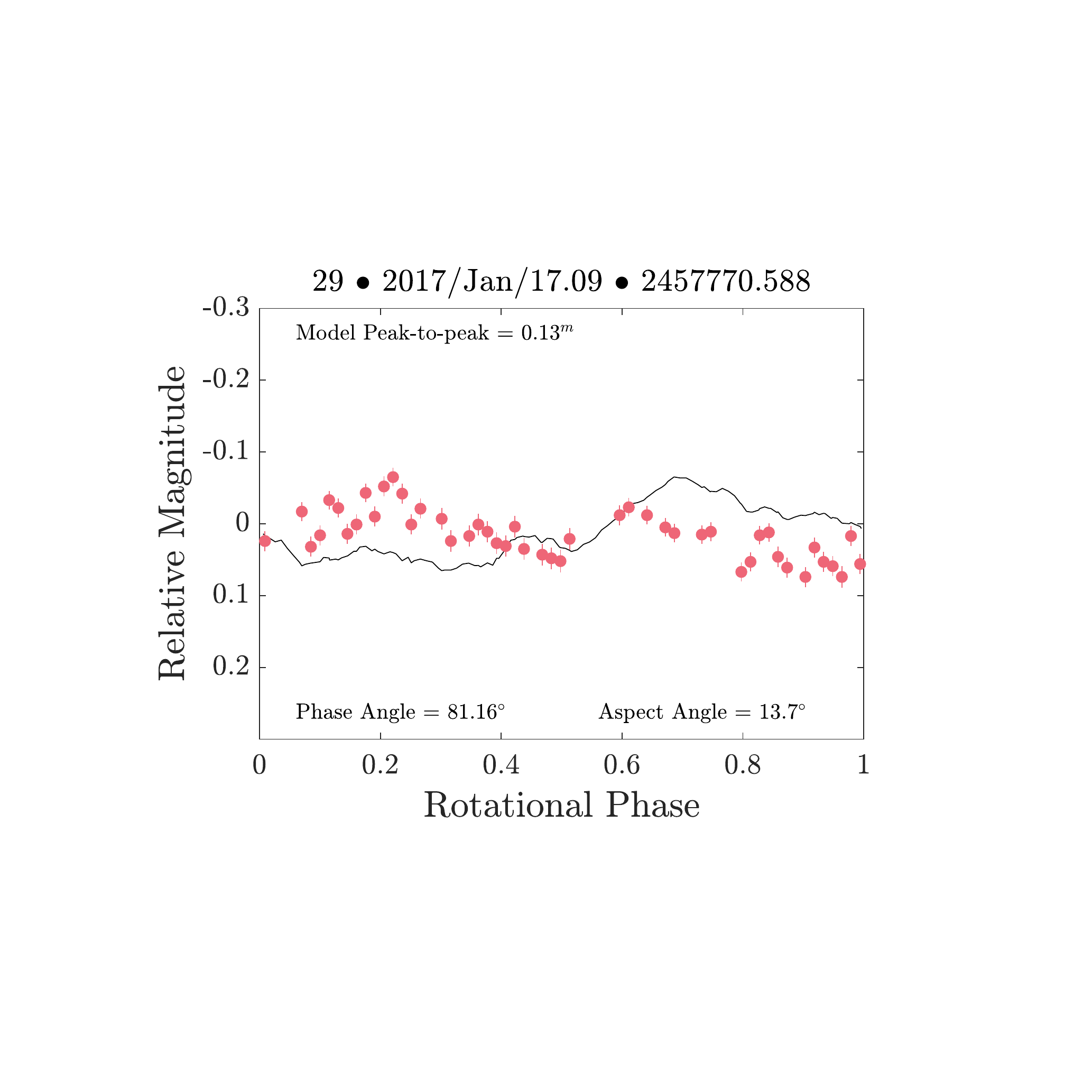} 
	}

			\resizebox{\hsize}{!}{
		\includegraphics[width=.48\textwidth, trim=2cm 4cm 3.8cm 4cm, clip=true]{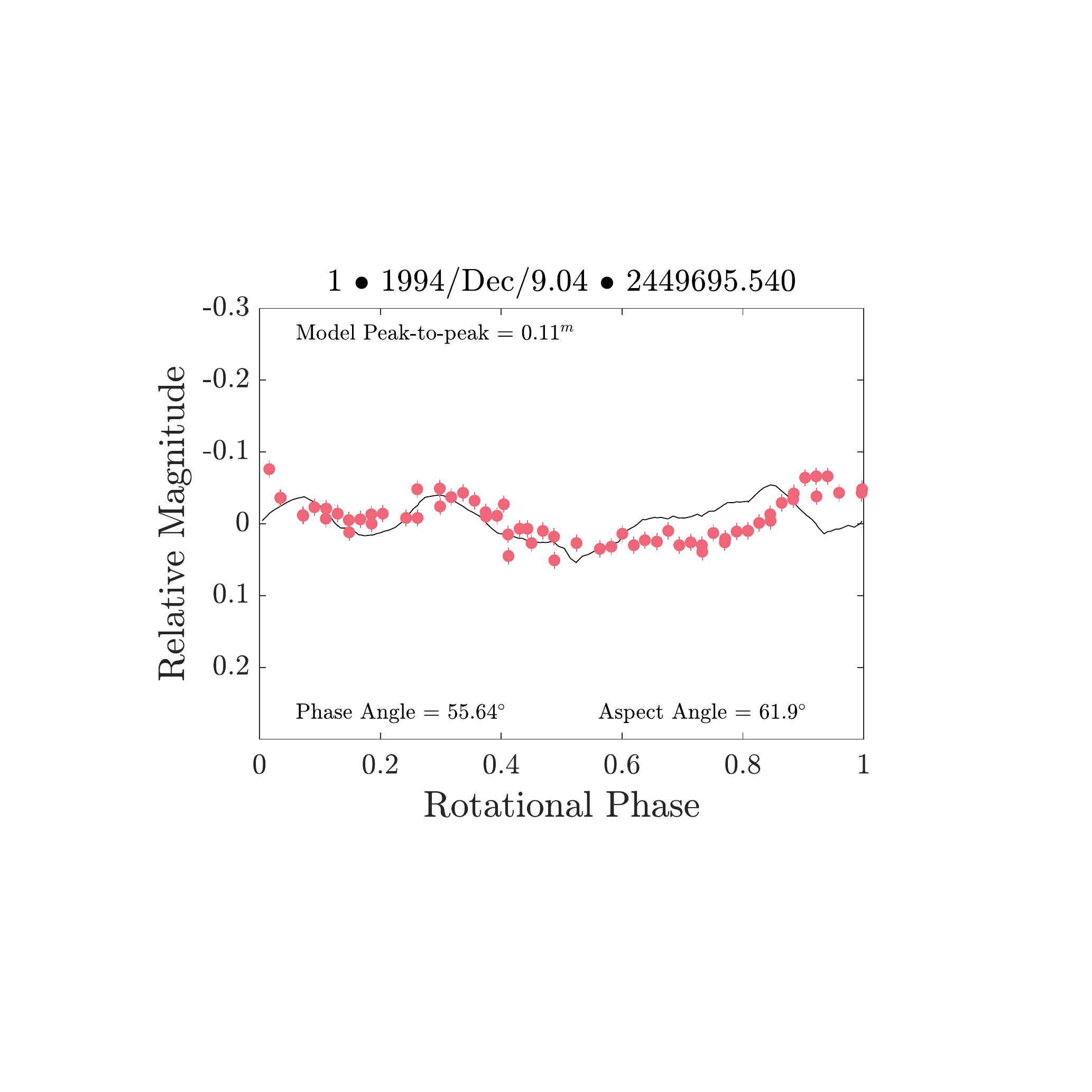} 
		\includegraphics[width=.48\textwidth, trim=2cm 4cm 3.8cm 4cm, clip=true]{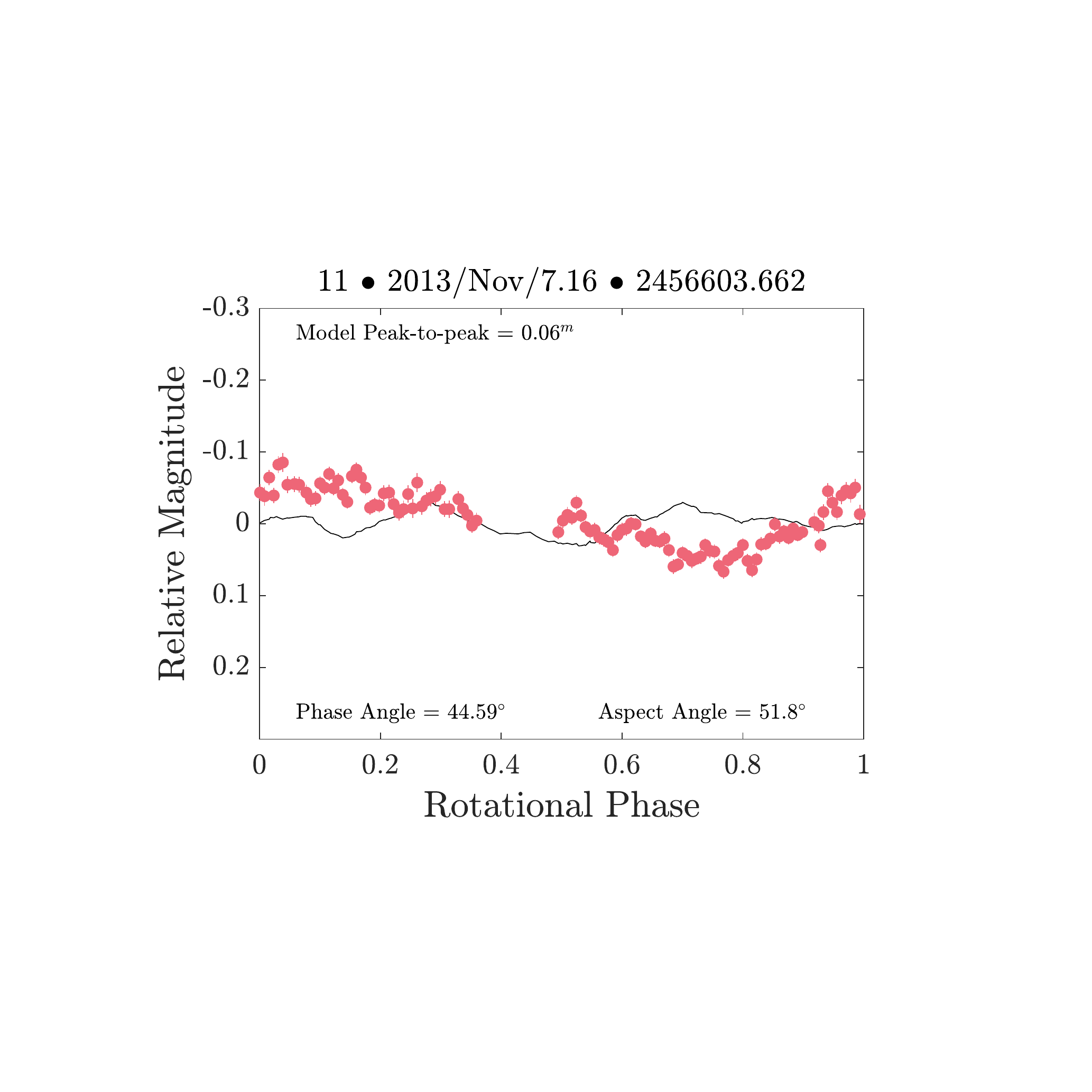} 		
		\includegraphics[width=.48\textwidth, trim=2cm 4cm 3.8cm 4cm, clip=true]{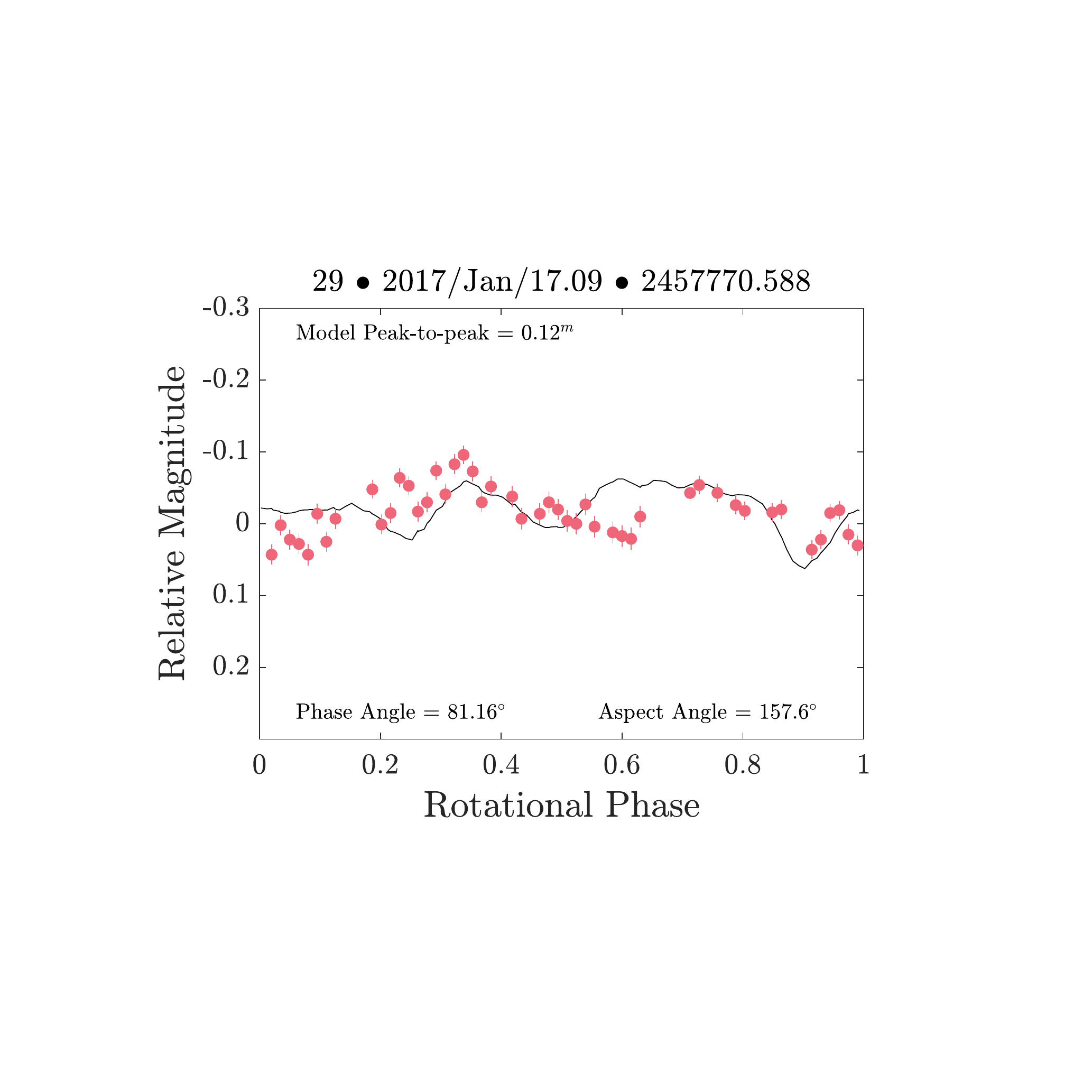} 
	}

	\caption{Example light curves collected for (2102) Tantalus. The left column of images corresponding to LC~1, middle column -- LC~11, and right column to LC~29 as listed in Table~\ref{tab:obs}. 
	The top row presents  the data plotted over synthetic light curves generated using the `full' constant-period light-curve inversion model (full set of synthetic light curves for this model is shown in Fig.~\ref{fig:conv-lcfit1}). 
     Second row includes example fits for the retrograde radar shape model (full set of light curves shown in \ref{fig:radar-retro1}).
	Finally, the bottom row show example fits for the prograde radar shape model (full set of light curves shown in \ref{fig:radar-pro1}).
		\label{fig:conv-lcfit:example}}
\end{figure*}

Fits of the synthetic light curves to the data for the best-fit constant-period shape model are shown in Figs.~\ref{fig:conv-lcfit:example} and \ref{fig:conv-lcfit1}. The light-curve fits are not perfect. While the amplitude of the synthetic light curves generally agrees with observations, the minima and maxima are misaligned. This effect can be sometimes seen in convex shape modelling if there are issues leading to the shape of the light curve being dominated by non-shape effects related for example to observing conditions, crowding of the star field, and quality of the detector, or timing errors.  We tested this hypothesis by removing the 2014 and 2017 unfiltered light curves \citep{Warner2015,Warner2017}, and using the light curves marked with black circles in the `LC-only' column of Table~\ref{tab:obs}. 

Dropping the subset of light curves and repeating the modelling procedure brings curious results. The rotation period is even less constrained with multiple periods having the quality of fit $\chi^2$ within $1\%$ of the best solution (Fig.~\ref{fig:periods}).  Despite having less constraint with considerable fraction of light-curves missing, the best solution is only $0.0002\,\textrm h$ away from the best-fit period obtained from the full light-curve data set, well within uncertainties in both measurements ($0.003\, \textrm{h}$ for the full period scan and $0.07\, \textrm{h}$ for restricted). Removing light curves has understandably made the pole determination even more ambiguous, but with the global minimum most likely in mid-negative ecliptic latitudes (as shown in Fig.~\ref{fig:chi:allLC}), and shifts the minimum for YORP-factor search towards negative values (Fig.~\ref{fig:chibowl}). The two YORP-factor searches give seemingly different results. However,  formal uncertainties would in both cases produce overlapping error bars,  encompassing constant-period, spin-up and slow-down solutions. The best-fit variable period solutions for either full or restricted light-curve data sets give synthetic light-curve fits of very similar quality to the constant-period solution. We conclude a detection of any period change is not possible with the current data set using the light-curve inversion method.

\begin{figure}
	\includegraphics[width=\columnwidth,trim=1.5cm 8.5cm 2cm 8.5cm,clip=true]{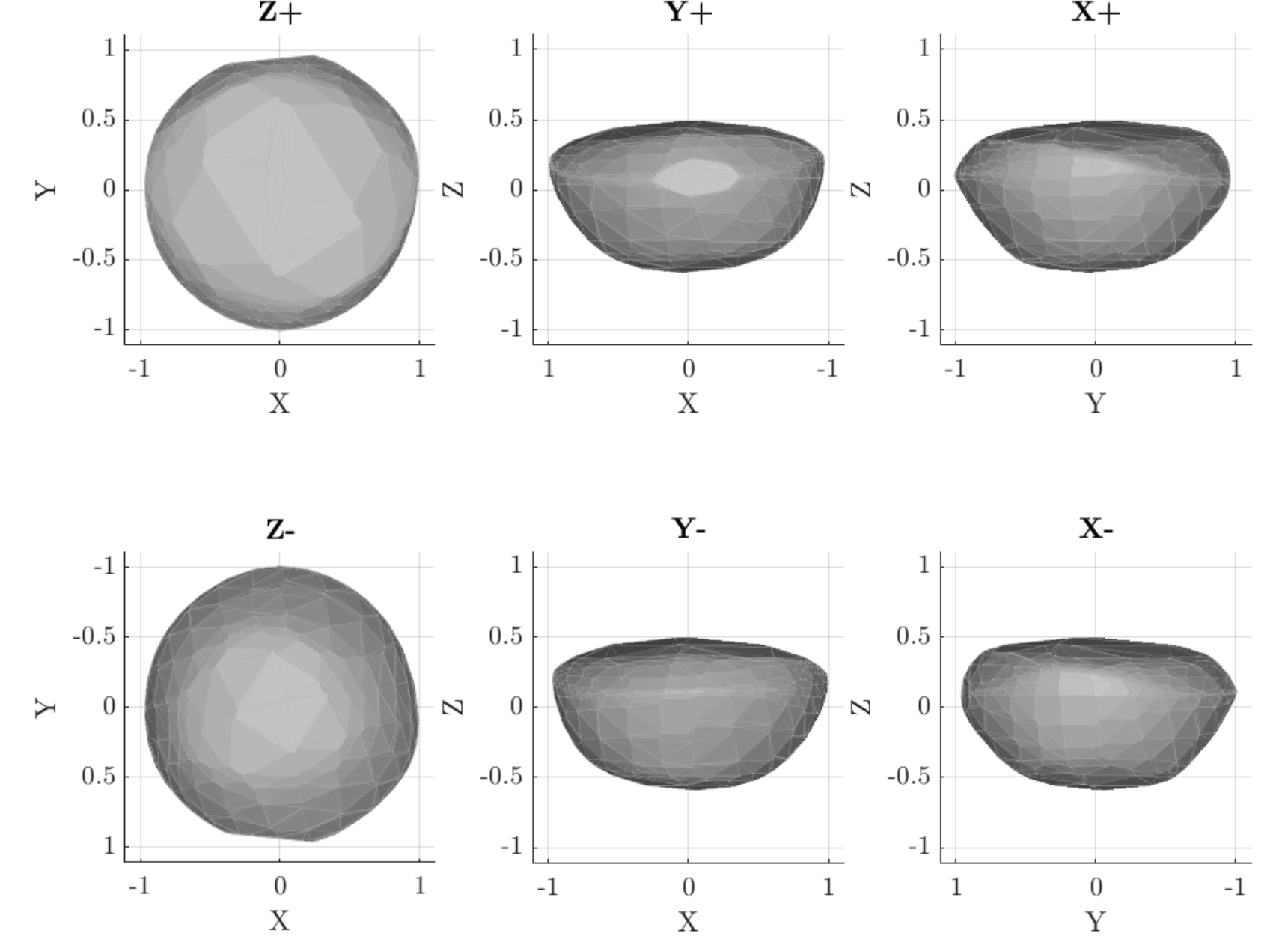} 
	\caption{Projections of the convex inversion shape model developed using the full light-curve data set. The XYZ-axes are aligned with the body principal moments of inertia. The units are arbitrarily selected so that the radius along the X-axis is 1. The flat face visible from the positive end of the Y-axis is an artefact of the modelling method. The presented shape model corresponds to the best-fit constant-period solution with $\lambda=210\degr$ and $\beta=-30\degr$. The best-fit variable-period solution with $\nu=2\times10^{-8}\,\textrm{rad/d}^2 $ has the same pole position and the shape is indistinguishable from the constant-period solution.}
	\label{fig:shape:allLC}
\end{figure}

The light-curve inversion produces a very symmetrical shape. Both the shape developed using the full light-curve set and a subset of light-curves have a nearly circular polar projection. This is consistent with the low amplitude of light curves observed at a wide range of observing geometries. The constant-period and varying-period best-fit solutions for the inversion of {the} full light-curve set are nearly identical with the same best-fit pole (shape model is shown in Fig.~\ref{fig:shape:allLC}).

\section{Radar modelling}
\label{sec:modelling}

\begin{figure}
	\centering

\resizebox{.95\hsize}{!}{\includegraphics{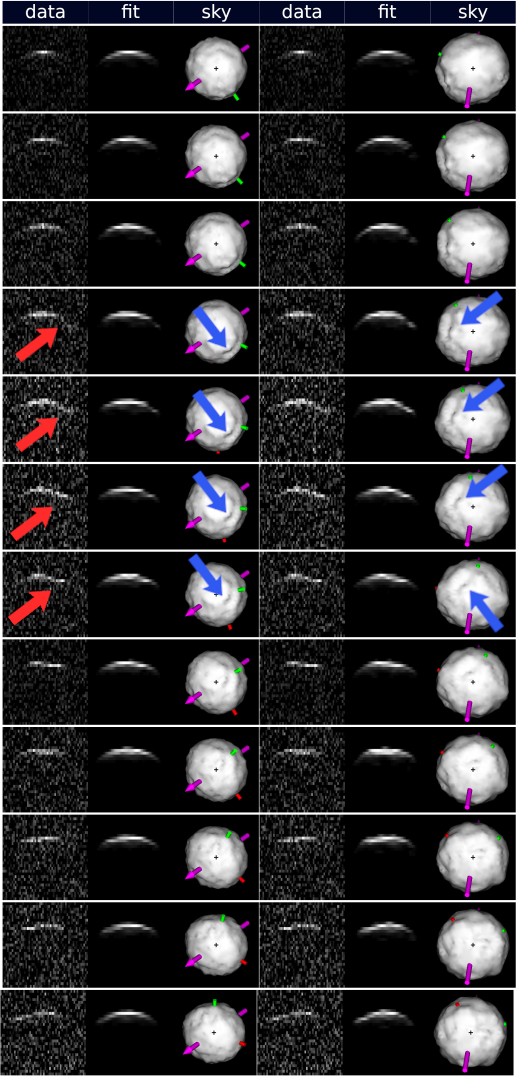}}

	\caption{ Comparison between best-fit retrograde (first three columns) and prograde (last three columns) radar shape models of asteroid (2102) and radar images (each row corresponds to the same radar image).
    The radar images were taken with Arecibo between 23:34 UTC on 1st January 2017 and 00:27 UTC on 2nd January 2017. Subsequent nights of observations are illustrated in the Appendix Figs.~\ref{fig:dd:retro:jan4}-\ref{fig:dd:pro:jan7}.
	Each three-image sub-panel is made of: the observational data (data), echo simulated from the radar model (fit), and plane-of-sky projection of the radar model (sky). On the data and synthetic-echo images the delay increases downwards and the frequency (Doppler) to the right. The plane-of-sky images are orientated with celestial north (in equatorial coordinate system) to the top and east to the left. The principal axes of inertia are marked with coloured rods (red for axis of minimum inertia, green for intermediate axis), and the rotation vector (Z-axis of body-fixed coordinate system, roughly aligned with axis of maximum inertia) is marked with a purple arrow. Note the rotation axis and Z-axis of the body overlap with the axis of maximum inertia. 
	Red arrows in selected images indicate asymmetry in the radar limb of the object, possibly corresponding to a surface indentation. The corresponding crater-like feature is marked with blue arrows in the plane-of-sky projections.
		\label{fig:dd:jan1}  }
	
\end{figure}

\begin{figure}
	\includegraphics[width=\columnwidth, trim=3cm 6cm 4cm 6.2cm, clip=true]{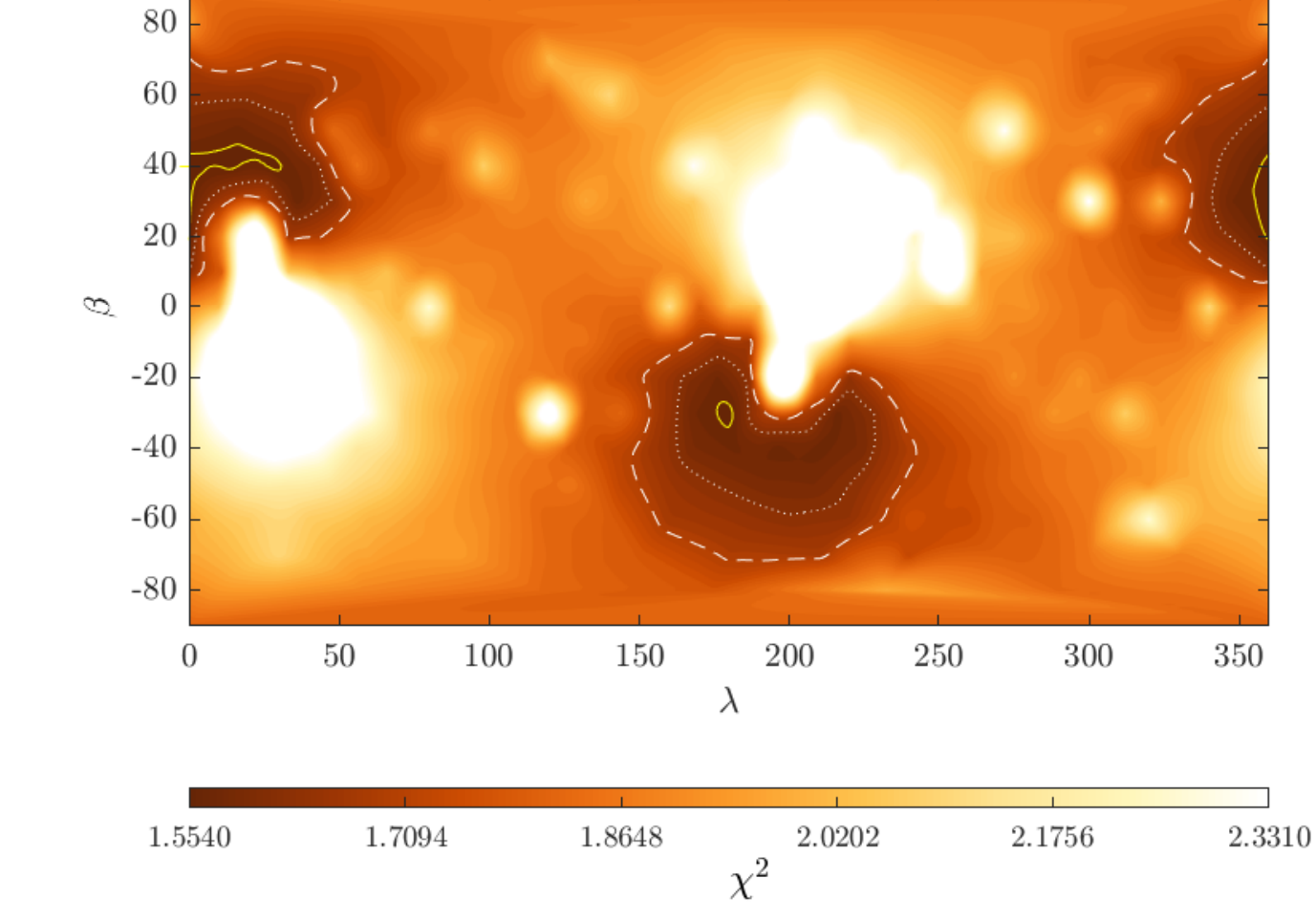} 	
	
	\caption{Similar to Fig.~\ref{fig:chi:allLC}, but for a constant period pole search using an ellipsoid model and a combination of radar and light curve data.
	Darker colours indicate lower $\chi^2$ values, and the lines mark gradual increases over the minimum $\chi^2$ value,  with solid line for 1\% increase, dotted line for 5\%, and dashed line for 10\%.
	Models with $\chi^2$ values smaller than $5\%$ above the minimum were selected for further improvement. The shape models were translated to the vertex representation and optimised using a combination of radar and light-curve data.
		 }
	\label{fig:chi:radar}
\end{figure}

\begin{figure*}
	\includegraphics[width=\columnwidth,trim=1.5cm 8.5cm 2cm 8.5cm,clip=true]{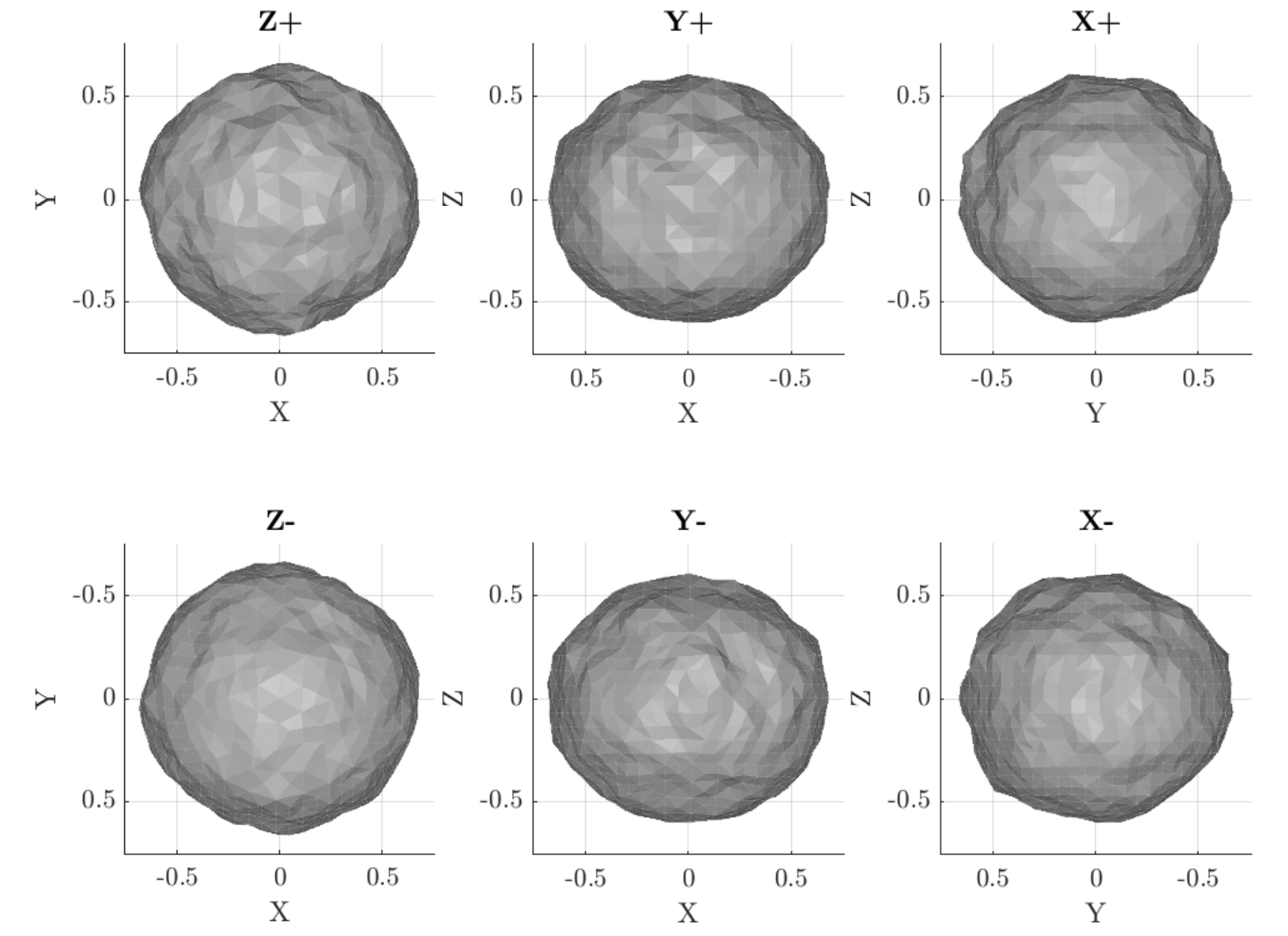} 
		\includegraphics[width=\columnwidth,trim=1.5cm 8.5cm 2cm 8.5cm,clip=true]{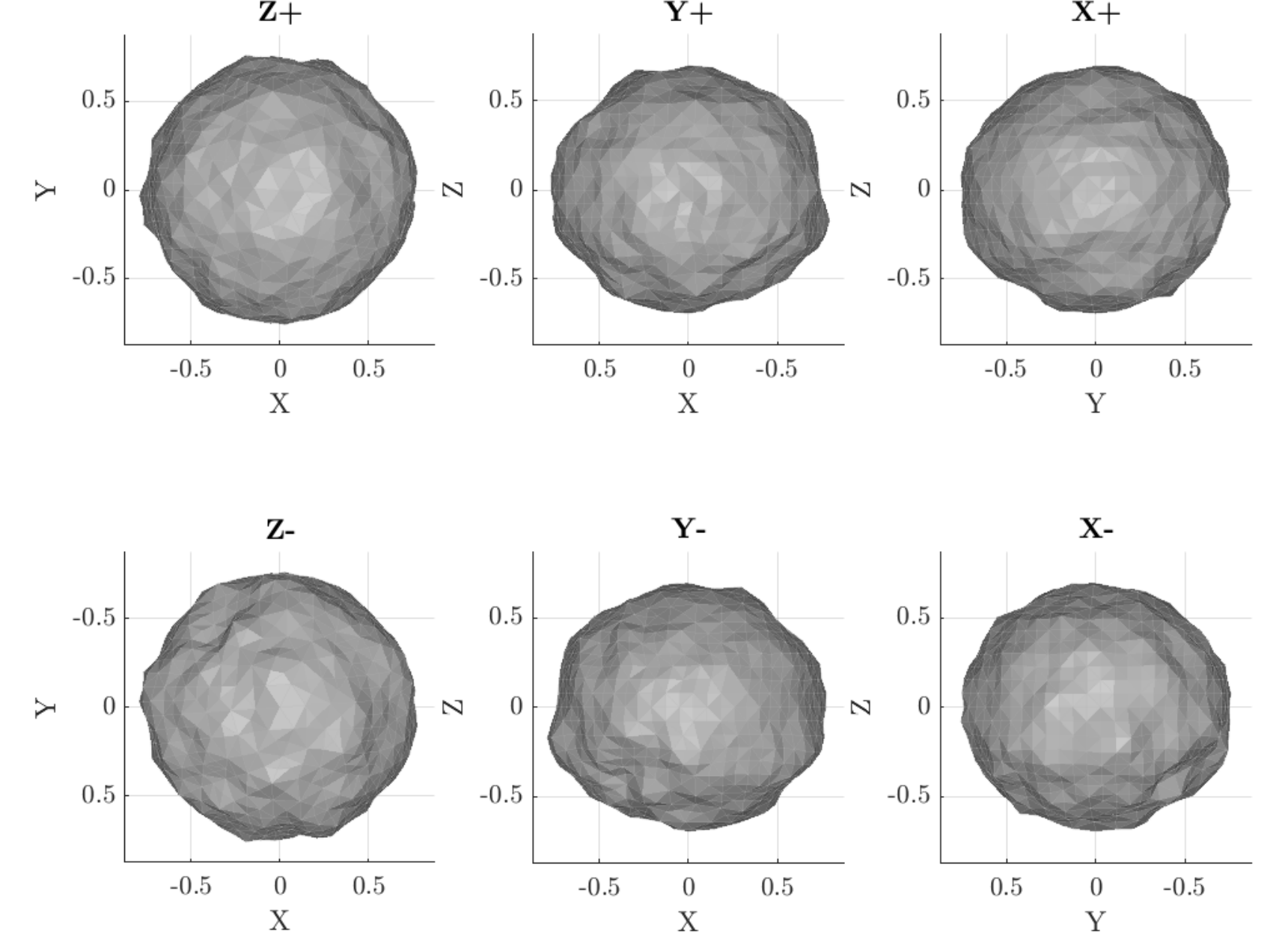} 
	\caption{Same as Fig.~\ref{fig:shape:allLC} but for the non-convex shape models developed with a subset of optical light-curves and the radar data. The six panels on the left illustrate the best-fit retrograde model with $\lambda=180\degr$ and $\beta=-30\degr$. The panels on the right illustrate the  best prograde solution with $\lambda=36\degr$ and $\beta=30\degr$. The difference of quality of fit of artificial light-curves and radar echos to data {between the two models} is negligible.}
	\label{fig:shape:radar}
\end{figure*}

\begin{table}
	\caption{Summary of spin-state parameters for (2102)  Tantalus from light curve inversion and radar {modelling}.  The presented sets of parameters correspond to the best-fit solutions from two approaches to shape modelling; in the `Light curve' column for the output of full convex-light-curve inversion, and in the `Radar retrograde' and `prograde' columns for a result of a global shape modelling that included both the radar and {light curve} data. The table lists: the ecliptic coordinates of the  rotation pole,  longitude ({$\lambda$}) and latitude ($\beta$), with associated uncertainties ({$\Delta_\lambda$} and $\Delta_\beta$), the model epoch ($T_0$), the sidereal rotation period ($P$) with uncertainty ($\Delta_P$). The uncertainties in pole and period are given as {the} standard deviation of the models within a  10\% increase of the minimum $\chi^2$ value (calculated from the ellipsoid models for the radar models).} 
	\label{tab:model:spin}
	\begin{tabular}[h]{l  l l l}
		\hline \hline
		{Parameter} & Light curve & Radar retrograde & Radar prograde \\ 	\hline	\noalign{\smallskip} 
		\noalign{\smallskip}
		$\lambda$         & $ 210 \degr $ & ${180} \degr$ & ${36} \degr$ \\
		$\Delta_\lambda$  & $ 41 \degr $ & $24 \degr$ & $23 \degr$ \\
		$\beta$           & $-30 \degr $  & ${-30} \degr$ & ${30} \degr$ \\
		$\Delta_\beta$    & $35 \degr $  & ${16} \degr$  & ${15} \degr$ \\
		\noalign{\smallskip}
		$T_0$ [JD]        & $2449695.50000$ & 
		$2457754.49022$
		&  $2457754.38451$  \\
		\noalign{\smallskip}
		$P$ [h]        &  2.385  & $2.391$ & $2.3901$       \\
		$\Delta_P$ [h]  & $0.003$ & $0.001$  & $0.0006$ \\ 
		\noalign{\smallskip} \hline                            
	\end{tabular}

\end{table}

\begin{table}
	\caption{Summary of shape parameters for asteroid (2102) Tantalus for the best-fit `retrograde' and `prograde' shape models developed from combined radar and light-curve data.  The extents are measured along the body-fixed coordinate system axis rather then principal axis of inertia. The D$_{eq}$ is the diameter of a sphere with volume equivalent to the model volume. The DEEVE stands for the dynamically equivalent equal-volume ellipsoid and `2a', `2b', and `2c' are its diameters. } 
	\label{tab:model:radar}

	\begin{tabular}[h]{l  l  l}
		\hline \hline \noalign{\smallskip} 
		Parameter & Retrograde & Prograde \\ 	\hline	\noalign{\smallskip}  
		Extent along \hfill X-axis [km] & 1.3 &  1.5 \\
		\hfill Y-axis [km] & 1.3 & 1.5 \\
		\hfill Z-axis [km] & 1.2 & 1.4 \\
	Surface area \hfill$[\textrm{km}^2]$ &   5.13 & 6.78\\
 Volume  \hfill $ [\textrm{km}^3]$ &   1.05 & 1.58\\
$\rm D_{eq}$ \hfill [km] & 1.3 & 1.5 \\
        DEEVE diameter \hfill 2a  [km] & 1.3 & 1.5\\
		\hfill 2b [km] &  1.3 & 1.5 \\
		\hfill 2c [km]&   1.2 & 1.4 \\
		\noalign{\smallskip} \hline	
	\end{tabular}

\end{table}

Observations of Tantalus are not restricted to optical photometry. Another source of shape information is the set of radar observations obtained in January 2017 (listed in Table~\ref{tab:radar}). Those observations reveal a quite symmetrical shape with a rounded limb and almost featureless surface. In only a few of the images, for example in the middle of the highest-resolution imaging sequence obtained on 1st January 2017 (illustrated in Fig.~\ref{fig:dd:jan1}),  some slight deformation of the asteroid can be noted. The limb of the radar echo appears asymmetrical  meaning there might be an indentation, like a crater, that causes the signal to travel a longer distance to this part of the surface.  

We assumed the sidereal rotation period derived from light-curve inversion as a starting point for the shape determination. This value was later refined as the model progressed. A subset of the highest quality light curves collected by us were selected (marked with a black circle in the `LC+radar model' column of Table~\ref{tab:obs}) and combined with the cw observations, and $75\,\textrm{m}$ and $150\,\textrm{m}$ resolution Doppler-delay imaging. We performed a pole search using a $5\degr\times5\degr$ grid of possible pole positions. The grid was constructed in such a way that the pole locations were evenly spaced in ecliptic latitude $\beta$, but in the longitudal dimension, $\lambda$, the search always started at $\lambda=0\degr$ and subsequent points were $5\degr$ as measured along a circle of latitude. At each point of the grid of possible pole positions we used the SHAPE modelling software \citep{Magri2007} to optimise the ellipsoidal shape and rotation rate, keeping the pole position fixed. 

The results of this pole search are illustrated in Fig.~\ref{fig:chi:radar}, similarly to Fig.~\ref{fig:chi:allLC}. The outcome is a slightly better constraint on the pole position, roughly in agreement with the convex light-curve inversion. 
The best solution corresponds to a retrograde solution with mid-negative latitude and longitude roughly between $150\degr$ and $250\degr$ (the mean $\beta=-39\degr\pm16 \degr$ and  $\lambda=190\degr\pm24\degr$). Due to inherent ambiguity between north and south in radar imaging \citep[the radar image is `folded' along the line of sight, e.g.][]{Ostro2002} there is a second, almost equally-good pole solution being a mirror with respect to the ecliptic plane and shifted by $180\degr$ in longitude (the mean $\beta=38\degr\pm15\degr$ and  $\lambda=11\degr\pm23\degr$). The sidereal rotation period is consistent between the mean value for retrograde, $P=2.391\pm0.001\,\textrm{h}$, and prograde, $P=2.3902\pm0.0006\,\textrm{h}$, families of solutions. Similarly, the average sizes of the triaxial ellipsoid axes are consistent, with  
 $ 1.2 \pm 0.3\, \rm km$, $ 1.2 \pm 0.3\, \rm km$, and $ 1.1 \pm 0.2\, \rm km$, a very slightly oblate spheroid,  for both retrograde and prograde models.

%

For further improvement and investigation of shape details, a family of 28 pole solutions, with the $\chi^2$ quality of fit as illustrated in Fig.~\ref{fig:chi:radar} within 5\% of the $\chi^2$ for best solution, were selected. For each of the selected pole positions the ellipsoid shape model was converted from triaxial-ellipsoid description to a triangular mesh by dividing the surface of the ellipsoid into 1000 vertices collected into 1996 triangular facets. At this stage the locations of individual vertices was optimised by shifting each individual vertex along the normal of initial ellipsoid surface. We show the final product of this stage in Fig.~\ref{fig:shape:radar}. The parameters of the best-fit retrograde and prograde solutions obtained after optimising those shape models are summarised in Table~\ref{tab:model:radar}. The average edge length for all 28 optimised models was $76\pm12\,\rm m$. The average equivalent volume sphere diameter is $ 1.2 \pm 0.2\, \rm km$, with the diameters of an ellipsoid with the same moments of inertia being $ 1.2 \pm 0.2\, \rm km$, $ 1.2 \pm 0.2\, \rm km$, and $ 1.1 \pm 0.2\, \rm km$, so the object is essentially spherical. 
The only notable feature is a crater-like indentation (close to the north pole of the retrograde model, and south pole of the prograde model). This is identifiable in the plane-of-sky projections {of both shape models} illustrated in Fig.~\ref{fig:dd:jan1} {(the feature is marked with red arrows in the radar images and blue arrows in the plane-of-sky projections)}.
Optical light curves, as illustrated in Figs.~\ref{fig:conv-lcfit:example}, \ref{fig:radar-retro1}, and \ref{fig:radar-pro1}, 
aren't sufficiently well reproduced from either radar model to enable reliable rotational phase offset measurement, as was done for example for (68346) 2001 KZ66 \citep{Zegmott2021}. We therefore conclude that no spin-state change can be dependably detected for Tantalus.

\section{Radar properties}
\label{sec:radarprop}

\begin{table*}
	\caption{ Radar-derived disc-integrated properties for the cw spectra of (2102) Tantalus.
	{UT Date} (in `year-month-day' format) and {UT Time} (in `hours:minutes:seconds' format)  is the universal-time date and time mid-receive. The body-fixed {longitude}, $\lambda_B$, and latitude, $\beta_B$,  of the radar line-of-sight, and rotation phase, $\varphi$, were determined using the spin-state for two radar shape models (see Table~\ref{tab:model:radar}). 
	The rotation phase is measured in degrees as an offset of X-axis relative to its position at $T_0$. The radar cross-sections (`$\sigma_{OC}$' column) were calculated from the OC power spectra. The `{Projected areas}' and radar OC albedos (`$\hat{\sigma}_{OC}$') were derived using the 3D shape of either of the two models. 
	The SC/OC ratio (`$\mu_{C} $') for each spectrum is also given. In the last row we list the means and standard deviations of OC albedos and SC/OC ratio. We note the difference in the mean radar albedo determination is primarily due to the size difference between the retrograde and prograde models.	
	\label{tab:radar:spec}} 
\resizebox{\hsize{}}{!}{  
		\begin{tabular}{c c | c  c c c  c c | c  c c c c c | c}
			\hline \hline \noalign{\smallskip} 
					&	&     \multicolumn{6}{c}{Retrograde model}                &  \multicolumn{6}{c}{Prograde model}\\
UT Date	&	UT Time	&	 $\lambda_{B}$ 	&	 $\beta_{B}$ 	&	 $\varphi$ 	&	   $\sigma_{OC}$ 	&	 Area 	&	  $\hat{\sigma}_{OC}$ 	&	 $\lambda_{B}$ 	&	 $\beta_{B}$ 	&	 $\varphi$ 	&	  $\sigma_{OC}$ 	&	 Area 	&	 $\hat{\sigma}_{OC}$ 	& $\mu_{C}$ \\
	&	  	&	 {[$\degr$]} 	&	{ [$\degr$] }	&	{ [$\degr$] }	&	{[$\rm km^2$]}	&	{[$\rm km^2$]}	&		&	 {[$\degr$]} 	&	{ [$\degr$] }	&	{ [$\degr$] }	&	{[$\rm km^2$]}	&	{[$\rm km^2$]}	&	  & 	\\ \hline
2017-01-01	&	22:56:30	&	182	&	44	&	178	&	0.370	&	1.300	&	0.284	&	254	&	-52	&	106	&	0.371	&	1.733	&	0.214	& $0.103 \pm 0.011$ \\
2017-01-04	&	21:33:13	&	339	&	61	&	21	&	0.234	&	1.326	&	0.177	&	36	&	-66	&	324	&	0.229	&	1.751	&	0.131	& $0.202 \pm 0.022$ \\
2017-01-05	&	21:08:24	&	22	&	65	&	338	&	0.181	&	1.320	&	0.137	&	74	&	-69	&	286	&	0.181	&	1.755	&	0.103	& 	$0.287 \pm 0.043$ \\
2017-01-06	&	21:06:11	&	8	&	69	&	352	&	0.233	&	1.322	&	0.176	&	54	&	-71	&	306	&	0.234	&	1.756	&	0.133	& 	$0.148 \pm 0.028$\\
2017-01-07	&	21:10:50	&	336	&	71	&	24	&	0.279	&	1.325	&	0.211	&	16	&	-73	&	344	&	0.287	&	1.757	&	0.163	& 			$0.202 \pm 0.003$\\
			&  &  &  &   &      \multicolumn{2}{r}{Mean values:}                &       $0.20\pm0.05$  & &  &  & & &        $ 0.15    \pm     0.04    $      & 			$0.19 \pm 0.06$ \\
						
			\noalign{\smallskip}
			\hline                                       
		\end{tabular}
}

\end{table*}

\begin{figure*}

	\resizebox{0.99\hsize}{!}{\includegraphics[trim=5cm 2.5cm 2cm 1cm, clip=true,angle=0]{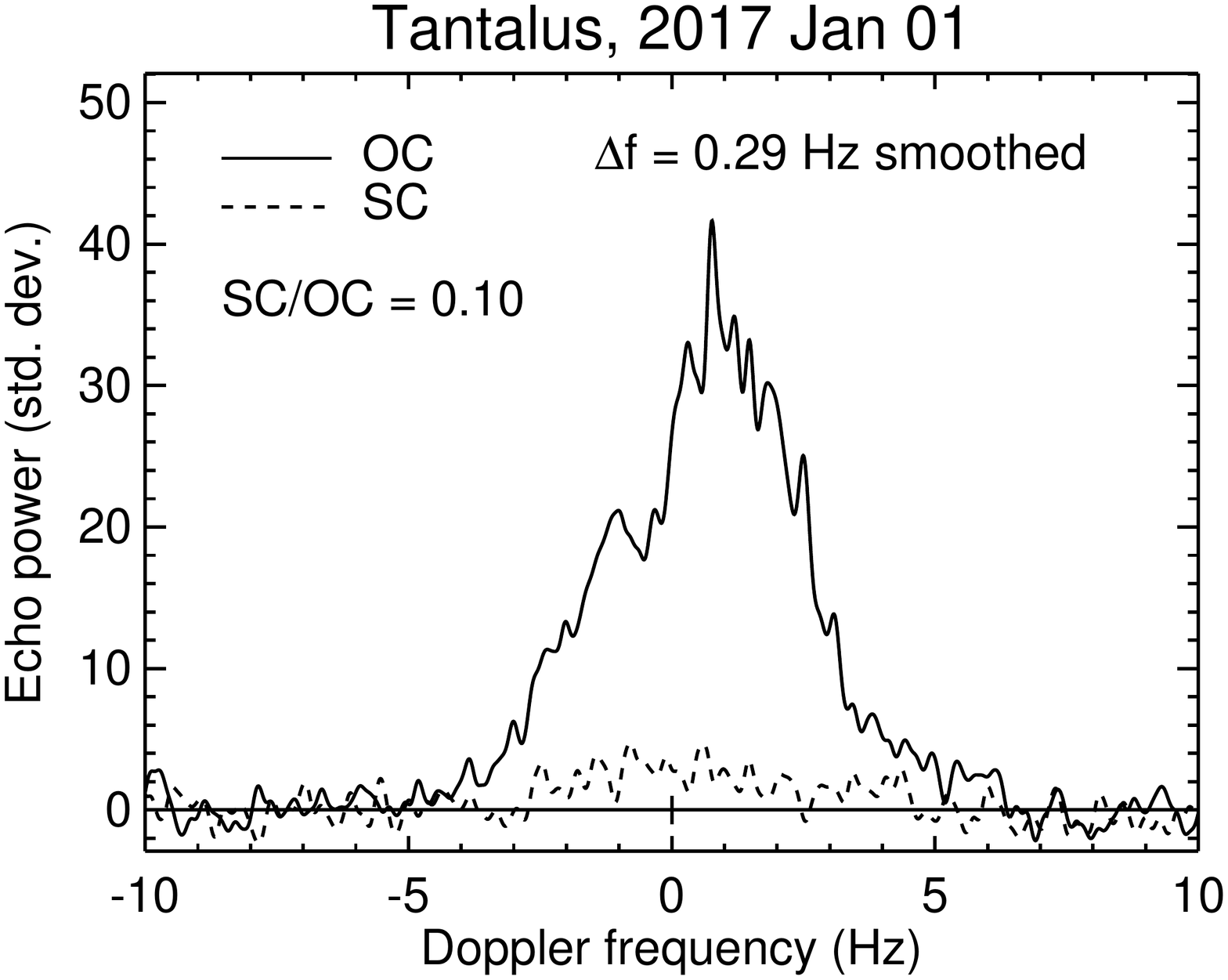} \includegraphics[trim=5cm 2.5cm 2cm 1cm, clip=true,angle=0]{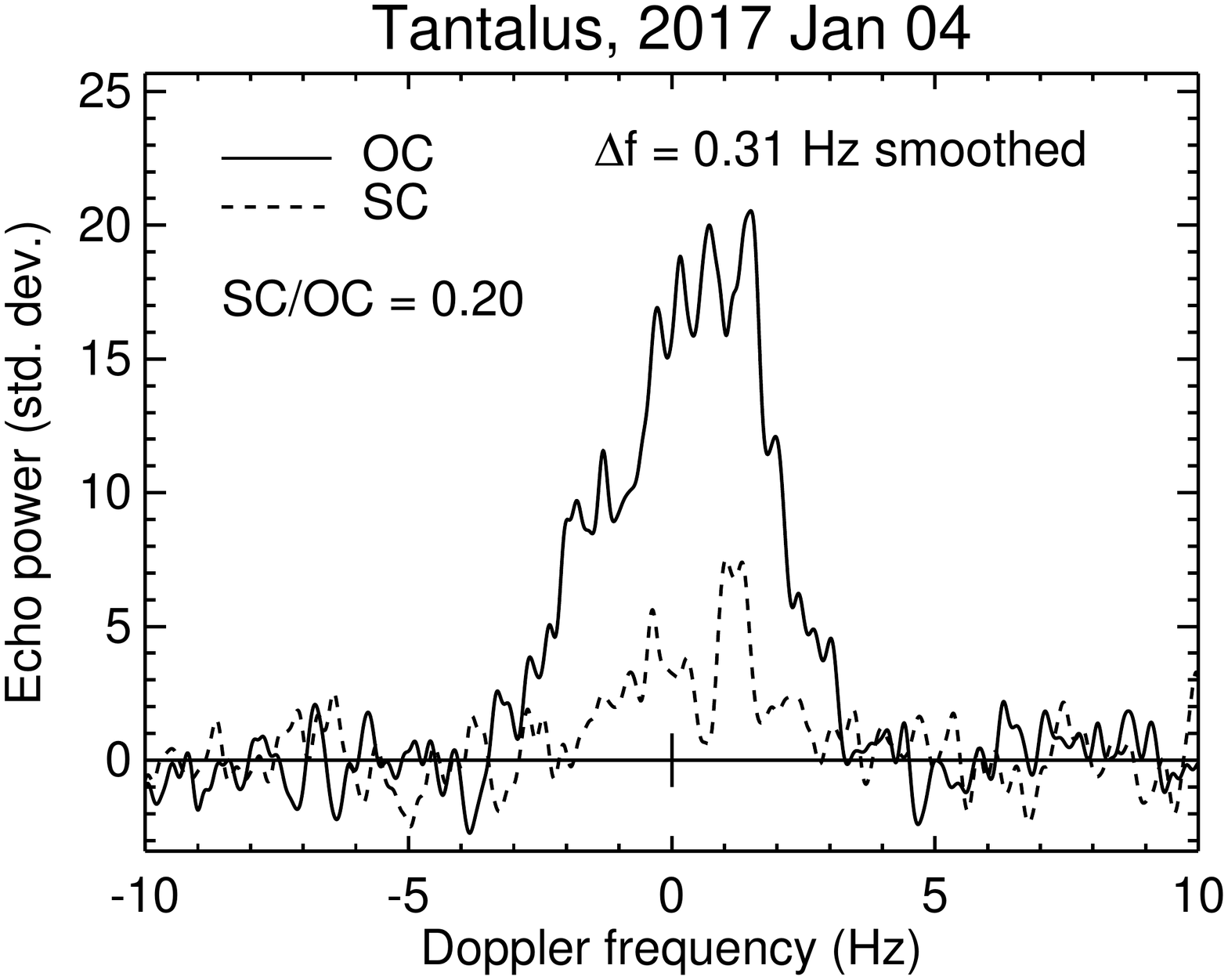} \includegraphics[trim=5cm 2.5cm 2cm 1cm, clip=true,angle=0]{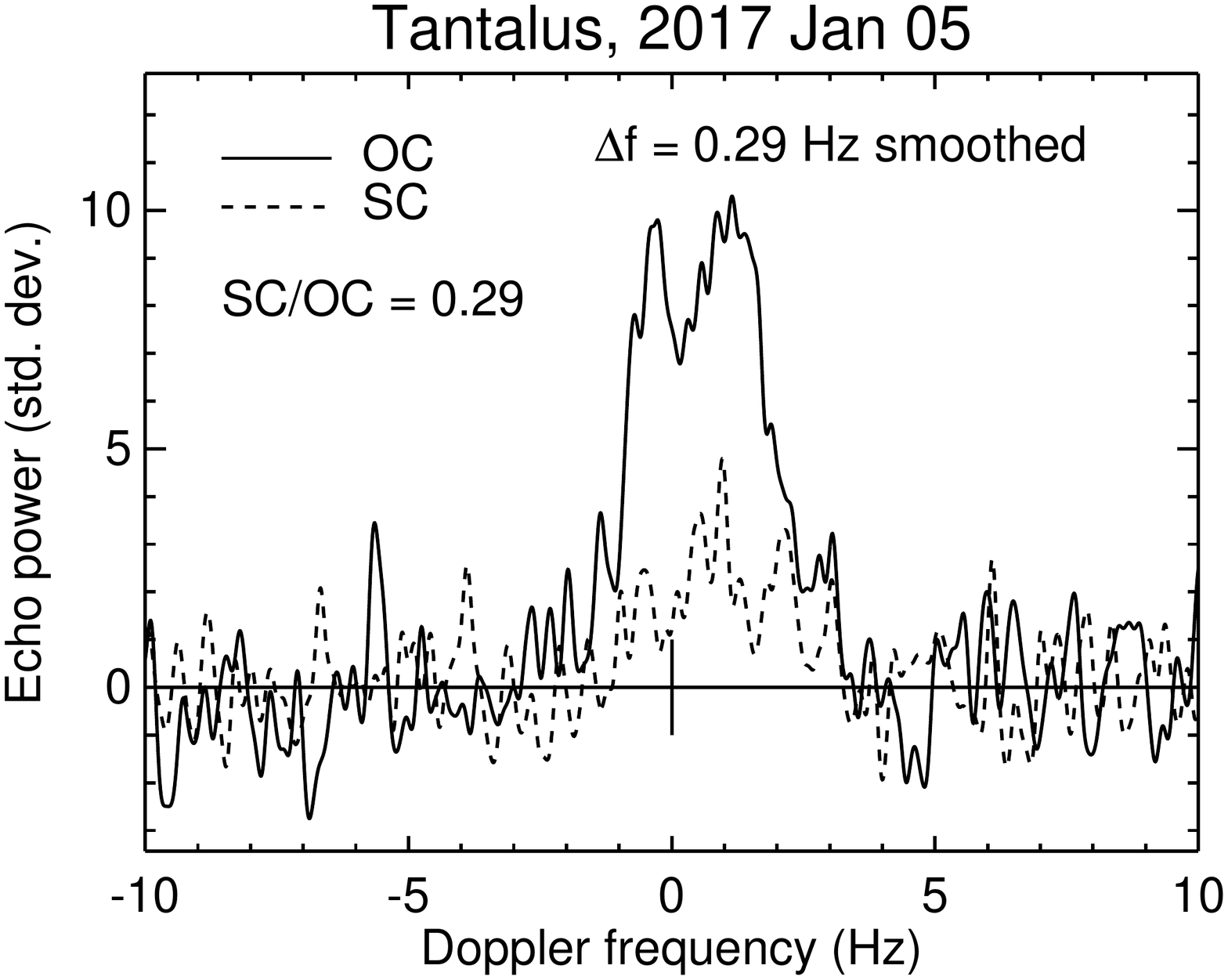}} 
	\resizebox{0.66\hsize}{!}{\includegraphics[trim=5cm 2.5cm 2cm 1cm, clip=true,angle=0]{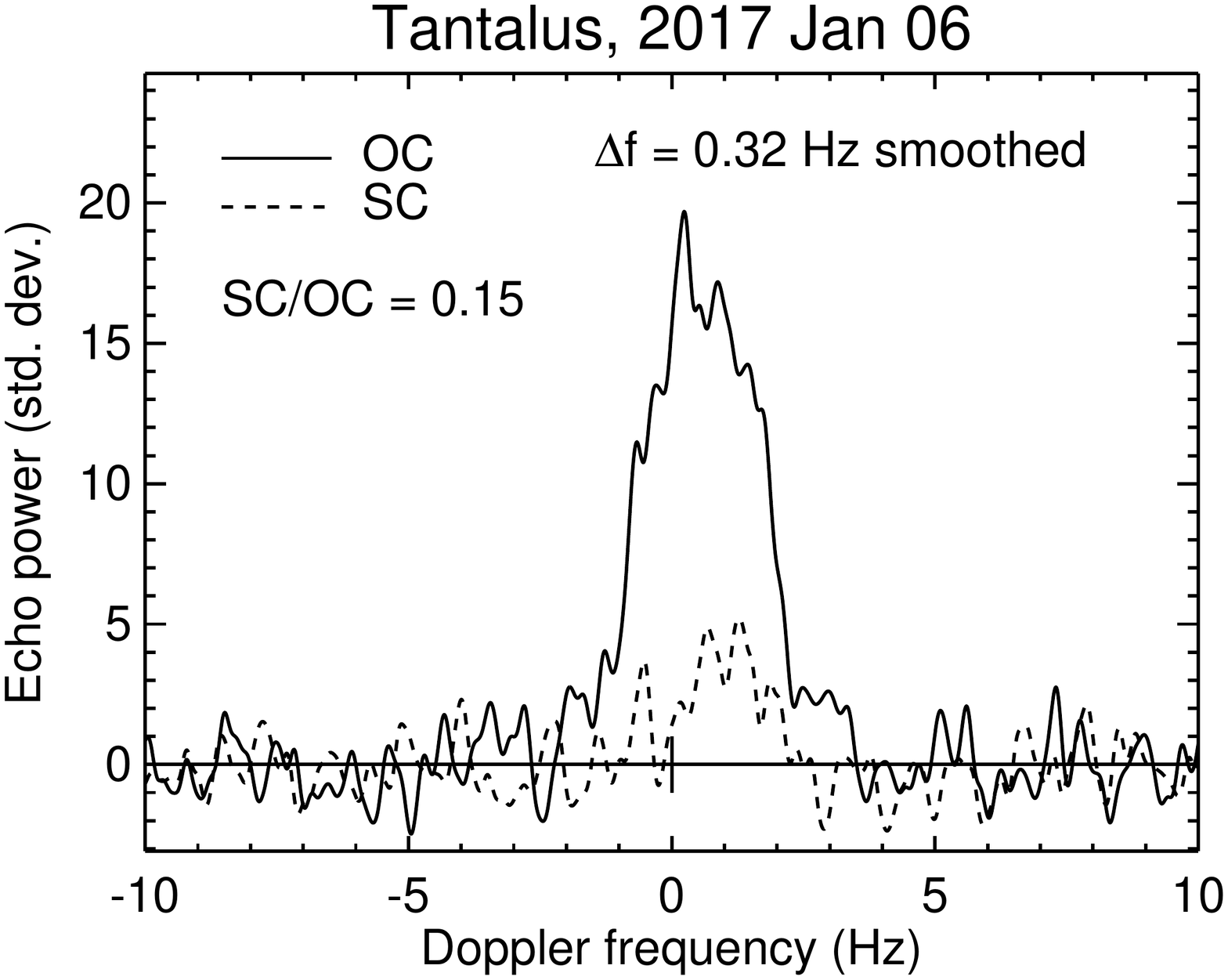} \includegraphics[trim=5cm 2.5cm 2cm 1cm, clip=true,angle=0]{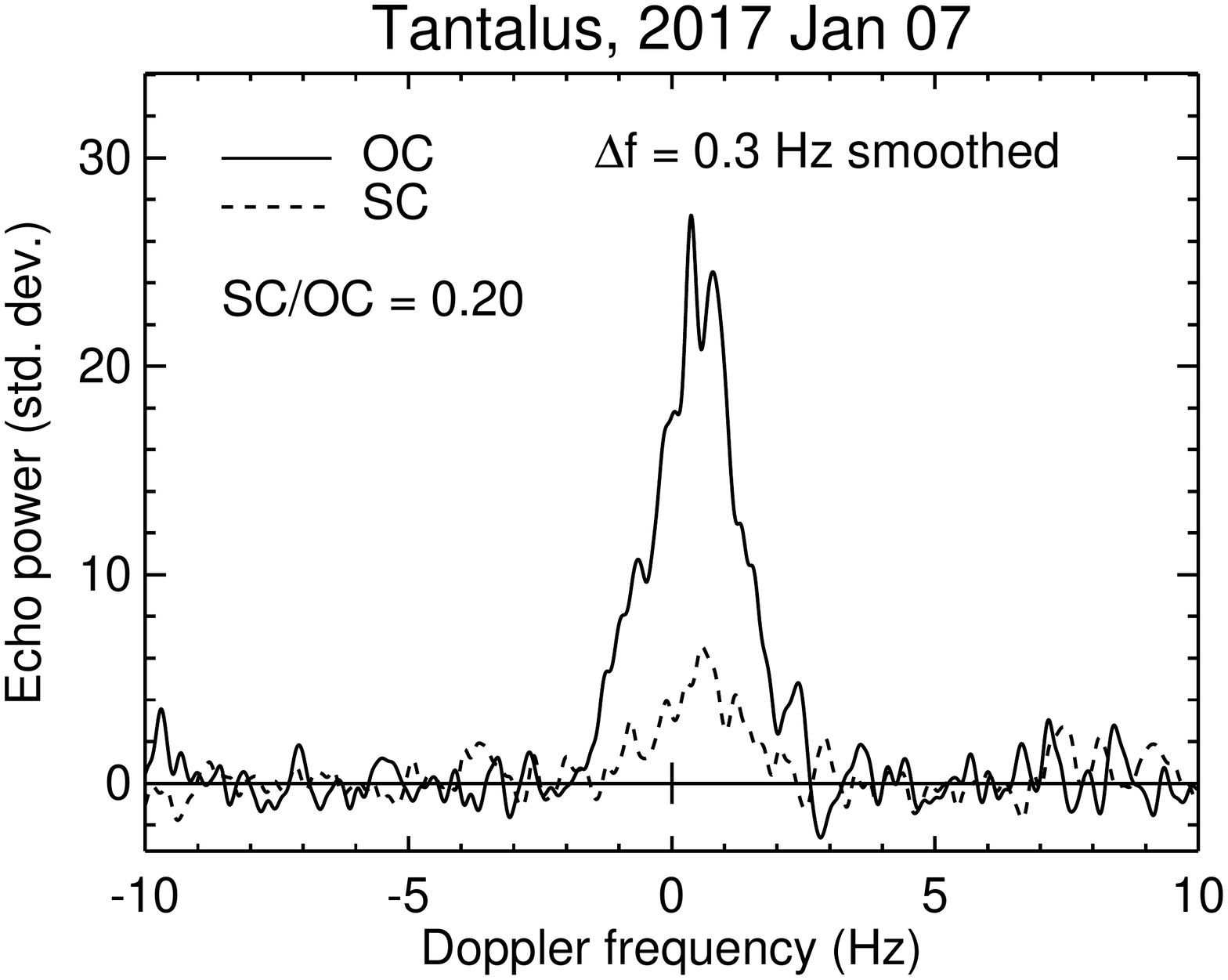}}

	\caption{The continuous-wave observations of (2102) Tantalus collected in January 2017 at Arecibo. The disc-integrated properties of Tantalus based on the cw spectra are gathered in Table~\ref{tab:radar:spec}. The signal is recorded in two channels, with the same circular (SC) polarisation as the transmitted radiation marked with a dashed line in each panel, and the opposite circular (OC) polarisation, marked with a solid line. The spectra show SC/OC ratios typical for an S-type NEA. \label{fig:cw}}
	
\end{figure*}

\begin{figure}
	\includegraphics[width=\columnwidth, trim=0cm 0cm 0cm 0cm, clip=true]{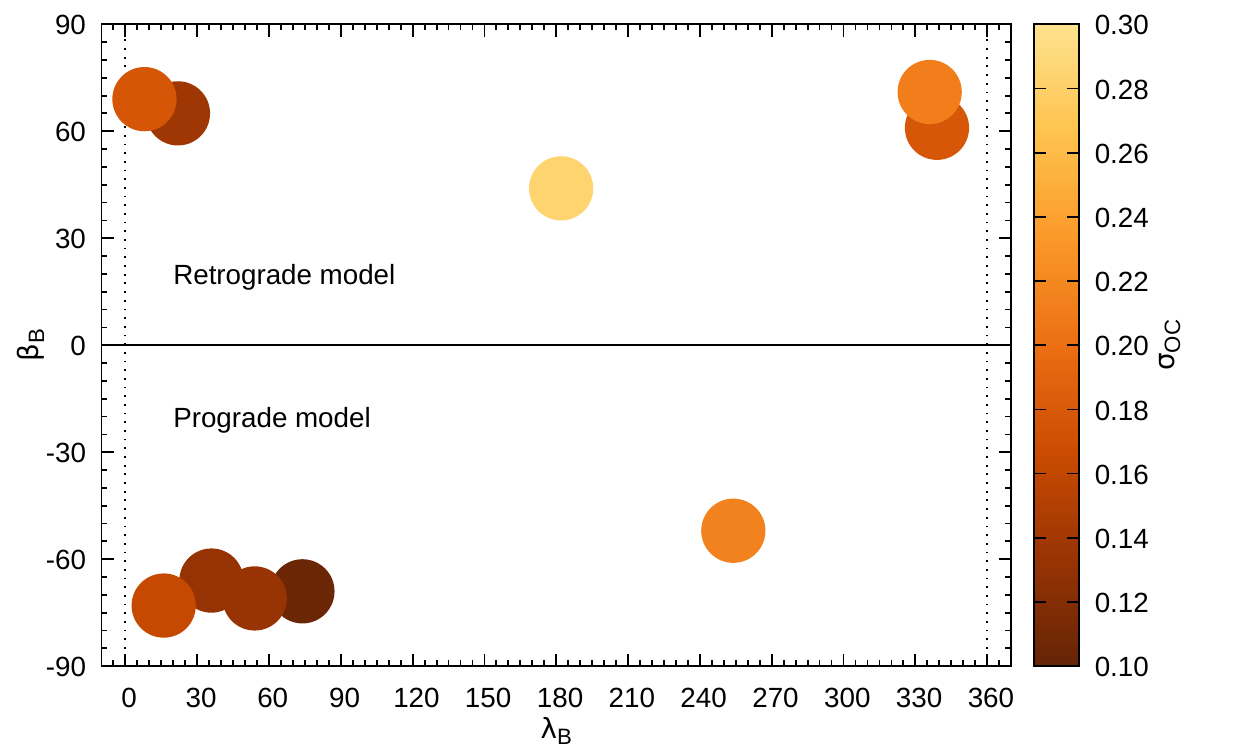}
	
	\includegraphics[width=\columnwidth, trim=0cm 0cm 0cm 0cm, clip=true]{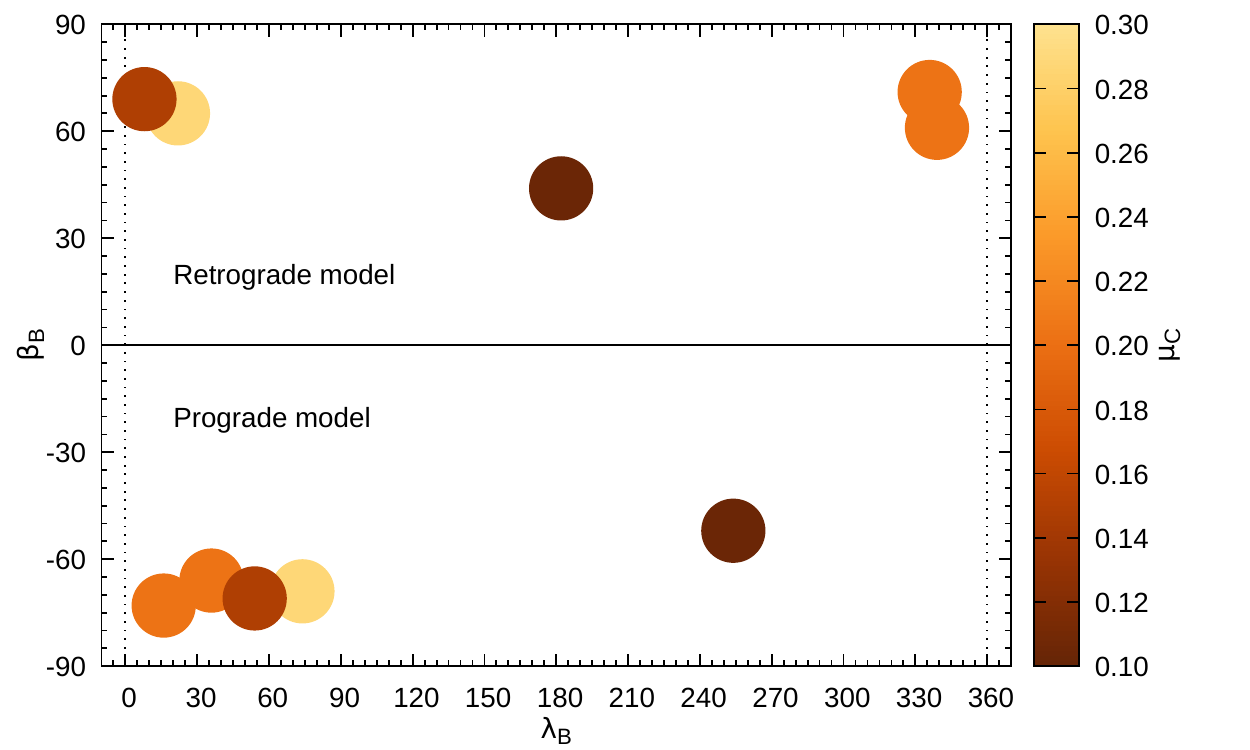}
	
	\caption{ A map of radar albedo ($\sigma_{OC}$, upper plot) and circular polarisation ratio ($\mu_C$) for asteroid 2102 Tantalus based on cw measurements. The figure is a visualisation of measurements listed in Table~\ref{tab:radar:spec} with darker colour in each plot corresponding to lower values of $\sigma_{OC}$ or $\mu_C$, and lighter to higher. The size of the spots of colour is arbitrary. Each plot is divided in two as measurements for the retrograde model were made in the northern hemisphere in body-centric coordinates (upper half of both plots), and in the southern hemisphere for the prograde model (lower half).}
	\label{fig:radarprop}
\end{figure}

Analysis of the continuous-wave (cw) spectra, summarised in Table~\ref{tab:radar:spec} and shown in Fig.~\ref{fig:cw}, shows some discrepancy between prograde and retrograde models due to the  difference in subradar latitude between the assumed poles and the radar line of sight, and the uncertainty in size determination. The best-fit prograde model is $230$, $150$, and $180\,\textrm m$ larger than the retrograde along each dimension. As a result of that, the radar albedo ($\hat{\sigma}_{OC}$) measurements are systematically lower for the prograde model. The values, $\hat{\sigma}_{OC}=0.20\pm0.05$ for the retrograde model and  $0.15\pm0.04$ for prograde, are  still consistent and agree with typical values determined for radar-observed S-type main-belt asteroids, which is $\hat{\sigma}_{OC}=0.14\pm	0.04$ \citep{Magri2007surv}. 
If we use equations in \citet{Shepard2008,Shepard2010,Shepard2015}, $\hat{\sigma}_{OC}$  would correspond to near-surface bulk densities between $2000$ and $3100\, \textrm{kg}\,\textrm{m}^{-3}$ for the retrograde model and $1800$ to $2600\, \textrm{kg}\,\textrm{m}^{-3}$ for prograde, which lie within the values expected for S-type asteroids \citep{Carry2012}.

The ratio of the same circular polarisation as the emitted signal (SC) to the opposite circular polarisation (OC) detected for the object is $\mu_C=0.19\pm0.06$. This measurement is based on the calibrated radar echo and is independent of shape information. This value is consistent with the typical  $\mu_C=0.270\pm0.079$ for S-type near-Earth objects \citep{Benner2008}, albeit at the lower limit. 
The polarisation ratio is
a zeroth-order gauge to the surface roughness to the extent that zero means a smooth surface in the wavelength scale  (13 cm for Arecibo), but beyond that it's not a linear scale of surface roughness because it's affected by different factors, such as the permittivity and particle-size frequency distribution in the near-surface \citep{Virkki2016}.
The $\mu_C$ for Tantalus is highest on 5th January, and lowest at 1st January, when the asteroid is viewed from almost opposite directions. Incidentally, the radar albedo is lowest on the 5th January and highest on the 1st. This might mean that there is some variation of surface material on Tantalus. 
One side would have a smoother, radar-reflective, so less porous, surface which could be exposed rock.
On the other side there could be a rough, radar-dark patch, perhaps a crater filled with fine-grained regolith (mapping of measured $\hat{\sigma}_{OC}$ and $\mu_C$ on the asteroid surface is presented in Fig.~\ref{fig:radarprop}).

Interestingly, the cw spectrum collected on the 5th January also shows bifurcation, which is a characteristic shape of the cw spectrum for contact binary objects. However, it is very clear from the images that the object is symmetrical and the dip in echo power is only at about $25\%$ level, rather than 50-100\% level typical for a contact binary. Similar bifurcation of radar echo was recently noted for (16) Psyche and attributed to a possible radar-dark spot surrounded by regions of higher albedo \citep{Shepard2021}. This explanation seems sensible as the same cw spectrum produced overall lowest value of $\hat{\sigma}_{OC}$ for Tantalus, suggesting less radar-reflective material. However, this particular cw spectrum also appears to be the noisiest, and we noted significant pointing errors in acquiring this observation. A cw spectrum taken on 6th January comes from a similar location on the surface but has a higher albedo and does not appear bifurcated, so we conclude there is insufficient evidence to claim that there is in fact a radar-dark spot on the surface in the specific location corresponding to the cw spectrum taken on 5th January. Still, the whole region probed between 4th and 7th January appears less reflective in radar wavelengths than the other side of the asteroid, corresponding to the spectrum from 1st January. On the other hand, the spectrum from 1st January is isolated, with no nearby measurement to confirm the high albedo, so further observations are needed to confirm the surface variation.

It might be worth considering here how a less radar-reflective spot would affect the optical light curves as the light-curve inversion method assumed uniform optical albedo. Radio signal penetrates the surface, so the radar albedo carries information about the physical properties of the top layer of material on a small body at the scale of the radar wavelength  \citep[e.g.][]{Virkki2016}, while optical light operates at much shorter wavelengths. In fact, analysis of a sample of radar-observed main-belt asteroids shows no correlation between radar and optical albedos for S-types \citep{Magri2007surv}.  While there is some mismatch between the synthetic light curves generated from the radar shape model and the data collected, it is mostly in rotation phase and not in the light-curve amplitude. A considerably darker spot on an otherwise symmetrical body would produce a dip in the asteroid brightness, but this is not evident in the data. Therefore, we conclude there is no convincing evidence of an optically dark spot.

\section{Thermophysical analysis}
\label{sec:atpm}

\begin{table}
	\caption{ Observational geometry used in the thermophysical modelling of Tantalus. For each set of dates in 2010 (column `Dates') the following information is given: 
	number of measurements in WISE W3 and W4 channels (`No. W3' and `No. W4' respectively),	
	heliocentric distance (`$R_h$'),
	heliocentric longitude (`$\lambda$') and latitude (`$\beta$'),  
	observer distance (`$\Delta$'), and observer longitude (`$\lambda_o$') and latitude (`$\beta_o$').
	\label{tab:X}} 
\resizebox{\hsize{}}{!}{  
		\begin{tabular}{c c c  c c c  c c c }
\hline \hline
Dates & No. & No. & $R_h$ & $\lambda$ & $\beta$ & $\Delta$ & $\lambda_o$ & $\beta_o$ \\ 
{[dd/mm]} & W3 & W4 & [AU] &  [$\degr$] & [$\degr$] &  [AU] & [$\degr$] & [$\degr$] \\ \hline
16-20/06 & 33 & 33 & 1.657 & 300.8 & -42.4 & 1.308 & 356.9 & -58.6 \\
4-19/07 & 74 & 74 & 1.673 & 308.9 & -49.2 & 1.322 & 17.7 & -73.6 \\
		\noalign{\smallskip}
			\hline  
		\end{tabular}
}
\end{table} 

\begin{table}
	\caption{ Derived thermophysical properties for the three different shape models of Tantalus: the full lightcurve-inversion model (`Lightcurve'), and the retrograde and prograde radar models. 
	\label{tab:Y}} 
\resizebox{\hsize{}}{!}{  
		\begin{tabular}{c c c  c }
\hline \hline
Shape Model & Lightcurve & Radar &
Radar  \\
 &  & Retrograde &
Prograde \\ \hline

Reduced-$\chi^2$ & 1.1 & 1.0 &  1.1 \\
Radar diameter [km] & {--} & $1.26 \pm 0.20$ & $1.45 \pm 0.20$ \\
Radiometric diameter [km] & $1.68 \pm 0.05$ & $1.66 \pm 0.05$ & 	$1.57 \pm 0.05$ \\
Geometric albedo & $0.25 \pm 0.01$ & $0.26 \pm 0.01$ & $0.29 \pm 0.02$ \\
Thermal inertia [$\textrm{J}\, \textrm{m}^{-2}\, \textrm{K}^{-1}\, \textrm{s}^{-1/2}$] & $100 \pm 20$ & $110 \pm 30$ & $270 \pm 80$ \\
Roughness fraction & $0.12 \pm 0.09$ & $0.49 \pm 0.29$ & $0.13\pm0.09$ \\
		\noalign{\smallskip}
			\hline  
		\end{tabular}
}

\end{table}

\begin{figure}
	\includegraphics[width=\columnwidth]{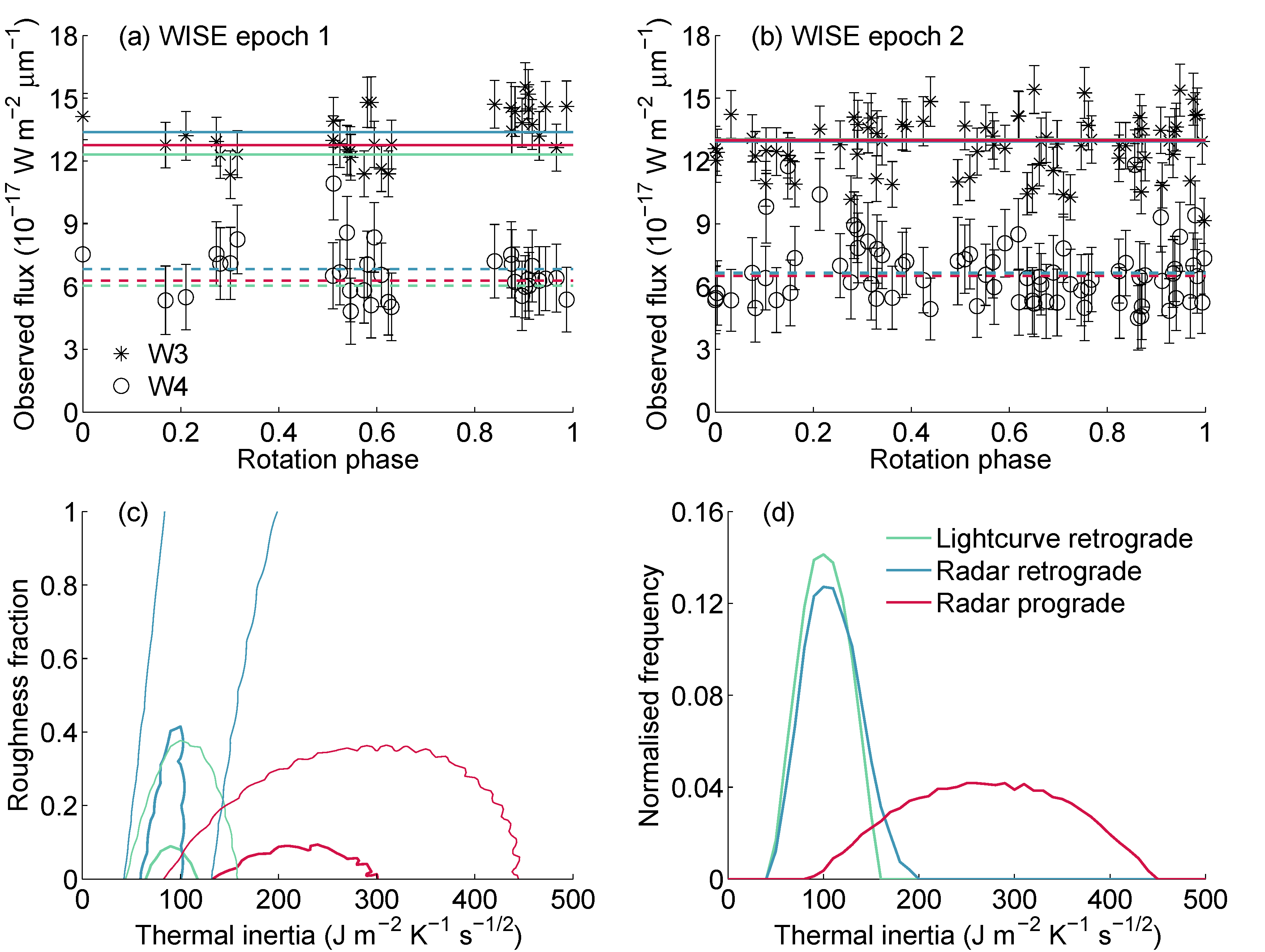}
	\caption{ Thermophysical modelling of the WISE data of Tantalus. The thermal lightcurves of Tantalus obtained by WISE on 16-20 June 2010 and 4-19 July 2010 are given in panels (a) and (b), respectively. Here, the W3 and W4 channel datapoints are given by the asterisks and circles, respectively, which have been rotationally phased relative to the first datapoint assuming a rotation period of 2.39 hours. The solid (W3) and dashed (W4) lines give the best model fits for the three shape models tested after rotational averaging. Panel (c) gives contours in $\chi^2$ of $1\sigma$ (thick lines) and $3\sigma$ (thin lines) confidence level in derived thermal inertia and roughness fraction for the three shape models tested. Finally, panel (d) shows the normalised frequency distributions of acceptable thermal inertia values contained within the $3\sigma$ confidence level $\chi^2$ contours.}
	\label{fig:X}
\end{figure}

To ascertain the preferred shape model solution of Tantalus, and its thermophysical properties, we modelled and fitted the infrared observations of Tantalus that were serendipitously acquired by WISE during its cryogenic operations in 2010 \citep[Table~\ref{tab:X};][]{Mainzer2011b}. 
Additional observations were also acquired in 2014, 2016, 2017, and 2020 
\citep{Masiero2017,Masiero2020}, 
but we did not include these datasets in our analyses because they were obtained during the non-cryogenic phase of the mission where reflected sunlight contributed significantly to the available near-infrared data. To retrieve the WISE data, the detections of Tantalus reported in the Minor Planet Center (MPC) database were used to query the WISE All-Sky Singe Exposure (L1b) source database via the NASA/IPAC Infrared Service Archive. Following 
\citet{Rozitis2018}, 
the queries were performed in the moving object search mode with a match radius of $5\arcsec$ for the range of dates provided by the MPC. The resulting detections were only kept in the instances where the measured flux levels of Tantalus were at the $3\sigma$ level or greater in both the W3 ($11.1\, \upmu\textrm{m}$) and W4 ($22.6\, \upmu\textrm{m}$) channels, and when the measured positions were within $1\arcsec$ of the predicted positions. The retrieved WISE magnitudes were converted to fluxes by accounting for the red-blue calibrator discrepancy reported by 
\citet{Wright2010}, 
and additional uncertainties of 4.5 and $5.7\%$ were added in quadrature to the retrieved uncertainties in the W3 and W4 channels, respectively, to account for other calibration issues 
\citep{Jarrett2011}. 
Finally, colour corrections were performed on the model fluxes, rather than the observed fluxes, using the WISE corrections provided by \citet{Wright2010}. This resulted in 33 W3 and 33 W4 usable flux measurements on 16-20 June 2010, and 74 W3 and 74 W4 flux measurements on 4-19 July 2010 (Table~\ref{tab:X}).

Thermophysical modelling was performed using the Advanced Thermophysical Model 
\citep[ATPM;][]{Rozitis2011,Rozitis2012,Rozitis2013}
in combination with the lightcurve- and radar-derived shape models of Tantalus. For a given shape model, the ATPM computes its surface temperature distribution by solving the 1-D heat conduction equation for each triangular facet with a surface boundary condition that accounts for direct solar illumination, shadowing, multiple scattered sunlight, and self-heating from thermal re-emission. The effects of rough surface thermal-infrared beaming (i.e., thermal re-direction of absorbed sunlight back towards the Sun) are incorporated by the fractional addition of hemispherical craters that represent the unresolved surface roughness. Surface temperatures were computed for thermal inertia ranging from $0$ to 
$1000\,\textrm{J}\, \textrm{m}^{-2}\, \textrm{K}^{-1}\, \textrm{s}^{-1/2}$
in steps of $10\,\textrm{J}\, \textrm{m}^{-2}\, \textrm{K}^{-1}\, \textrm{s}^{-1/2}$ for the observational geometries given in Table~\ref{tab:X} assuming an emissivity of $0.9$ and a Bond albedo of $0.15$
\citep[i.e., calculated using an H of 16.0 and a G of 0.15 with the radar-derived diameter;][]{Rozitisetal2013}.
The predicted model fluxes were then calculated as a function of wavelength, thermal inertia, roughness fraction, and rotation phase by applying and summing the Planck function across all facets that were visible to the observer at the instances of the observations. Finally, $\chi^2$-minimisation was performed between the data and thermophysical models using the rotational averaging technique described in 
\citet{Rozitis2018}
to obtain the best-fit diameter, thermal inertia, and roughness fraction for each possible shape model of Tantalus. 

Figure~\ref{fig:X} and Table~\ref{tab:Y} summarise the resulting thermophysical fits and properties, respectively, for the three different shape models of Tantalus tested. As indicated, the thermophysical fits did not prefer one shape model over the others based purely on their derived $\chi^2$ values, but only the radar prograde shape model produced a radiometric diameter that was consistent with its radar-derived diameter within the uncertainties (Table~\ref{tab:Y}). Therefore, we concluded that the prograde sense of rotation was the most likely orientation for Tantalus. The preferred solution had a thermal inertia of $270 \pm 80 \,\textrm{J}\, \textrm{m}^{-2}\, \textrm{K}^{-1}\, \textrm{s}^{-1/2}$, which was rather typical for a kilometre-sized near-Earth asteroid 
\citep[i.e., ${\sim}200 \,\textrm{J}\, \textrm{m}^{-2}\, \textrm{K}^{-1}\, \textrm{s}^{-1/2}$;][]{Delbo2007,Delbo2015}.
It also had a low roughness fraction of $0.13\pm 0.09$, equivalent to $16 \pm 7\degr$ RMS slope, which was consistent with its slightly lower than average radar circular polarisation ratio (i.e., another qualitative measure of surface roughness). Additionally, the uniform thermal lightcurves shown in 
Figures~\ref{fig:X}a and \ref{fig:X}b
indicated that no large hemispherical differences in thermal inertia or roughness were present on Tantalus’s surface. 

Our derived thermal inertia value was about half that of $670 \pm 240 \,\textrm{J}\, \textrm{m}^{-2}\, \textrm{K}^{-1}\, \textrm{s}^{-1/2}$ obtained by 
\citet{Koren2015}
who used a similar subset of WISE data. However, the rotation period of Tantalus was not well constrained at the time of their work, and thus 
\citet{Koren2015}
allowed it to vary between 2 and 24 hours in their thermophysical model. Therefore, their obtained thermal inertia value was centred on a median rotation period of 13 hours. If we scale their thermal inertia value by $(2.39/13)^{0.5}$ to ensure that the non-dimensional thermal parameter is conserved, then a corrected thermal inertia of ${\sim}290 \,\textrm{J}\, \textrm{m}^{-2}\, \textrm{K}^{-1}\, \textrm{s}^{-1/2}$ is obtained, which is consistent with our value.  

Traditionally, asteroid thermal inertia values have  been interpreted in terms of regolith grain size 
\citep[e.g.][]{Gundlach2013}
but the recent Hayabusa2 and OSIRIS-REx missions have demonstrated that rock porosity can also dictate the thermal inertia for some asteroids 
\citep{Okada2020,Rozitis2020}.
Therefore, Tantalus’s moderately low thermal inertia value could either imply the presence of mm- to cm-sized regolith grains or highly porous rocks. However, 
\citet{Cambioni2021}
suggests that the grain size interpretation is appropriate for S-type asteroids, such as Tantalus, because they are expected to produce more fine-grained regolith from impact cratering and thermal fracturing than other spectral types. Interestingly, this would imply that Tantalus’s surface is dominated by small grains despite its relatively fast spin-rate.

\section{Geophysical analysis} 
\label{sec:geo}

\begin{figure}
	\includegraphics[width=\columnwidth]{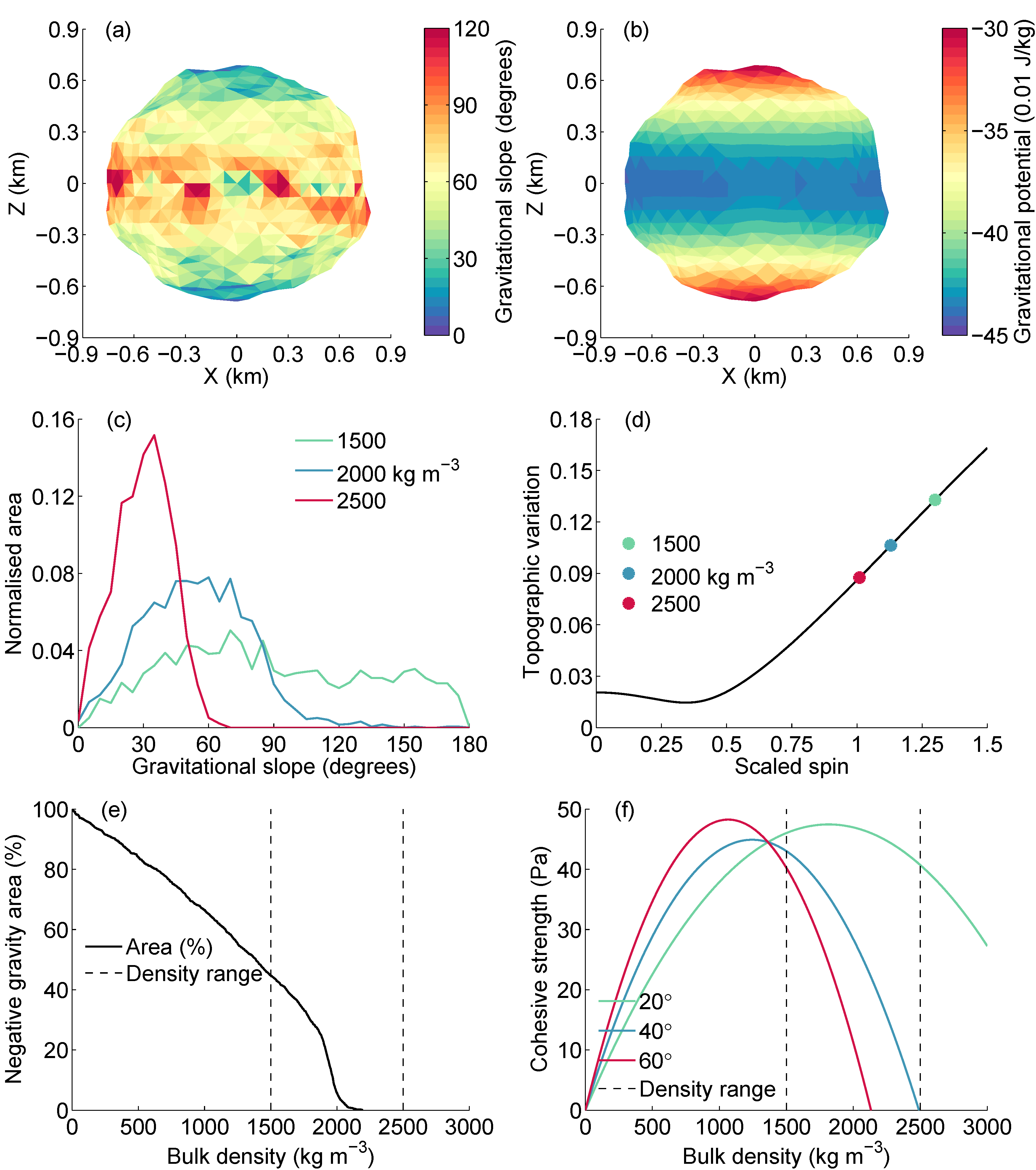}
	\caption{Geophysical analysis of Tantalus. (a) Gravitational slopes computed with the radar prograde shape model assuming a bulk density of $2000\,\textrm{kg\, m}^{-3}$. (b) Same as (a) but for gravitational potential. (c) Areal distribution of gravitational slope computed for three different values of bulk density. (d) Topographic variation in gravitational potential as a function of scaled spin, i.e., $\omega/(G\pi\rho)^{0.5}$. The current topographic variation of Tantalus is identified for three different values of bulk density. (e) Fractional area experiencing negative effective gravity as a function of bulk density. The bulk density range for an S-type rubble-pile asteroid is shown for comparison. (f) Minimum cohesive strength required to prevent the rotational breakup of Tantalus as a function of bulk density. The different lines show the cohesive strengths required for different assumed values of the angle of friction.}
	\label{fig:Y}
\end{figure}

To investigate the spin-stability of Tantalus, we applied several geophysical analyses to the prograde shape model that was preferred by the combined analysis of the radar and infrared data. In particular, we applied a polyhedron gravity field model modified for rotational centrifugal forces 
\citep{Werner1997,Rozitis2014}
to determine the gravitational slopes, gravitational potential, and topographic variation 
\citep[i.e., the relative standard deviation of gravitational potential variations across the surface of the asteroid;][]{Richardson2014,Richardson2019}
of Tantalus as a function of bulk density 
(Figure~\ref{fig:Y}).
Additionally, we also applied the Drucker-Prager failure criterion 
\citep{Holsapple2007}
to determine if structural cohesive forces are required to prevent the rotational breakup of Tantalus 
(Figure~\ref{fig:Y}f).
For these calculations, the appropriate bulk density range for Tantalus was $1500$ to $2500\,\textrm{kg}\,\textrm{m}^{-3}$ given that it is likely to be an S-type rubble-pile asteroid 
\citep{Carry2012}
which happens to be consistent with that inferred previously from the radar albedo measurements. 

As shown in 
Figures~\ref{fig:Y}a and \ref{fig:Y}b, 
there was a strong dependence of gravitational slope and potential with latitude because of Tantalus’s fast spin-rate. Some slopes on the equator are greater than $90\degr$ at the lower end of the bulk density range where centrifugal forces exceed its self-gravity 
(Figure~\ref{fig:Y}c), 
and there are also large topographic variations regardless of the precise bulk density value 
(Figure~\ref{fig:Y}d). 
{The geophysical analysis is more sensitive to the large-scale shape features rather than the small-scale topography. The small-scale features produce the small `spikes' seen in Figure~\ref{fig:Y}c, but they do not strongly influence the overall slope distributions.}
A minimum bulk density of ${\sim}2200 \,\textrm{kg}\,\textrm{m}^{-3}$ is required to prevent surface mass shedding 
(Figure~\ref{fig:Y}e), 
and a cohesive strength of up to ${\sim}45\,\textrm{Pa}$ is required to prevent structural failure depending on the assumed angle of friction (Figure~\ref{fig:Y}f). Therefore, Tantalus could be exceeding its critical spin-rate via cohesive forces like the rapidly spinning near-Earth asteroid (29075) 1950 DA 
\citep{Rozitis2014}.

If Tantalus is exceeding its critical spin-rate, then it would be expected to undergo frequent landslide and mass shedding events 
\citep{Scheeres2015}.
Although not conclusive, the possible variations in radar surface properties noted previously could be evidence of a past landslide and/or shedding event. Additionally, Tantalus has historically been classified as a Q-type asteroid 
\citep{Bus2002},
which has traditionally been interpreted as exposure of fresh un-weathered material by a recent re-surfacing event 
\citep{Binzel2010}.
Theoretical modelling of various space weathering processes has demonstrated that re-surfacing of asteroids by YORP spin-up could be a significant mechanism for producing Q-types in the near-Earth asteroid population 
\citep{Graves2018}.
However, the most recent spectrum of Tantalus indicates an Sr-type classification 
\citep{Thomas2014},
and therefore it is possible that only parts of Tantalus’s surface, if any, have undergone recent re-surfacing.

\section{Conclusions}

The asteroid (2102) Tantalus has a very symmetrical shape. Shape modelling using standard methods applied to radar data and optical light curves have shown no sign of an equatorial bulge typical for fast-spinning NEAs, but rather an almost spherical object, which is consistent with rotationally-driven evolution of a body with low effective friction angle \citep{Sugiura2021}.
The radar imaging indicates a crater-like feature close to one of the poles. The radar pole search shows two families of possible pole solutions, but with no clear preference for either based on the available optical and radar data. Including thermophysical analysis based on WISE data we were able to constrain this to prograde rotation. 

Surface properties of Tantalus are rather typical for an S-class object. Combined radar and thermophysical analysis suggests a surface covered in low-porosity fine-grained regolith. There is an indication of differentiation across the surface of Tantalus in radar spectra, but the spectrum showing the highest radar albedo and lowest $\mu_C$ (which  indicates a more solid surface) is isolated. Additional data would be needed to confirm that this is not just an anomalous measurement. The cw radar spectrum showing the lowest albedo and highest $\mu_C$, when combined with the peculiar shape of this spectrum might indicate a very localised rough radar-dark spot, like a rock-filled crater, surrounded by more reflective material. However, this conclusion is only tentative due to low SNR for this particular spectrum and inconsistency with albedo measurements made at similar locations on the surface.

Our analysis of optical light curves shows that caution should be advised when searching for signatures of period change using low-amplitude and low-SNR light curves. 
A detailed search shows a preference for slow-down of rotation for the whole available data set and excluding some of the data coming from the smaller telescopes produced a slightly better fit for a rotational spin-up. However, with the level of analysis applied here, we do not yet firmly conclude that we are seeing an actual spin-change for Tantalus. 
Close inspection of synthetic light curves generated with the radar shape model combined with the available light-curve data shows that the uncertainty in rotational phase determination might be driving the ambiguity in the spin-rate change estimates. Additionally, our modelling indicated retrograde rotation from lightcurve inversion alone. However, to reach an agreement between radar and thermophysical effective diameter determination, prograde rotation was preferred. Therefore, {care should be taken when} determining pole orientation for symmetrical objects from light curves alone.

Earlier photometric studies suggested a presence of a $16\,\rm{h}$-rotation-period satellite \citep{Warner2015,Warner2017,Vaduvescu2017}. That periodicity is not apparent in the full lightcurve data set. Furthermore,
the radar data show no sign of a companion larger than $75\,\rm{m}$. With a primary diameter of $1.3\,\rm{km}$, the difference in photometric signal from a secondary of diameter $75\,\rm{m}$ would be 0.0036 magnitudes. This cannot explain the light curve effects observed.

The radar-derived shape models of Tantalus are consistent with radar observations, however all three presented models  
demonstrate issues with fits to the optical data. When it comes to the light-curve-inversion model this might be due to the convex model's limitations; from radar observations we know that there is a crater-like feature on the surface that would not be reproduced by convex inversion. Meanwhile, the radar modelling procedure can over-fit for the noise in radar data, producing bump-like artefacts in the 3D model. These might cast shadows producing  mismatch with optical light curves. Some of the light-curve effects might be also due to albedo variegation, which is possible given the variation suggested by analysis of radar-derived surface properties, but not very well constrained with presented data.

 Tantalus makes a much closer approach to Earth in December 2038, coming at about $0.044\,\textrm{AU}$, as compared to the December 2016 approach when its minimum distance from Earth was $0.137\,\textrm{AU}$. It would be useful to obtain additional measurements of surface properties using planetary radar to confirm any surface variation during that approach. A smaller geocentric distance means a higher SNR for the radar echo and would provide an opportunity to investigate the surface topography in more detail, for example to confirm the presence of any surface craters. Sadly, the Arecibo telescope used to obtain results presented here collapsed on 1st December 2020, considerably reducing observational capabilities in the planetary radar domain. Hopefully, the Goldstone Solar System Radar facility along with alternative instrumentation (to be commissioned) can be used for continued detailed characterisation of NEAs.

\section*{Acknowledgements}

We thank the anonymous referee for their detailed comments.
We thank all the staff at the observatories involved in this study for their support. %
AR, SCL, BR, LRD, SFG, CS and AF acknowledge support from the UK Science and Technology Facilities Council. %
Based in part on observations collected at the European Organisation for Astronomical Research in the Southern Hemisphere under ESO programme 185.C-1033.  
Based in part on the Arecibo Planetary Radar observations collected under programme R3037.
During the time of the radar observations, The Arecibo Planetary Radar Program was fully supported by NASA’s Near-Earth Object Observations Program through grants no. NNX12AF24G and NNX13AF46G awarded to Universities Space Research Association (USRA). The Arecibo Observatory is an NSF facility.
This publication uses data products from WISE, a project of the Jet Propulsion Laboratory/California Institute of Technology, funded by the Planetary Science Division of NASA. We made use of the NASA/IPAC Infrared Science Archive, which is operated by the Jet Propulsion Laboratory/California Institute of Technology under a contract with NASA.
This research has received support from the National Research Foundation (NRF; 2019R1I1A1A01059609), 
the European Union  H2020-SPACE-2018-2020 research and innovation programme under grant agreement No. 870403 (NEOROCKS), 
the European Union H2020-MSCA-ITN-2019 under grant No. 860470, 
and the Novo Nordisk Foundation
Interdisciplinary Synergy Programme grant no. NNF19OC0057374.
For the purpose of open access, the author has applied a Creative Commons Attribution (CC BY) licence to any Author Accepted Manuscript version arising from this submission.

\section*{Data Availability}


The light-curve data published by \citet{Pravec1997b} are publicly available from NASA Planetary Data System at \url{https://sbnapps.psi.edu/ferret/datasetDetail.action?dataSetId=EAR-A-3-RDR-NEO-light-curveS-V1.1}. 
The light-curve data published by \citet{Warner2015} and \citet{Warner2017} are publicly available at the Asteroid light-curve Photometry Database \citep{Warner2009}, accessible through \url{http://alcdef.org/}. 
The light curve data first published here can be accessed in an on-line Table~{A1} available at the CDS via anonymous ftp to cdsarc.u-strasbg.fr (130.79.128.5) or via \url{http://cdsarc.u-strasbg.fr/viz-bin/qcat?J/MNRAS/vol/page}.
The shape models discussed here can be downloaded from the Database of Asteroid Models from Inversion Techniques (DAMIT) accessible through \url{https://astro.troja.mff.cuni.cz/projects/damit/}.
The radar CW data is available online at  \url{https://www.lpi.usra.edu/resources/asteroids/asteroid/?asteroid_id=1975YA}
and the radar imaging data is available from the authors upon request.



\bibliographystyle{mnras}
\bibliography{ARbib} 



\newpage

\appendix

\section{Additional figures}



\begin{figure*}
	\resizebox{.99\hsize}{!}{
		\includegraphics[width=.48\textwidth, trim=0.8cm 0cm 0.2cm 0cm, clip=true]{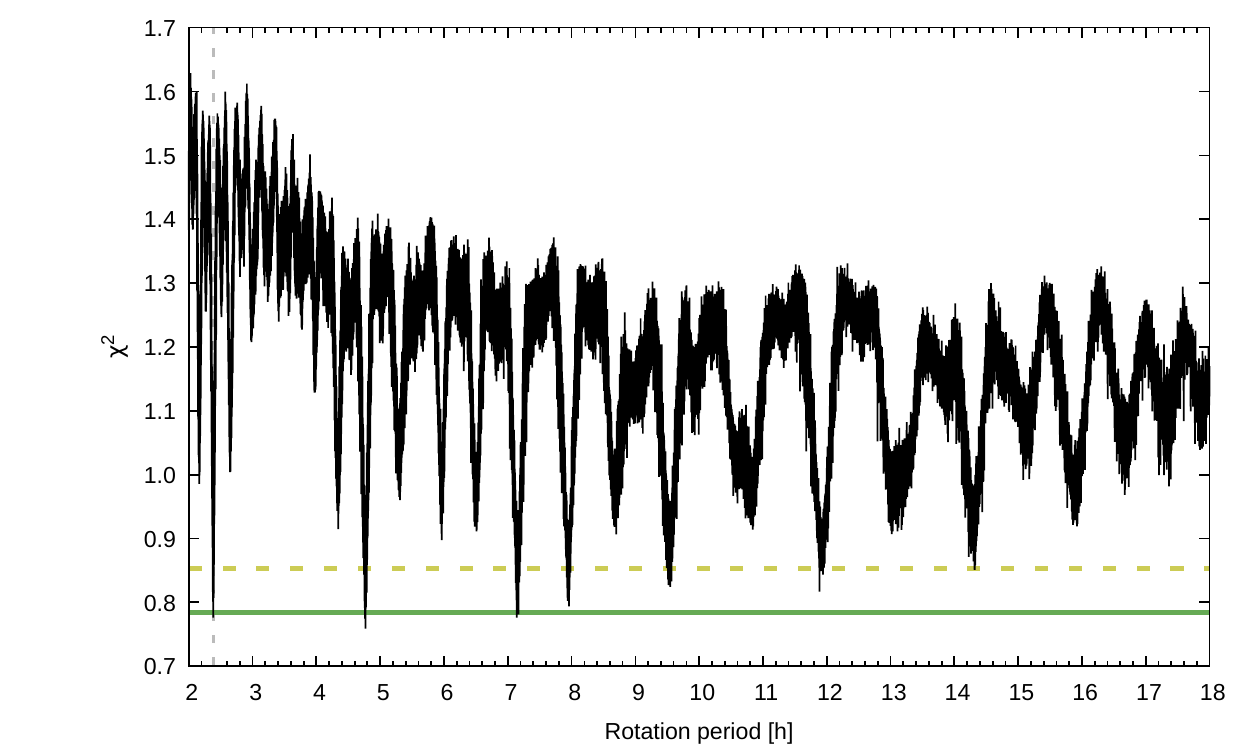} 
	}

	\caption{Periodogram obtained using convex light-curve inversion with a wide search range ($2-18\,\rm{h}$). All available light curves with the vertical dashed line marking $P = 2.385455\, \textrm{h}$ corresponding to the adopted period,  the continuous horizontal line indicating 1\% increase in $\chi^2$ value above the value for that period, and dashed line -- $10\%$ increase. 
		\label{fig:periodogram}
	}
\end{figure*}

\begin{figure*}
		\resizebox{\hsize}{!}{
		\includegraphics[width=.48\textwidth, trim=2cm 4cm 3.8cm 4cm, clip=true]{LC/2102_210_-30_0_v190906_20190911_initial_spinstate_01_fix.pdf} 
		\includegraphics[width=.48\textwidth, trim=2cm 4cm 3.8cm 4cm, clip=true]{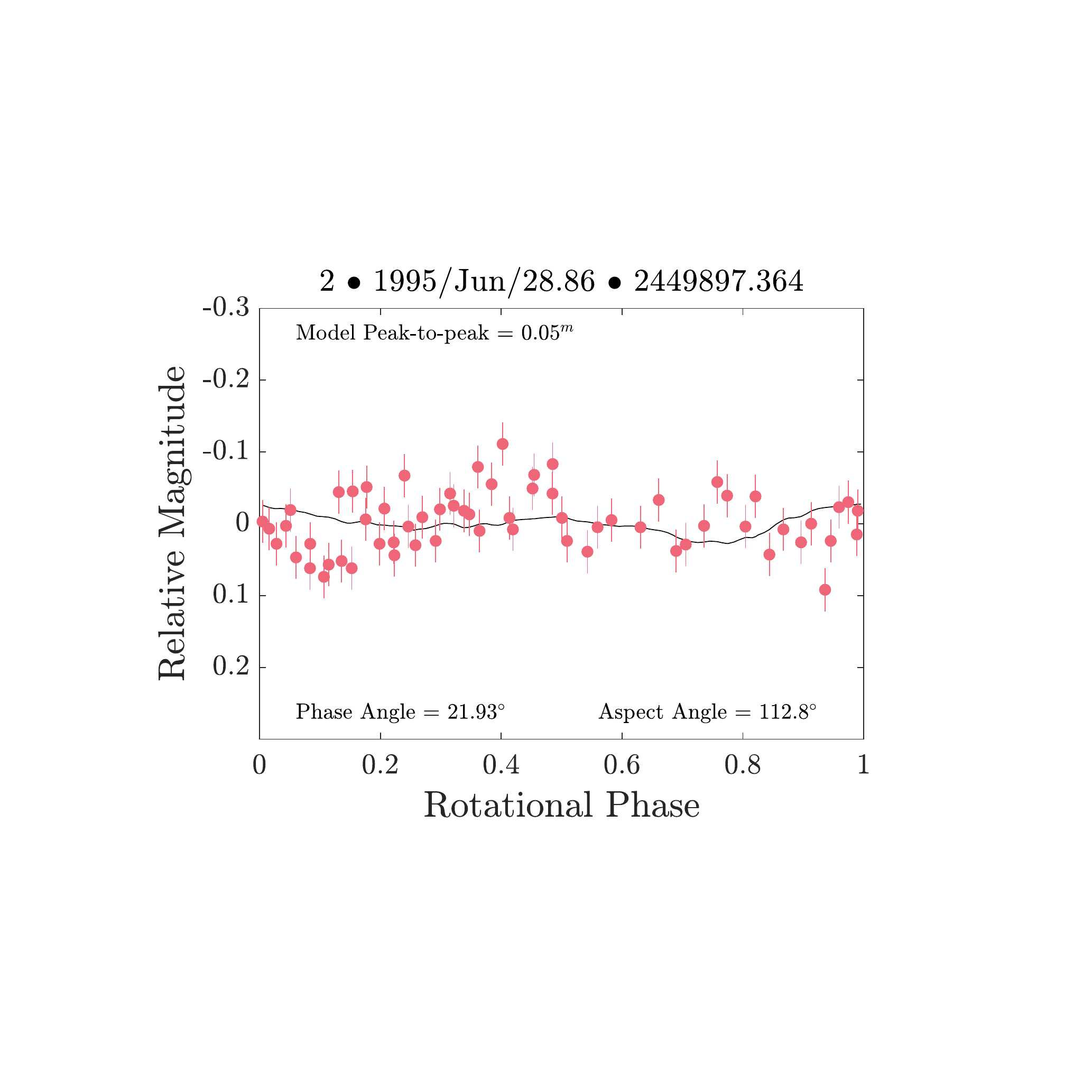} 		
		\includegraphics[width=.48\textwidth, trim=2cm 4cm 3.8cm 4cm, clip=true]{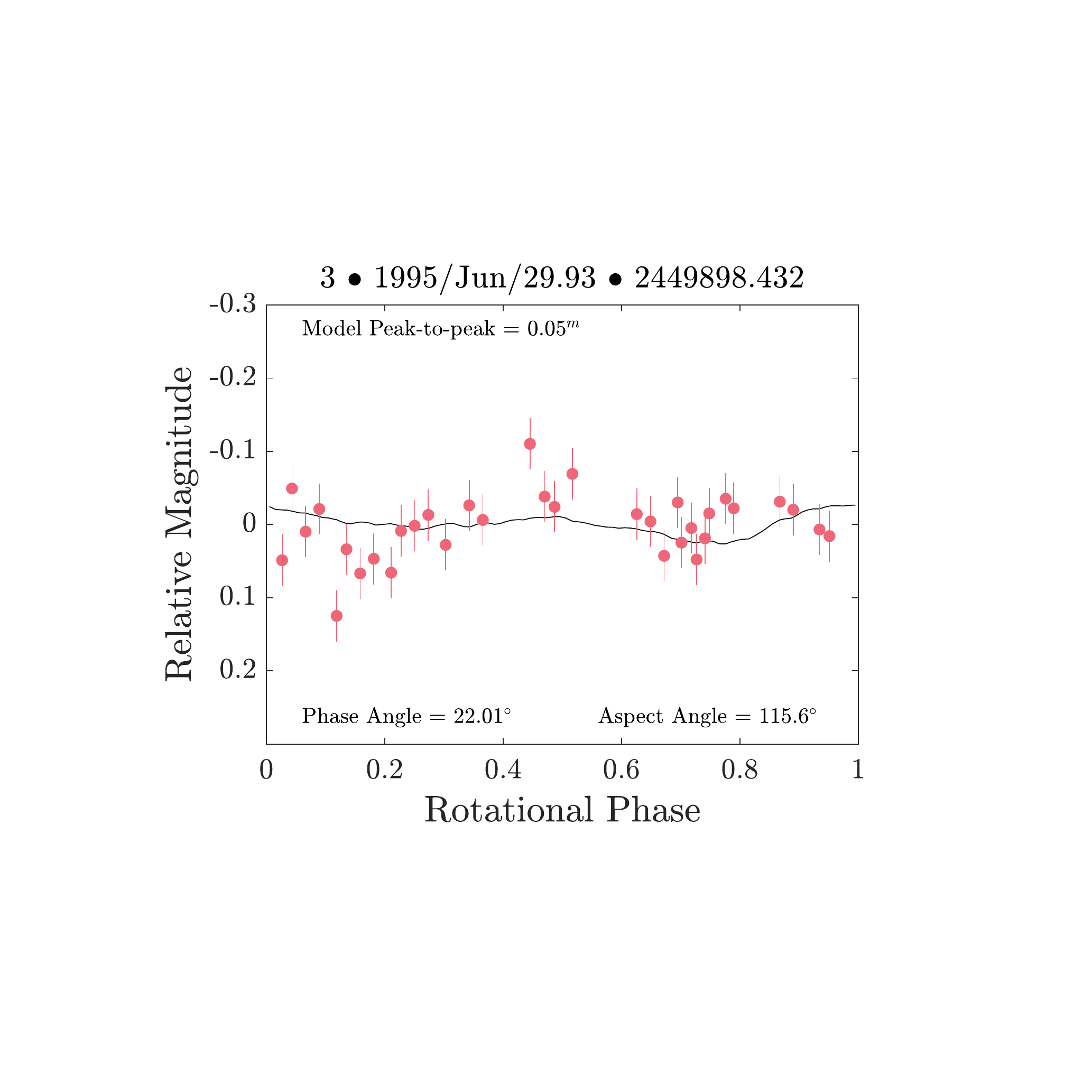} 
	}
	
		\resizebox{\hsize}{!}{
		\includegraphics[width=.48\textwidth, trim=2cm 4cm 3.8cm 4cm, clip=true]{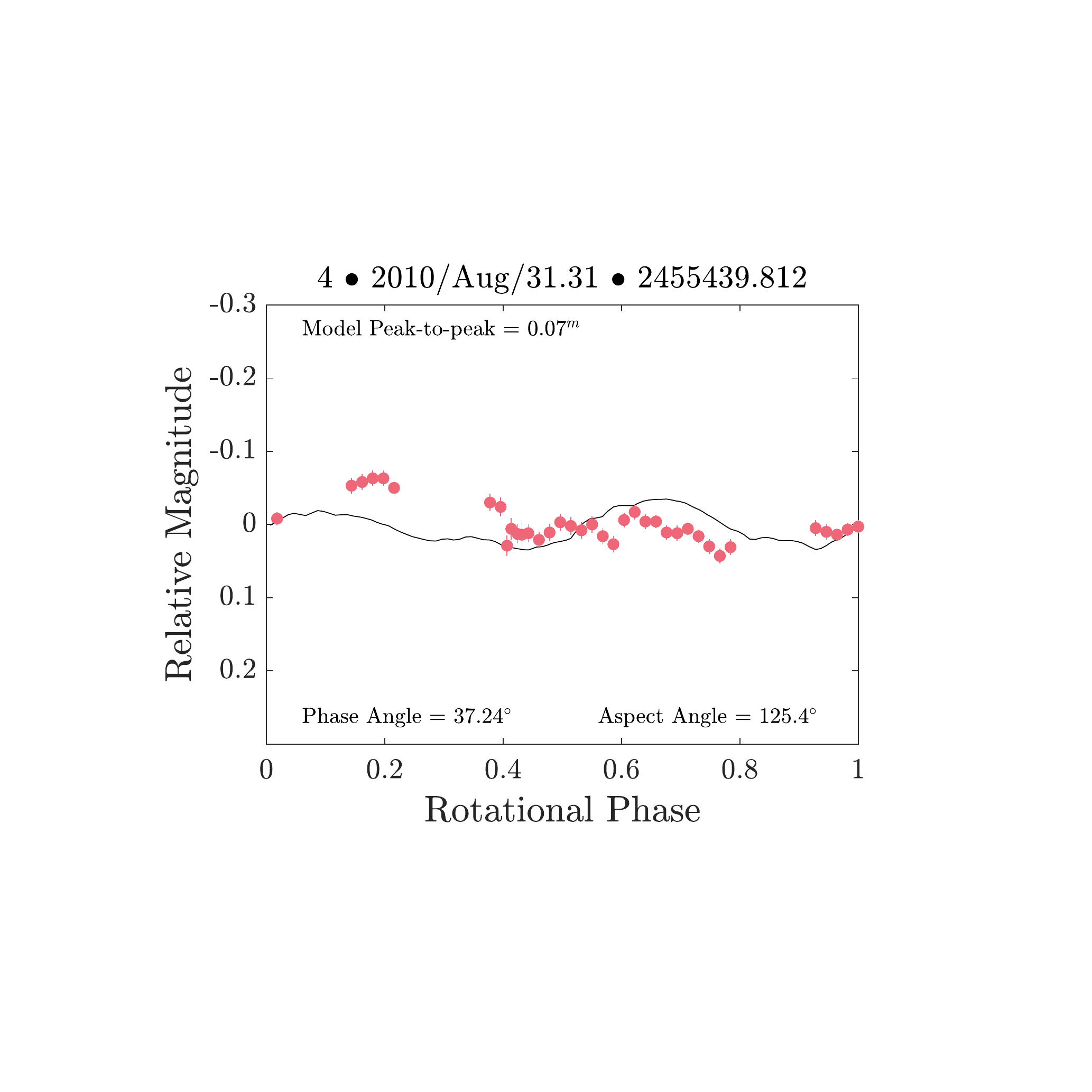} 
		\includegraphics[width=.48\textwidth, trim=2cm 4cm 3.8cm 4cm, clip=true]{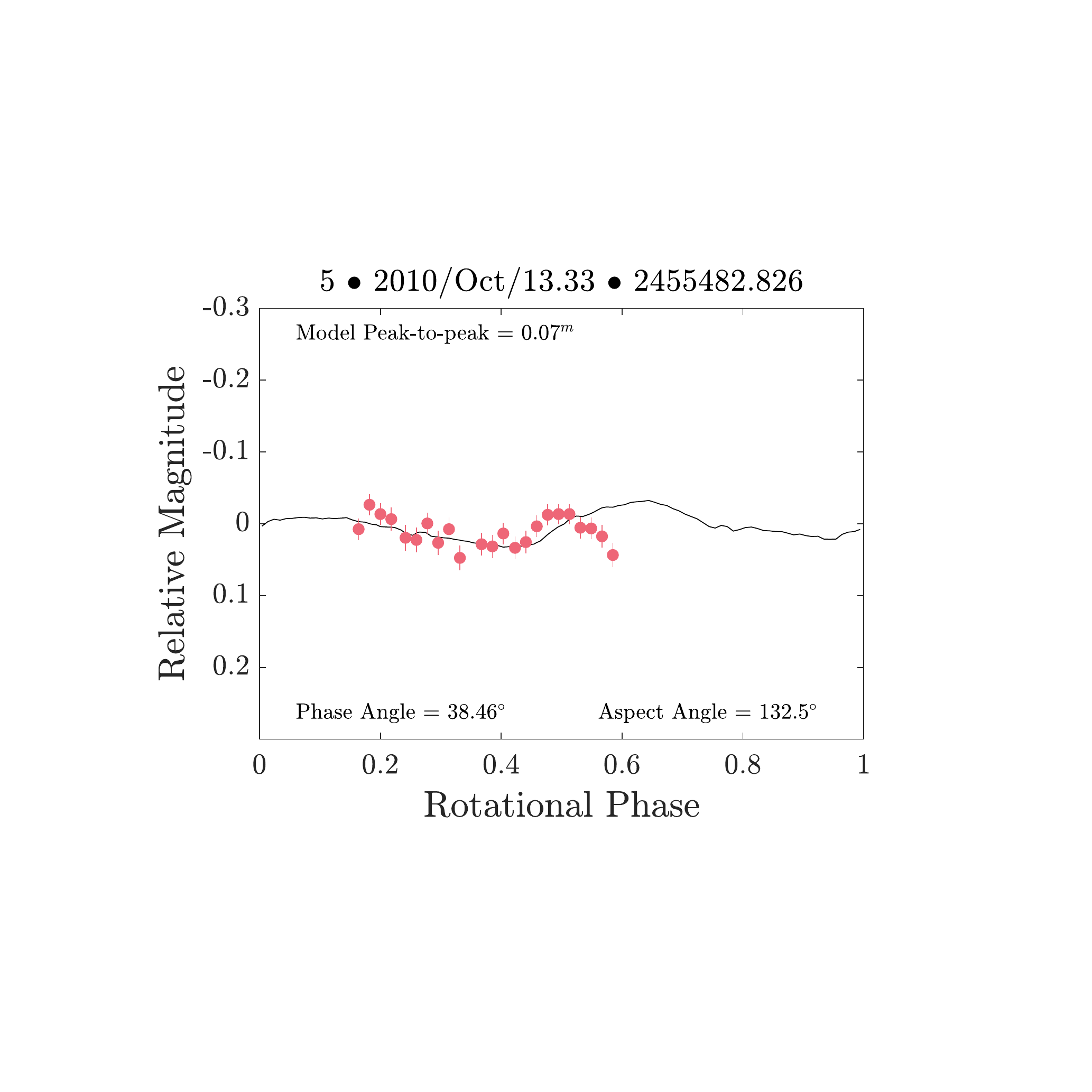} 		
		\includegraphics[width=.48\textwidth, trim=2cm 4cm 3.8cm 4cm, clip=true]{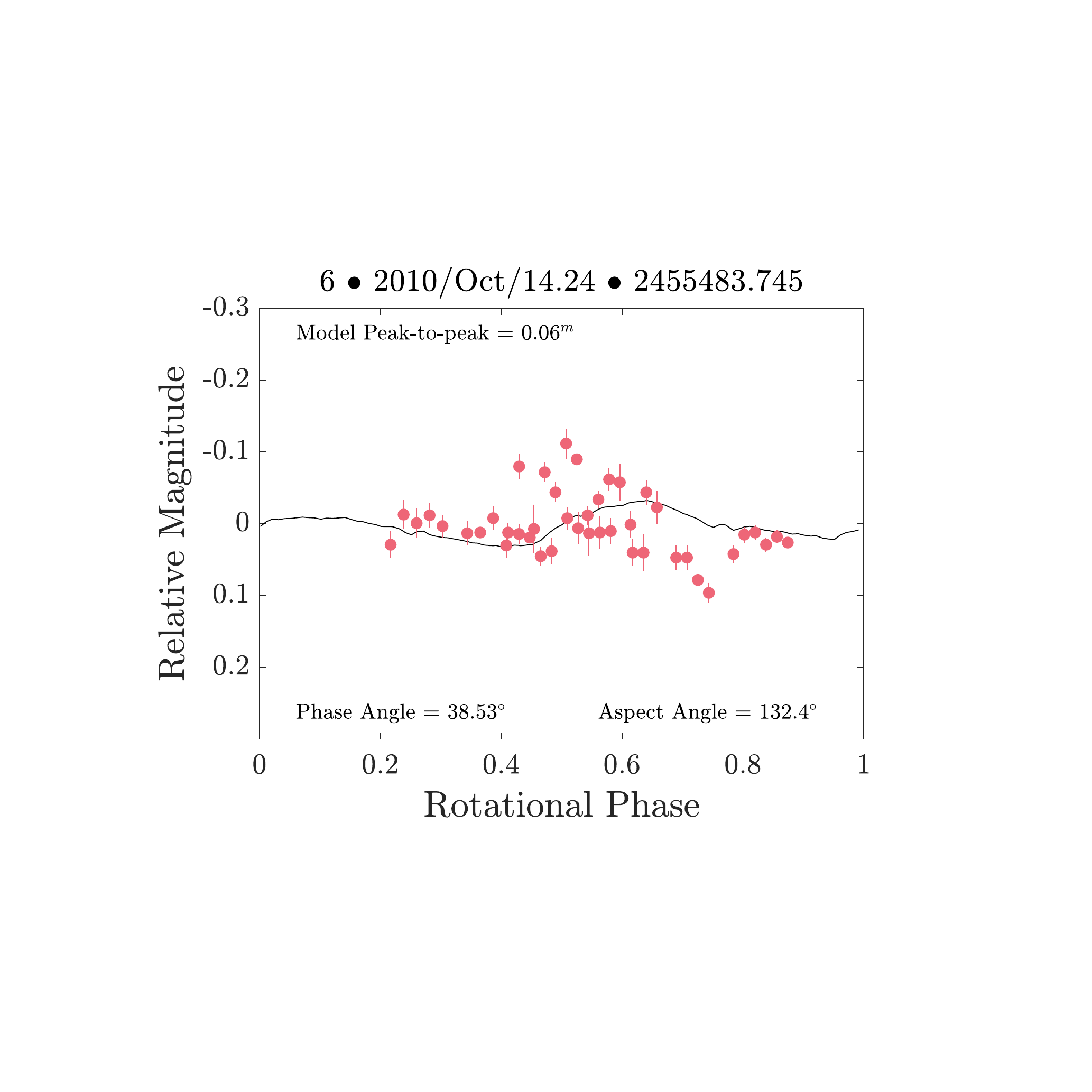} 
	}

	\resizebox{\hsize}{!}{
	\includegraphics[width=.48\textwidth, trim=2cm 4cm 3.8cm 4cm, clip=true]{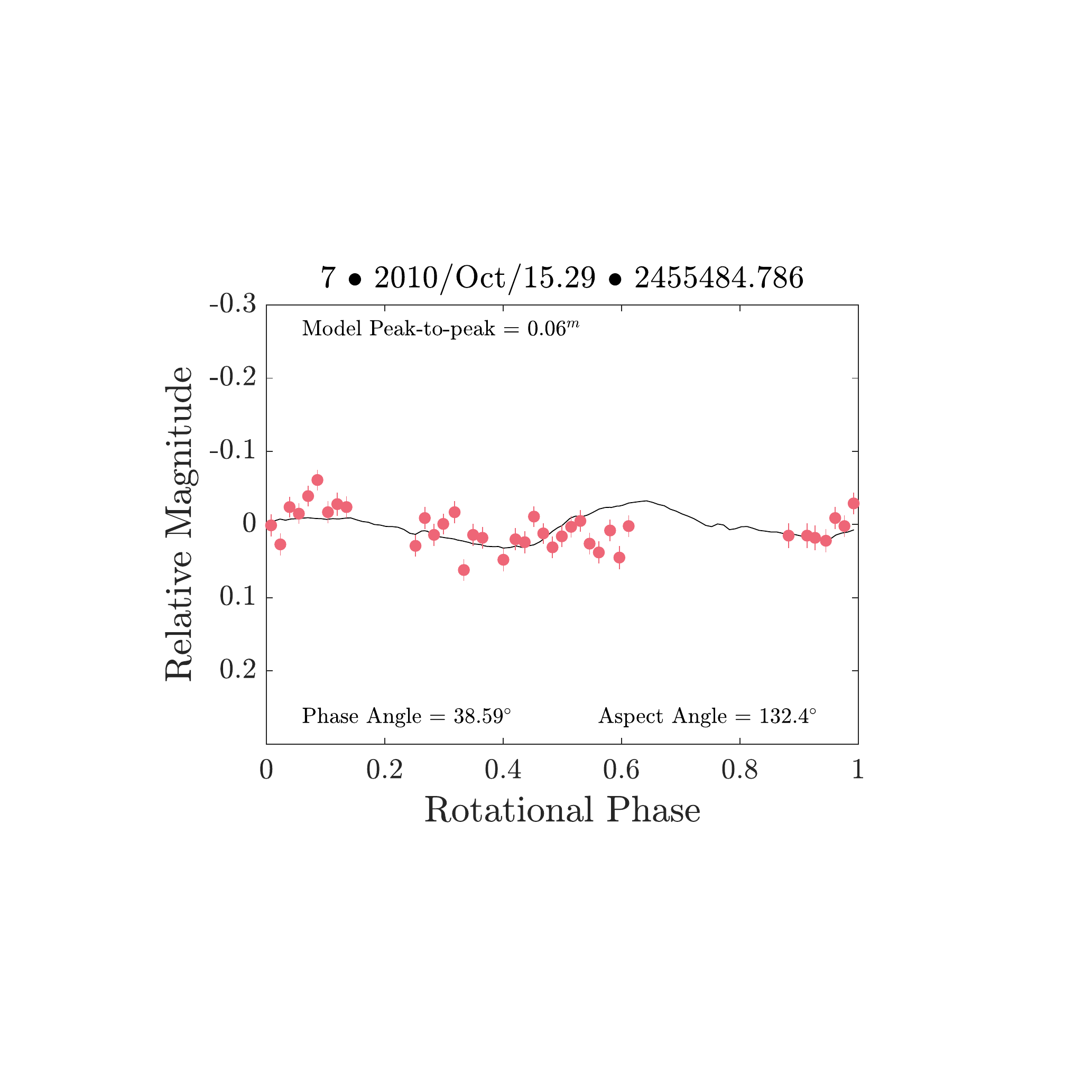} 
	\includegraphics[width=.48\textwidth, trim=2cm 4cm 3.8cm 4cm, clip=true]{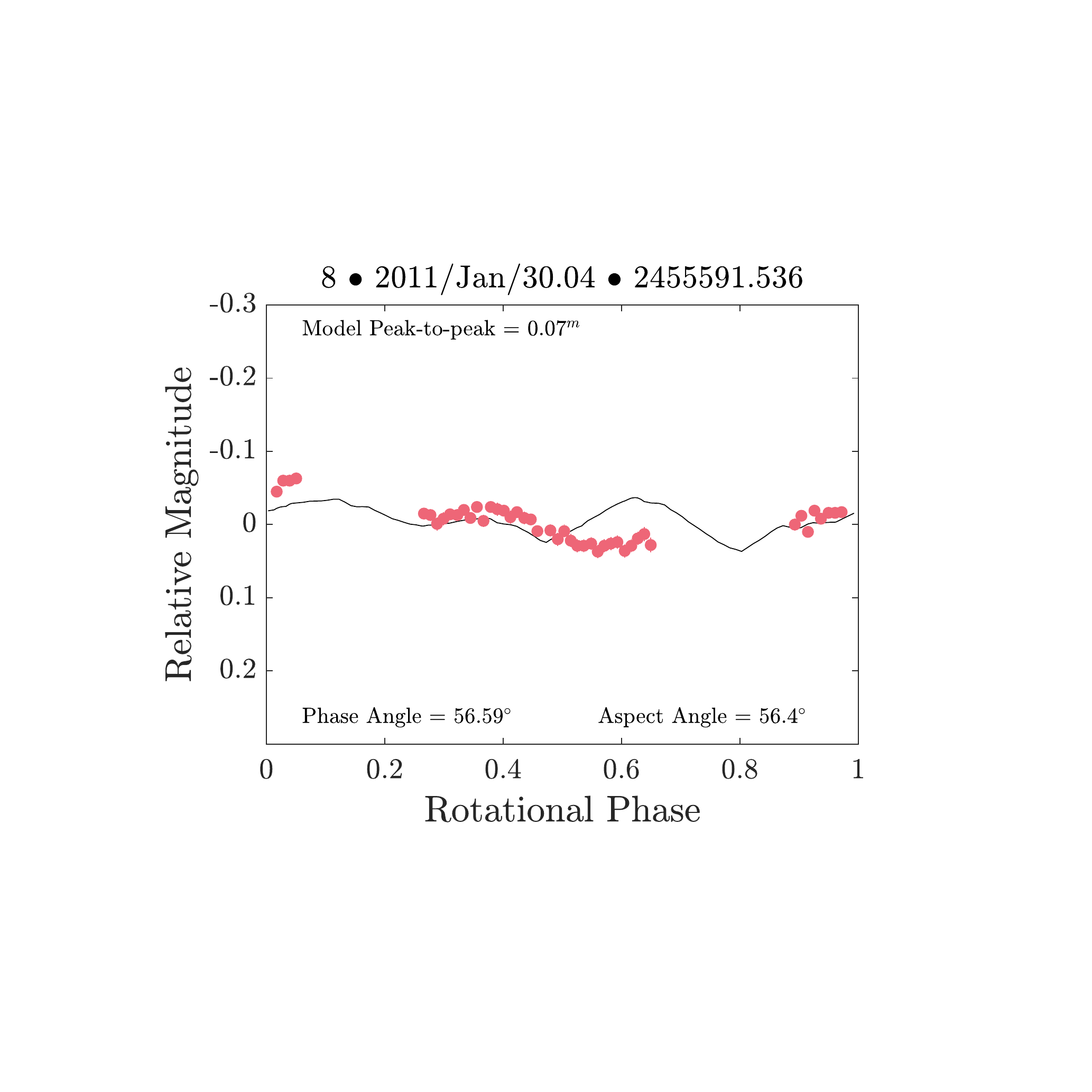} 		
	\includegraphics[width=.48\textwidth, trim=2cm 4cm 3.8cm 4cm, clip=true]{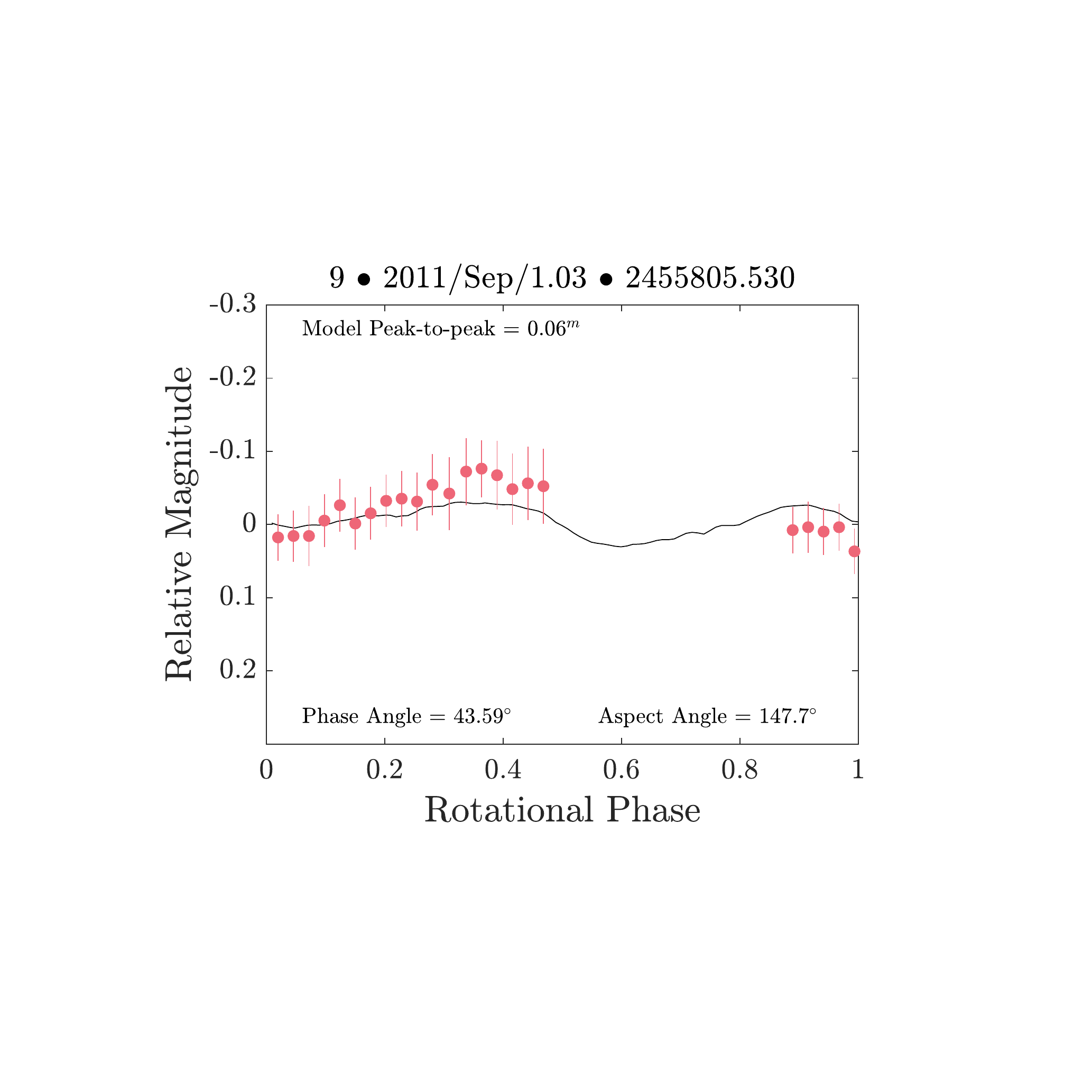} 
}

	\resizebox{\hsize}{!}{
		\includegraphics[width=.48\textwidth, trim=2cm 4cm 3.8cm 4cm, clip=true]{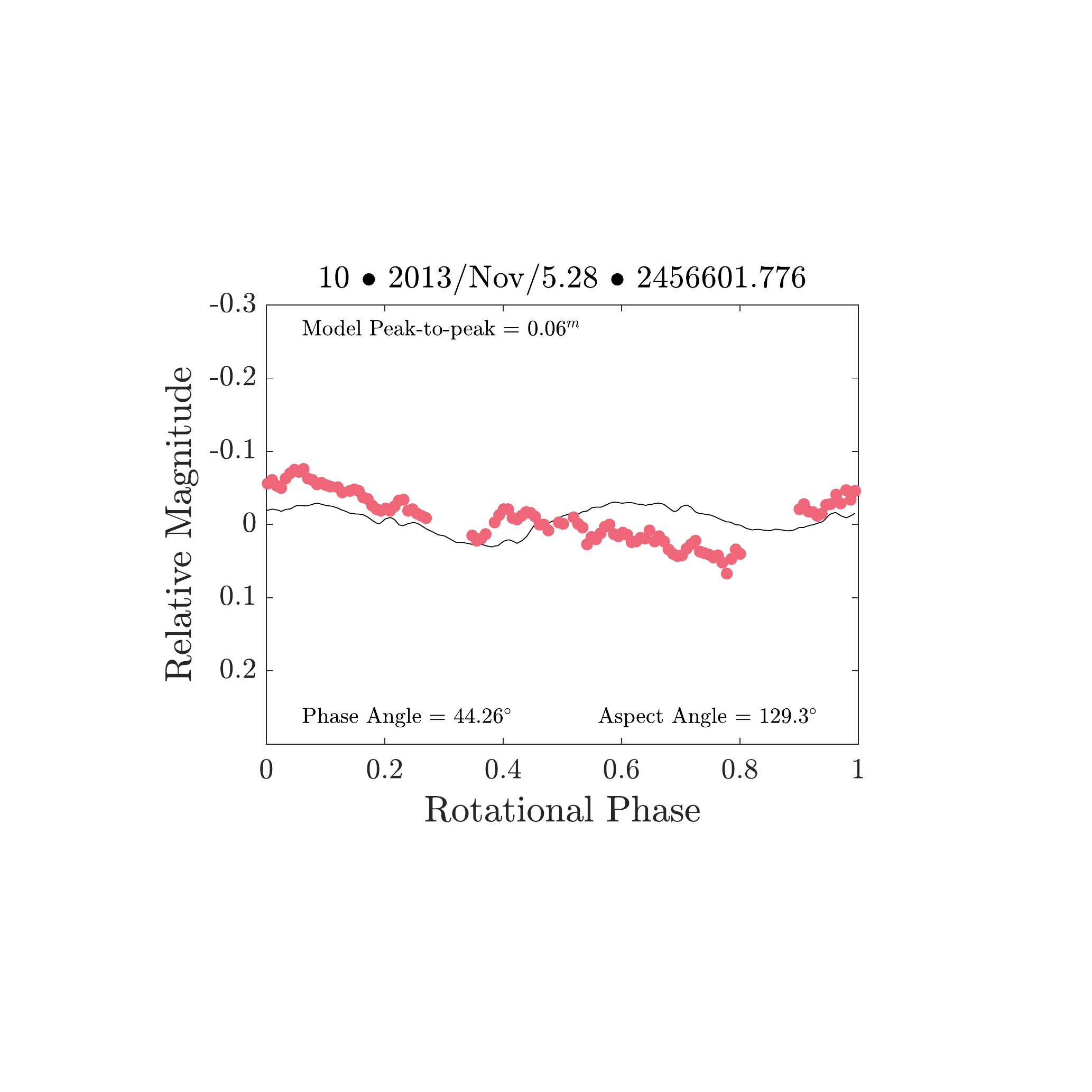} 
		\includegraphics[width=.48\textwidth, trim=2cm 4cm 3.8cm 4cm, clip=true]{LC/2102_210_-30_0_v190906_20190911_initial_spinstate_11_fix.pdf} 		
		\includegraphics[width=.48\textwidth, trim=2cm 4cm 3.8cm 4cm, clip=true]{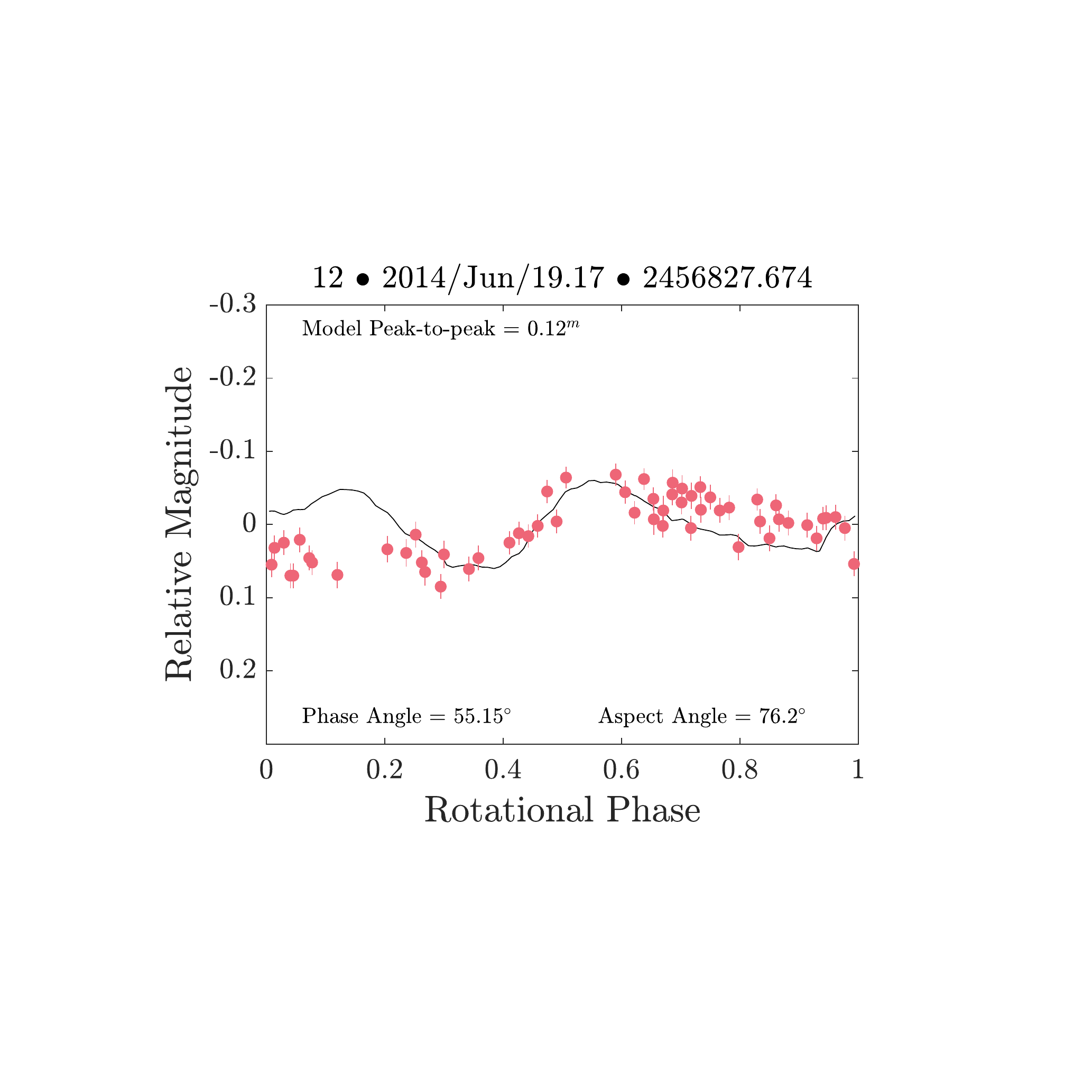} 
	}

	\caption{All available data plotted over synthetic light curves generated with the convex-inversion shape model of asteroid (2102) Tantalus. This shape model was generated using all available light-curve data  listed in Table~\ref{tab:obs}. 
		\label{fig:conv-lcfit1}}
\end{figure*}

\addtocounter{figure}{-1}

\begin{figure*}
	\resizebox{\hsize}{!}{
		\includegraphics[width=.48\textwidth, trim=2cm 4cm 3.8cm 4cm, clip=true]{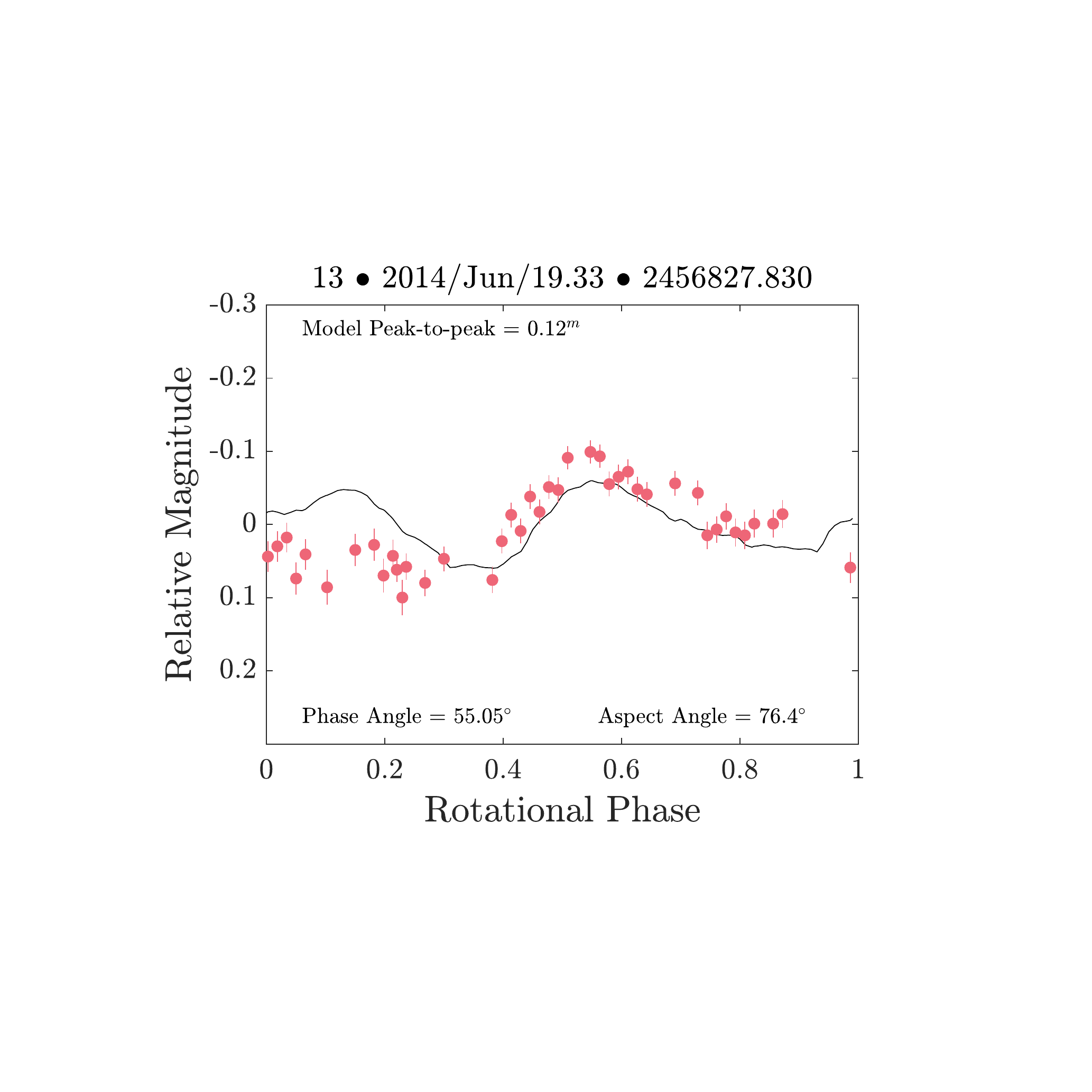} 
		\includegraphics[width=.48\textwidth, trim=2cm 4cm 3.8cm 4cm, clip=true]{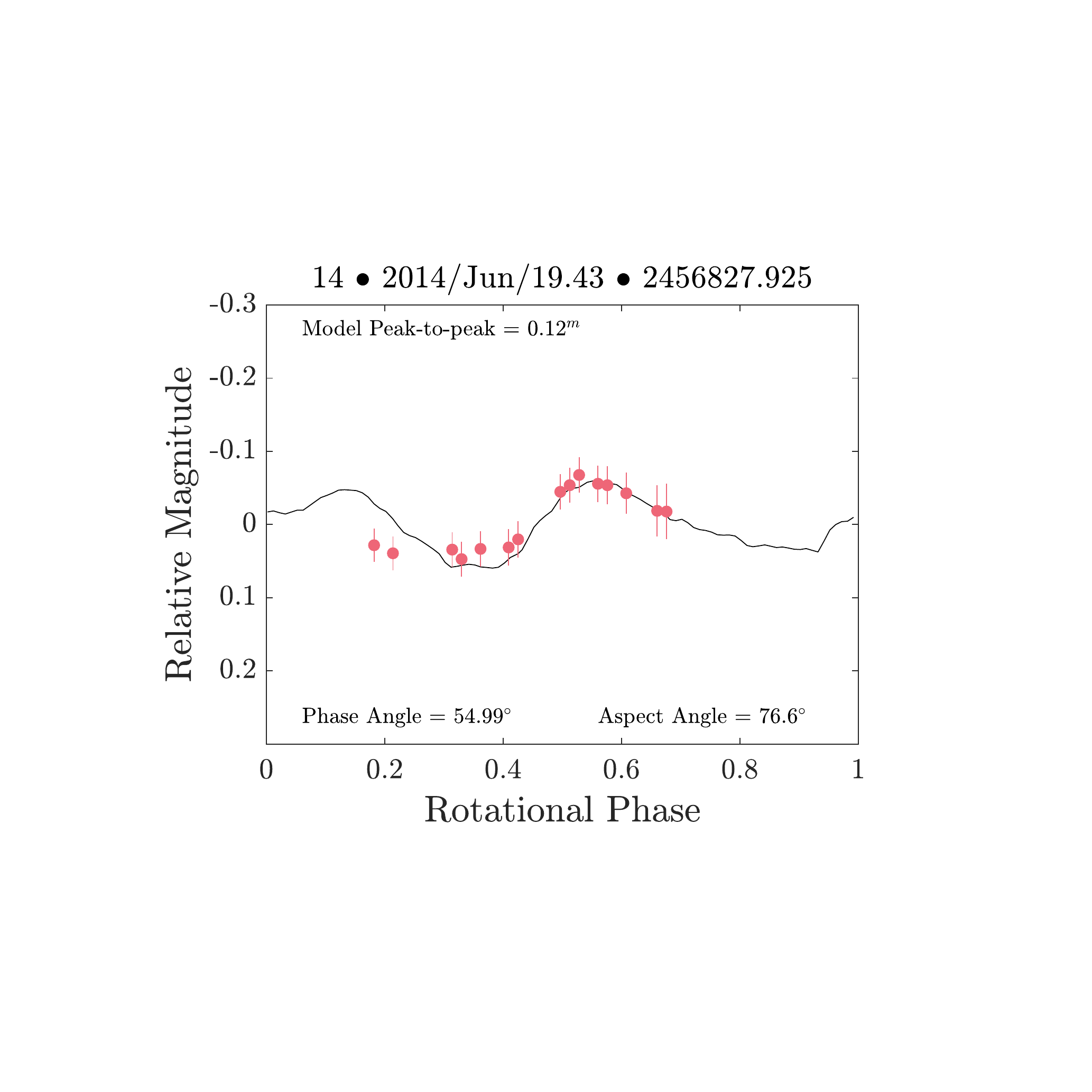} 		
		\includegraphics[width=.48\textwidth, trim=2cm 4cm 3.8cm 4cm, clip=true]{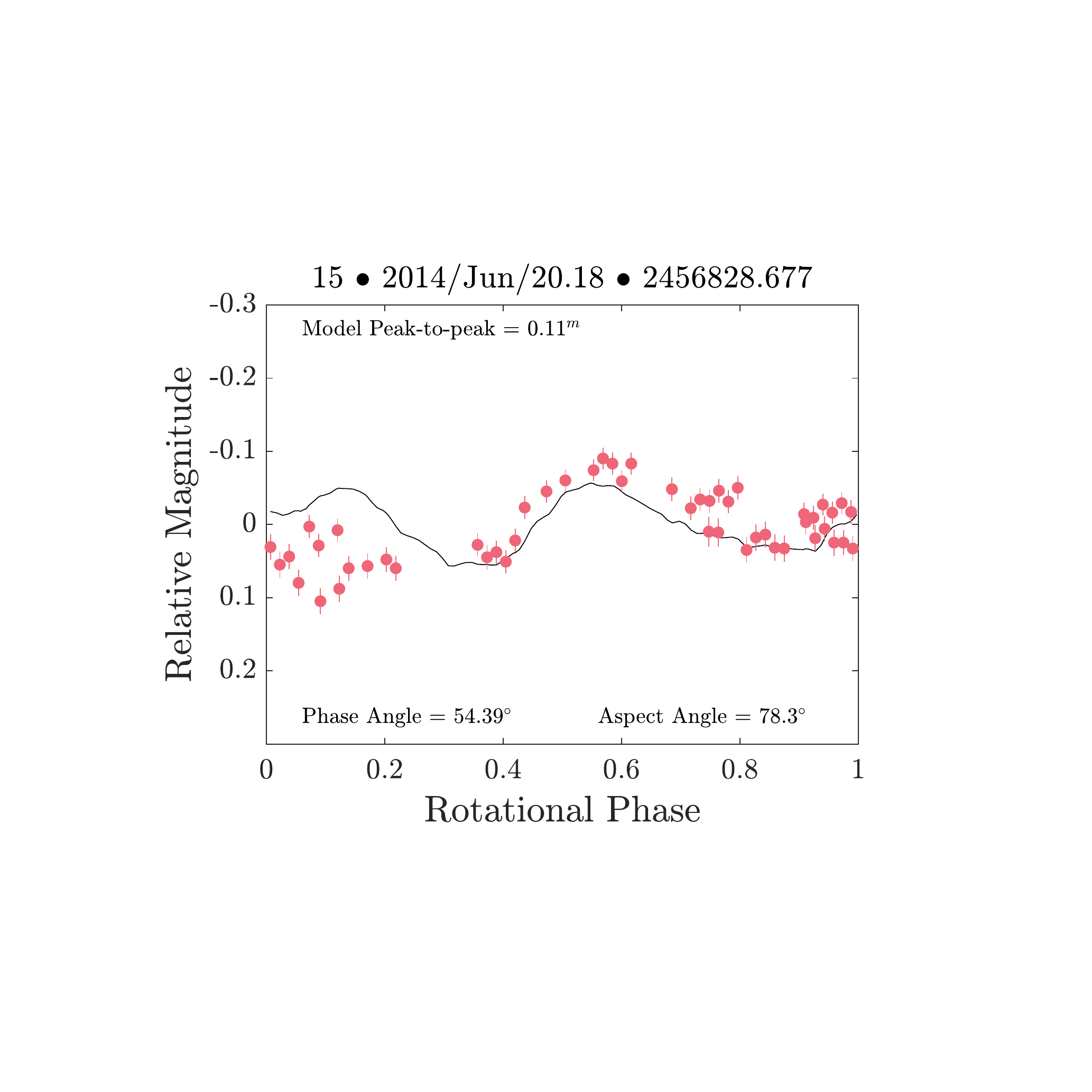} 
	}
	
	\resizebox{\hsize}{!}{
		\includegraphics[width=.48\textwidth, trim=2cm 4cm 3.8cm 4cm, clip=true]{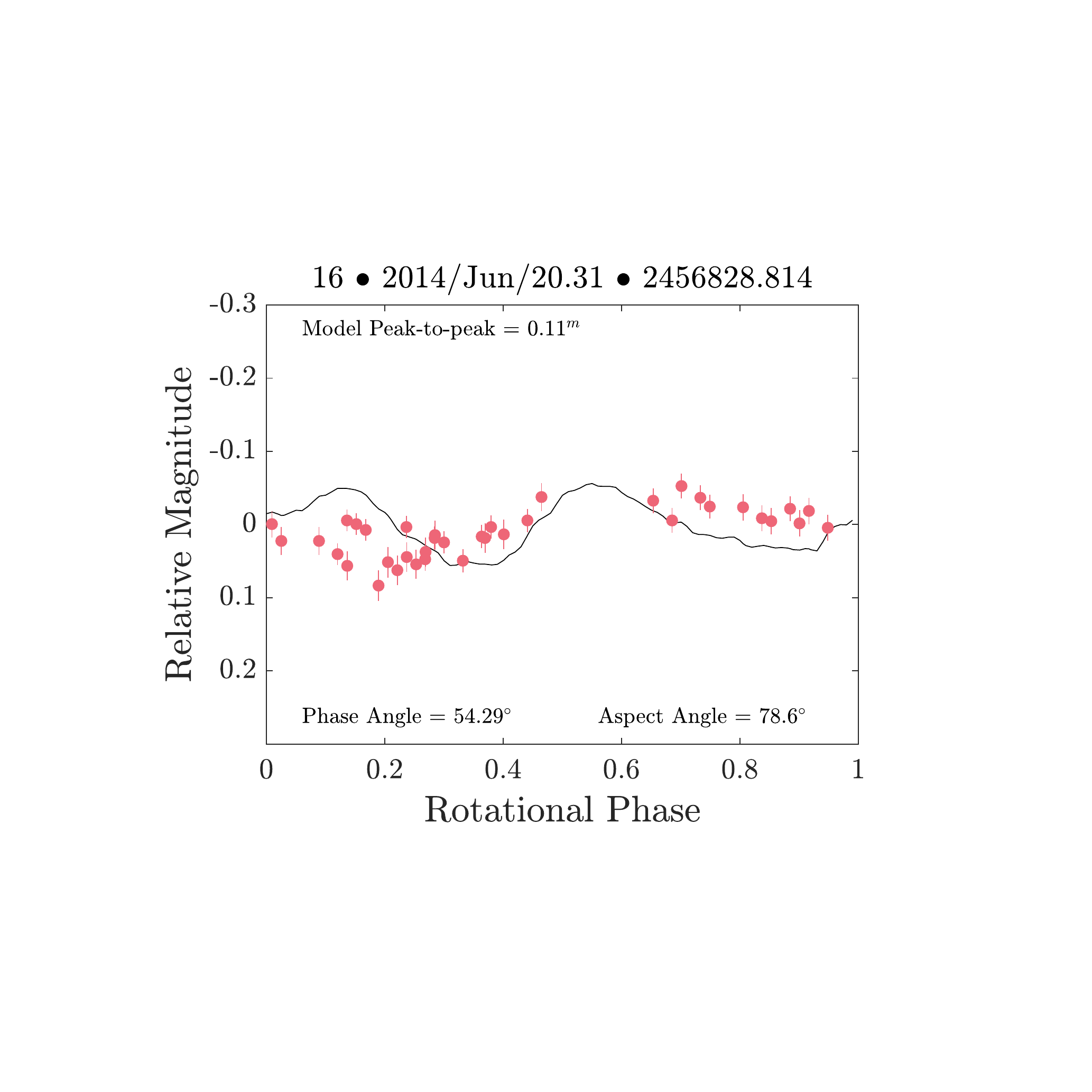} 
		\includegraphics[width=.48\textwidth, trim=2cm 4cm 3.8cm 4cm, clip=true]{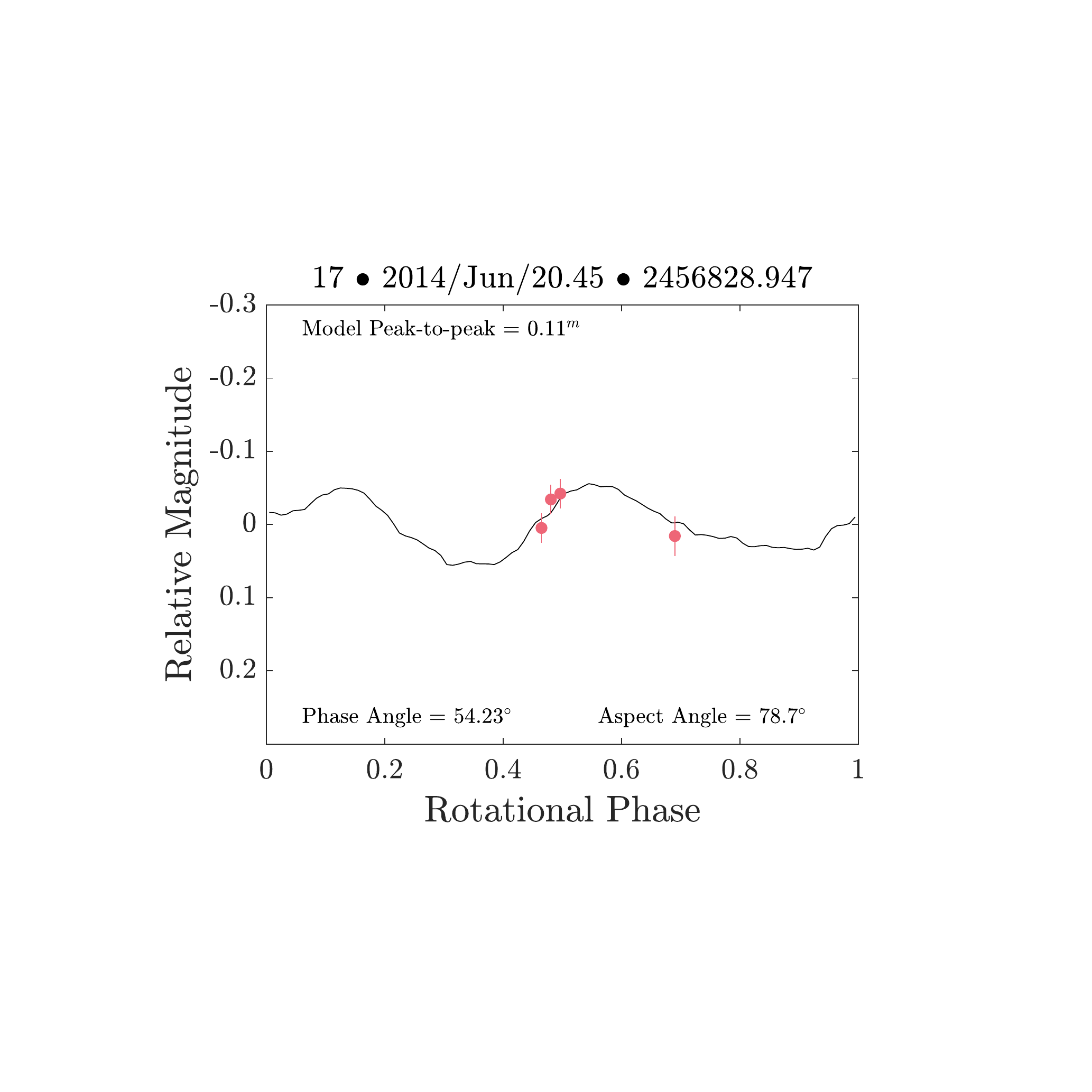} 		
		\includegraphics[width=.48\textwidth, trim=2cm 4cm 3.8cm 4cm, clip=true]{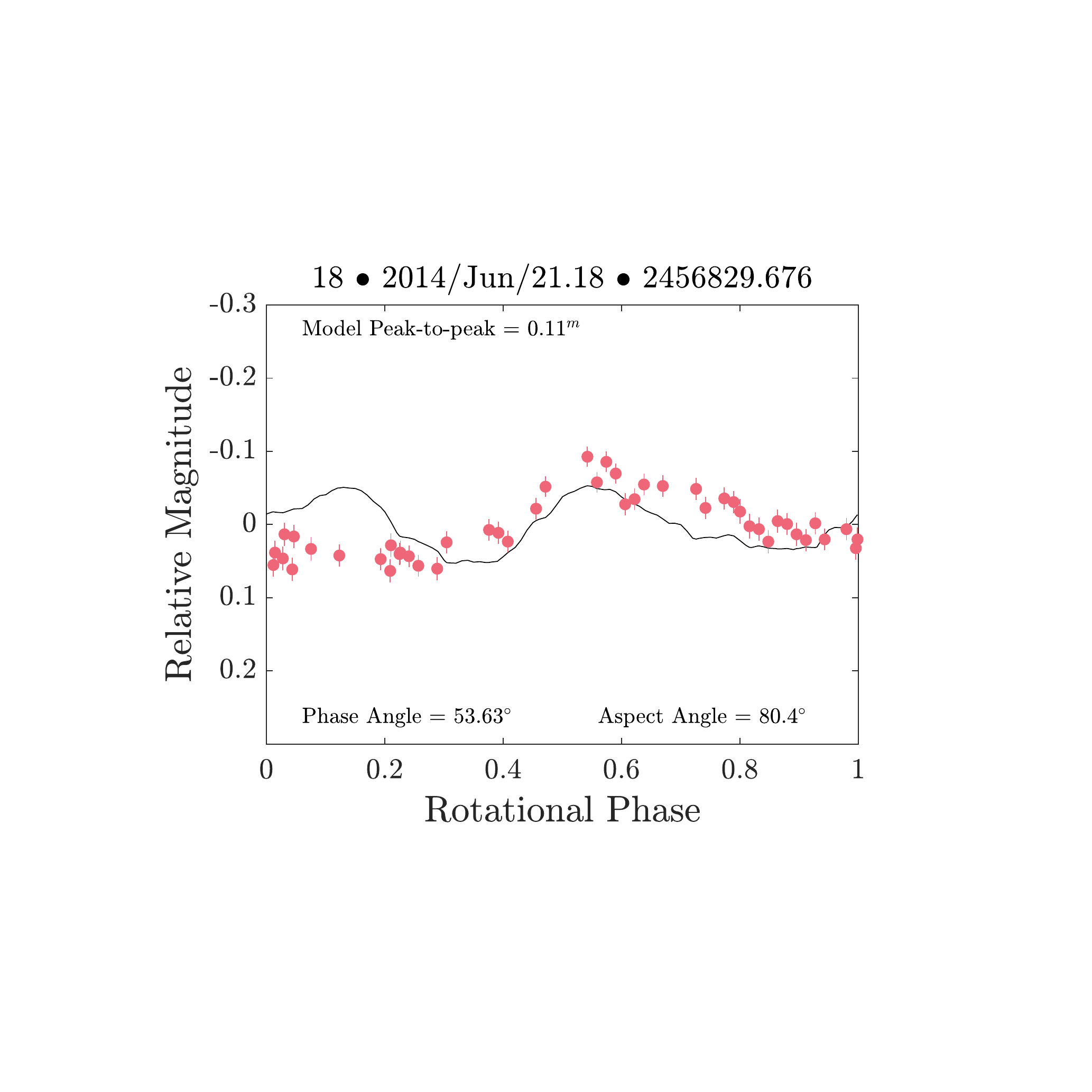} 
	}
	
	\resizebox{\hsize}{!}{
		\includegraphics[width=.48\textwidth, trim=2cm 4cm 3.8cm 4cm, clip=true]{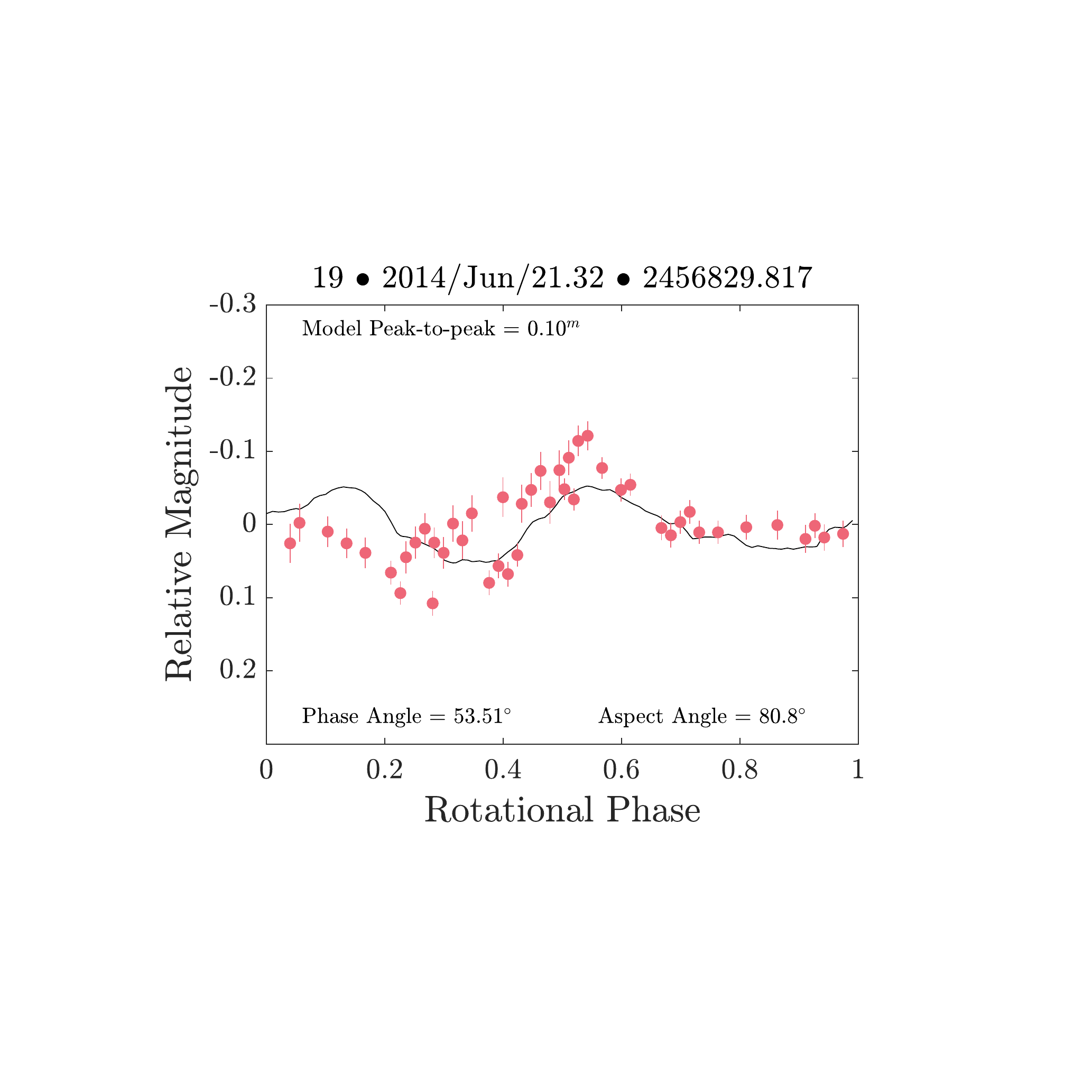} 
		\includegraphics[width=.48\textwidth, trim=2cm 4cm 3.8cm 4cm, clip=true]{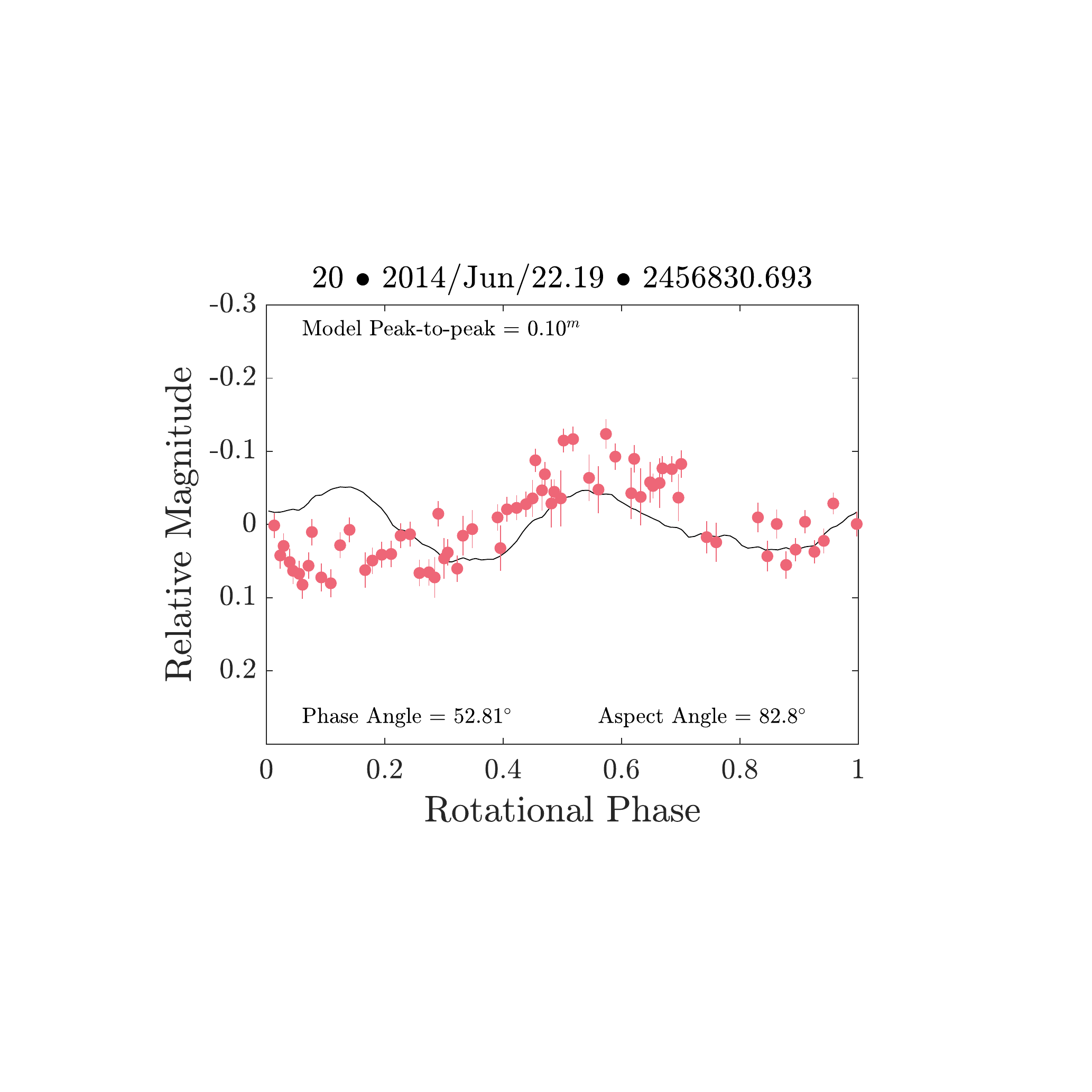} 		
		\includegraphics[width=.48\textwidth, trim=2cm 4cm 3.8cm 4cm, clip=true]{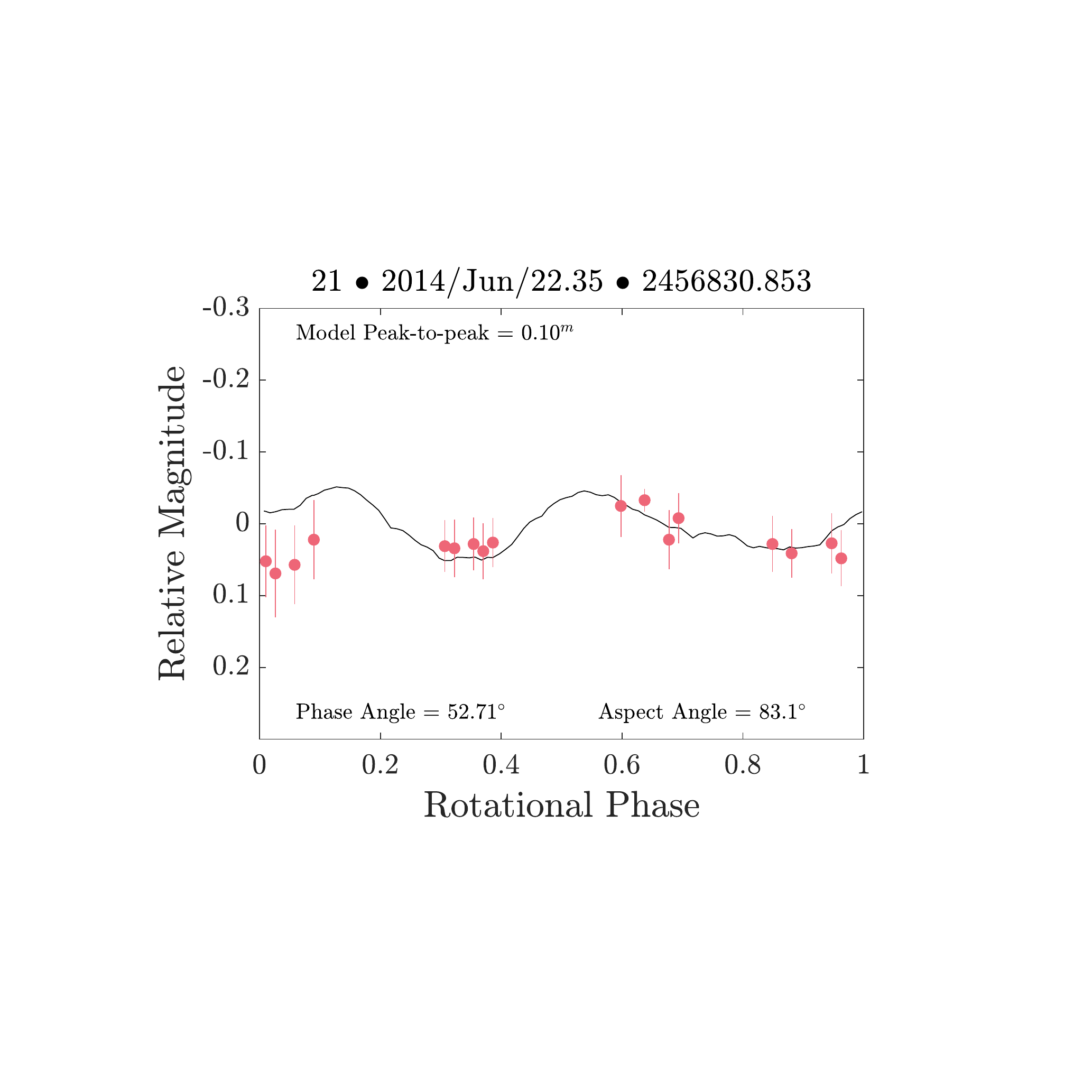} 
	}

	\resizebox{\hsize}{!}{
		\includegraphics[width=.48\textwidth, trim=2cm 4cm 3.8cm 4cm, clip=true]{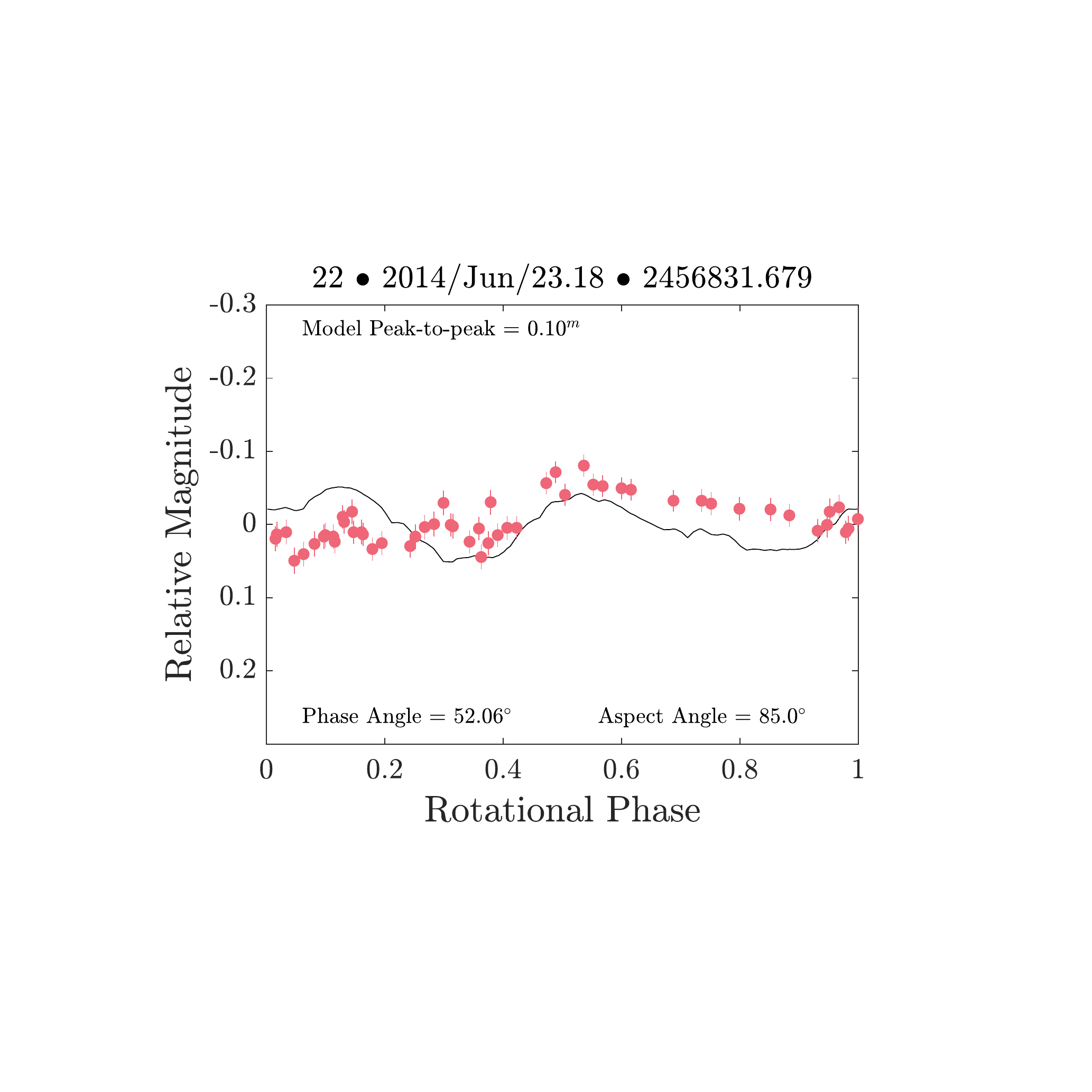} 
		\includegraphics[width=.48\textwidth, trim=2cm 4cm 3.8cm 4cm, clip=true]{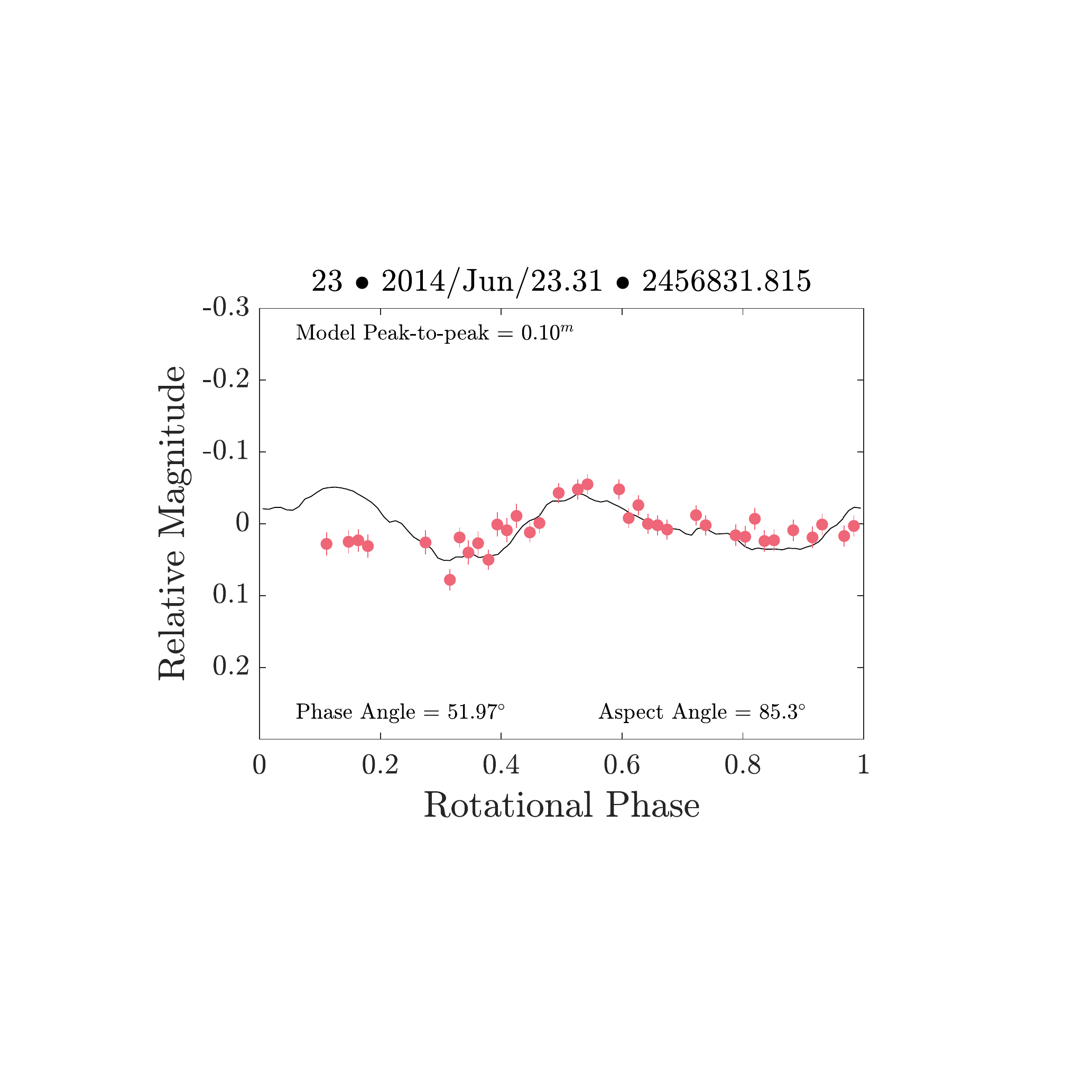} 		
		\includegraphics[width=.48\textwidth, trim=2cm 4cm 3.8cm 4cm, clip=true]{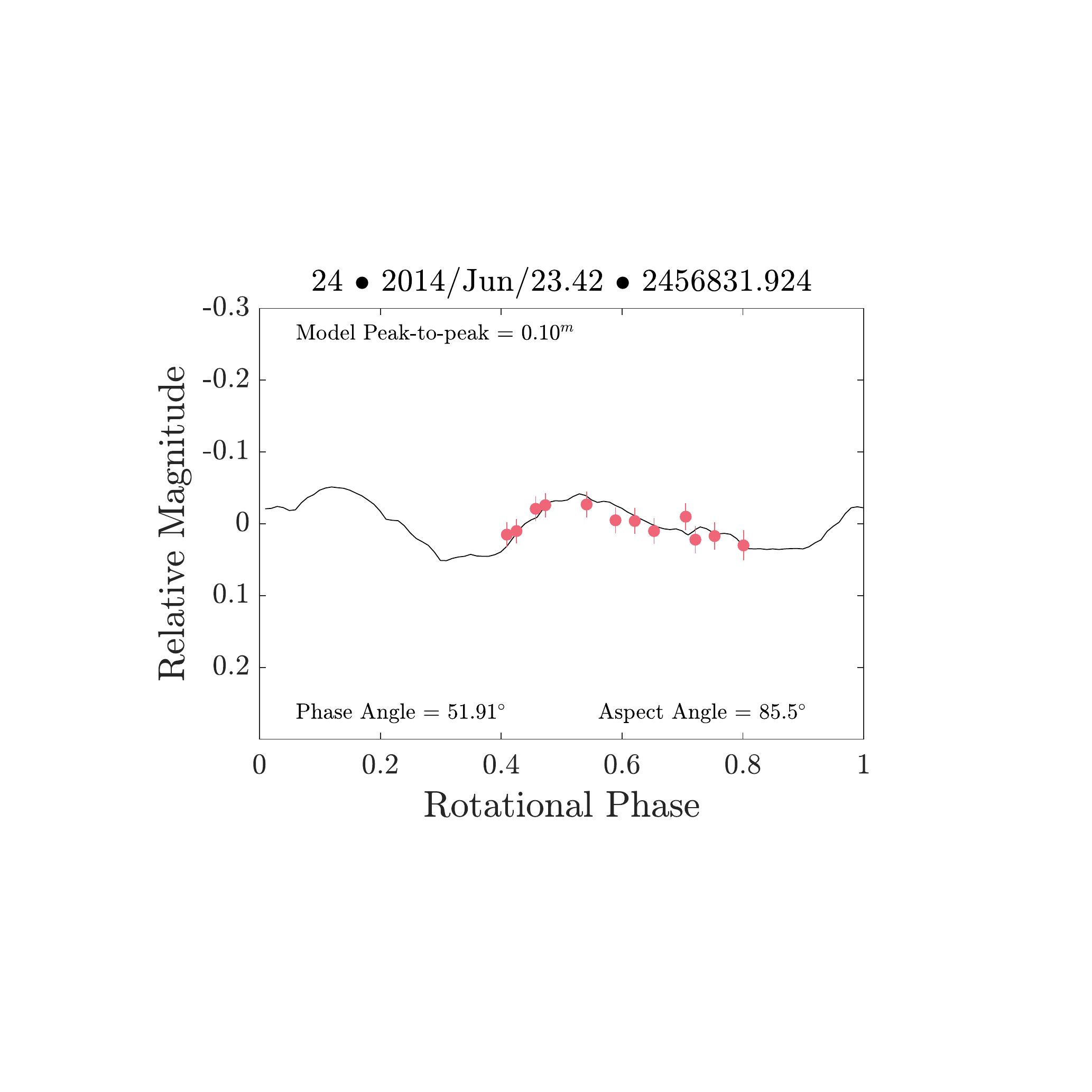} 
	}
	\caption{[Continued]
		\label{fig:conv-lcfit2}}
\end{figure*}

\addtocounter{figure}{-1}

\begin{figure*}
	\resizebox{\hsize}{!}{
		\includegraphics[width=.48\textwidth, trim=2cm 4cm 3.8cm 4cm, clip=true]{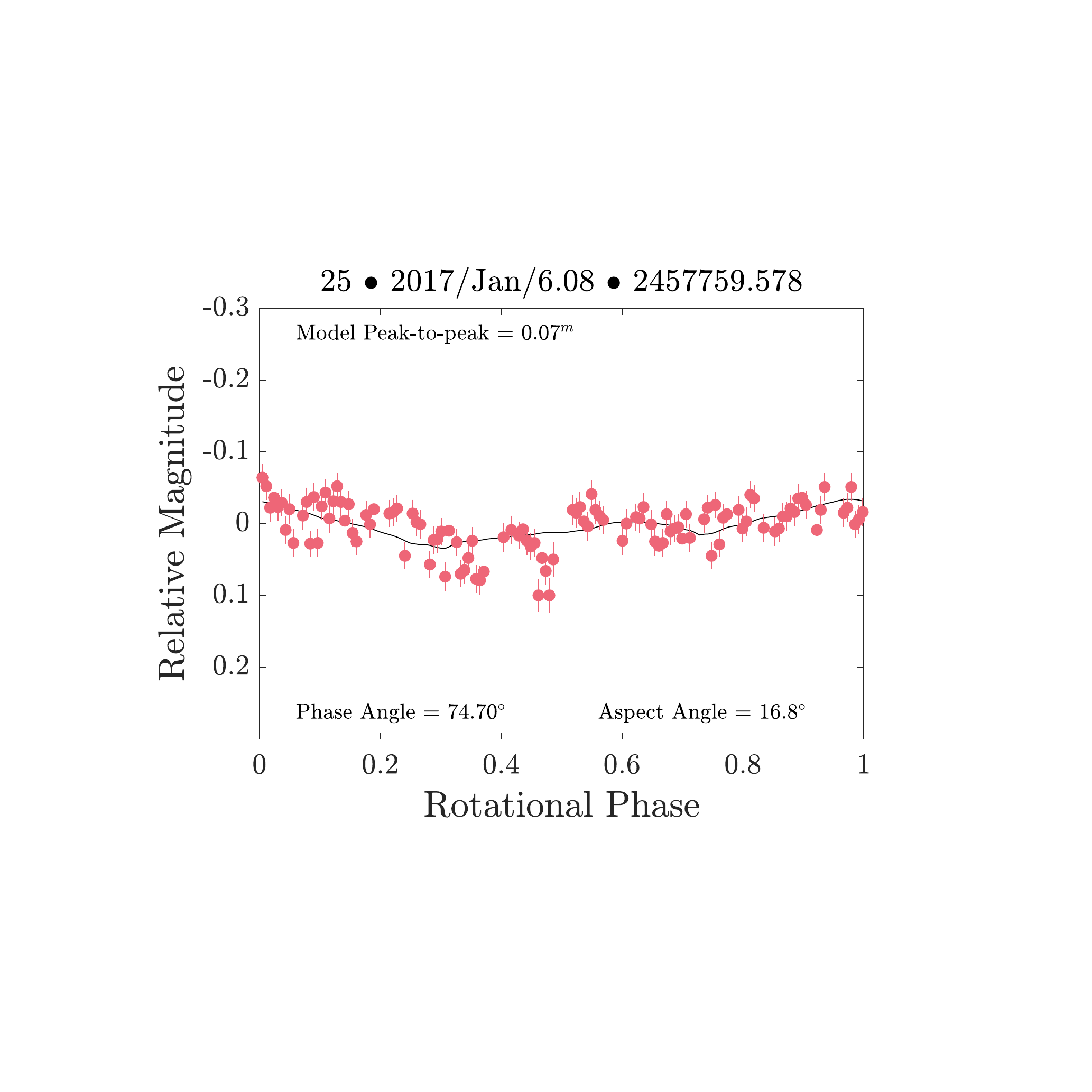} 
		\includegraphics[width=.48\textwidth, trim=2cm 4cm 3.8cm 4cm, clip=true]{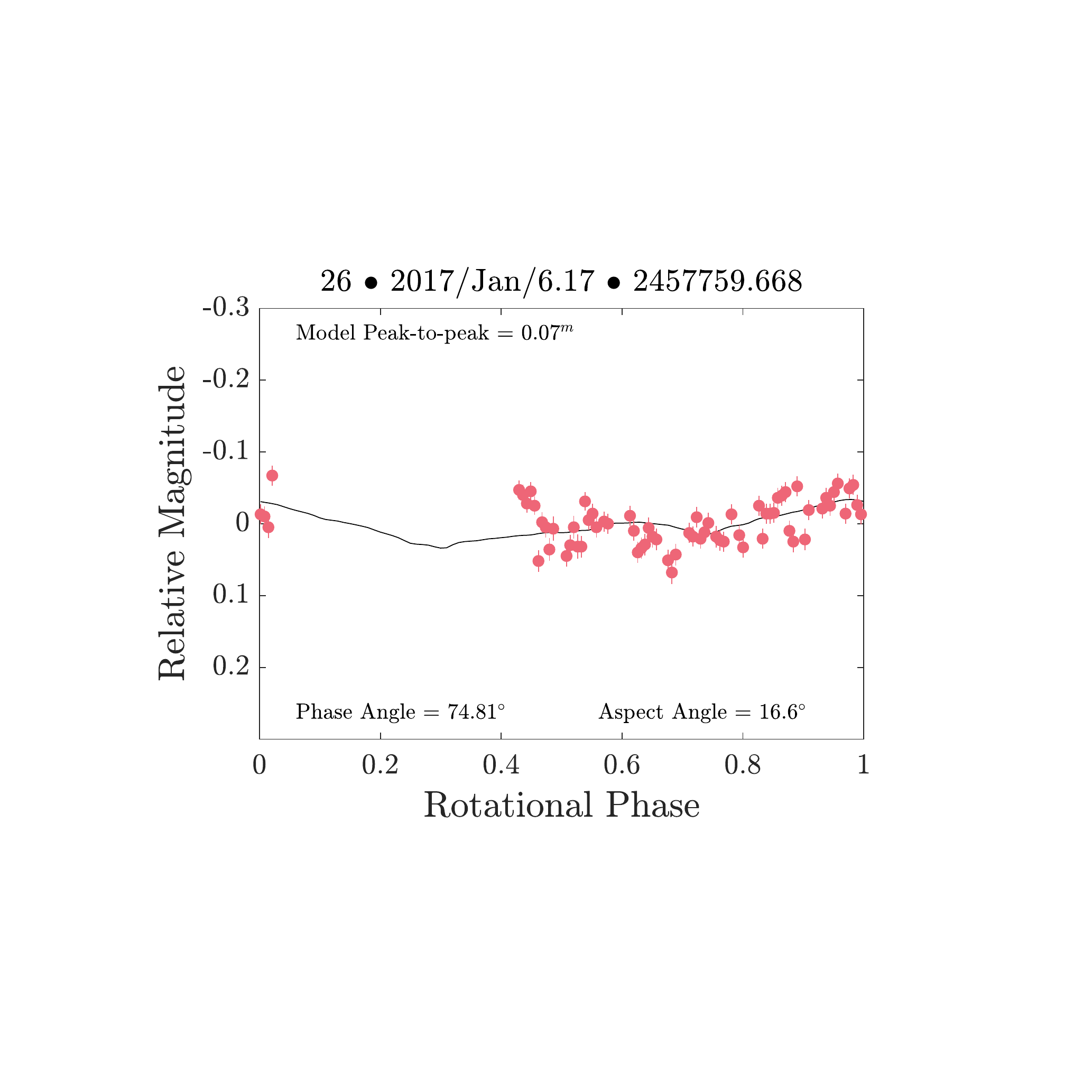} 		
		\includegraphics[width=.48\textwidth, trim=2cm 4cm 3.8cm 4cm, clip=true]{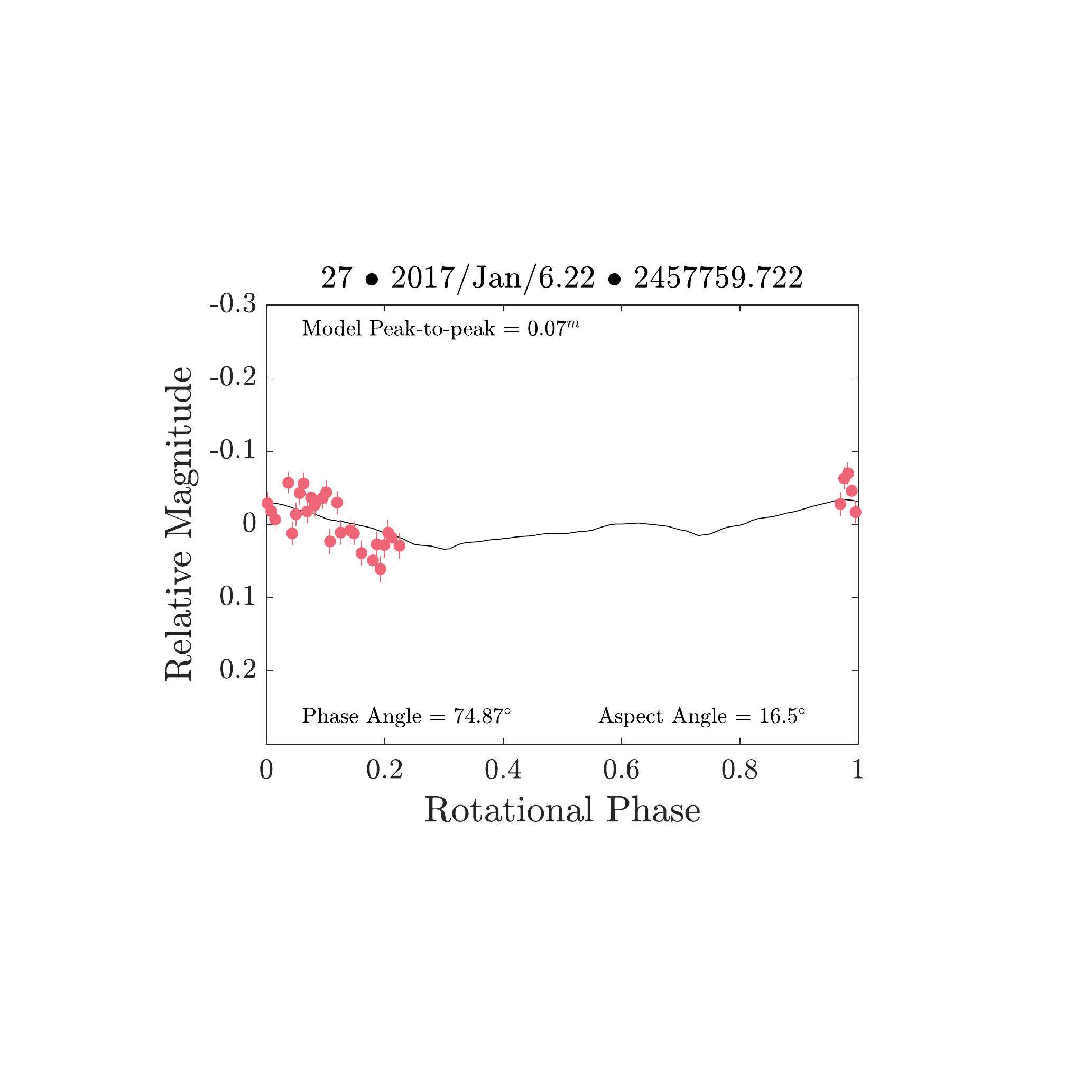} 
	}
	
	\resizebox{\hsize}{!}{
		\includegraphics[width=.48\textwidth, trim=2cm 4cm 3.8cm 4cm, clip=true]{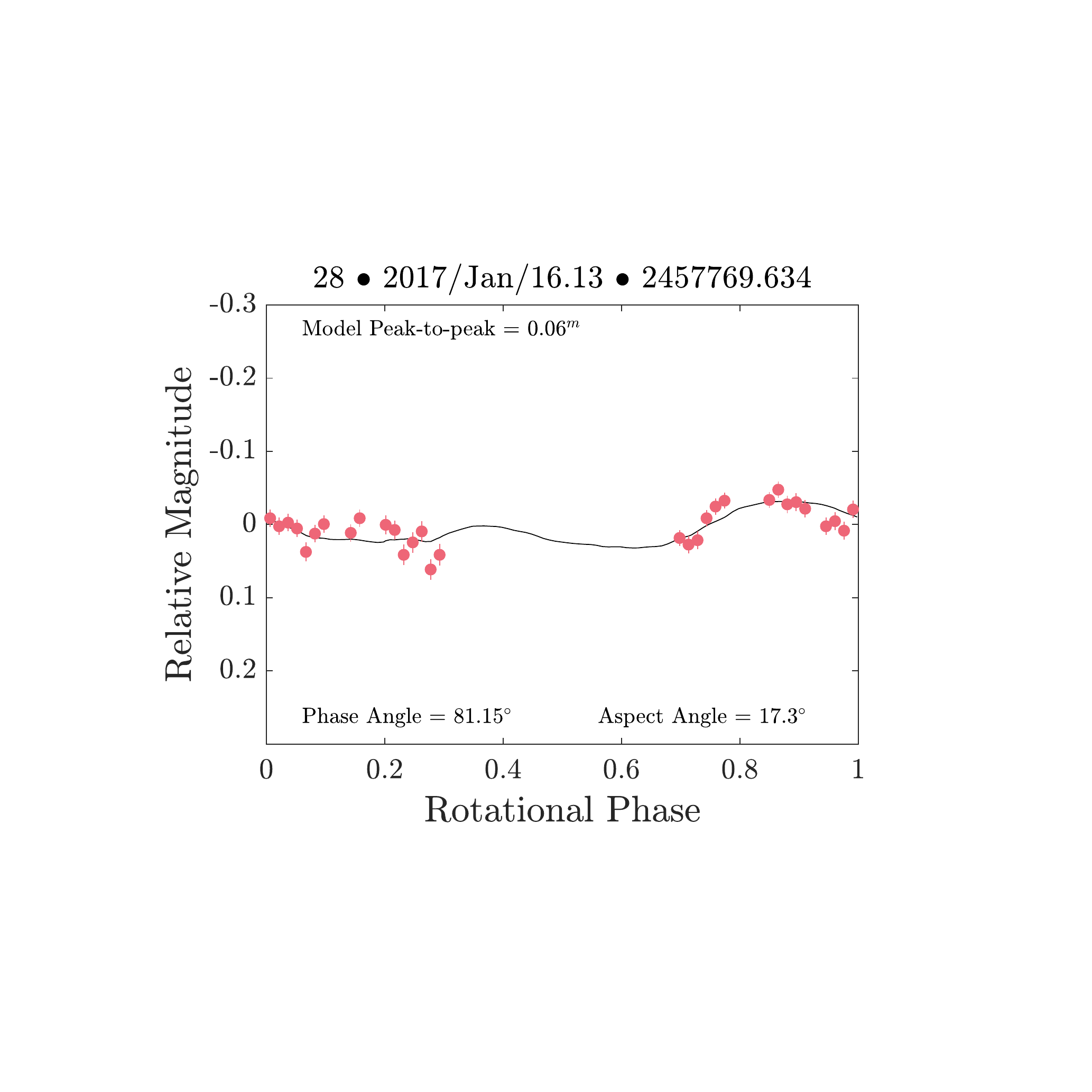} 
		\includegraphics[width=.48\textwidth, trim=2cm 4cm 3.8cm 4cm, clip=true]{LC/2102_210_-30_0_v190906_20190911_initial_spinstate_29_fix.pdf} 		
		\includegraphics[width=.48\textwidth, trim=2cm 4cm 3.8cm 4cm, clip=true]{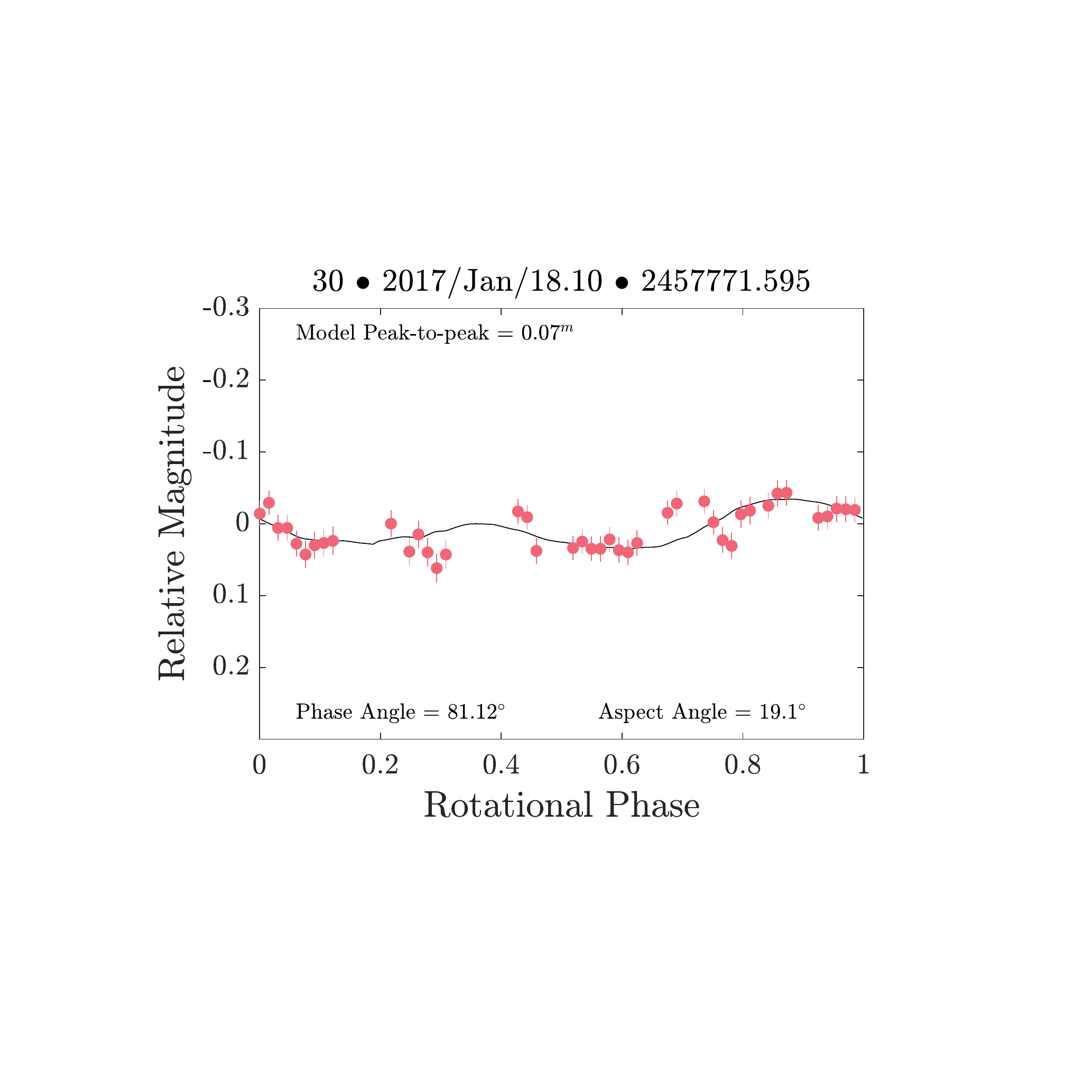} 
	}
	
	\resizebox{.6666\hsize}{!}{
		\includegraphics[width=.48\textwidth, trim=2cm 4cm 3.8cm 4cm, clip=true]{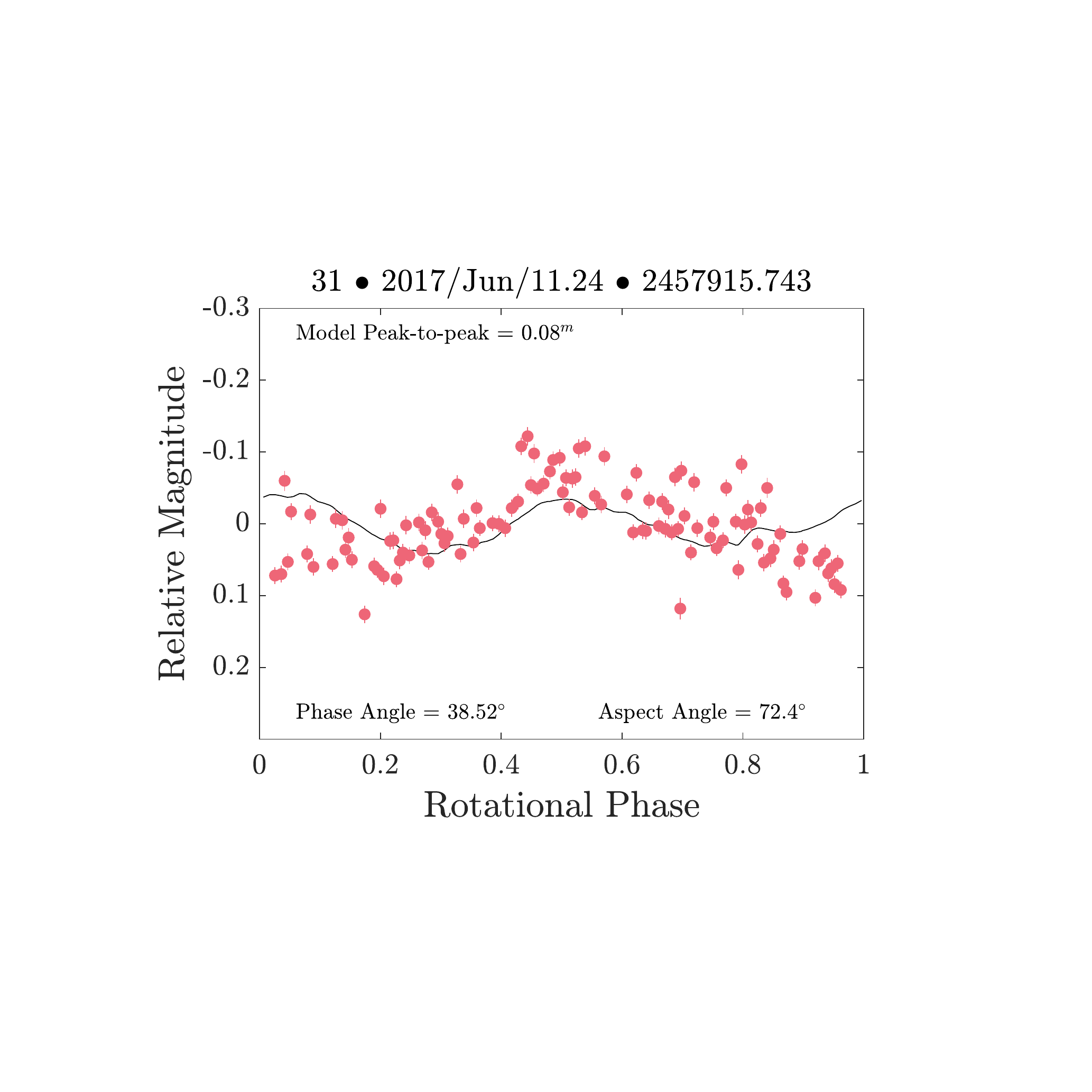} 
		\includegraphics[width=.48\textwidth, trim=2cm 4cm 3.8cm 4cm, clip=true]{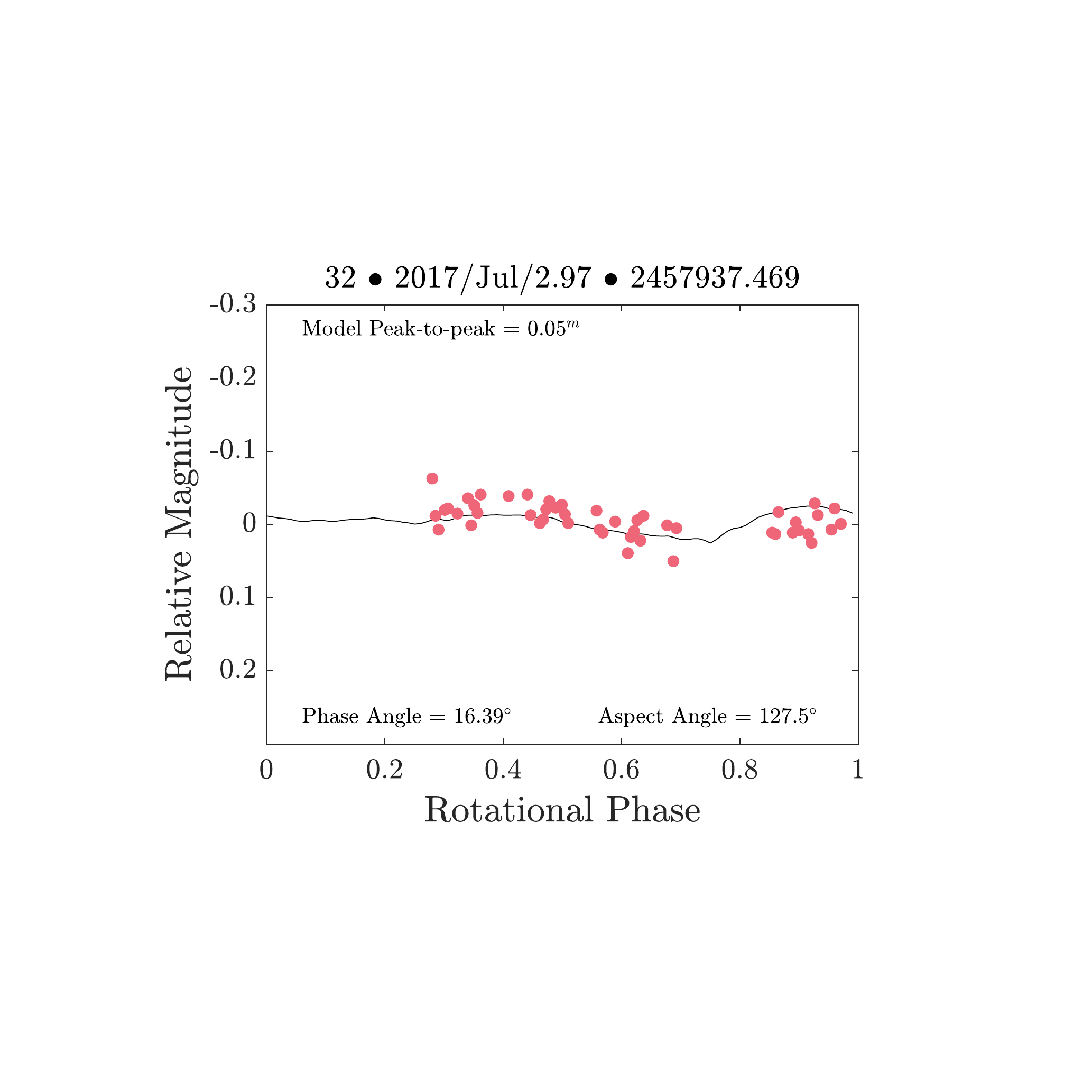} 	
	
	}

	\caption{[Continued]
		\label{fig:conv-lcfit3}}
\end{figure*}

\begin{figure}

	\resizebox{\hsize}{!}{\includegraphics{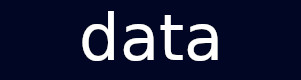}\includegraphics{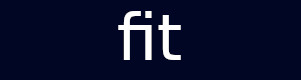}\includegraphics{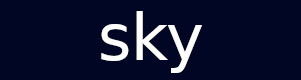}\includegraphics{radar/radar_data.jpg}\includegraphics{radar/radar_fit.jpg}\includegraphics{radar/radar_sky.jpg}}
    \resizebox{\hsize}{!}{\includegraphics{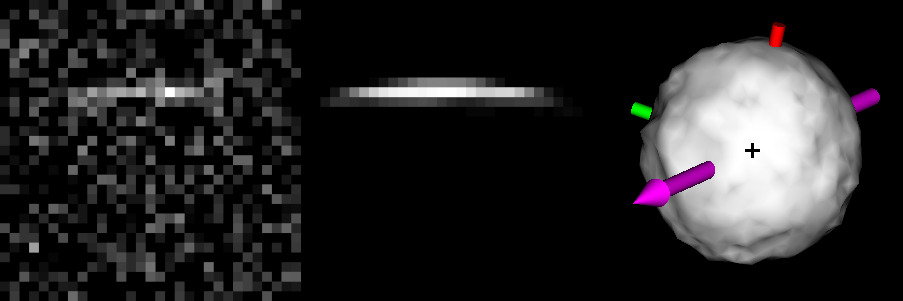}\includegraphics{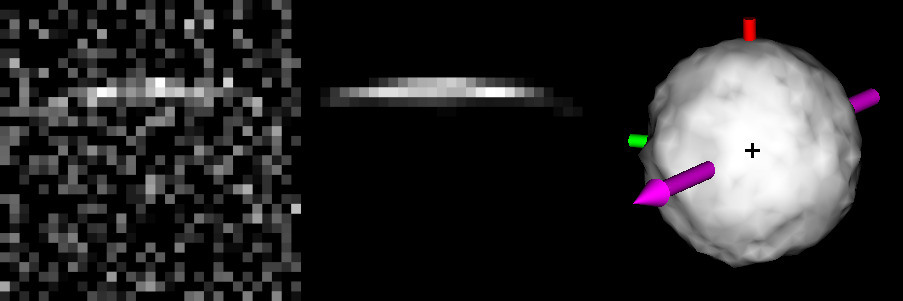}}
    \resizebox{\hsize}{!}{\includegraphics{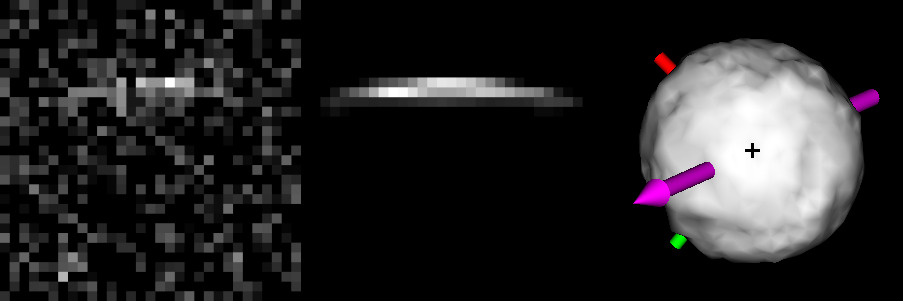}\includegraphics{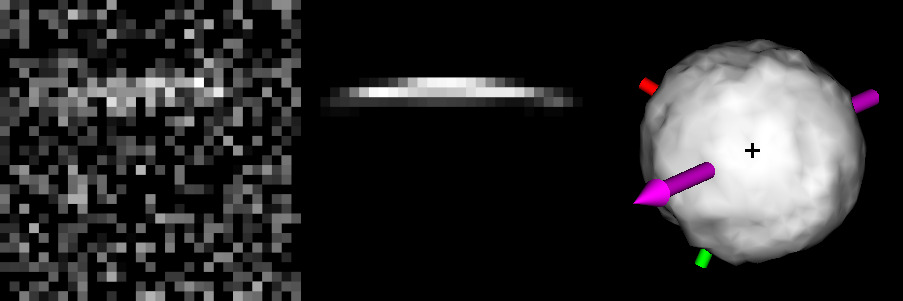}}
    \resizebox{\hsize}{!}{\includegraphics{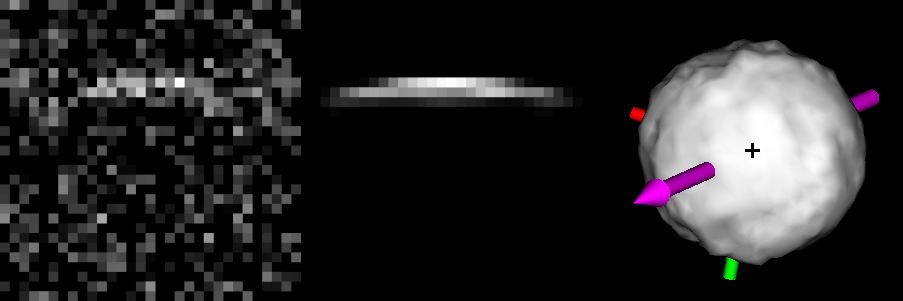}\includegraphics{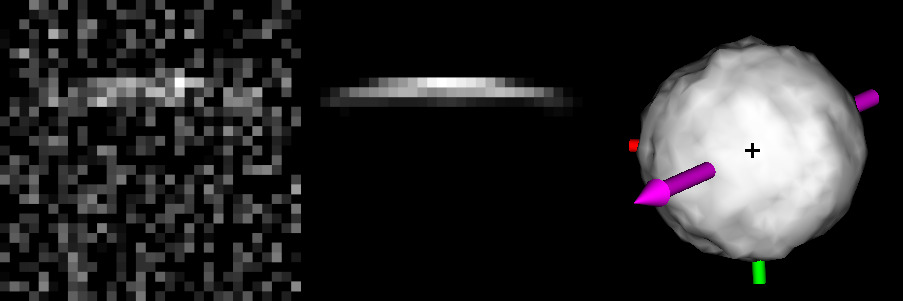}}
    \resizebox{\hsize}{!}{\includegraphics{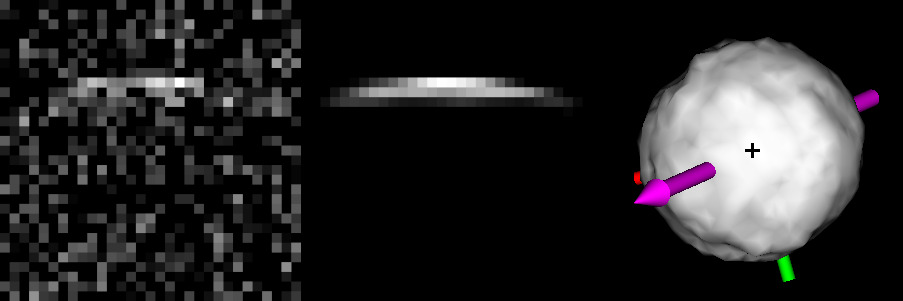}\includegraphics{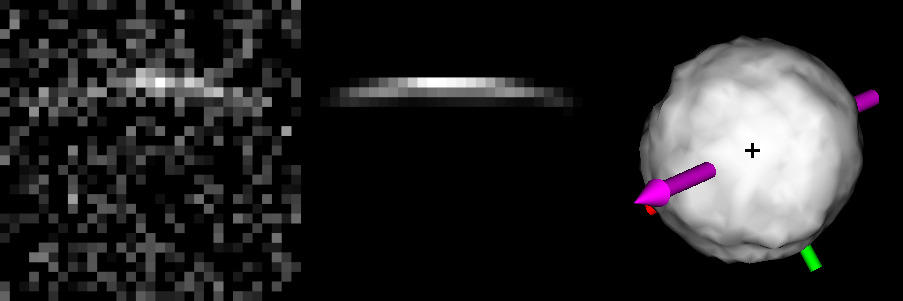}}
    \resizebox{\hsize}{!}{\includegraphics{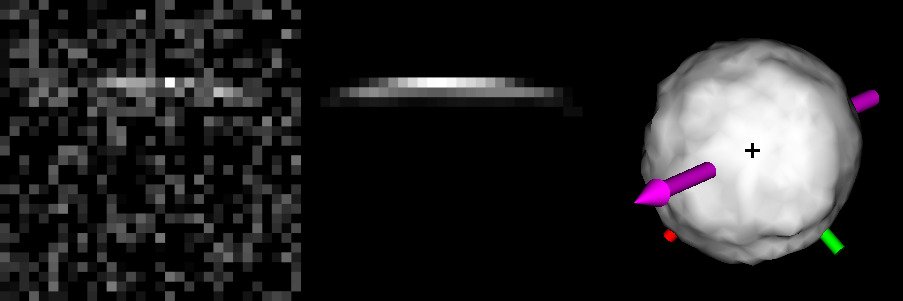}\includegraphics{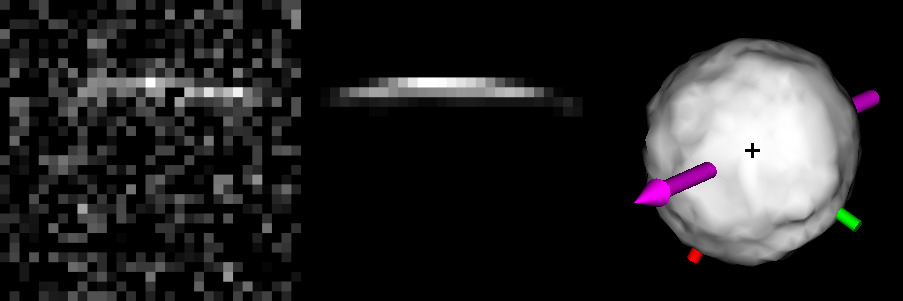}}
    \resizebox{\hsize}{!}{\includegraphics{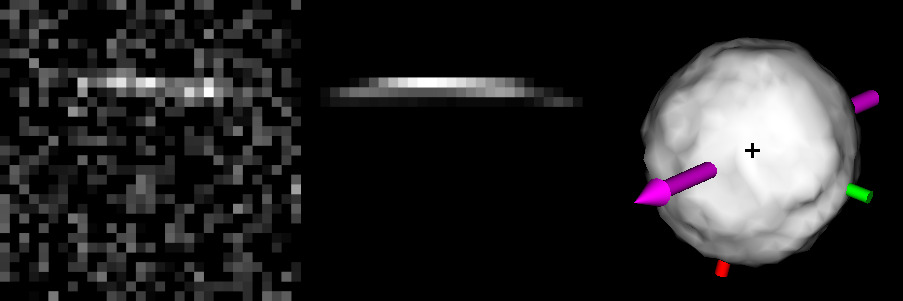}\includegraphics{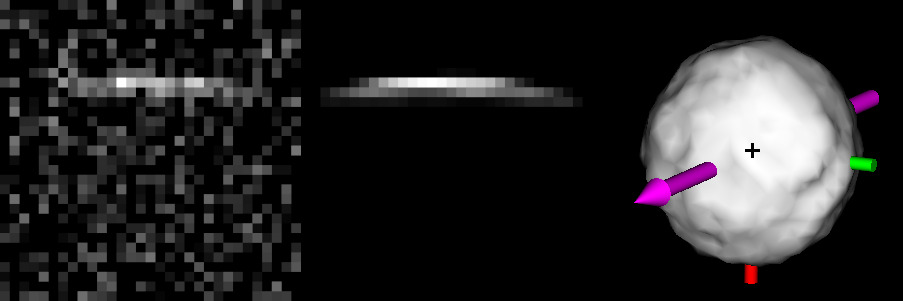}}
    \resizebox{\hsize}{!}{\includegraphics{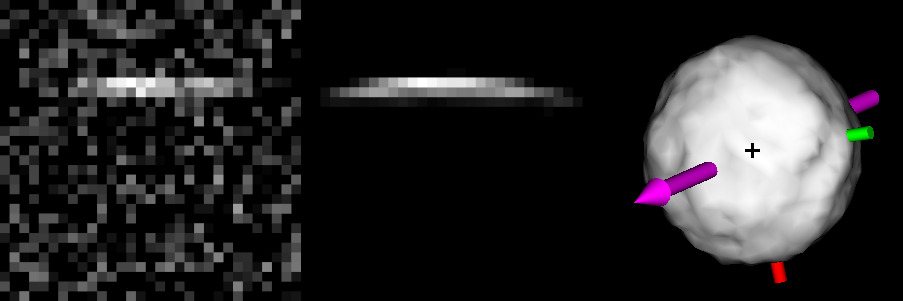}\includegraphics{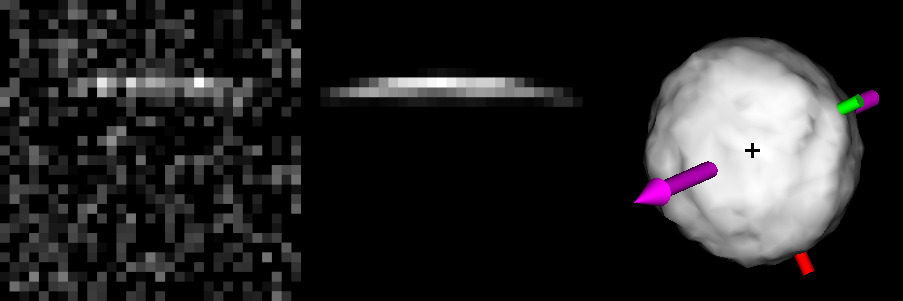}}
    \resizebox{0.5\hsize}{!}{\includegraphics{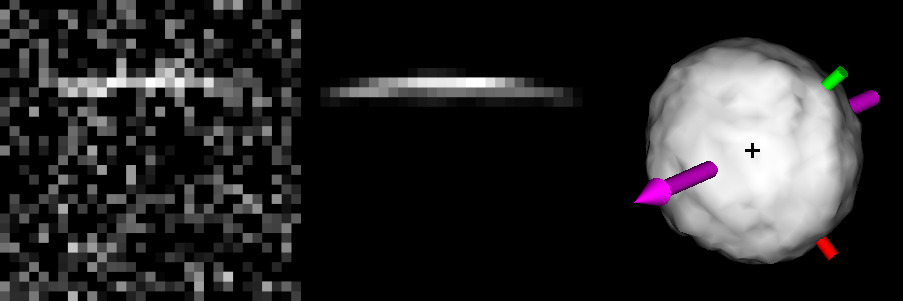}}
	\caption{Same as Fig.~\ref{fig:dd:jan1}, but for observations taken with Arecibo on the night beginning on 4th January 2017 and retrograde model only.
		\label{fig:dd:retro:jan4}  }
	
\end{figure}

\begin{figure}

	\resizebox{\hsize}{!}{\includegraphics{radar/radar_data.jpg}\includegraphics{radar/radar_fit.jpg}\includegraphics{radar/radar_sky.jpg}\includegraphics{radar/radar_data.jpg}\includegraphics{radar/radar_fit.jpg}\includegraphics{radar/radar_sky.jpg}}
    \resizebox{\hsize}{!}{\includegraphics{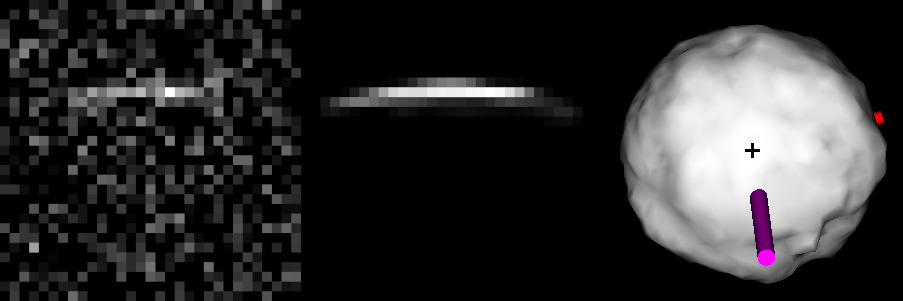}\includegraphics{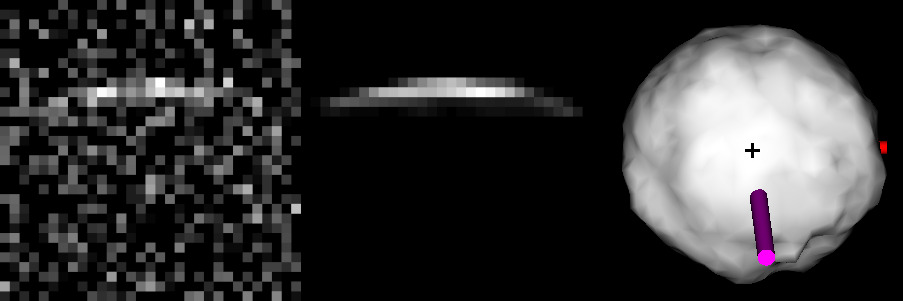}}
    \resizebox{\hsize}{!}{\includegraphics{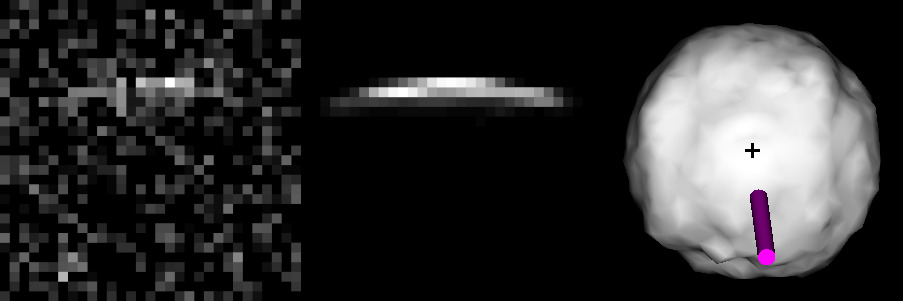}\includegraphics{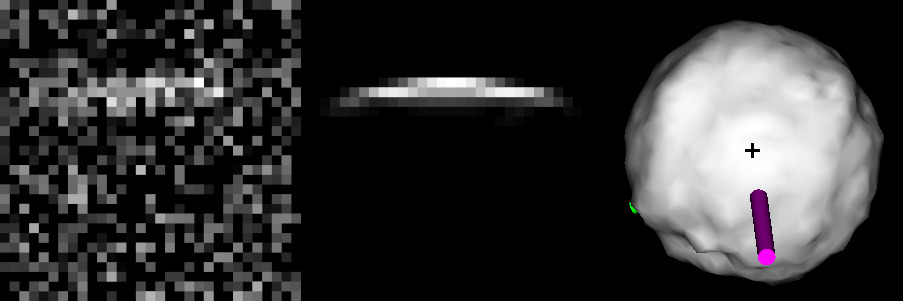}}
    \resizebox{\hsize}{!}{\includegraphics{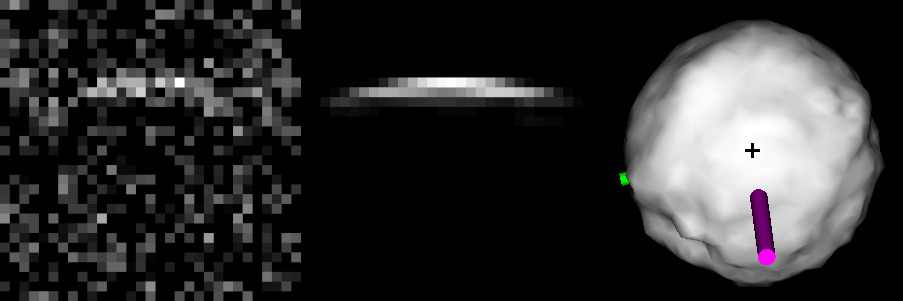}\includegraphics{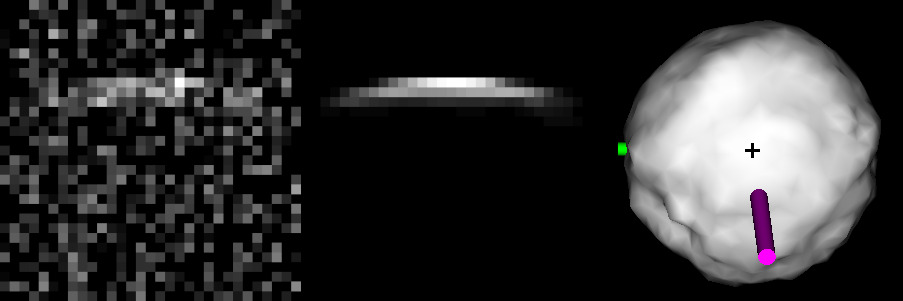}}
    \resizebox{\hsize}{!}{\includegraphics{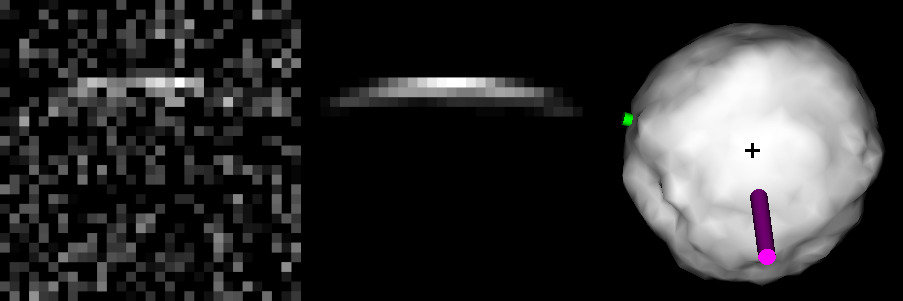}\includegraphics{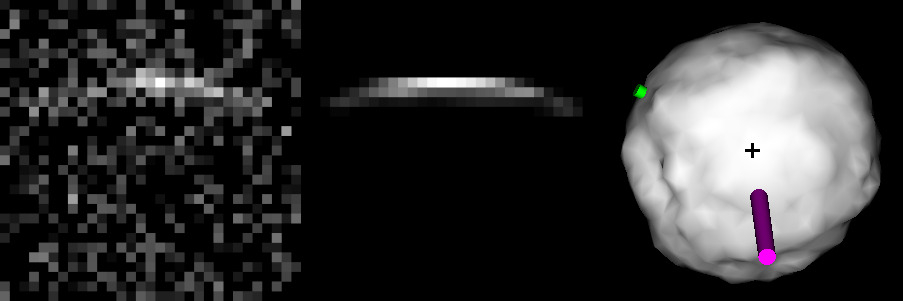}}
    \resizebox{\hsize}{!}{\includegraphics{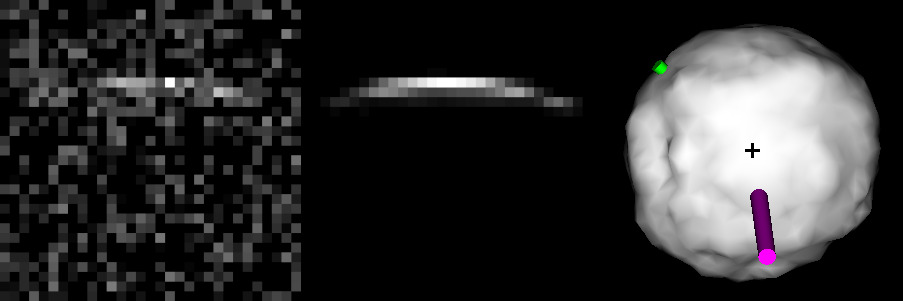}\includegraphics{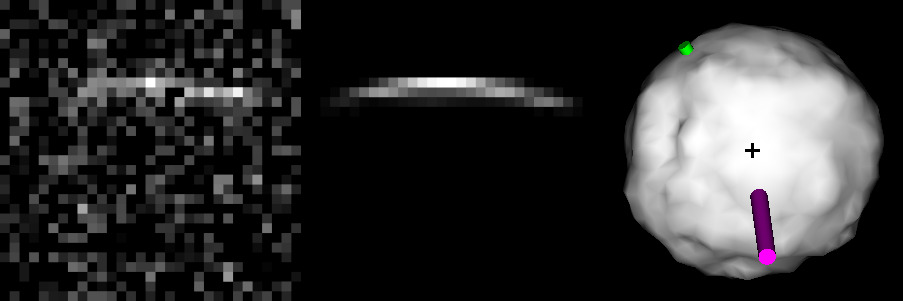}}
    \resizebox{\hsize}{!}{\includegraphics{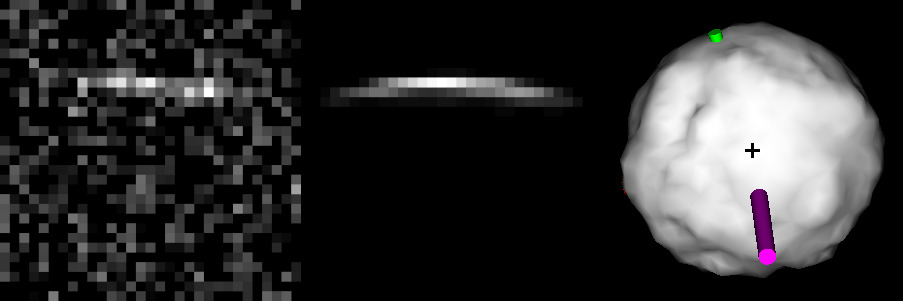}\includegraphics{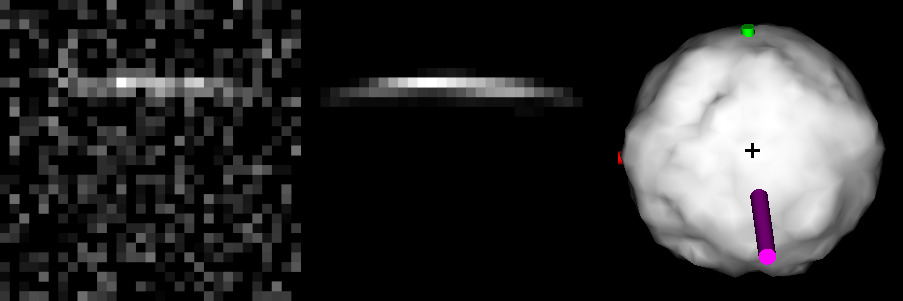}}
    \resizebox{\hsize}{!}{\includegraphics{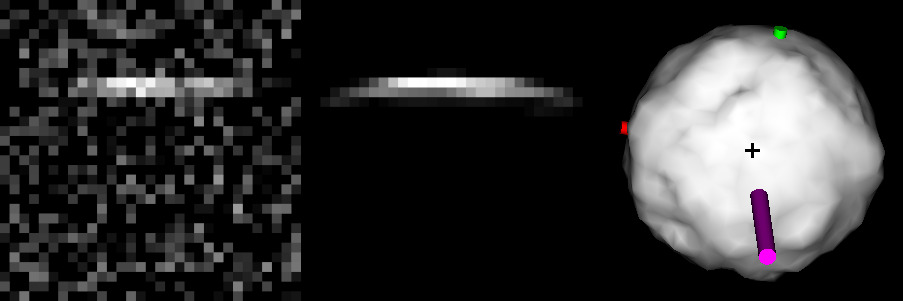}\includegraphics{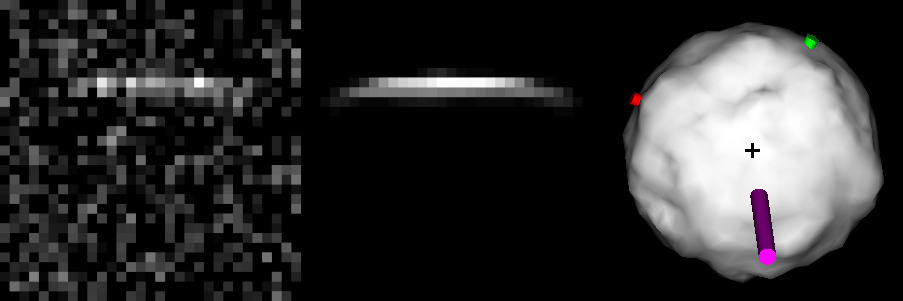}}
    \resizebox{0.5\hsize}{!}{\includegraphics{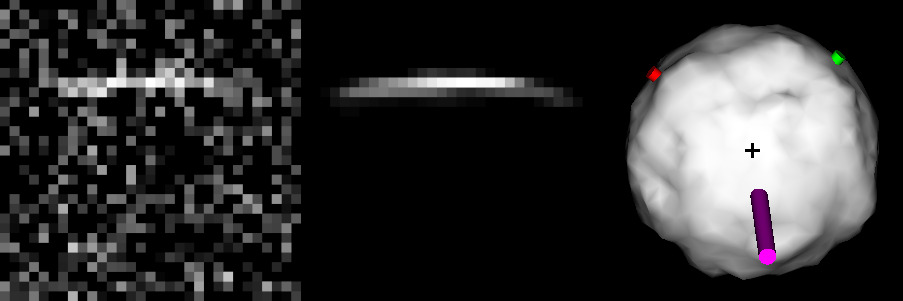}}
	\caption{Same as Fig.~\ref{fig:dd:jan1}, but for observations taken with Arecibo on the night beginning on 4th January 2017 and prograde model only.
		\label{fig:dd:pro:jan4}  }
	
\end{figure}

\begin{figure}

	\resizebox{\hsize}{!}{\includegraphics{radar/radar_data.jpg}\includegraphics{radar/radar_fit.jpg}\includegraphics{radar/radar_sky.jpg}\includegraphics{radar/radar_data.jpg}\includegraphics{radar/radar_fit.jpg}\includegraphics{radar/radar_sky.jpg}}
    \resizebox{\hsize}{!}{\includegraphics{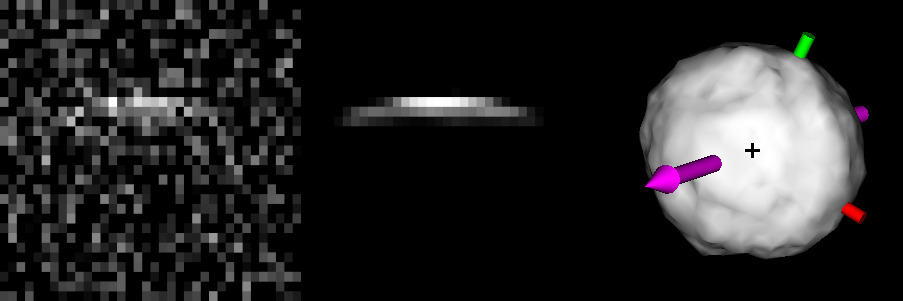}\includegraphics{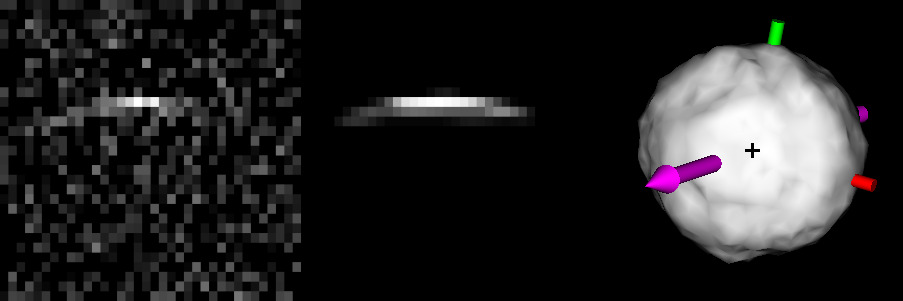}}
    \resizebox{\hsize}{!}{\includegraphics{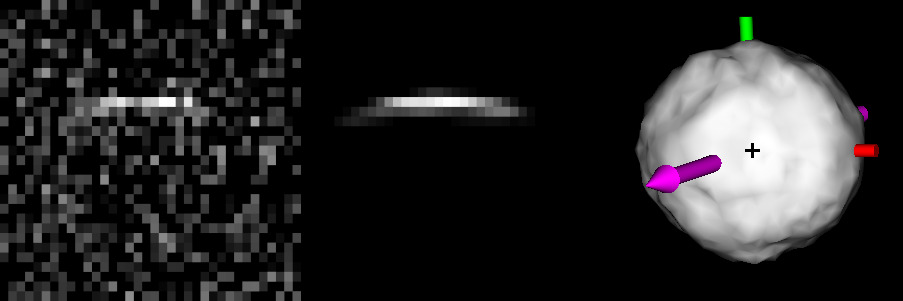}\includegraphics{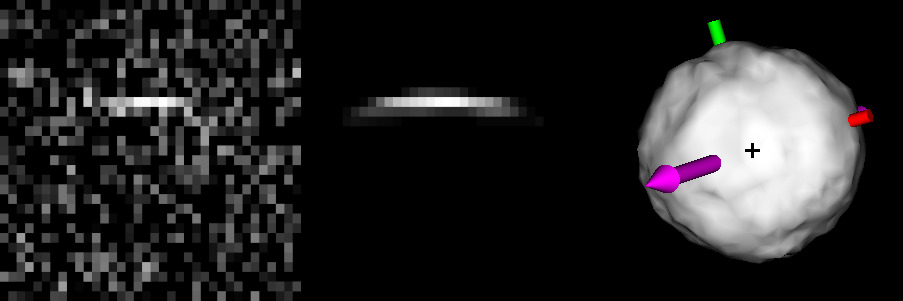}}
    \resizebox{\hsize}{!}{\includegraphics{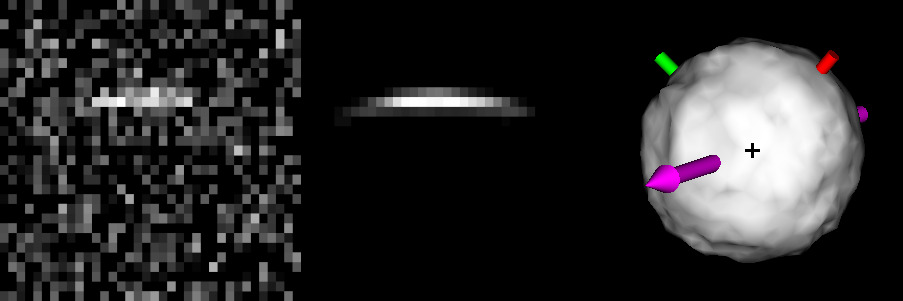}\includegraphics{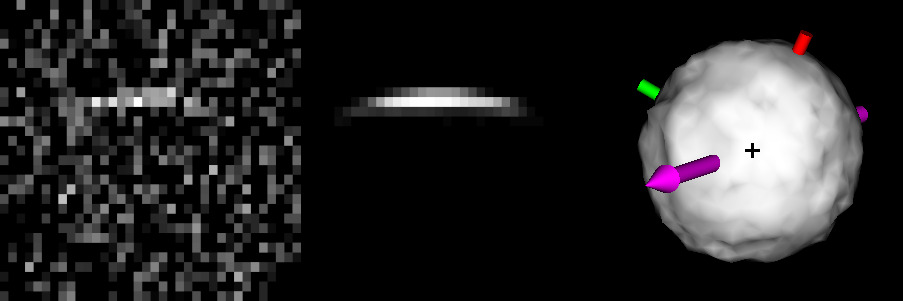}}
    \resizebox{\hsize}{!}{\includegraphics{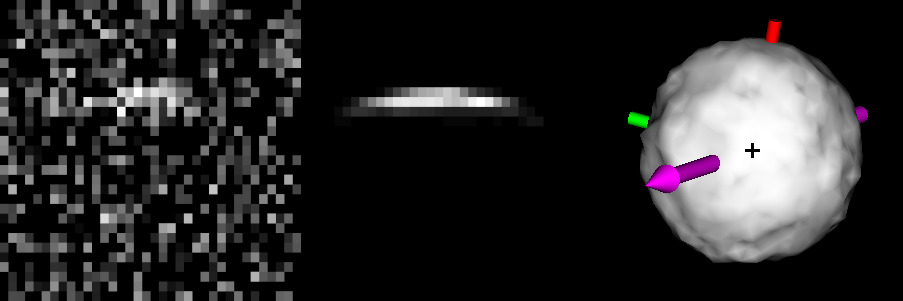}\includegraphics{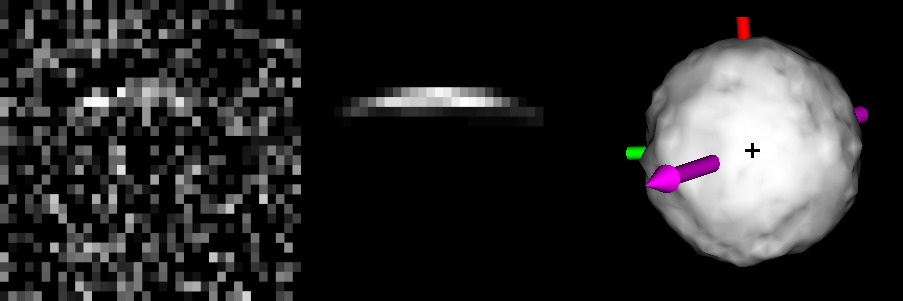}}
    \resizebox{\hsize}{!}{\includegraphics{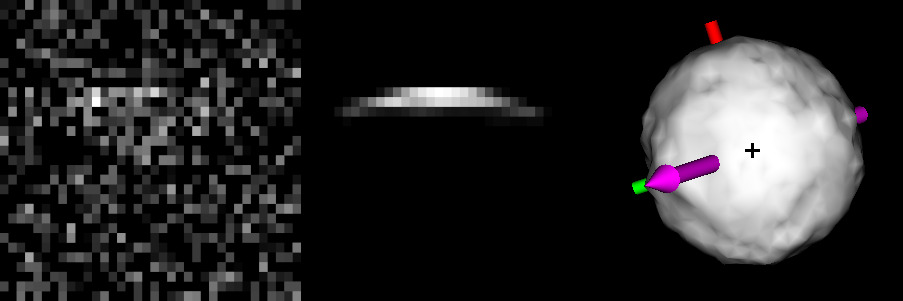}\includegraphics{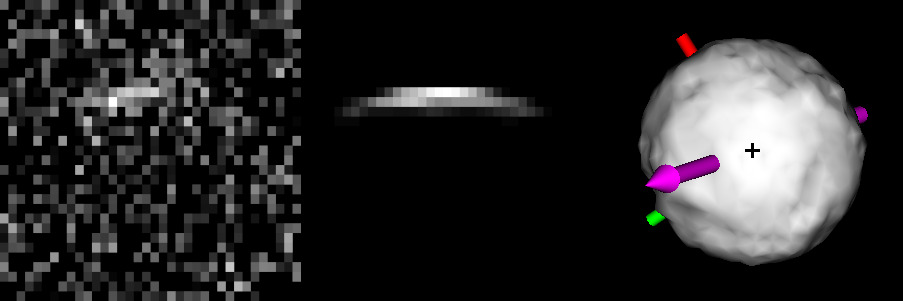}}
    \resizebox{\hsize}{!}{\includegraphics{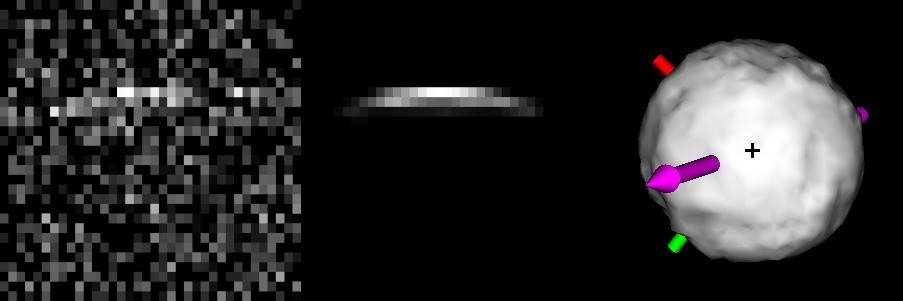}\includegraphics{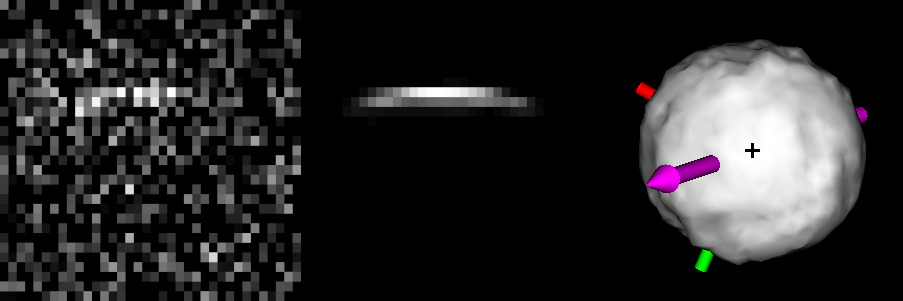}}
    \resizebox{\hsize}{!}{\includegraphics{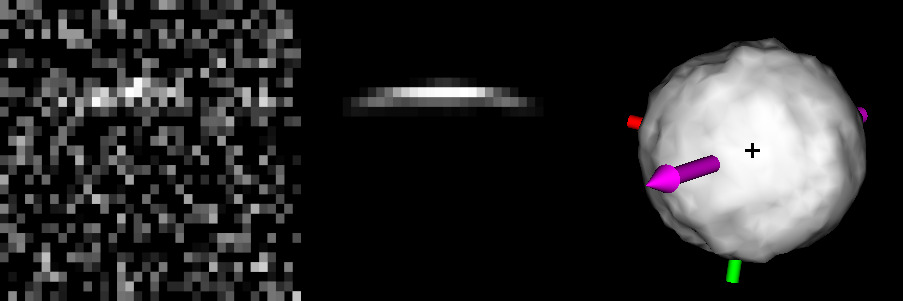}\includegraphics{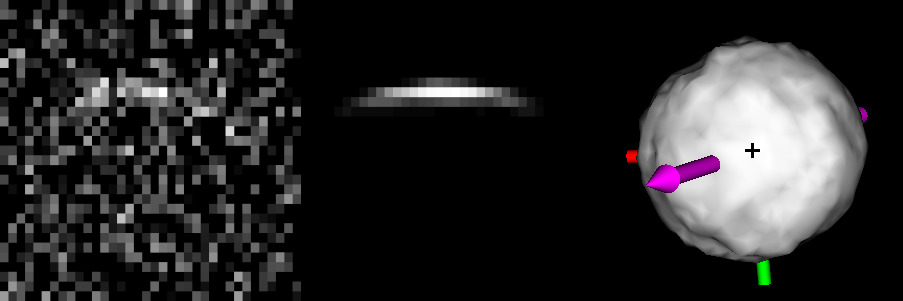}}
    \resizebox{\hsize}{!}{\includegraphics{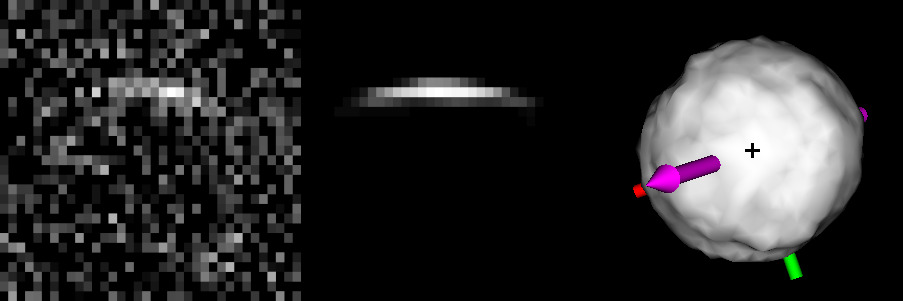}\includegraphics{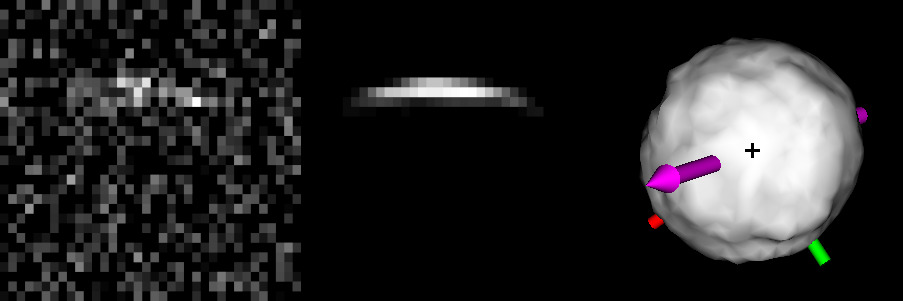}}
    \resizebox{\hsize}{!}{\includegraphics{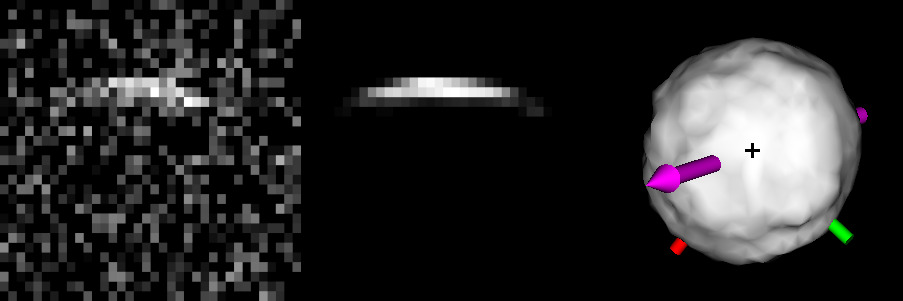}\includegraphics{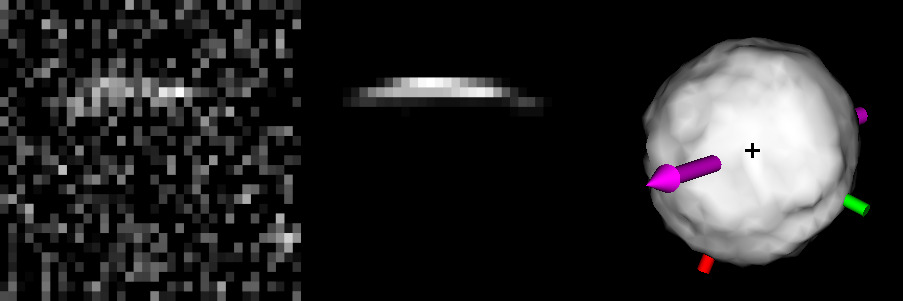}}
    \resizebox{\hsize}{!}{\includegraphics{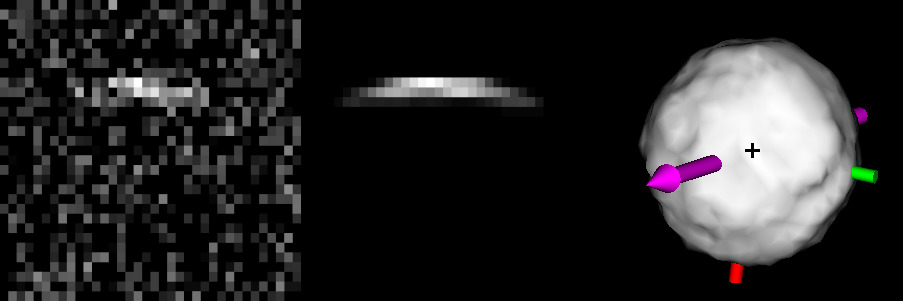}\includegraphics{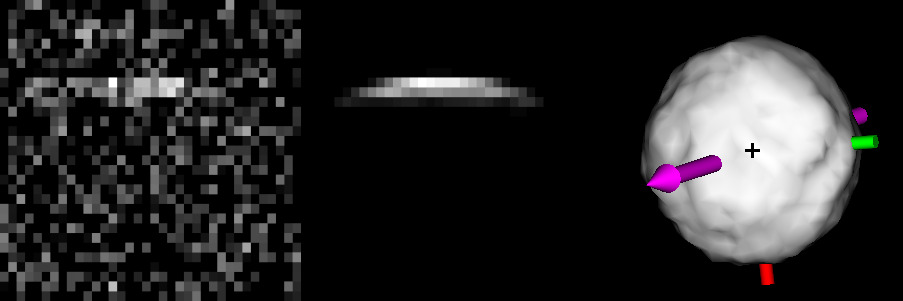}}
 
	\caption{Same as Fig.~\ref{fig:dd:jan1}, but for observations taken with Arecibo on the night beginning on 5th Januray 2017 and retrograde model only.
		\label{fig:dd:retro:jan5}  }
\end{figure}

\begin{figure}

	\resizebox{\hsize}{!}{\includegraphics{radar/radar_data.jpg}\includegraphics{radar/radar_fit.jpg}\includegraphics{radar/radar_sky.jpg}\includegraphics{radar/radar_data.jpg}\includegraphics{radar/radar_fit.jpg}\includegraphics{radar/radar_sky.jpg}}
    \resizebox{\hsize}{!}{\includegraphics{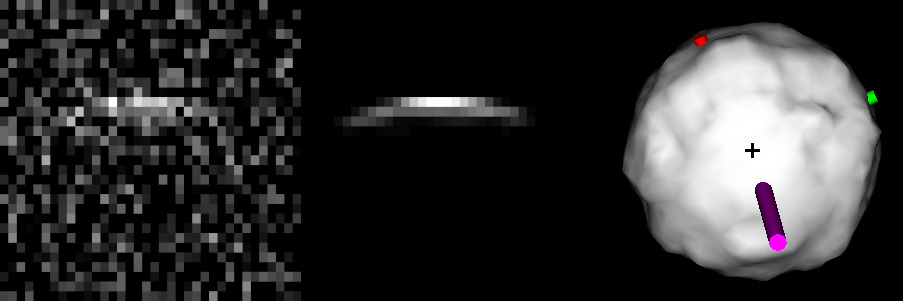}\includegraphics{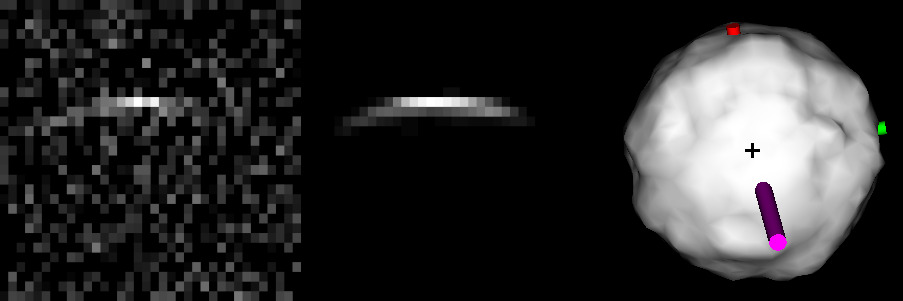}}
    \resizebox{\hsize}{!}{\includegraphics{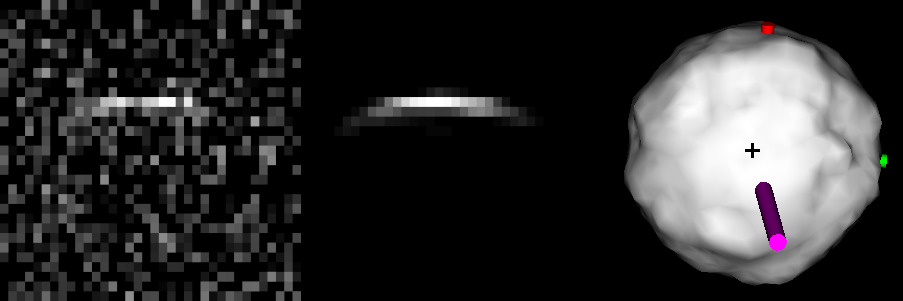}\includegraphics{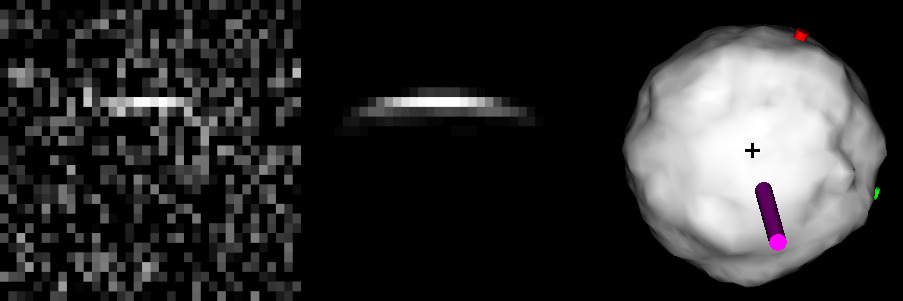}}
    \resizebox{\hsize}{!}{\includegraphics{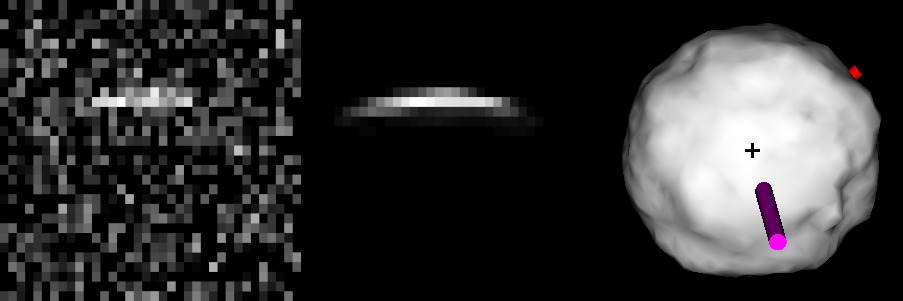}\includegraphics{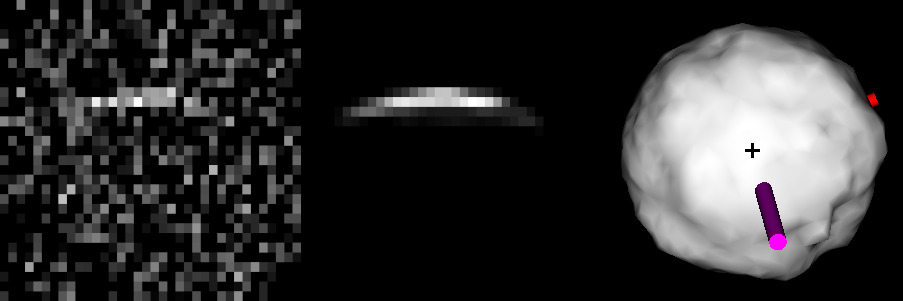}}
    \resizebox{\hsize}{!}{\includegraphics{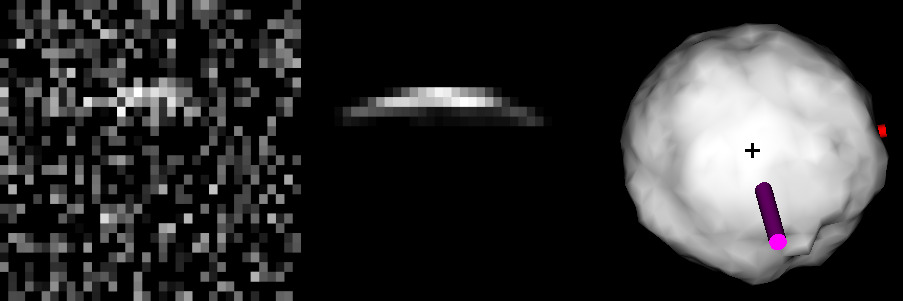}\includegraphics{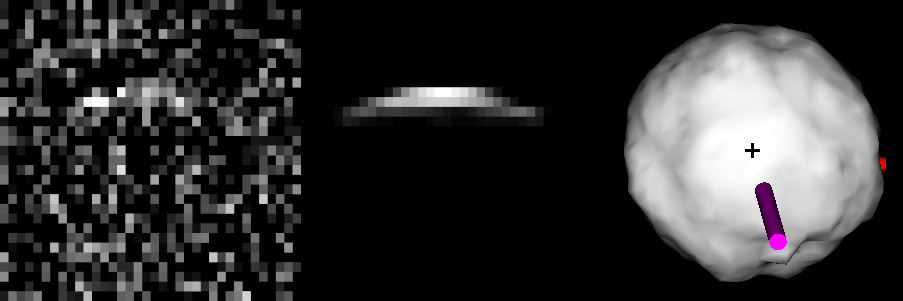}}
    \resizebox{\hsize}{!}{\includegraphics{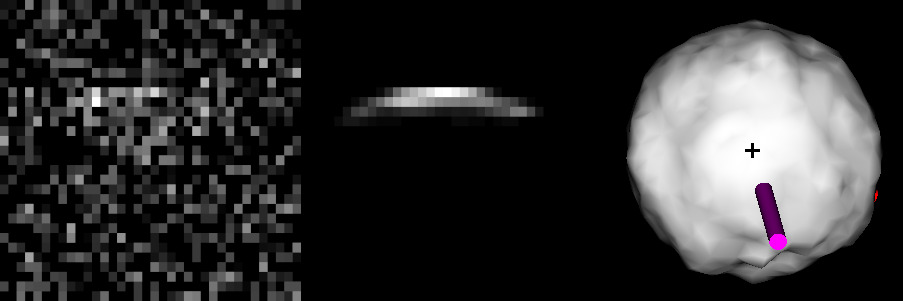}\includegraphics{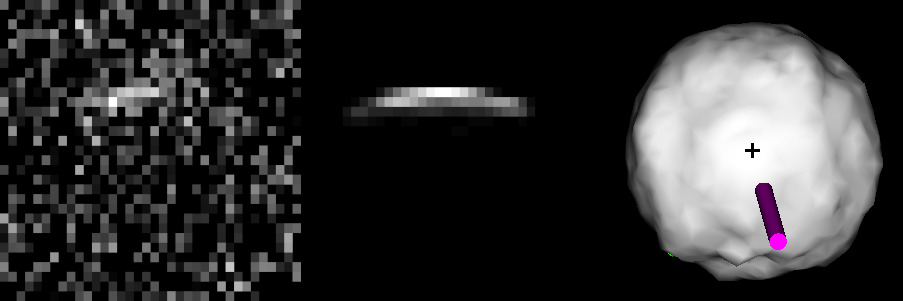}}
    \resizebox{\hsize}{!}{\includegraphics{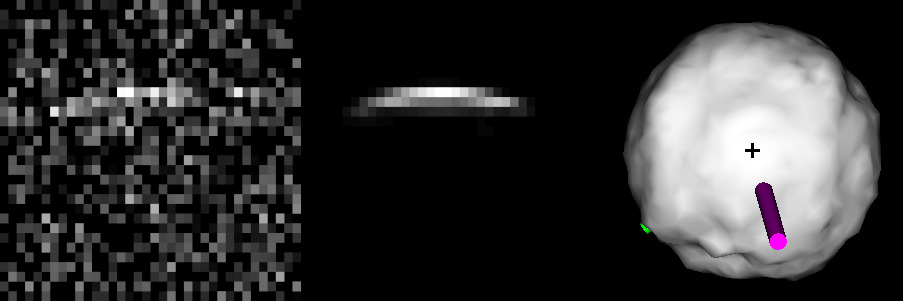}\includegraphics{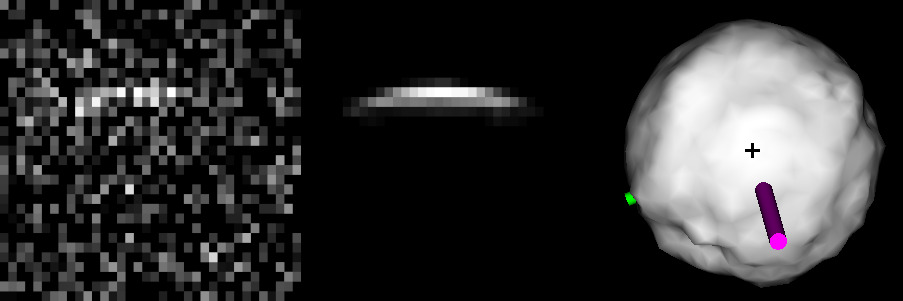}}
    \resizebox{\hsize}{!}{\includegraphics{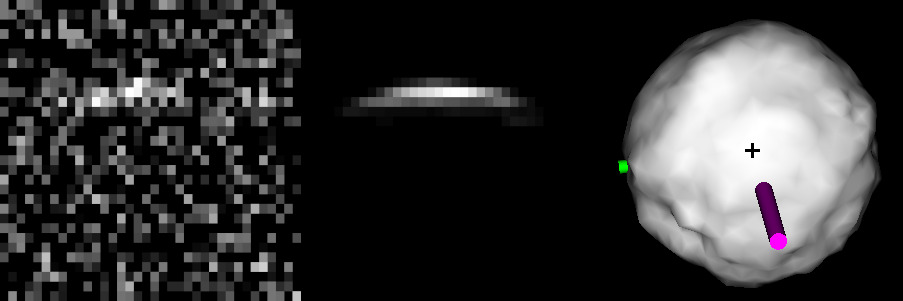}\includegraphics{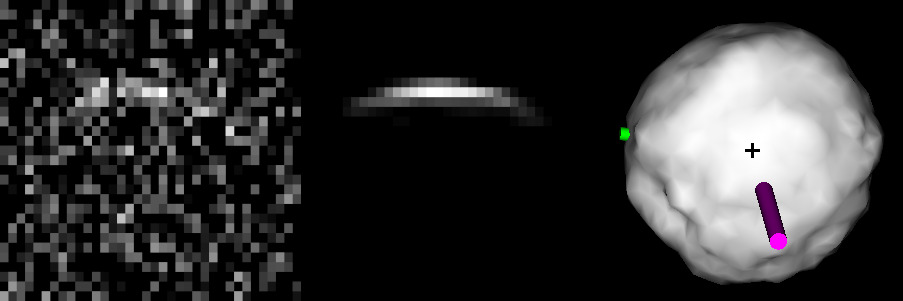}}
    \resizebox{\hsize}{!}{\includegraphics{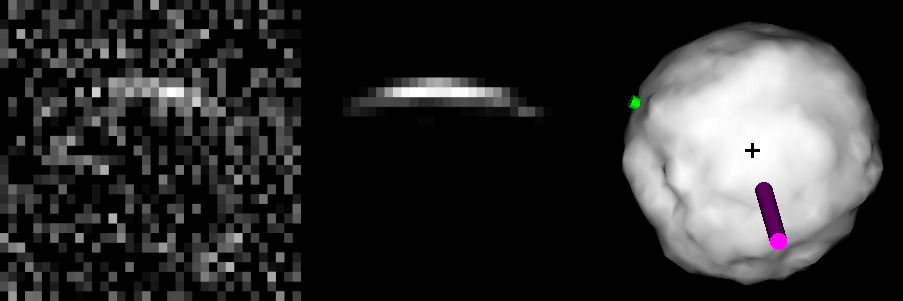}\includegraphics{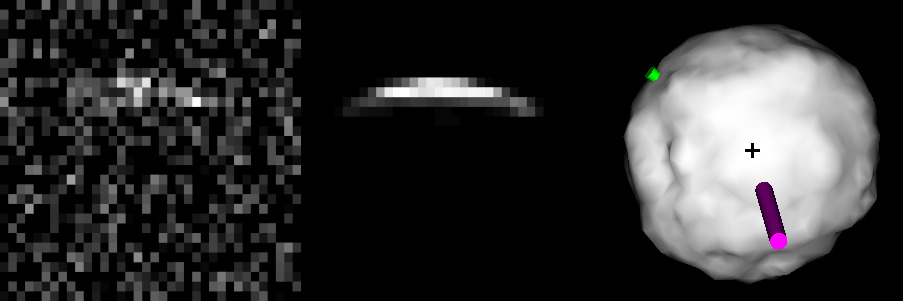}}
    \resizebox{\hsize}{!}{\includegraphics{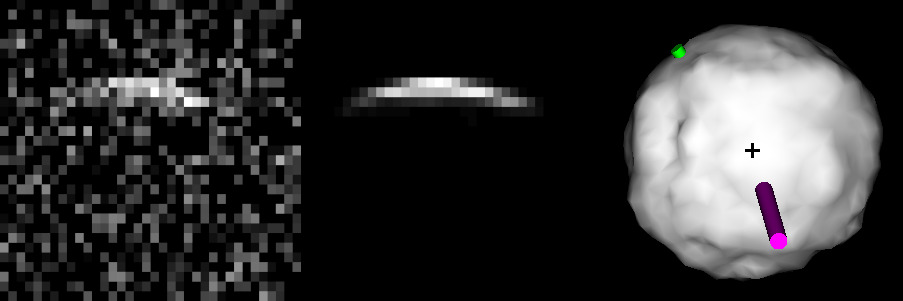}\includegraphics{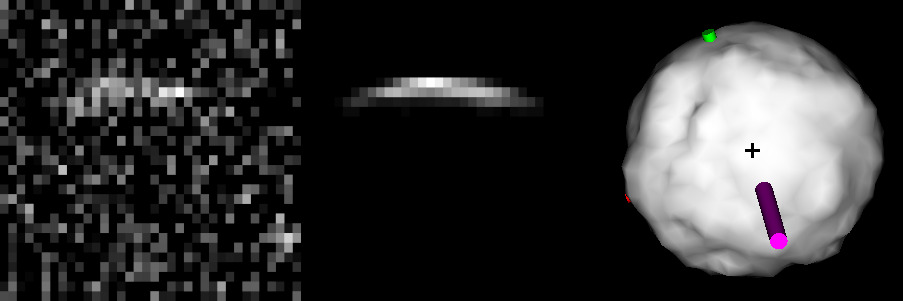}}
    \resizebox{\hsize}{!}{\includegraphics{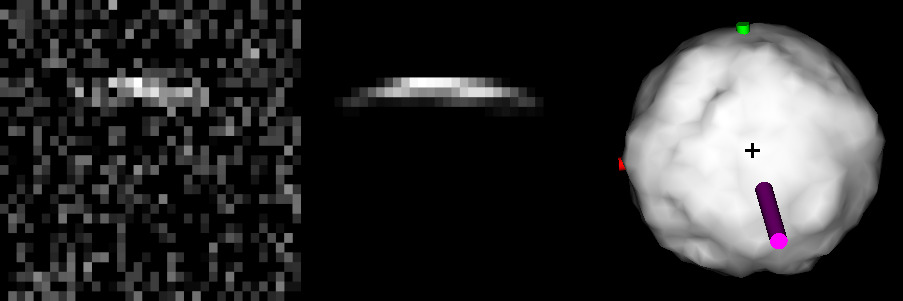}\includegraphics{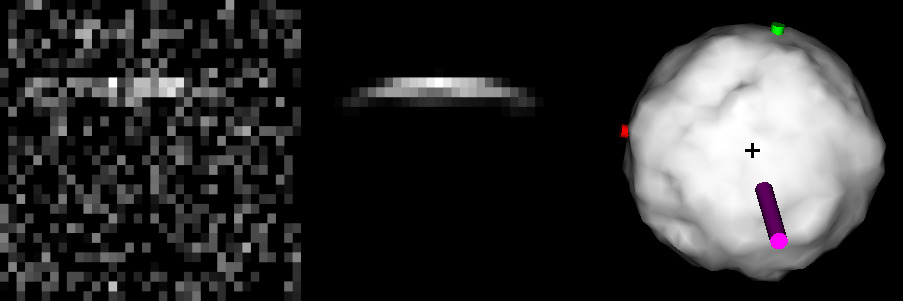}}
 
	\caption{Same as Fig.~\ref{fig:dd:jan1}, but for observations taken with Arecibo on the night beginning on 5th Januray 2017 and prograde model only.
		\label{fig:dd:pro:jan5}  }
\end{figure}

\begin{figure}
	\resizebox{\hsize}{!}{\includegraphics{radar/radar_data.jpg}\includegraphics{radar/radar_fit.jpg}\includegraphics{radar/radar_sky.jpg}\includegraphics{radar/radar_data.jpg}\includegraphics{radar/radar_fit.jpg}\includegraphics{radar/radar_sky.jpg}}
    \resizebox{\hsize}{!}{\includegraphics{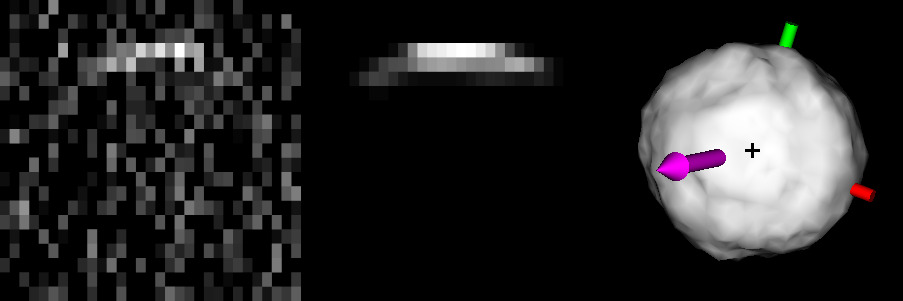}\includegraphics{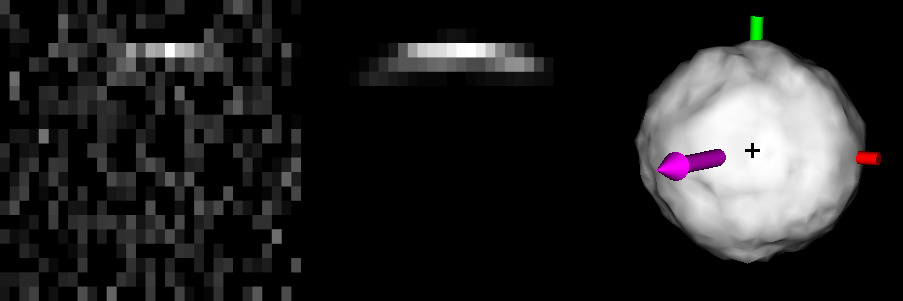}}
    \resizebox{\hsize}{!}{\includegraphics{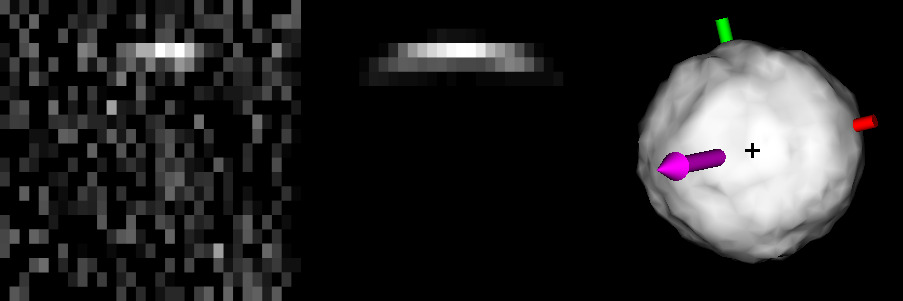}\includegraphics{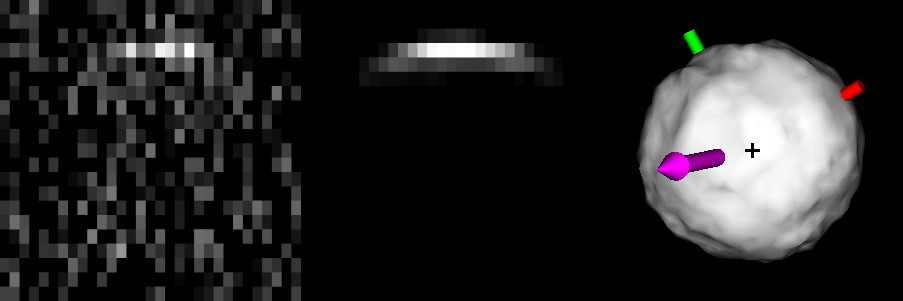}}
    \resizebox{\hsize}{!}{\includegraphics{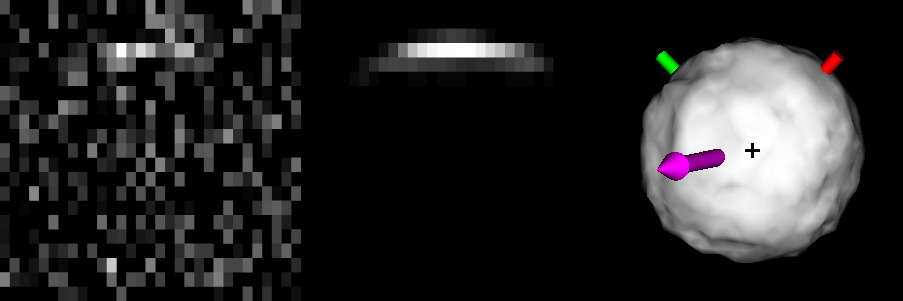}\includegraphics{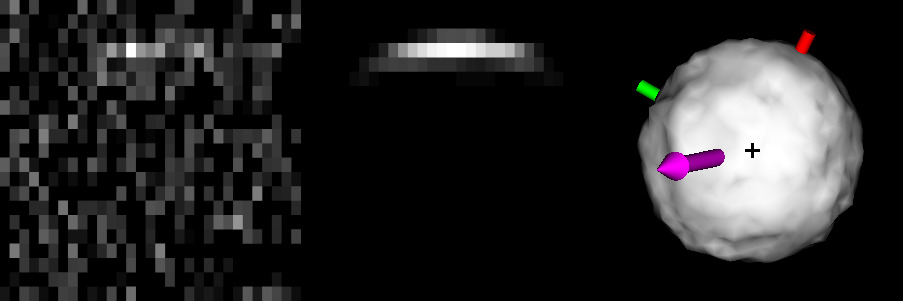}}
    \resizebox{\hsize}{!}{\includegraphics{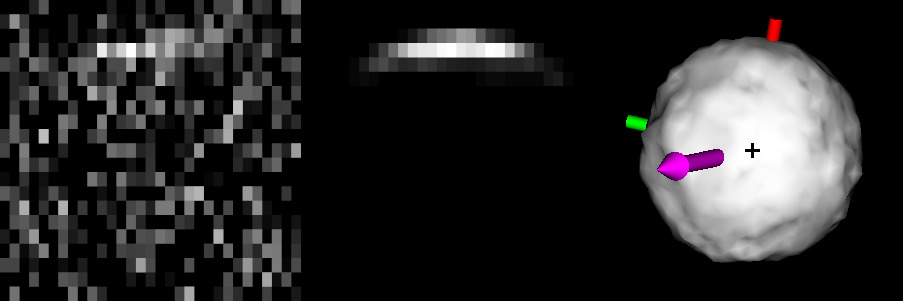}\includegraphics{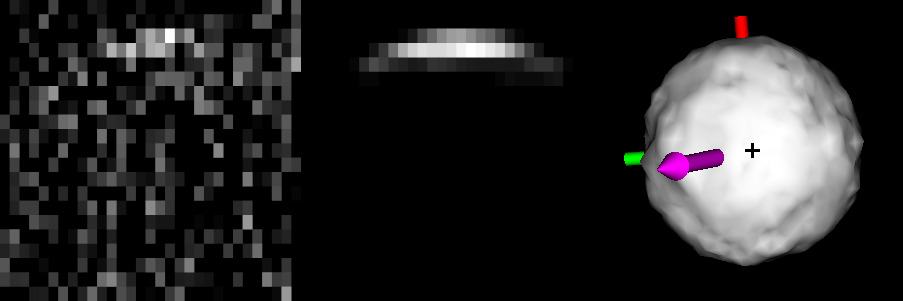}}
    \resizebox{\hsize}{!}{\includegraphics{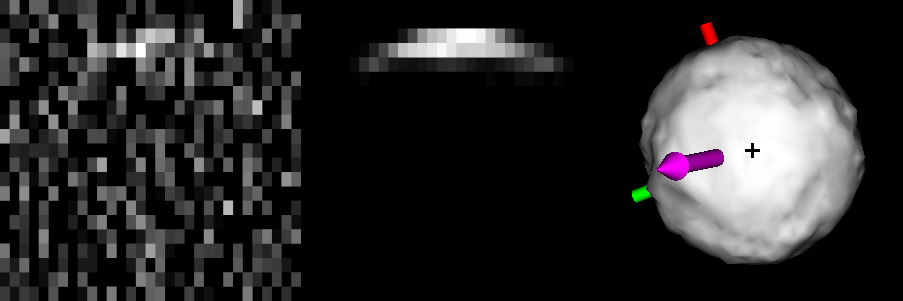}\includegraphics{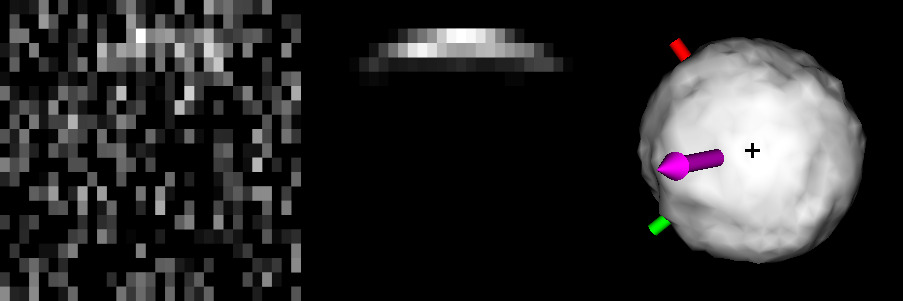}}
    \resizebox{\hsize}{!}{\includegraphics{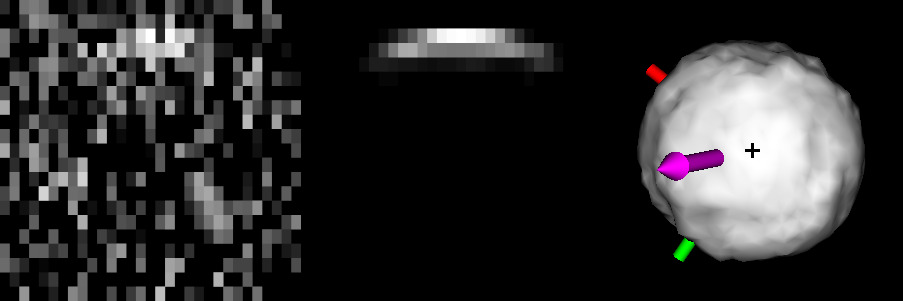}\includegraphics{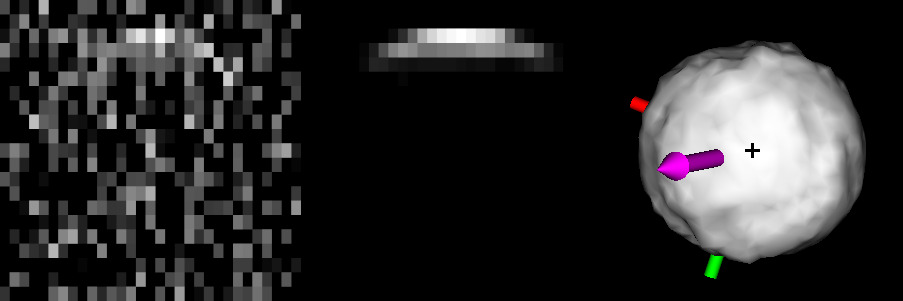}}
 
	\caption{Same as Fig.~\ref{fig:dd:jan1}, but for observations taken with Arecibo on the night beginning on 6th January 2017 and retrograde model only.
		\label{fig:dd:retro:jan6}  }
\end{figure}

\begin{figure}
	\resizebox{\hsize}{!}{\includegraphics{radar/radar_data.jpg}\includegraphics{radar/radar_fit.jpg}\includegraphics{radar/radar_sky.jpg}\includegraphics{radar/radar_data.jpg}\includegraphics{radar/radar_fit.jpg}\includegraphics{radar/radar_sky.jpg}}
    \resizebox{\hsize}{!}{\includegraphics{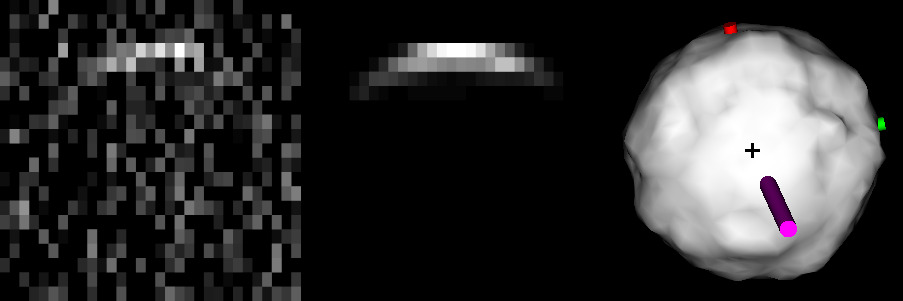}\includegraphics{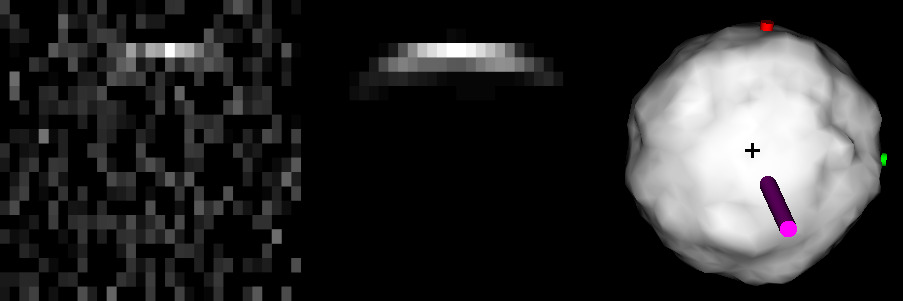}}
    \resizebox{\hsize}{!}{\includegraphics{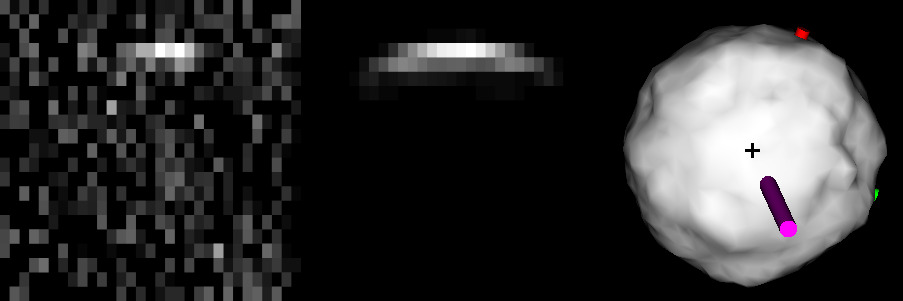}\includegraphics{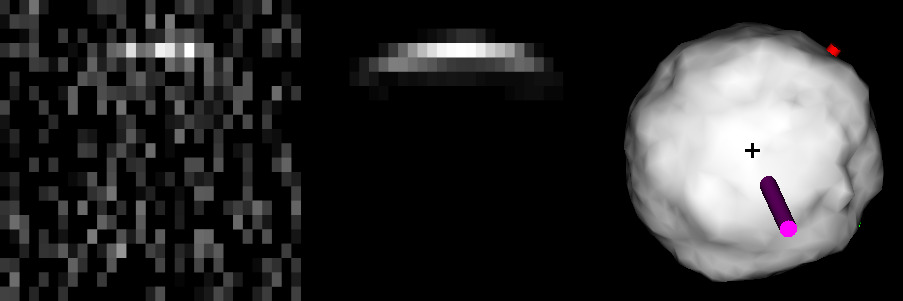}}
    \resizebox{\hsize}{!}{\includegraphics{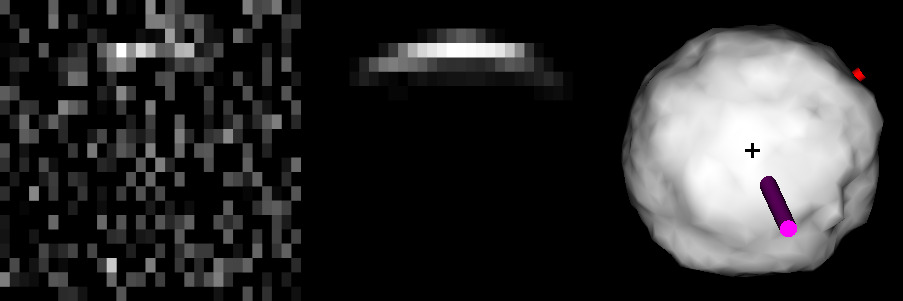}\includegraphics{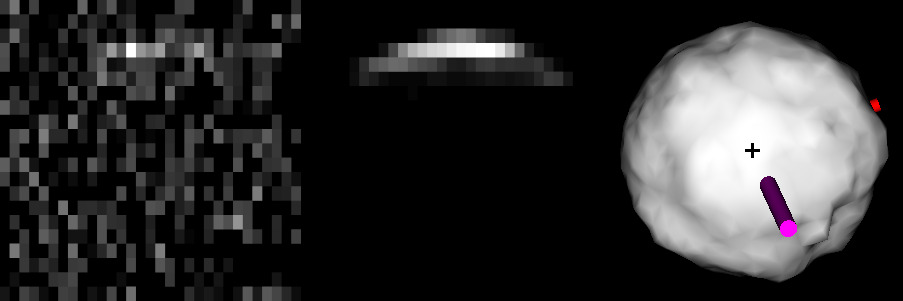}}
    \resizebox{\hsize}{!}{\includegraphics{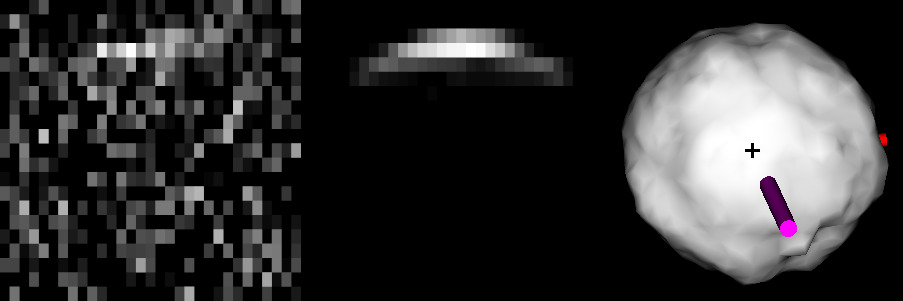}\includegraphics{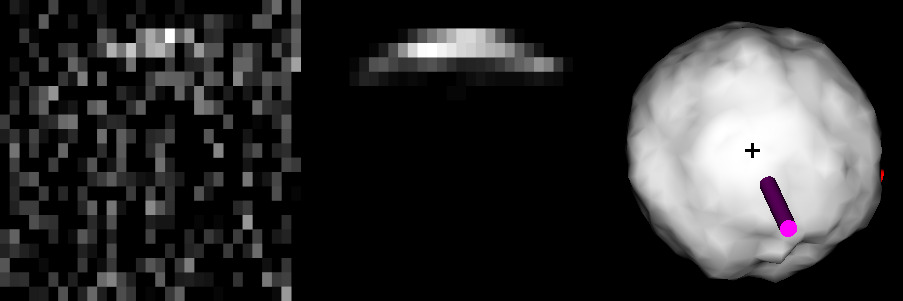}}
    \resizebox{\hsize}{!}{\includegraphics{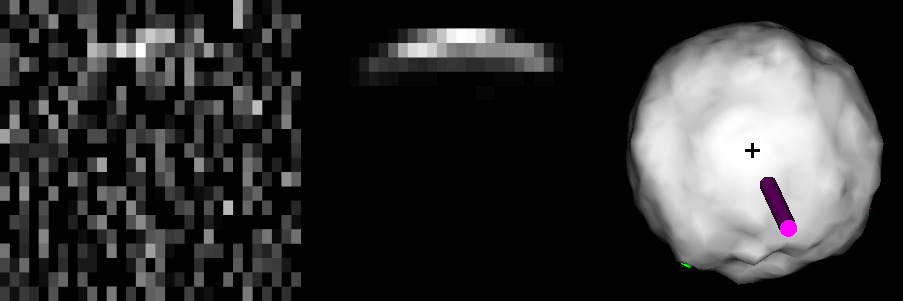}\includegraphics{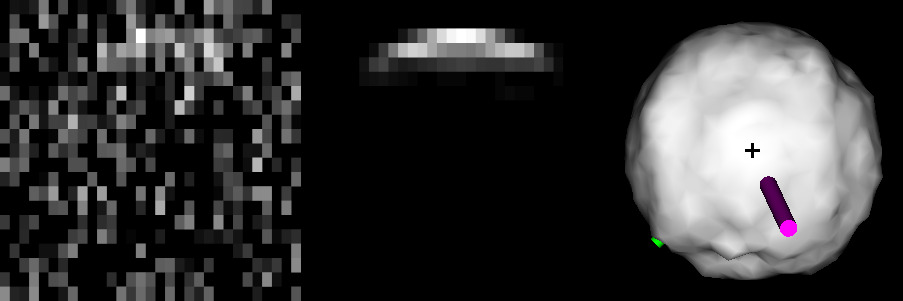}}
    \resizebox{\hsize}{!}{\includegraphics{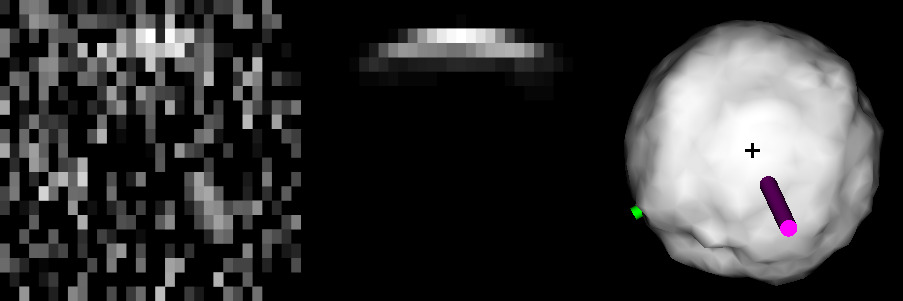}\includegraphics{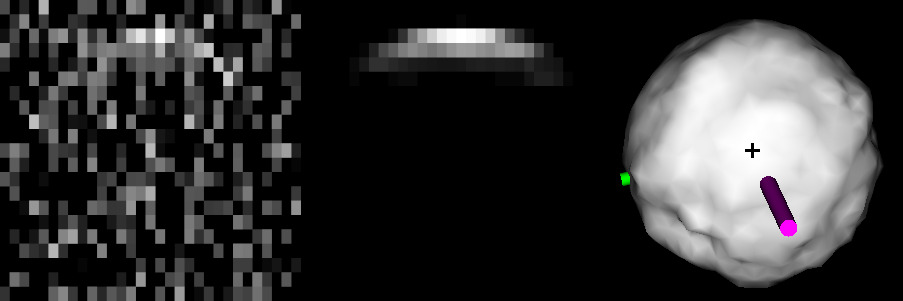}}
 
	\caption{Same as Fig.~\ref{fig:dd:jan1}, but for observations taken with Arecibo on the night beginning on 6th January 2017 and prograde model only.
		\label{fig:dd:pro:jan6}  }
\end{figure}

\begin{figure}
	\resizebox{\hsize}{!}{\includegraphics{radar/radar_data.jpg}\includegraphics{radar/radar_fit.jpg}\includegraphics{radar/radar_sky.jpg}\includegraphics{radar/radar_data.jpg}\includegraphics{radar/radar_fit.jpg}\includegraphics{radar/radar_sky.jpg}}
    \resizebox{\hsize}{!}{\includegraphics{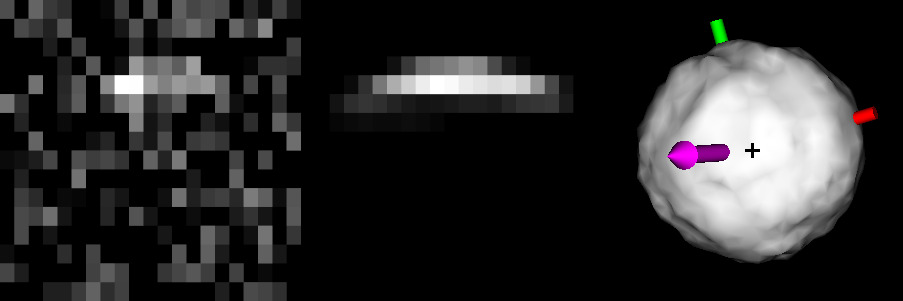}\includegraphics{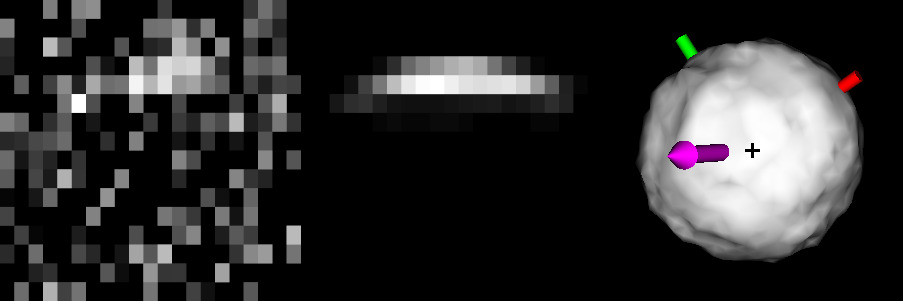}}
    \resizebox{\hsize}{!}{\includegraphics{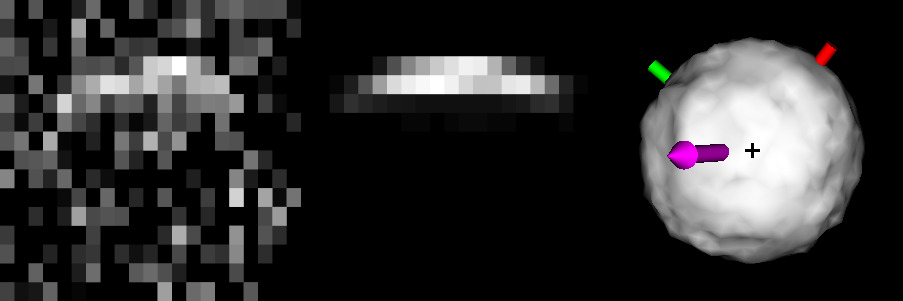}\includegraphics{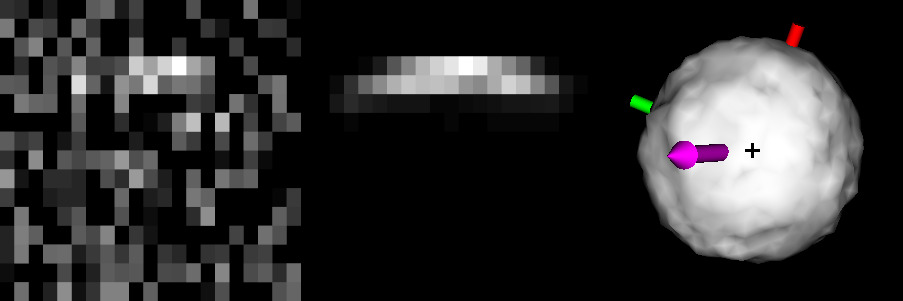}}
    \resizebox{\hsize}{!}{\includegraphics{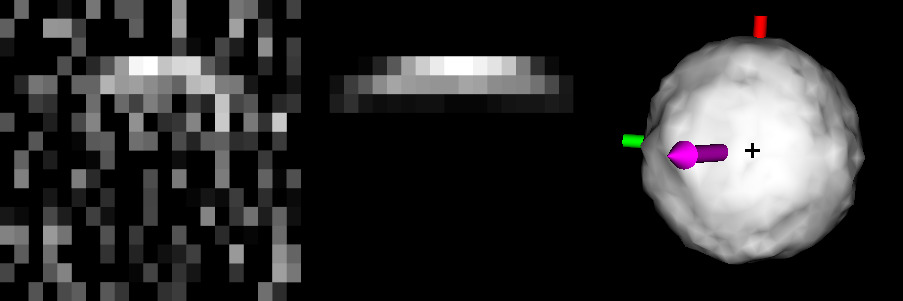}\includegraphics{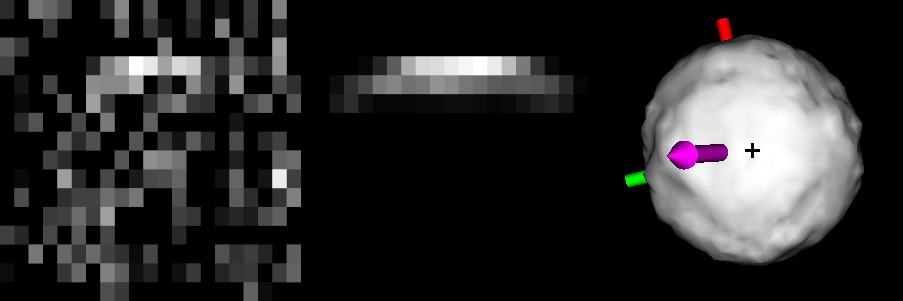}}
 
	\caption{Same as Fig.~\ref{fig:dd:jan1}, but for observations taken with Arecibo on the night beginning on 7th January 2017 and retrograde model only. As the object moved away from Earth the Doppler resolution of images decreased significantly relative to earlier observations. 
		\label{fig:dd:retro:jan7}  }
\end{figure}

\begin{figure}
	\resizebox{\hsize}{!}{\includegraphics{radar/radar_data.jpg}\includegraphics{radar/radar_fit.jpg}\includegraphics{radar/radar_sky.jpg}\includegraphics{radar/radar_data.jpg}\includegraphics{radar/radar_fit.jpg}\includegraphics{radar/radar_sky.jpg}}
    \resizebox{\hsize}{!}{\includegraphics{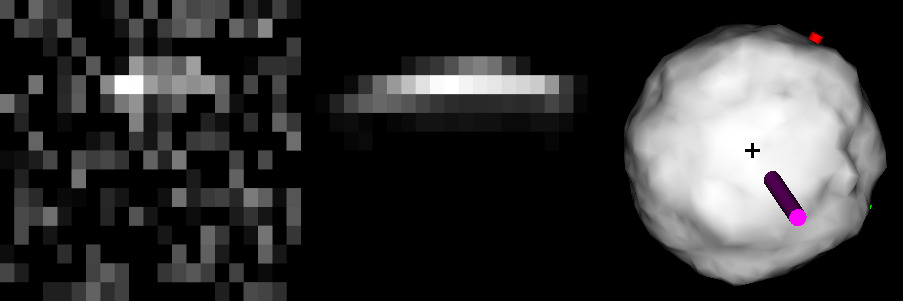}\includegraphics{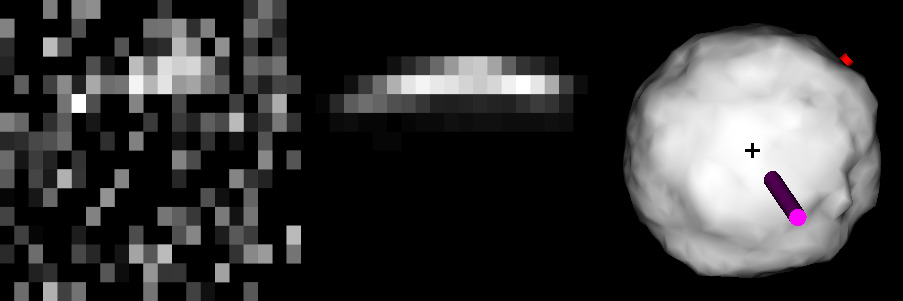}}
    \resizebox{\hsize}{!}{\includegraphics{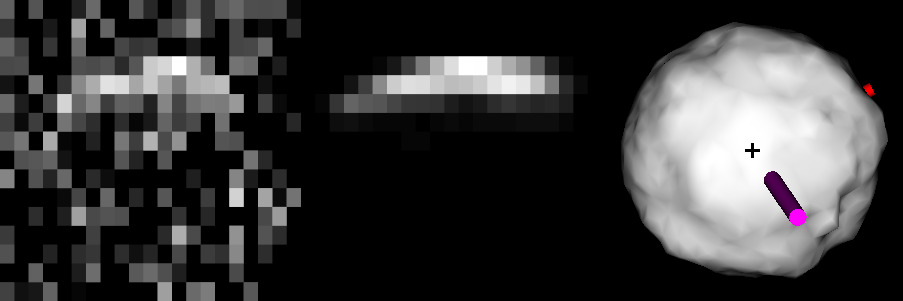}\includegraphics{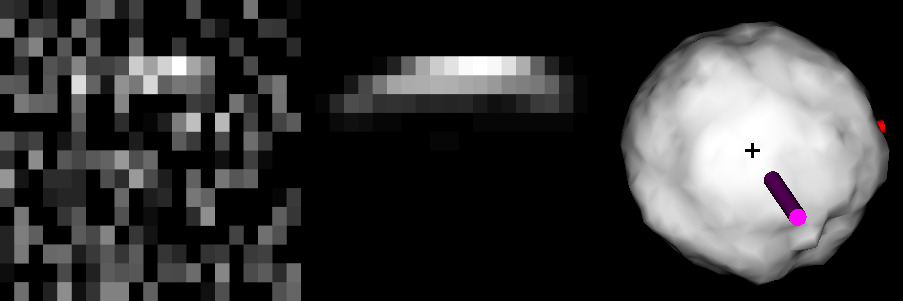}}
    \resizebox{\hsize}{!}{\includegraphics{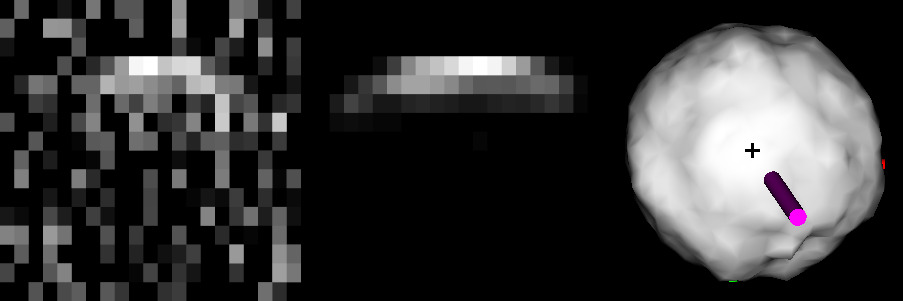}\includegraphics{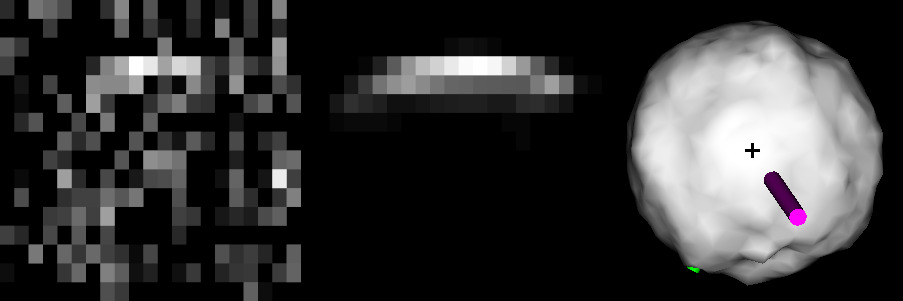}}
 
	\caption{Same as Fig.~\ref{fig:dd:jan1}, but for observations taken with Arecibo on the night beginning on 7th January 2017 and prograde model only.
		\label{fig:dd:pro:jan7}  }
\end{figure}

\begin{figure*}
	\resizebox{\hsize}{!}{
		\includegraphics[width=.48\textwidth, trim=2cm 4cm 3.8cm 4cm, clip=true]{LC/2102_lat-30lon180_v190906_20220204_newnewT0_01_fix.pdf} 
		\includegraphics[width=.48\textwidth, trim=2cm 4cm 3.8cm 4cm, clip=true]{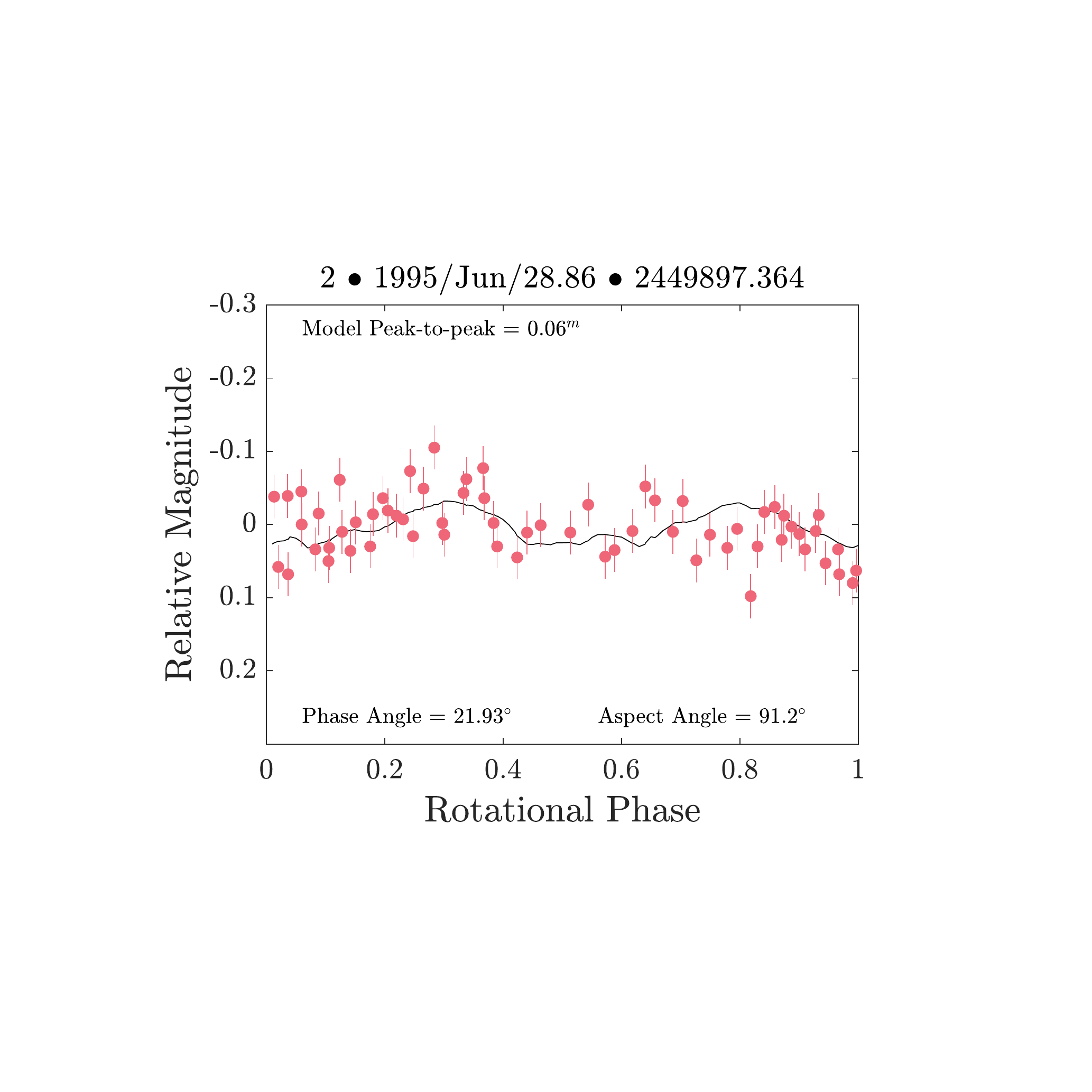} 		
		\includegraphics[width=.48\textwidth, trim=2cm 4cm 3.8cm 4cm, clip=true]{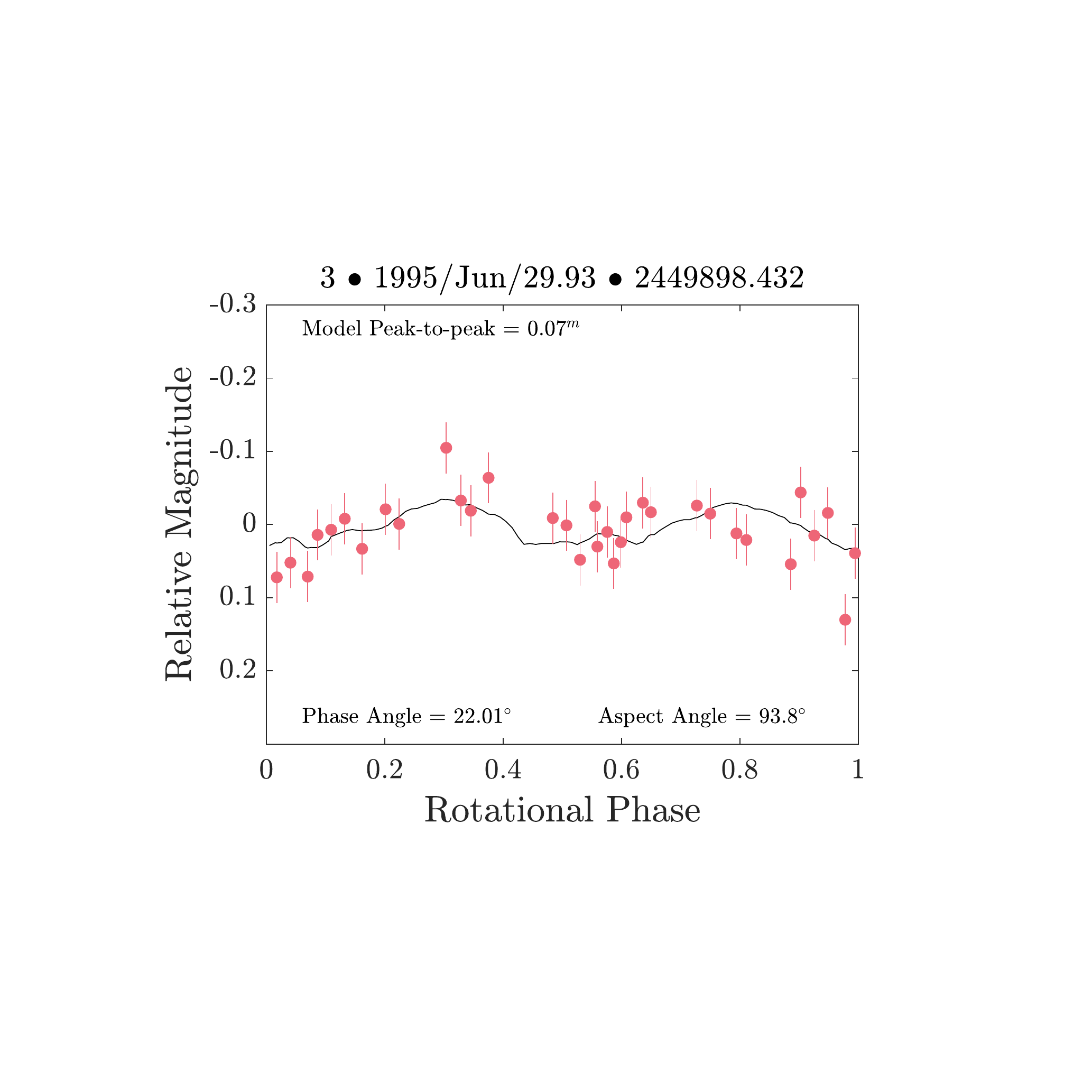} 
	}
	
	\resizebox{\hsize}{!}{
		\includegraphics[width=.48\textwidth, trim=2cm 4cm 3.8cm 4cm, clip=true]{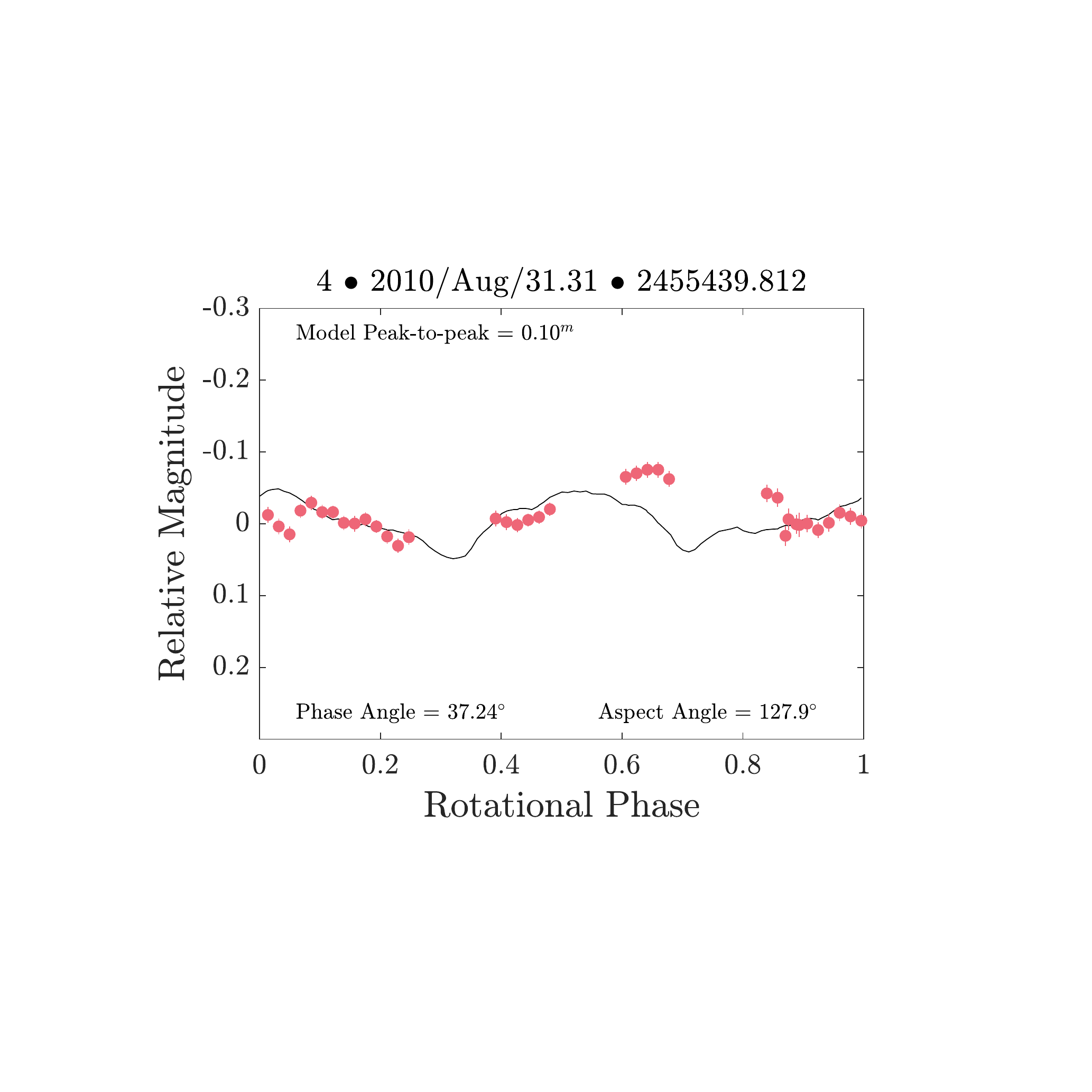} 
		\includegraphics[width=.48\textwidth, trim=2cm 4cm 3.8cm 4cm, clip=true]{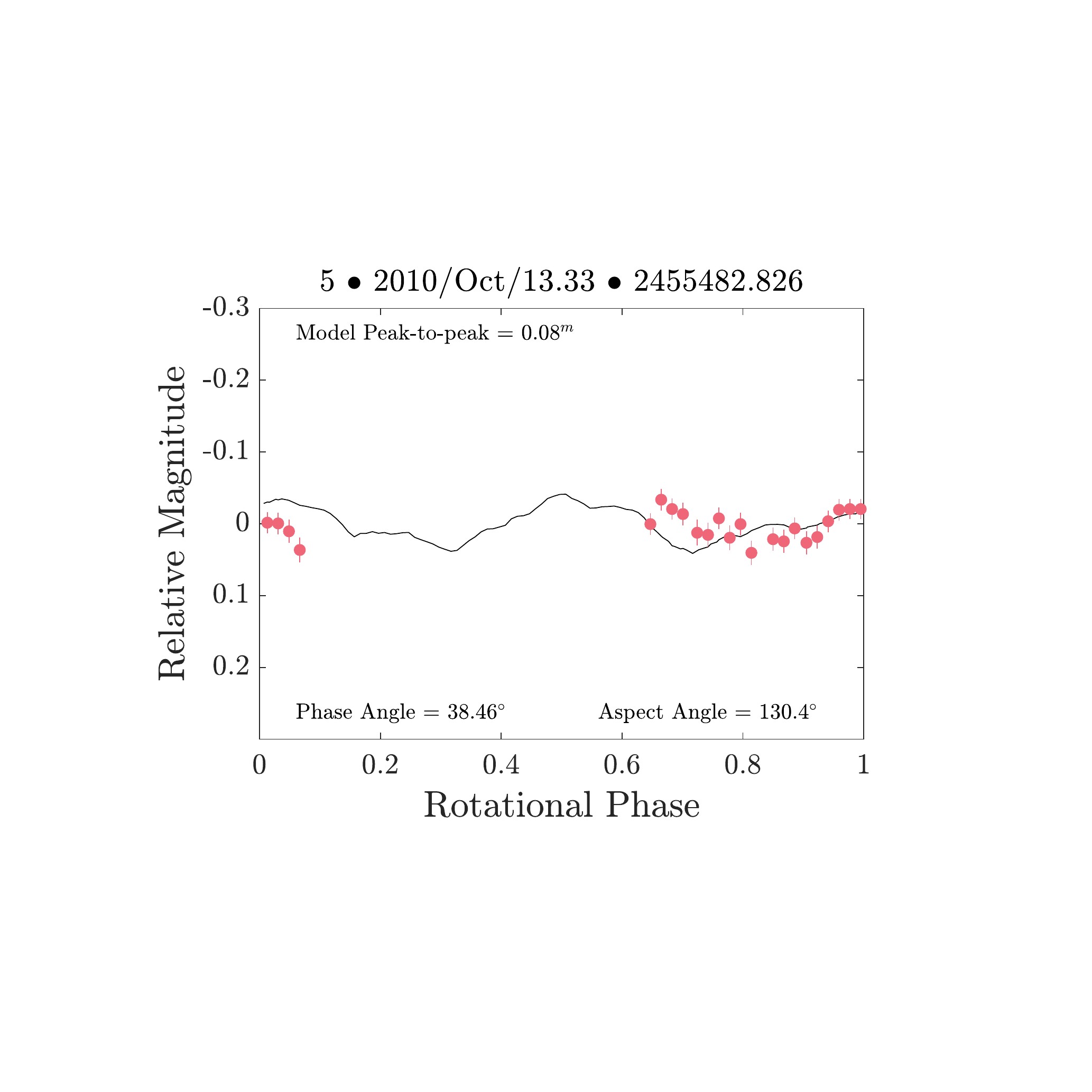} 		
		\includegraphics[width=.48\textwidth, trim=2cm 4cm 3.8cm 4cm, clip=true]{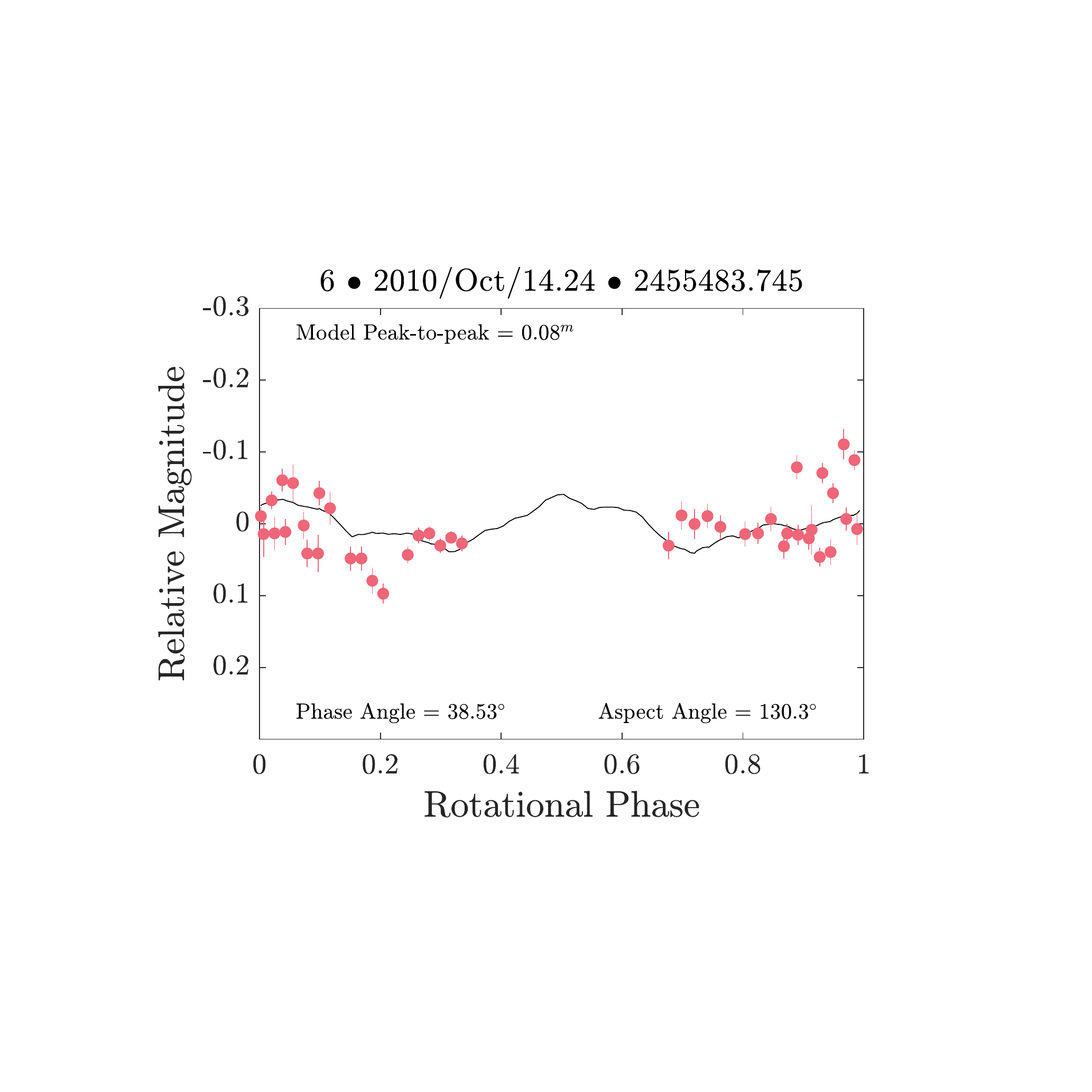} 
	}
	
	\resizebox{\hsize}{!}{
		\includegraphics[width=.48\textwidth, trim=2cm 4cm 3.8cm 4cm, clip=true]{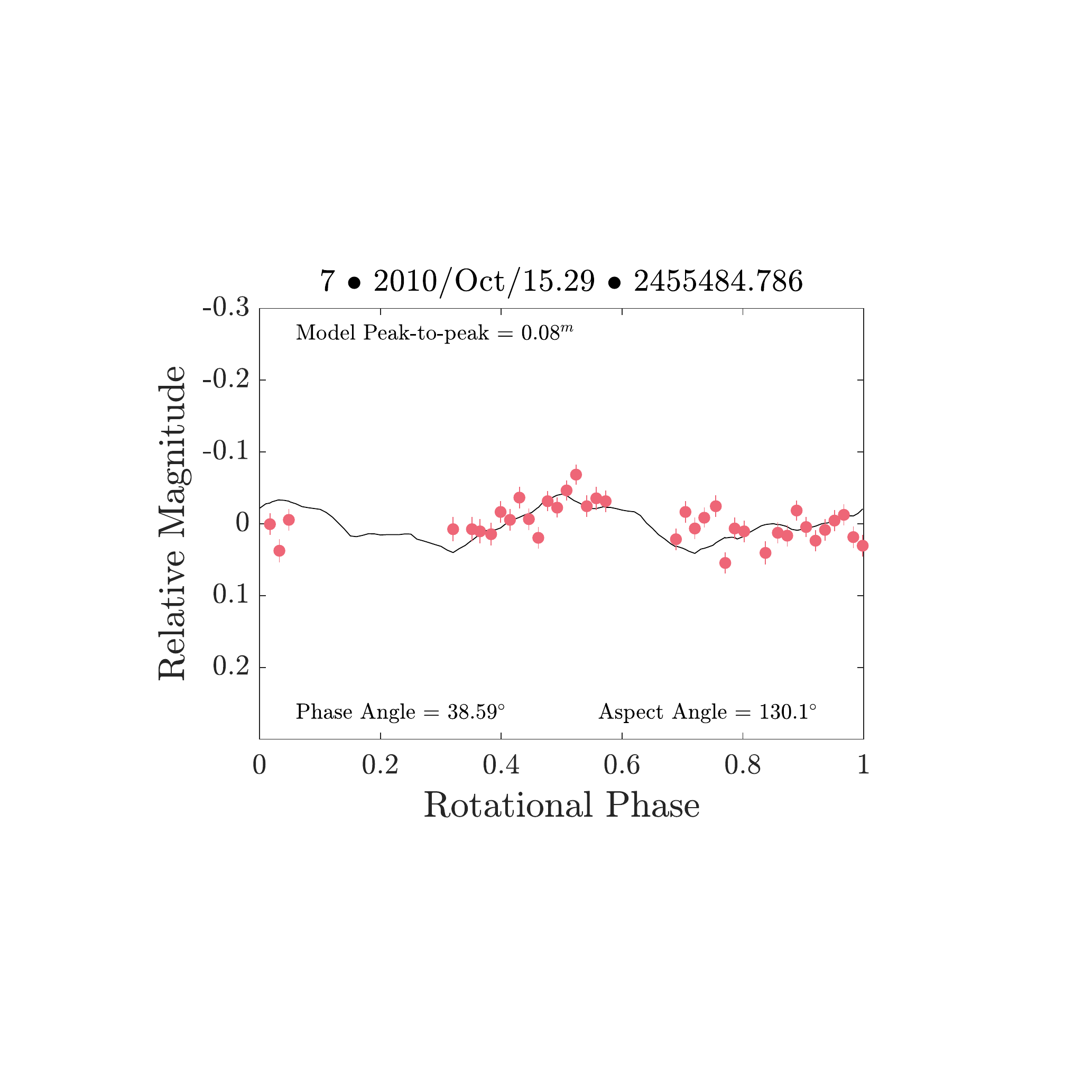} 
		\includegraphics[width=.48\textwidth, trim=2cm 4cm 3.8cm 4cm, clip=true]{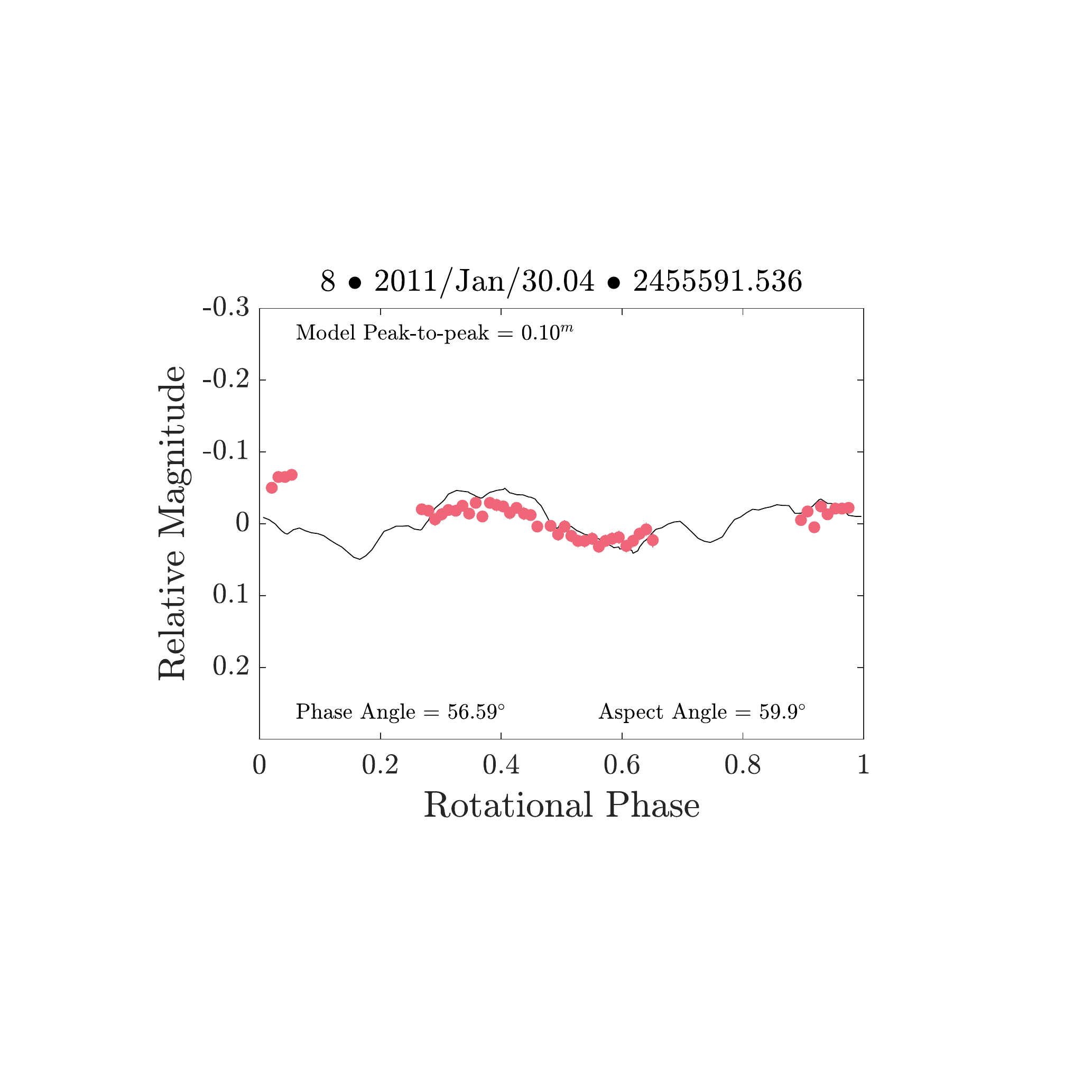} 		
		\includegraphics[width=.48\textwidth, trim=2cm 4cm 3.8cm 4cm, clip=true]{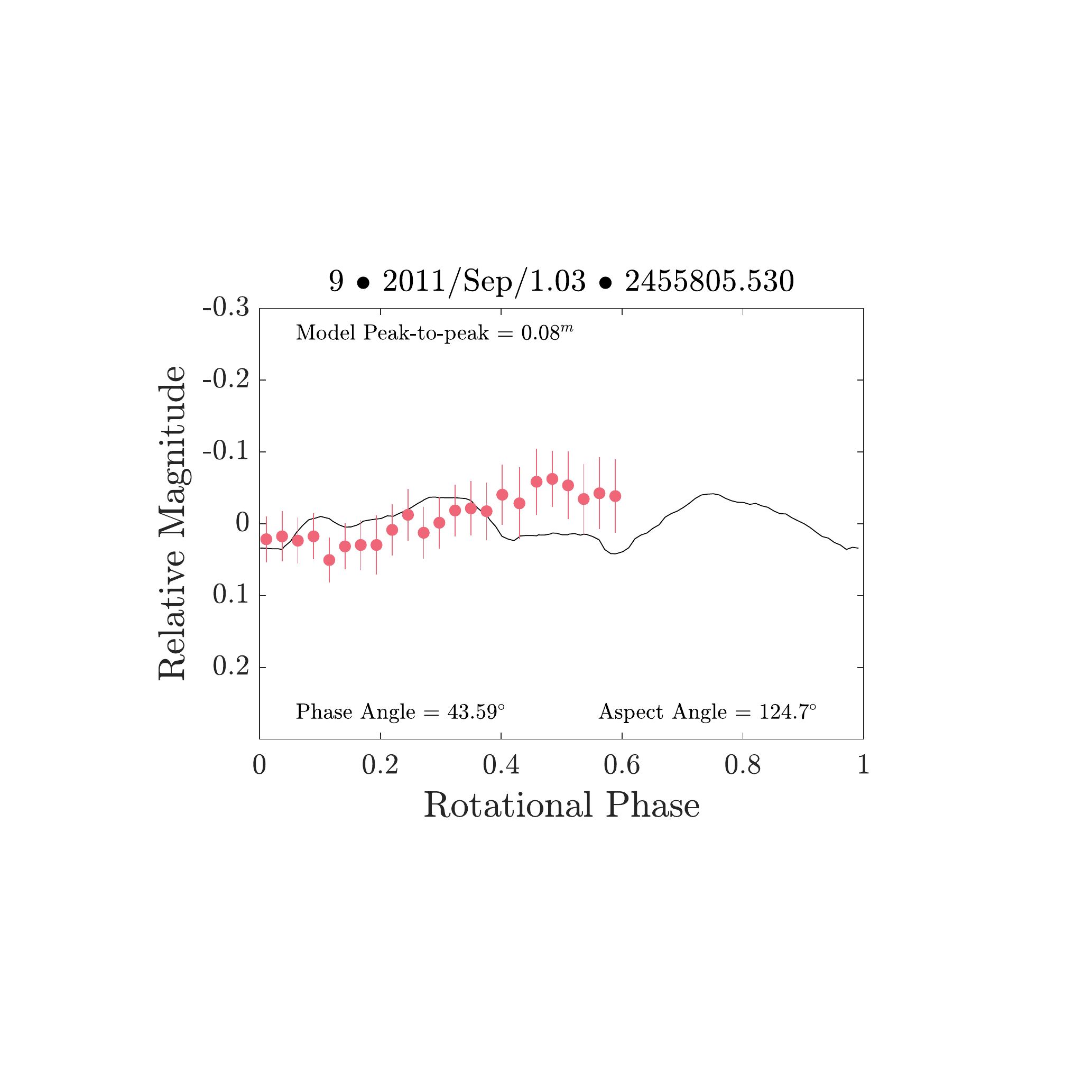} 
	}

	\resizebox{\hsize}{!}{
		\includegraphics[width=.48\textwidth, trim=2cm 4cm 3.8cm 4cm, clip=true]{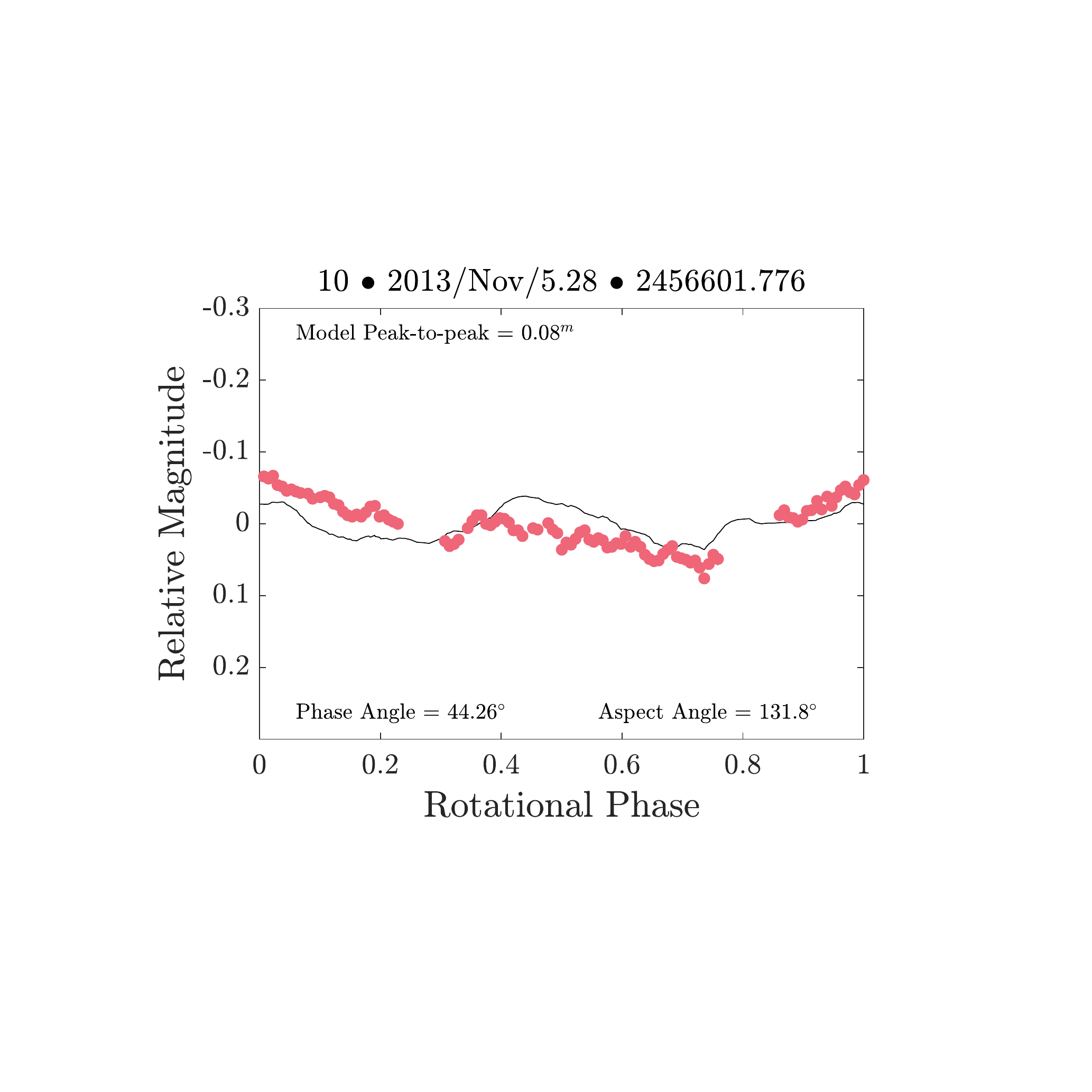} 
		\includegraphics[width=.48\textwidth, trim=2cm 4cm 3.8cm 4cm, clip=true]{LC/2102_lat-30lon180_v190906_20220204_newnewT0_11_fix.pdf} 		
		\includegraphics[width=.48\textwidth, trim=2cm 4cm 3.8cm 4cm, clip=true]{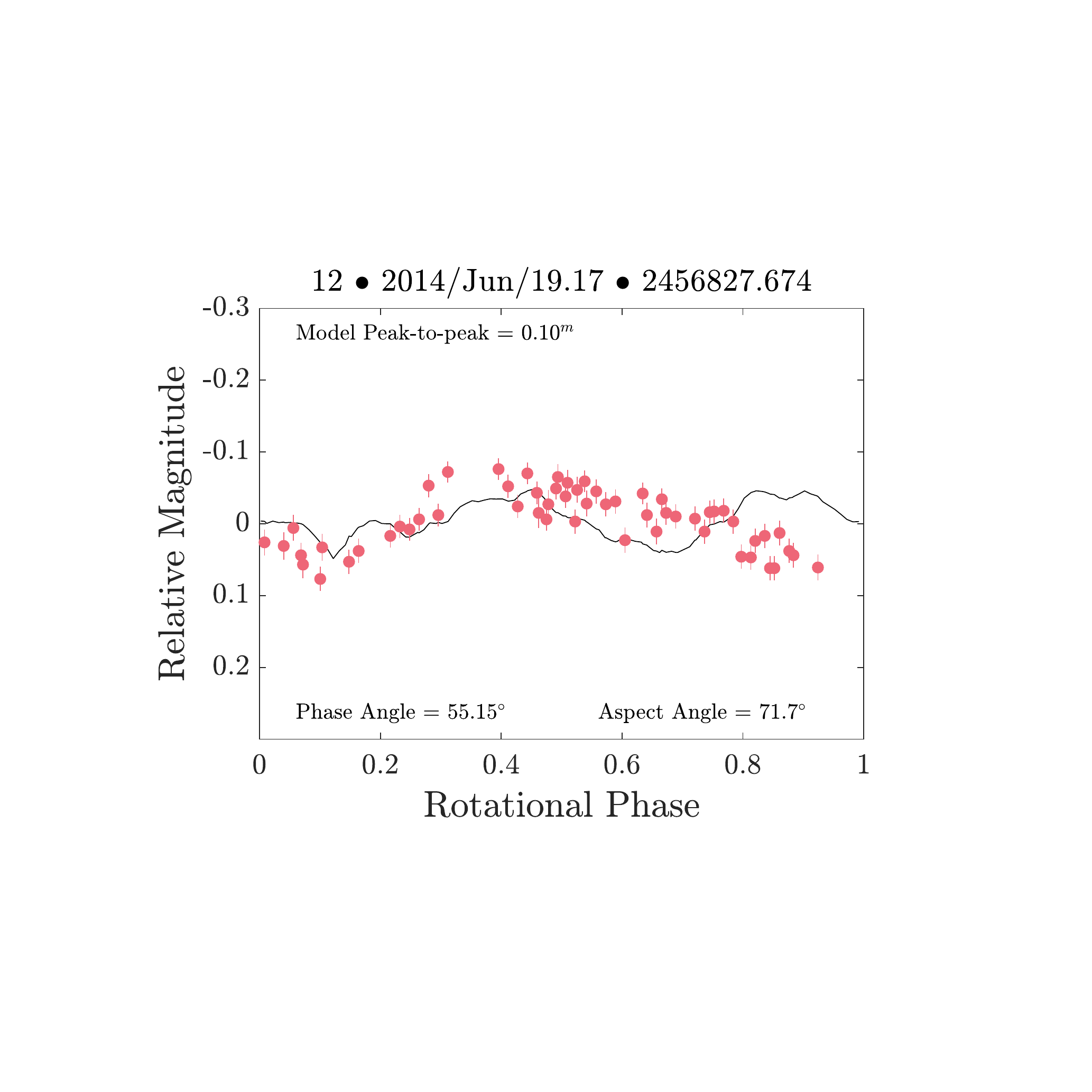} 
	}
	
				\caption{All available data plotted over synthetic light curves generated with the retrograde radar shape model of asteroid (2102) Tantalus. This shape model utilised a subset of available optical light-curves and the radar data. 
		\label{fig:radar-retro1}}
		
\end{figure*}

\addtocounter{figure}{-1}

\begin{figure*}
	\resizebox{\hsize}{!}{
		\includegraphics[width=.48\textwidth, trim=2cm 4cm 3.8cm 4cm, clip=true]{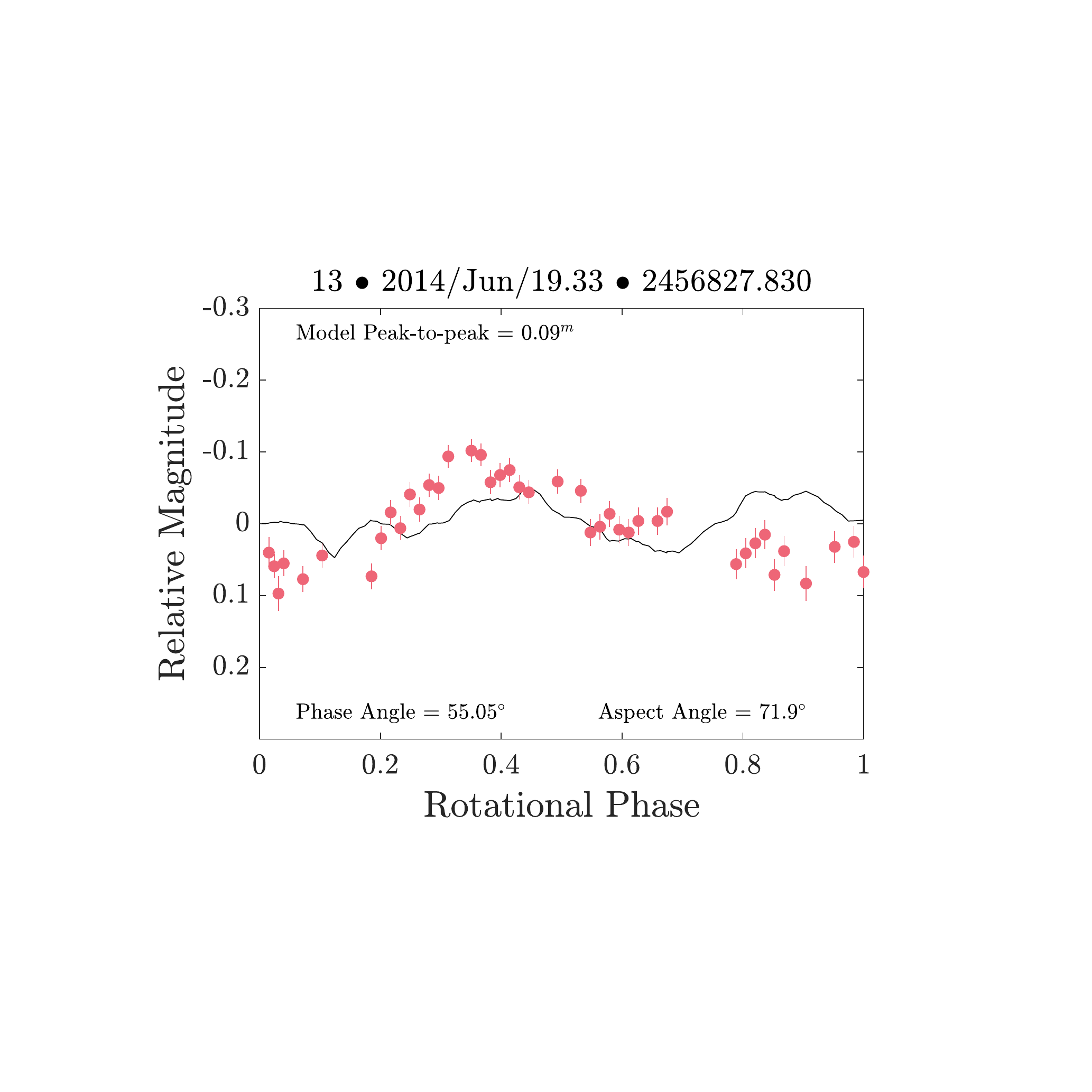} 
		\includegraphics[width=.48\textwidth, trim=2cm 4cm 3.8cm 4cm, clip=true]{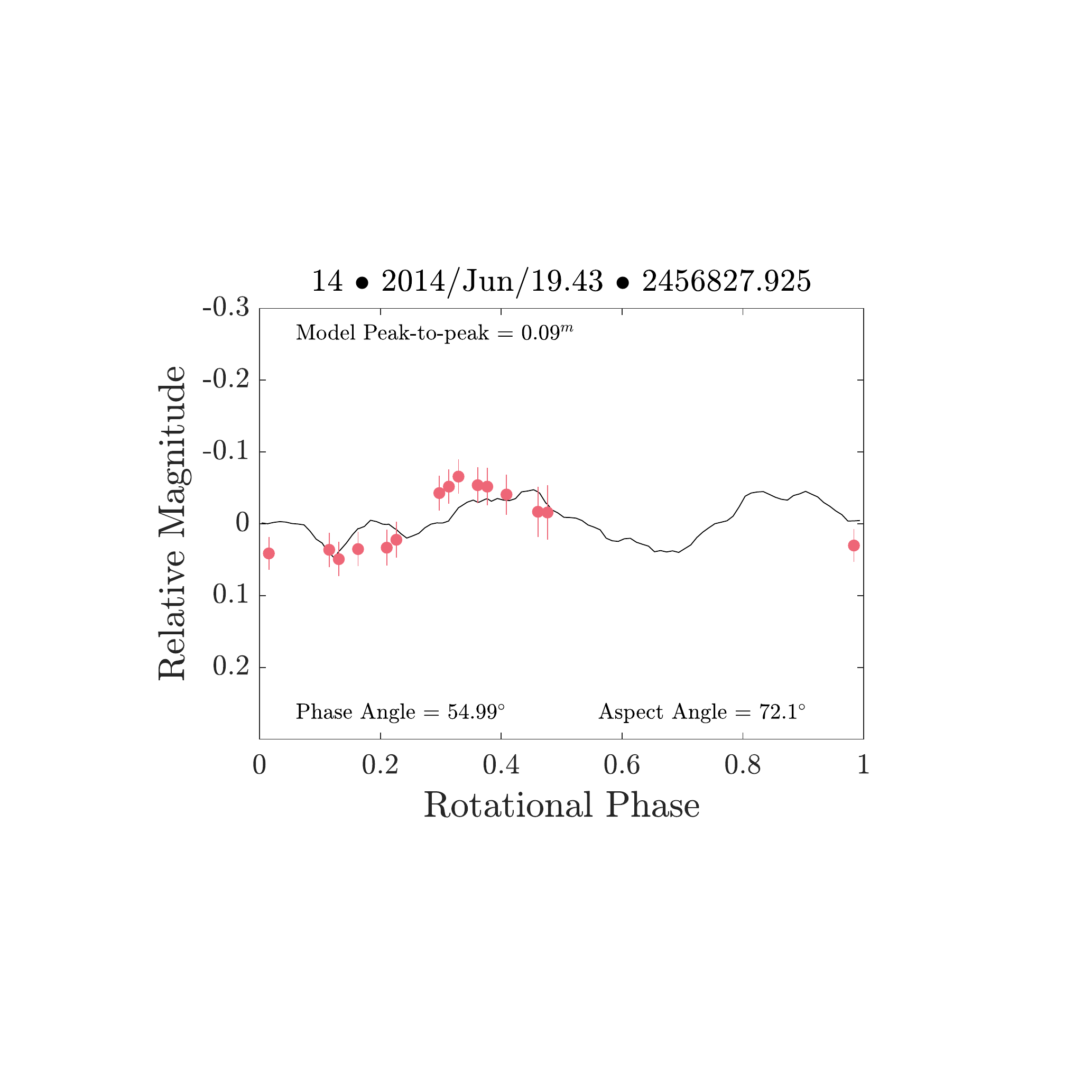} 		
		\includegraphics[width=.48\textwidth, trim=2cm 4cm 3.8cm 4cm, clip=true]{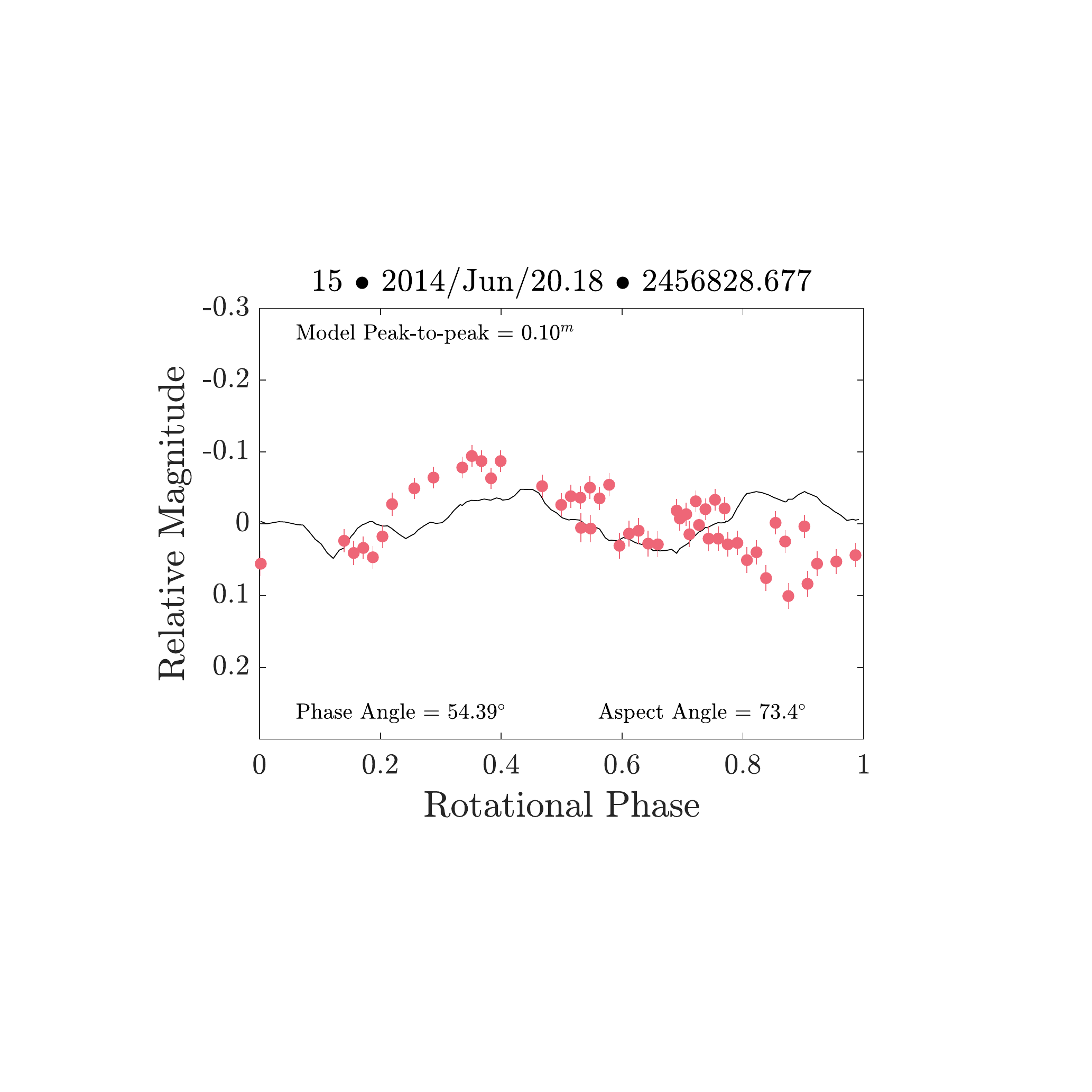} 
	}
	
	\resizebox{\hsize}{!}{
		\includegraphics[width=.48\textwidth, trim=2cm 4cm 3.8cm 4cm, clip=true]{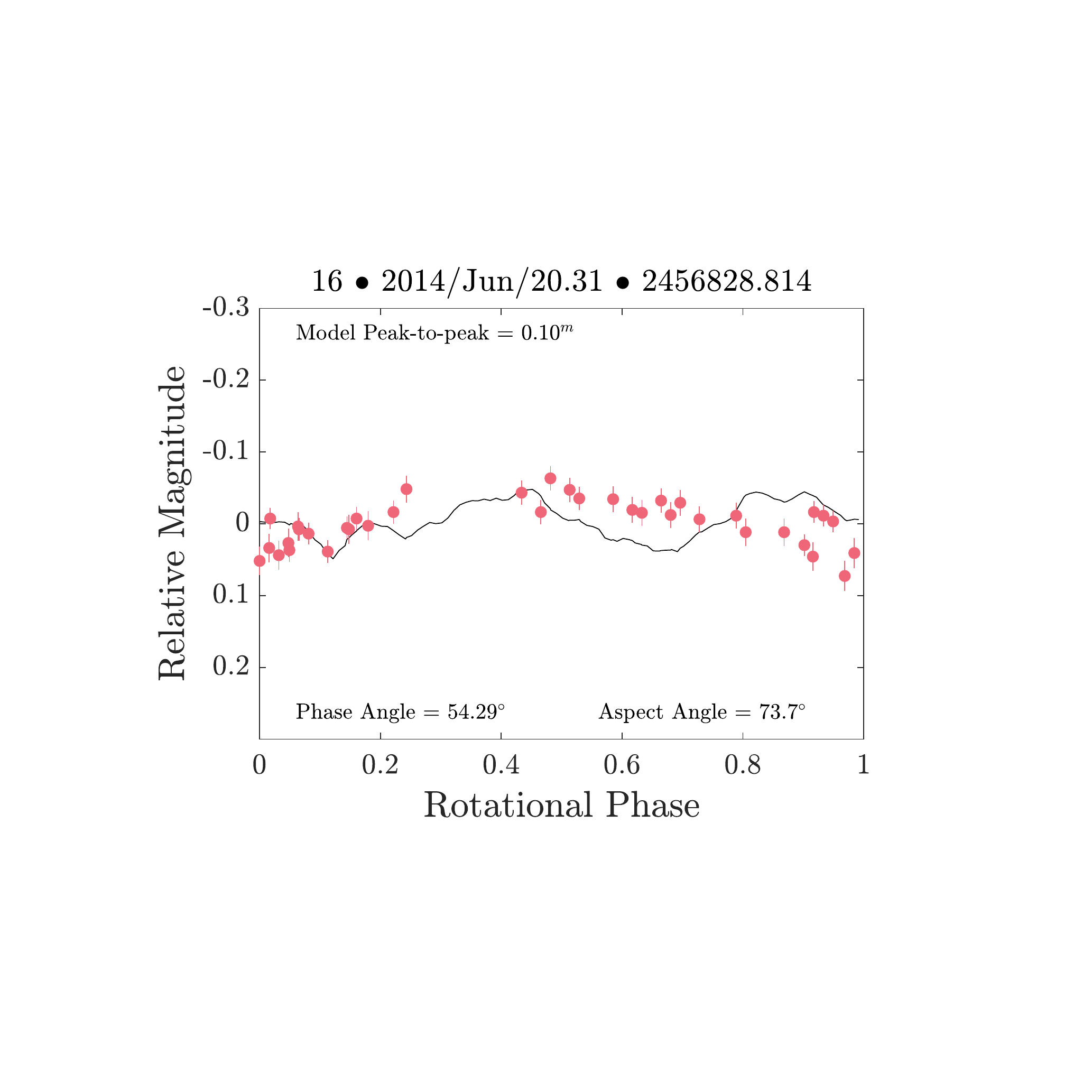} 
		\includegraphics[width=.48\textwidth, trim=2cm 4cm 3.8cm 4cm, clip=true]{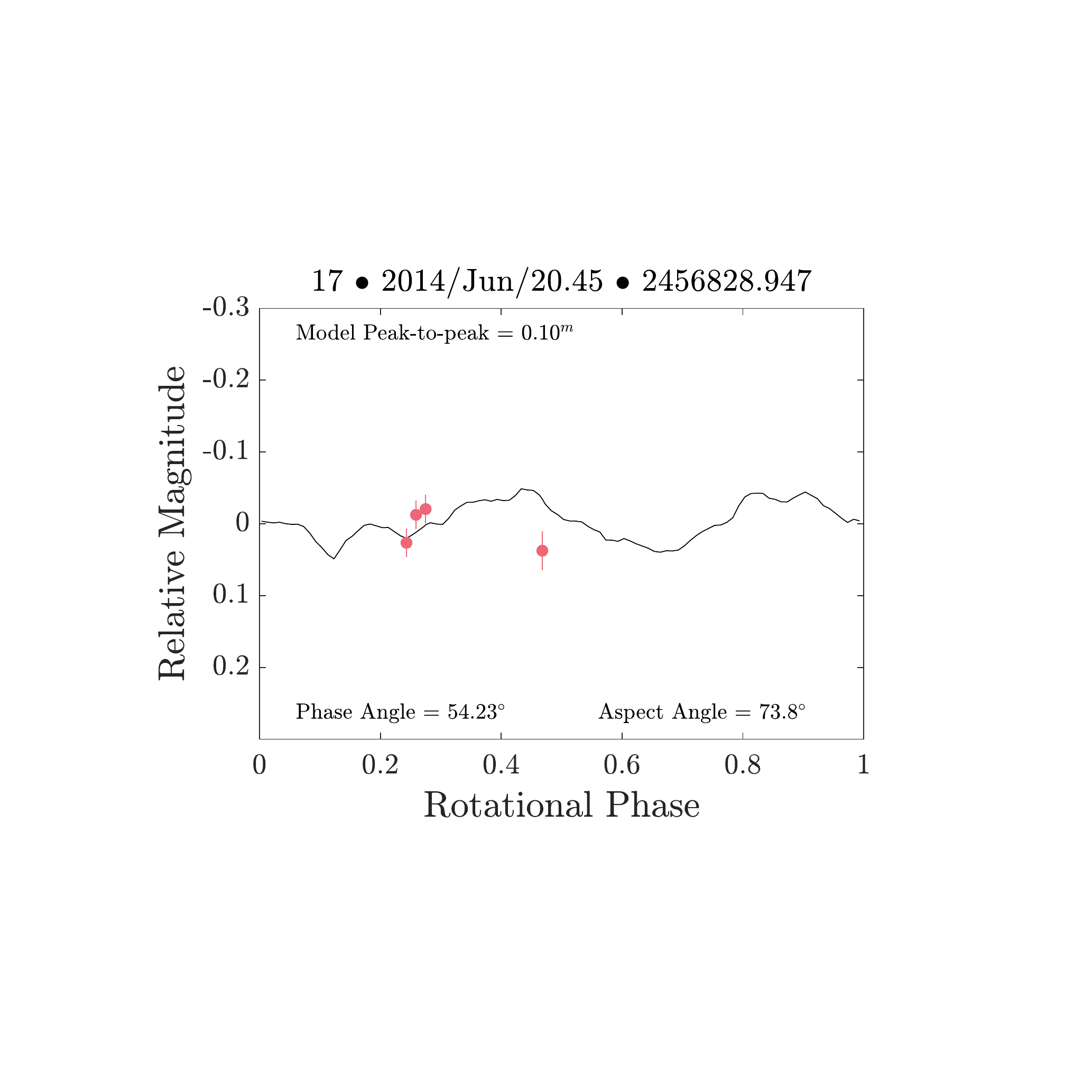} 		
		\includegraphics[width=.48\textwidth, trim=2cm 4cm 3.8cm 4cm, clip=true]{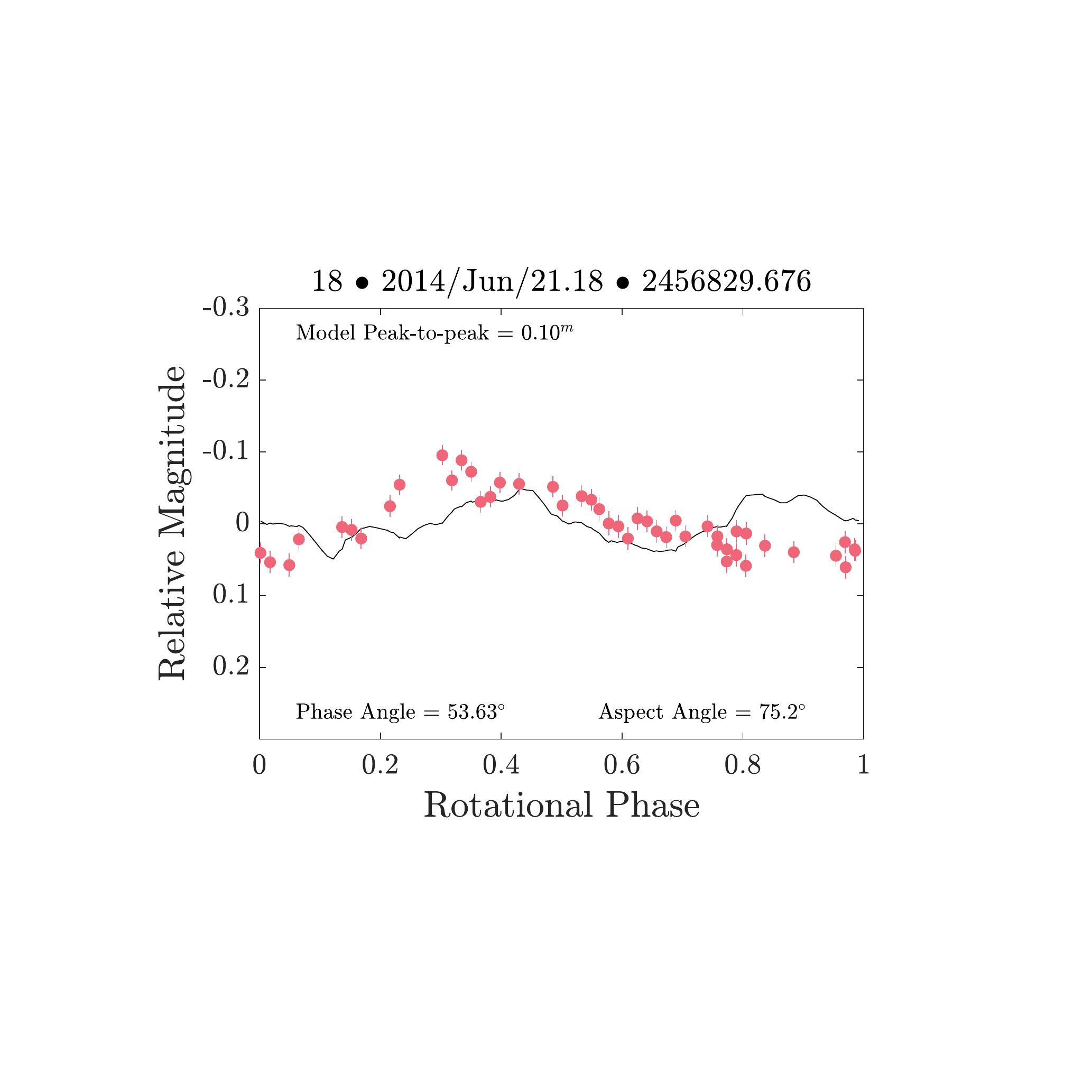} 
	}
	
	\resizebox{\hsize}{!}{
		\includegraphics[width=.48\textwidth, trim=2cm 4cm 3.8cm 4cm, clip=true]{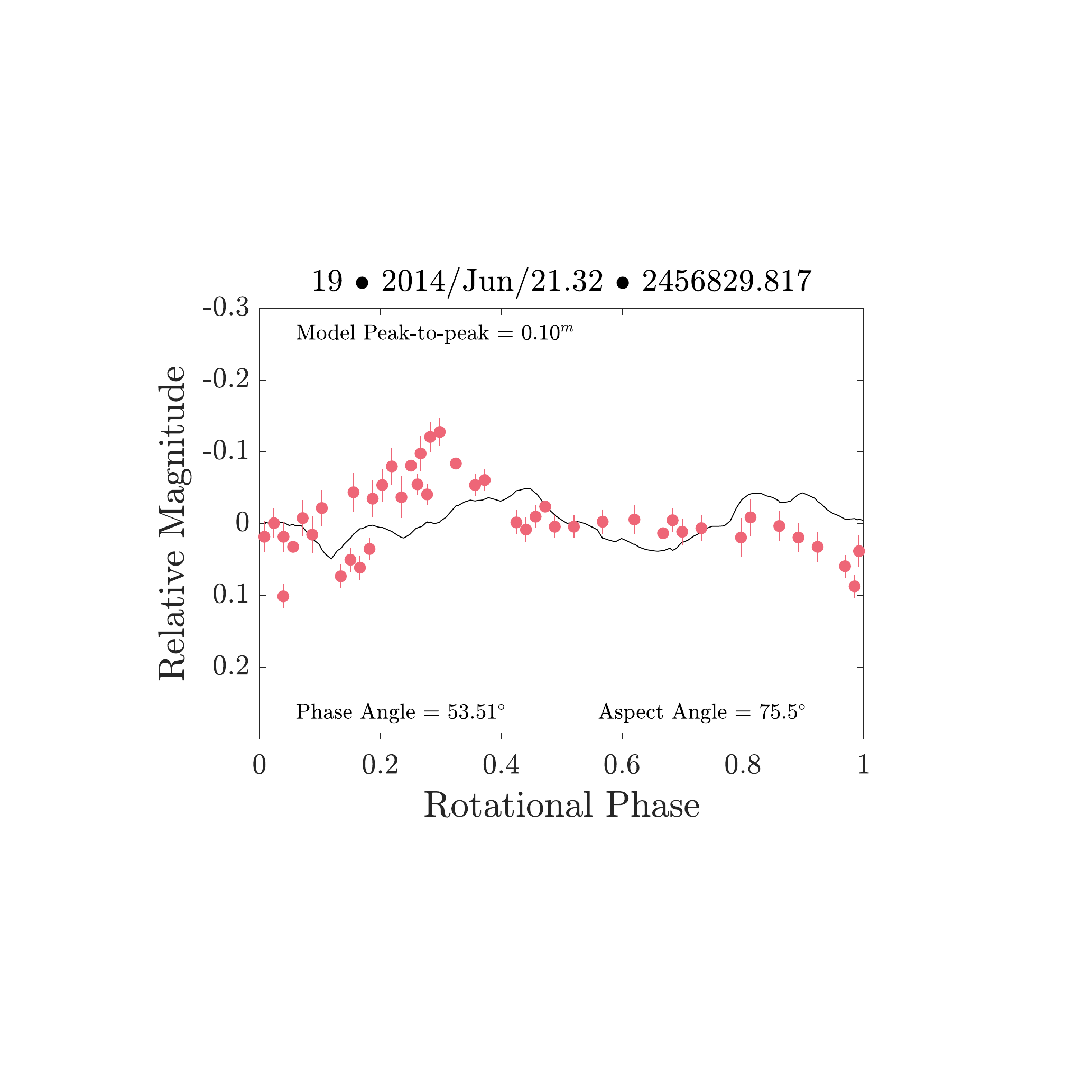} 
		\includegraphics[width=.48\textwidth, trim=2cm 4cm 3.8cm 4cm, clip=true]{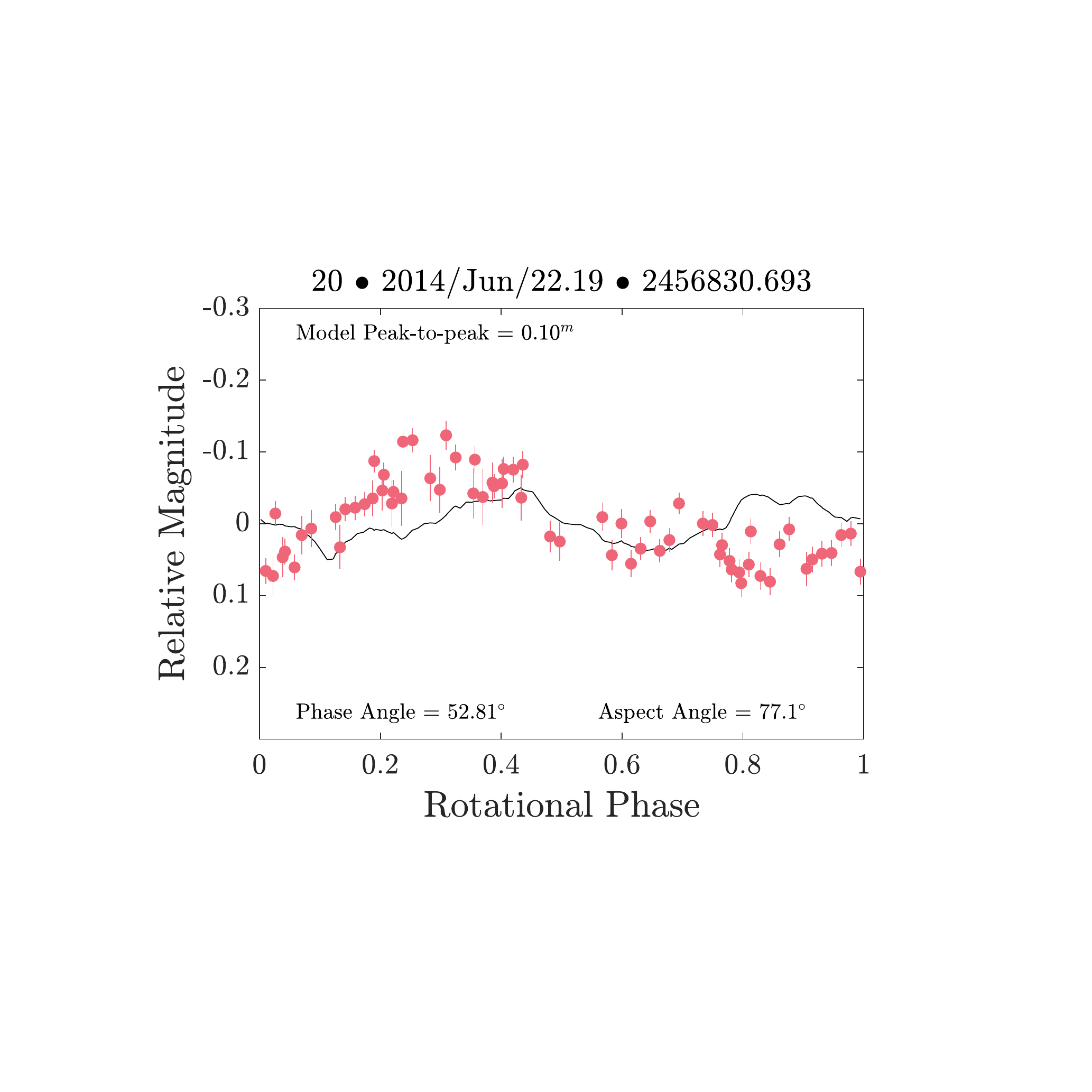} 		
		\includegraphics[width=.48\textwidth, trim=2cm 4cm 3.8cm 4cm, clip=true]{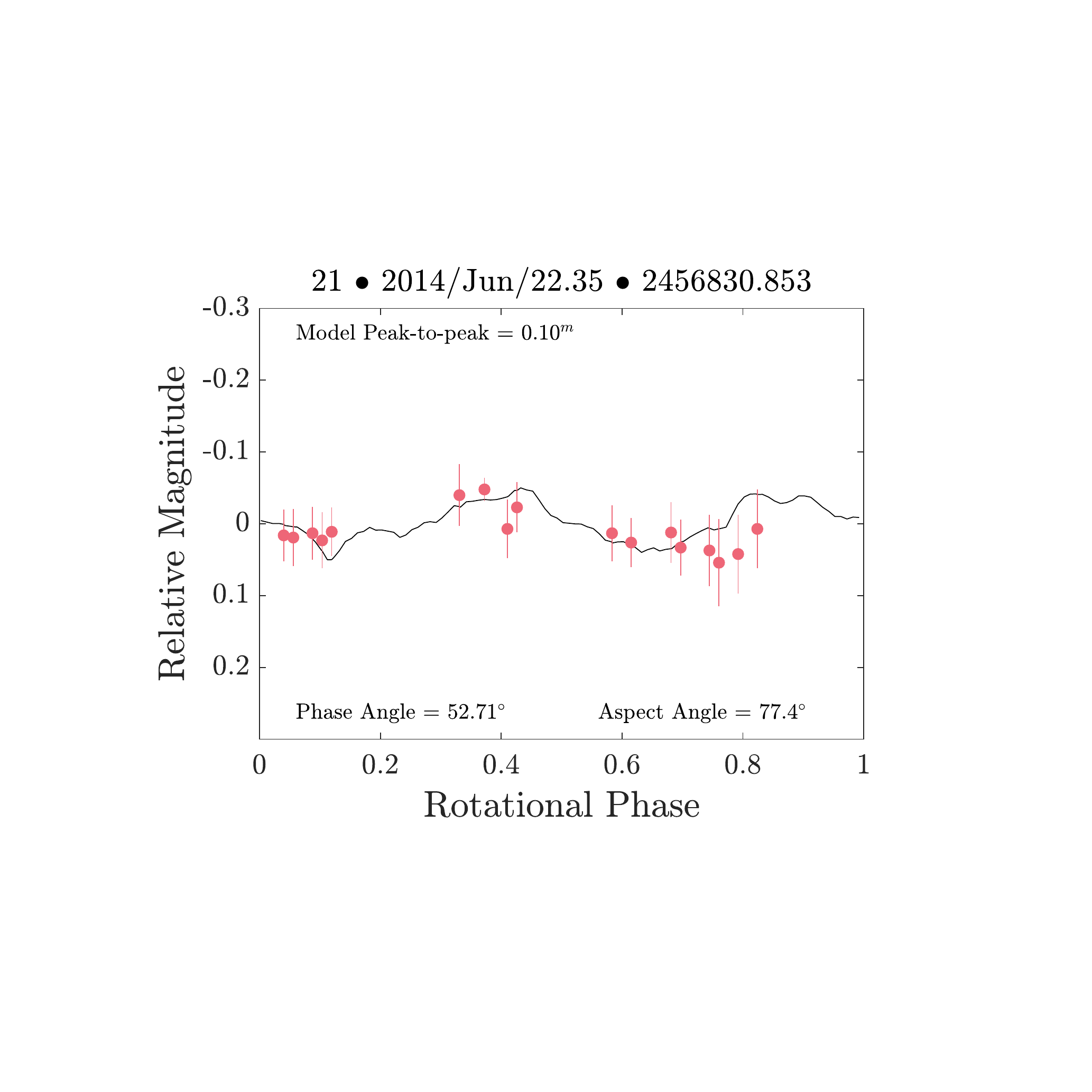} 
	}

	\resizebox{\hsize}{!}{
		\includegraphics[width=.48\textwidth, trim=2cm 4cm 3.8cm 4cm, clip=true]{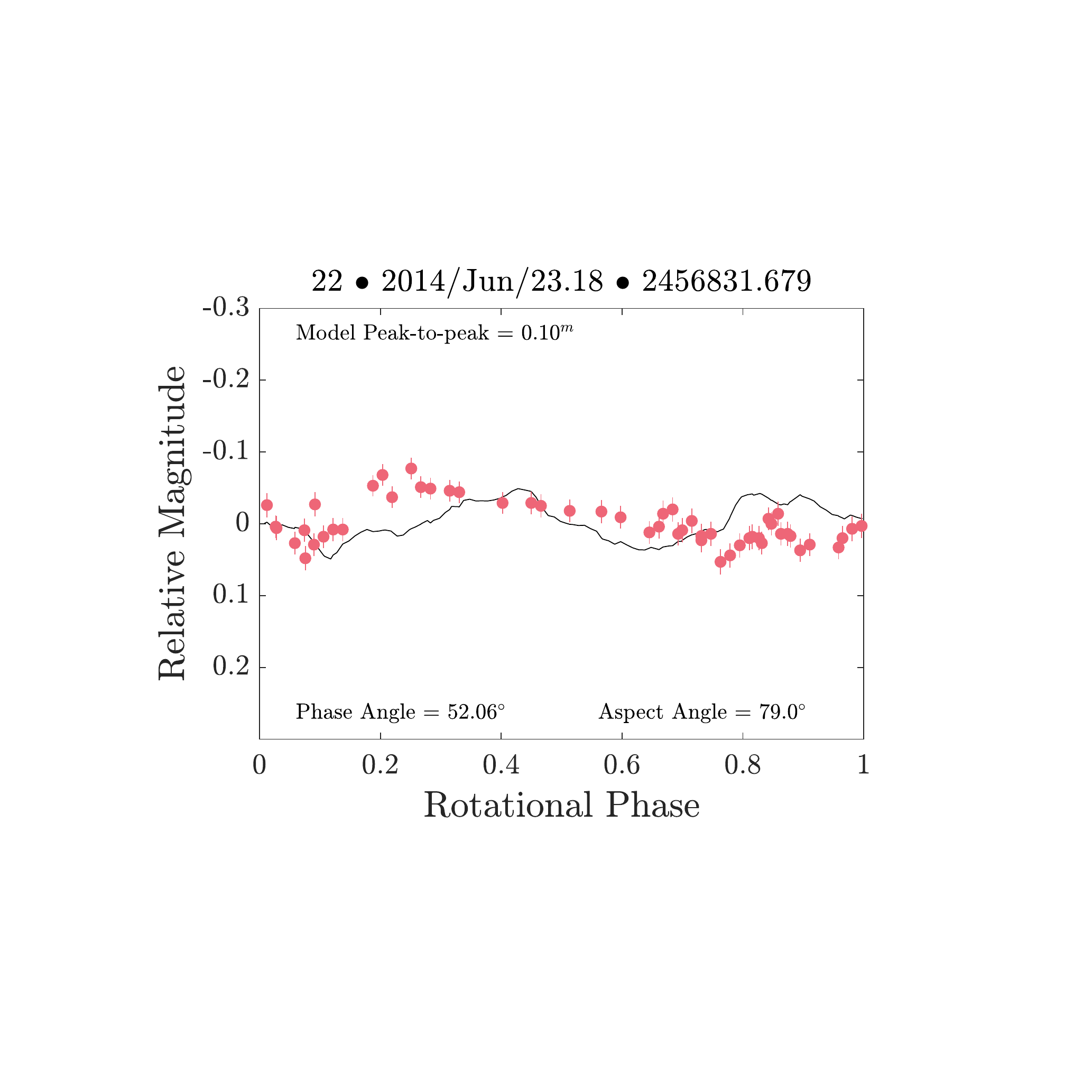} 
		\includegraphics[width=.48\textwidth, trim=2cm 4cm 3.8cm 4cm, clip=true]{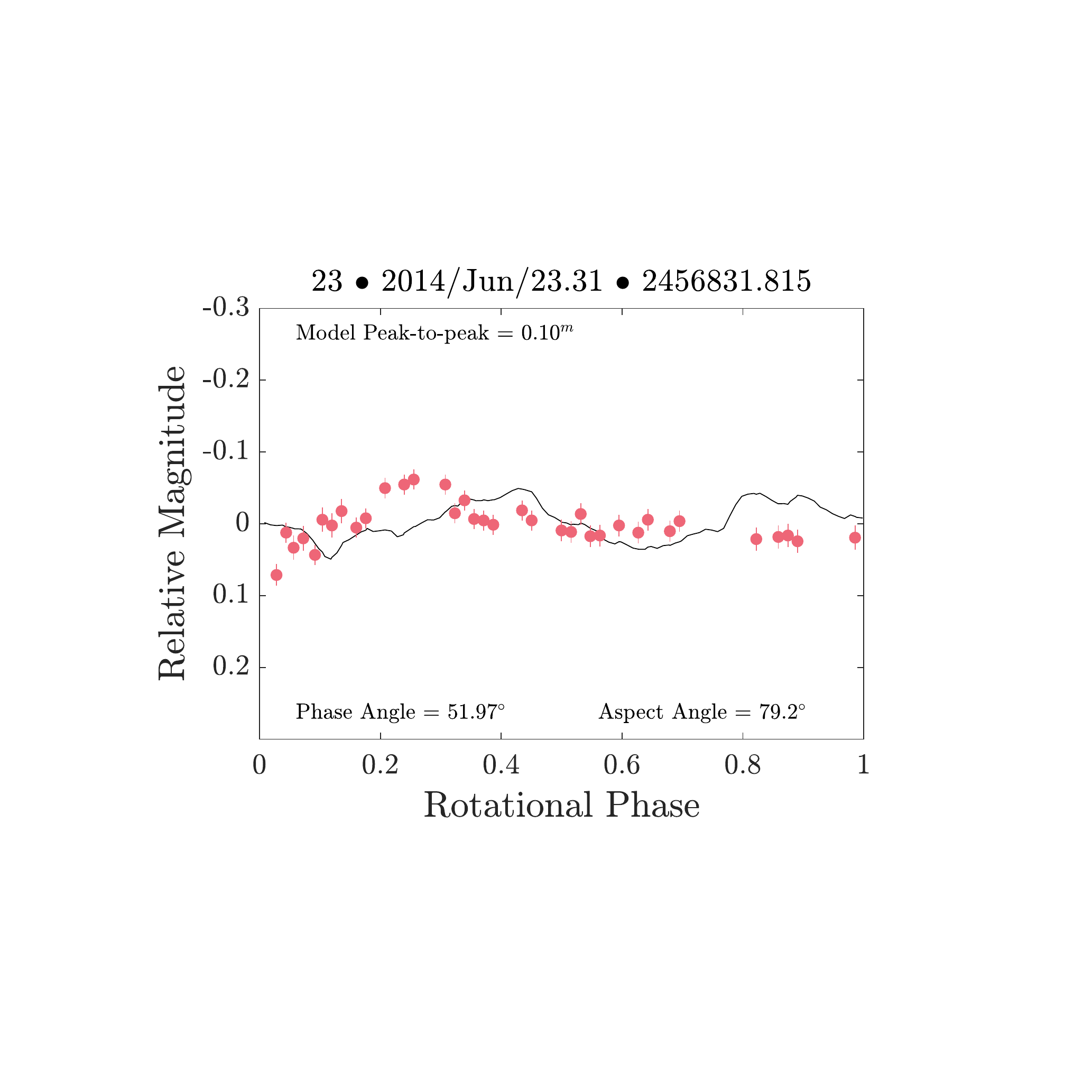} 		
		\includegraphics[width=.48\textwidth, trim=2cm 4cm 3.8cm 4cm, clip=true]{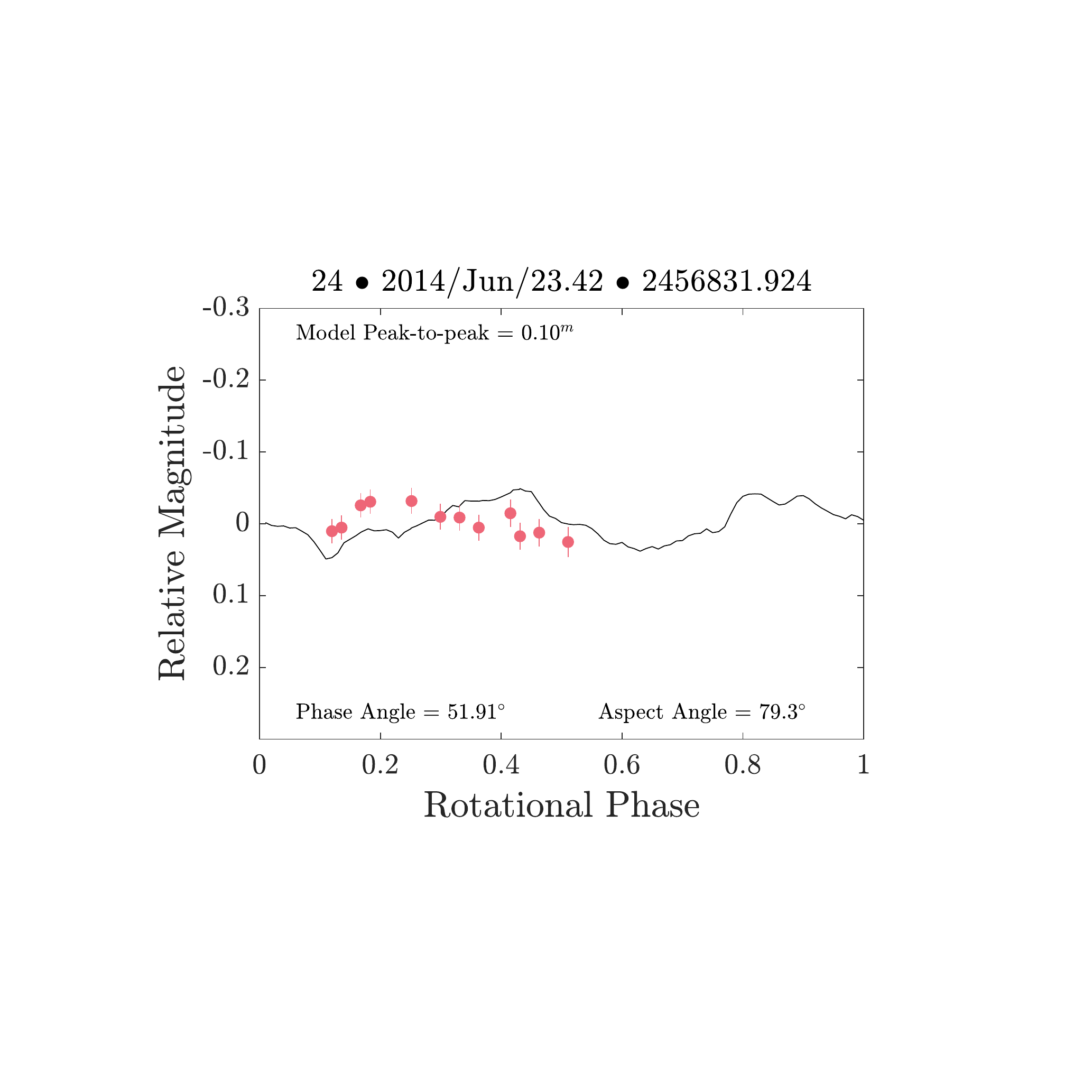} 
	}
	\caption{[Continued]
		\label{fig:radar-retro2}}
\end{figure*}

\addtocounter{figure}{-1}

\begin{figure*}
	\resizebox{\hsize}{!}{
		\includegraphics[width=.48\textwidth, trim=2cm 4cm 3.8cm 4cm, clip=true]{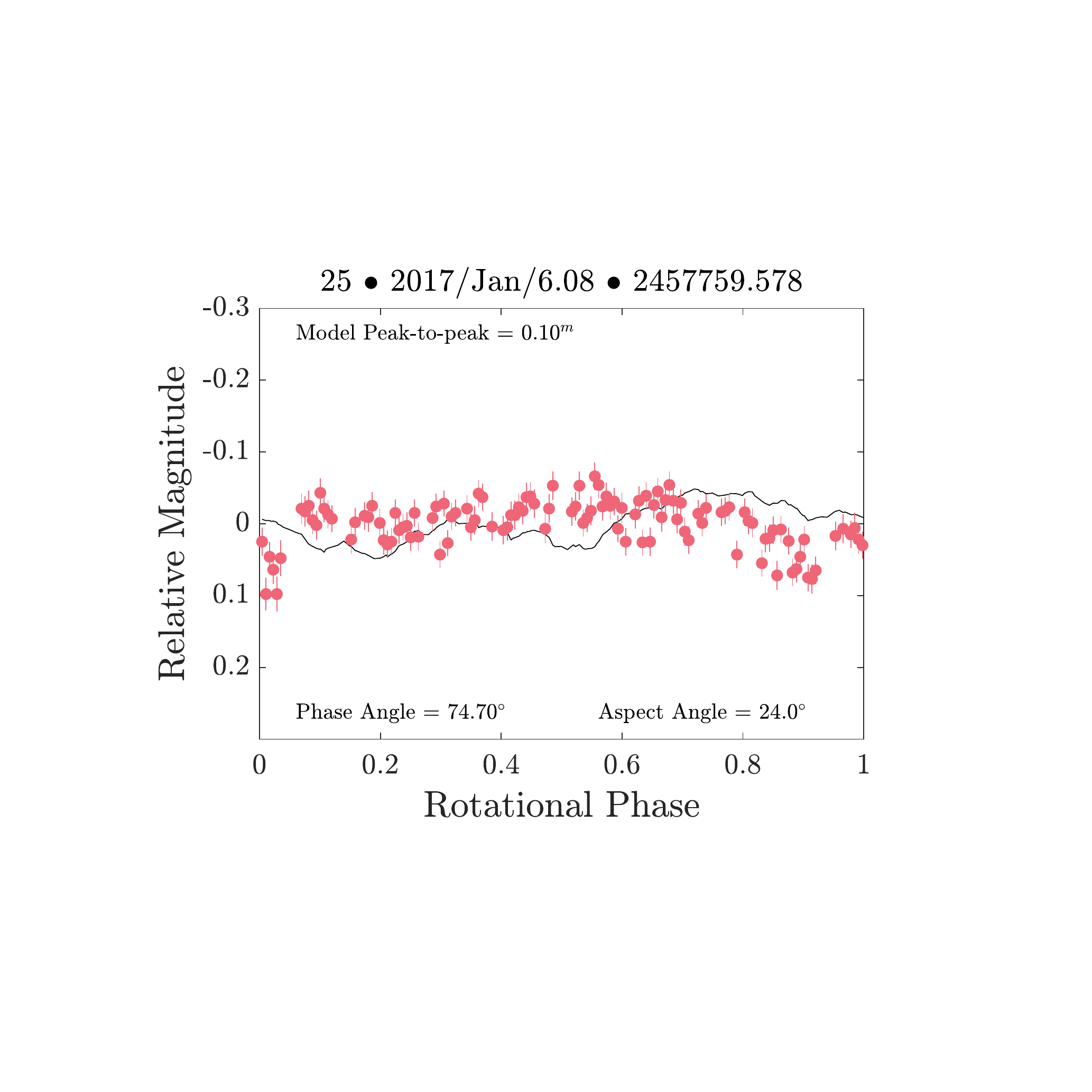} 
		\includegraphics[width=.48\textwidth, trim=2cm 4cm 3.8cm 4cm, clip=true]{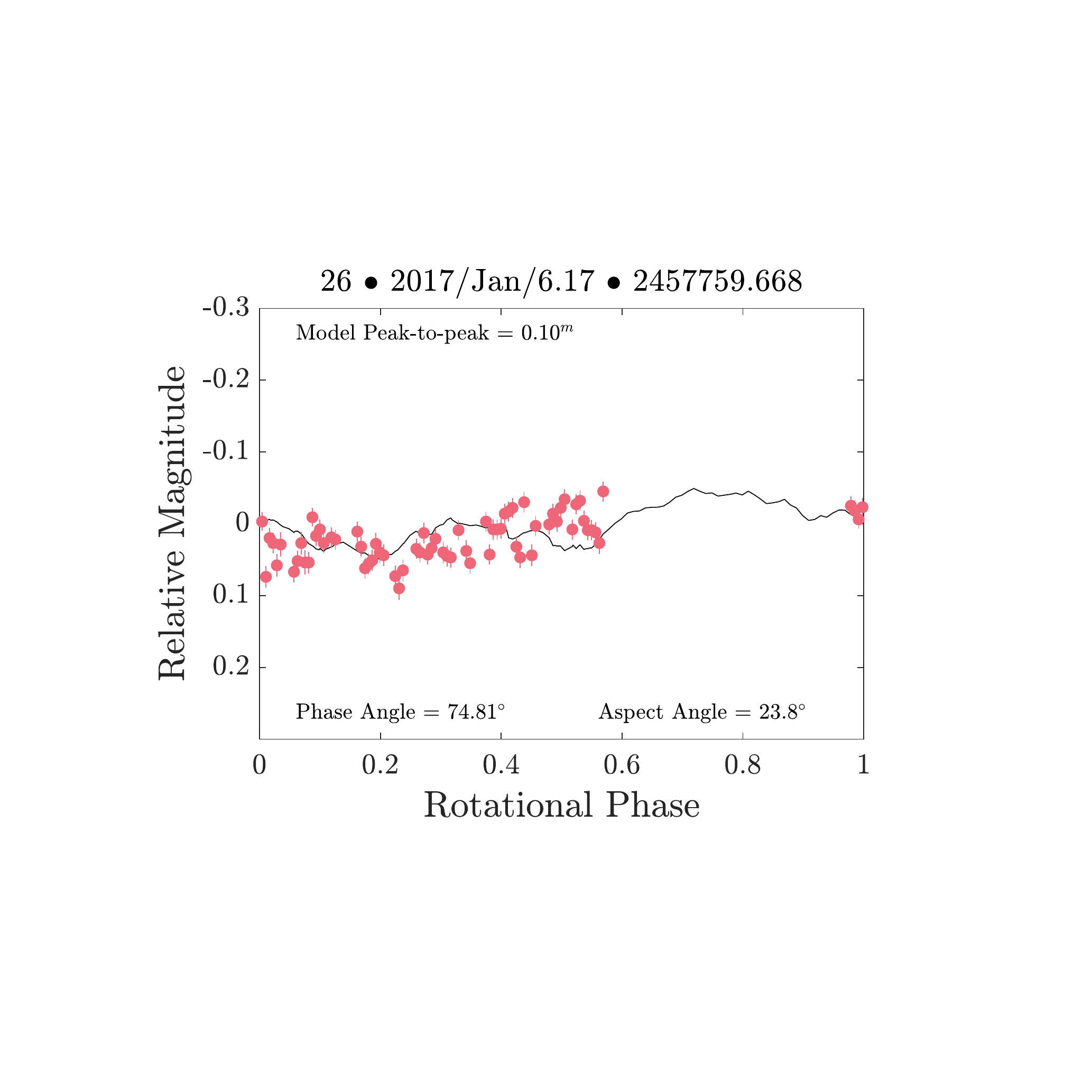} 		
		\includegraphics[width=.48\textwidth, trim=2cm 4cm 3.8cm 4cm, clip=true]{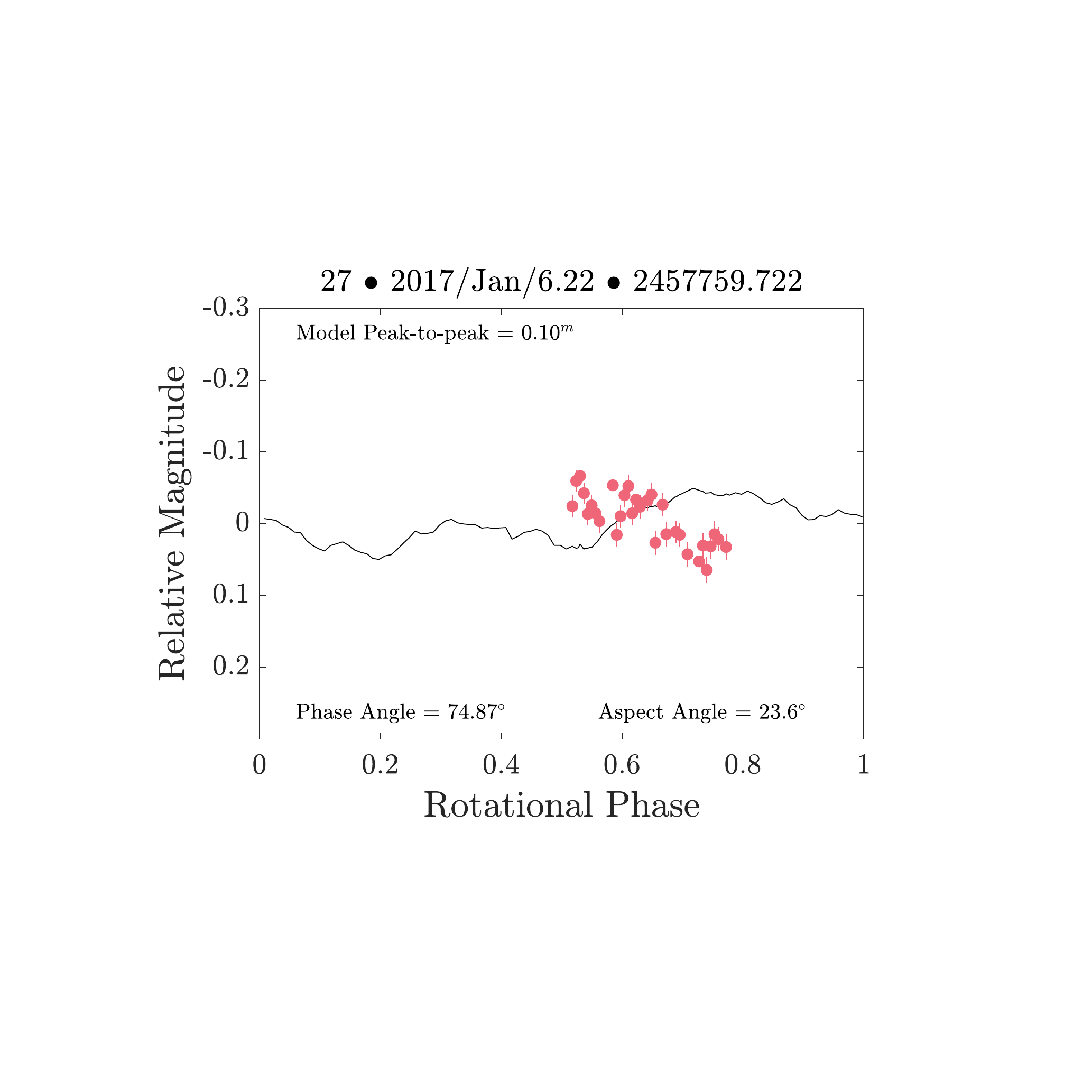} 
	}
	
	\resizebox{\hsize}{!}{
		\includegraphics[width=.48\textwidth, trim=2cm 4cm 3.8cm 4cm, clip=true]{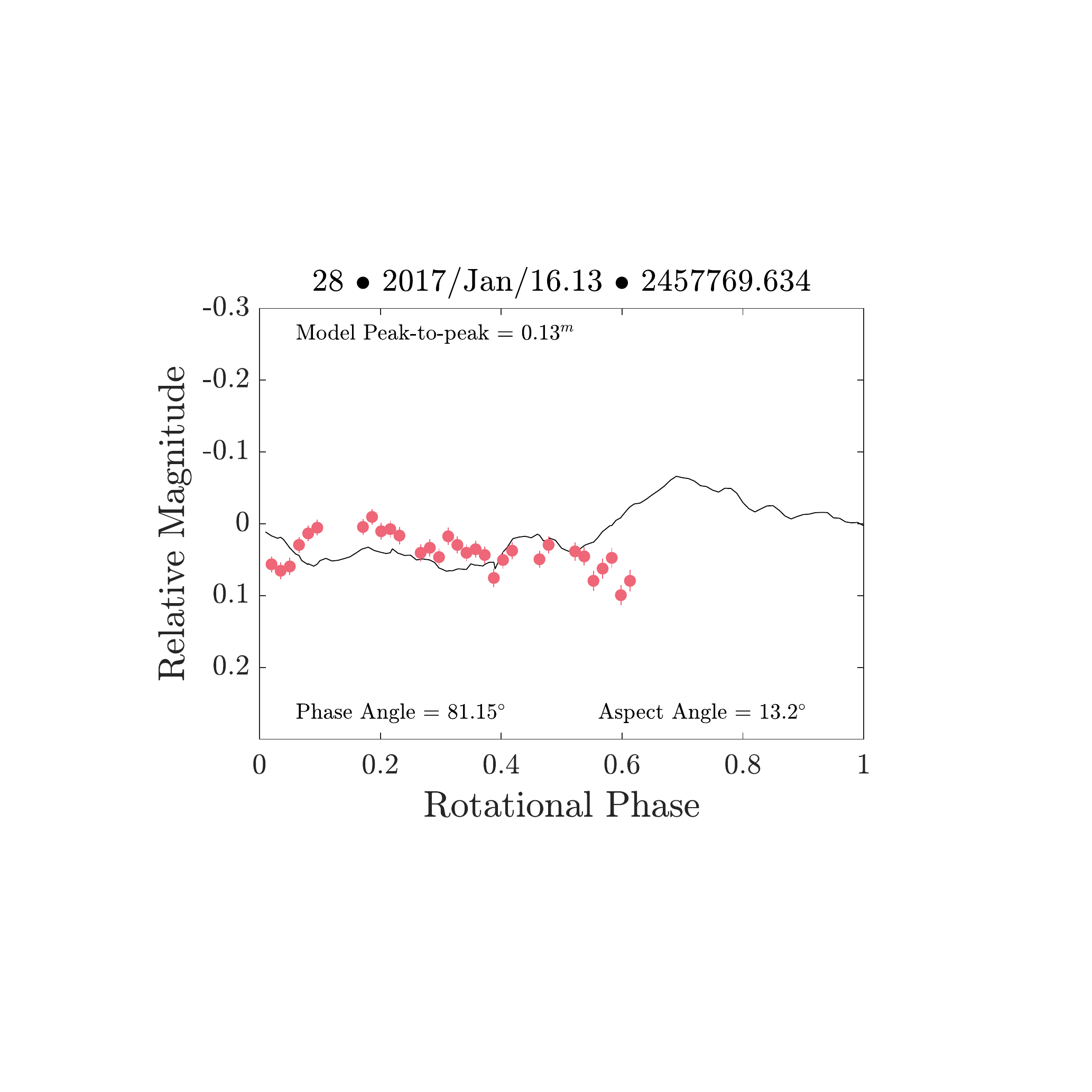} 
		\includegraphics[width=.48\textwidth, trim=2cm 4cm 3.8cm 4cm, clip=true]{LC/2102_lat-30lon180_v190906_20220204_newnewT0_29_fix.pdf} 		
		\includegraphics[width=.48\textwidth, trim=2cm 4cm 3.8cm 4cm, clip=true]{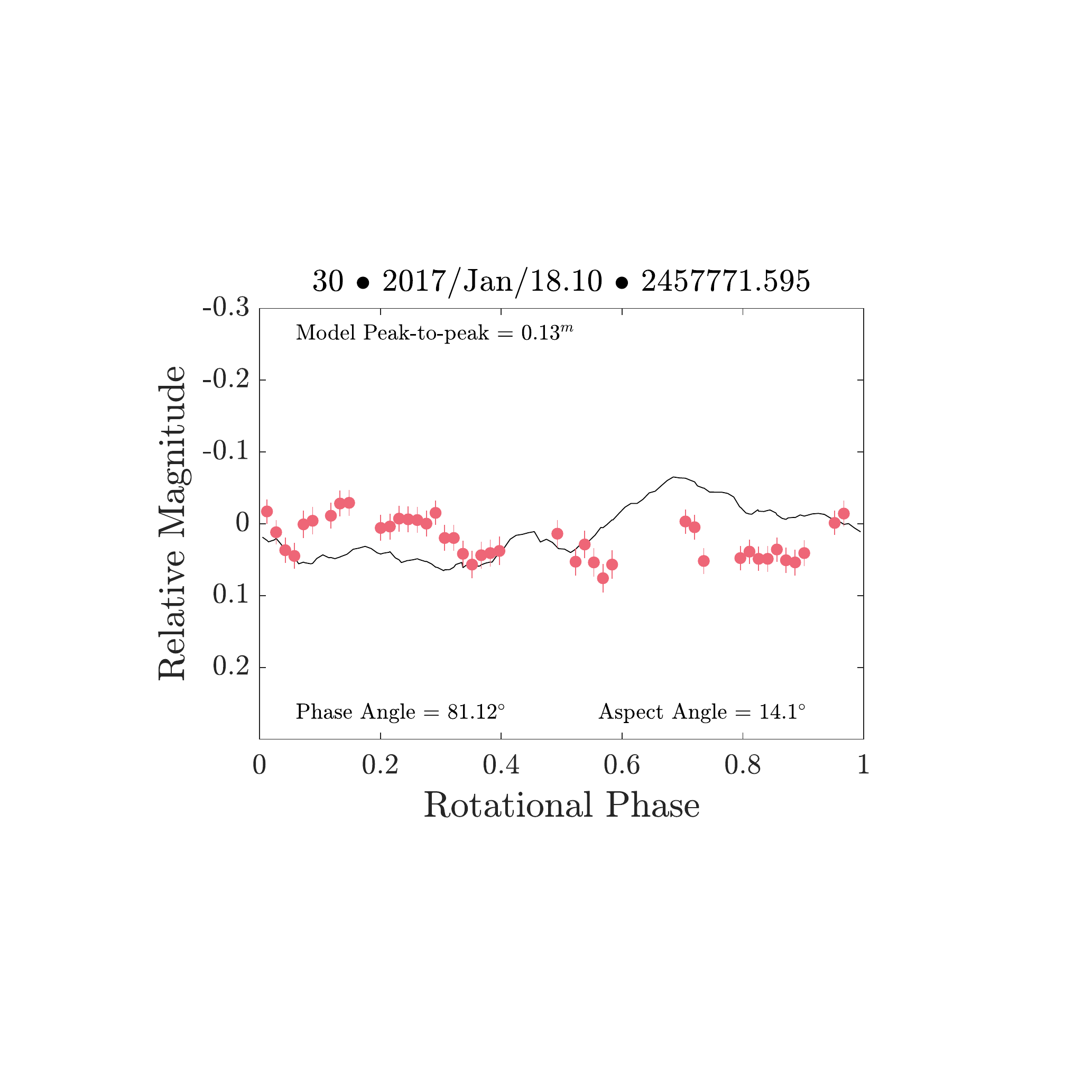} 
	}
	
	\resizebox{.6666\hsize}{!}{
		\includegraphics[width=.48\textwidth, trim=2cm 4cm 3.8cm 4cm, clip=true]{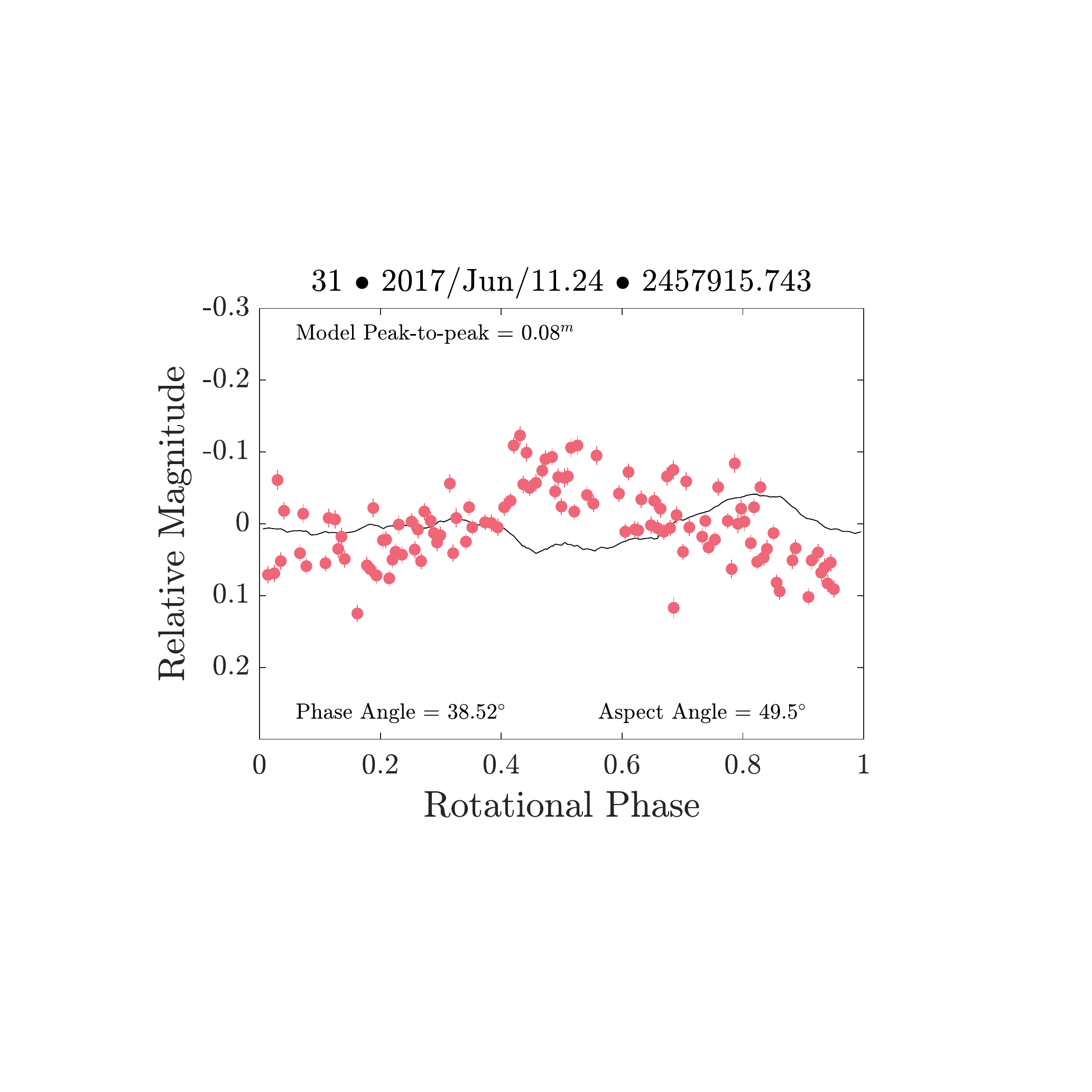} 
		\includegraphics[width=.48\textwidth, trim=2cm 4cm 3.8cm 4cm, clip=true]{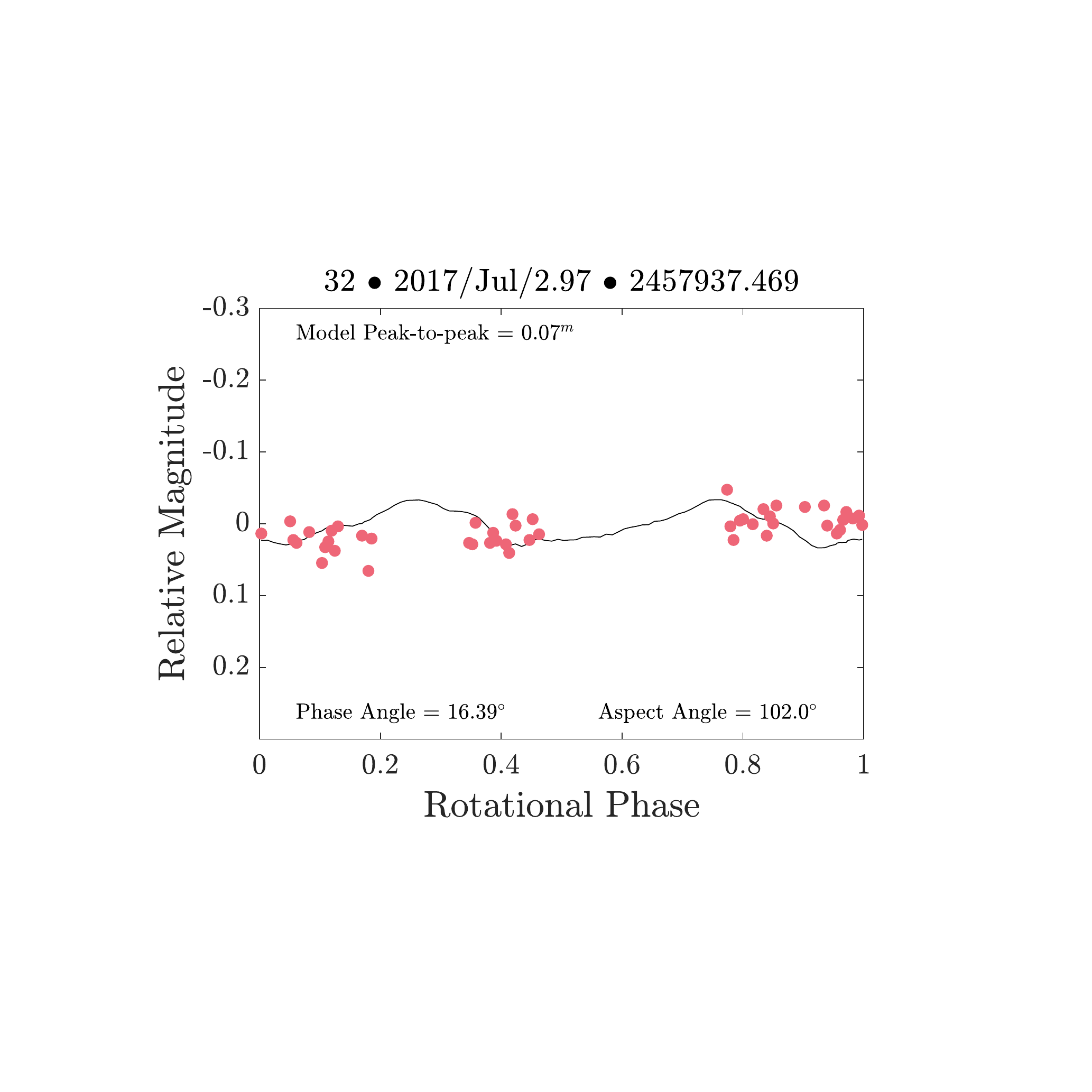} 	
	}
	
	\caption{[Continued]
		\label{fig:radar-retro3}}
\end{figure*}

\begin{figure*}
	\resizebox{\hsize}{!}{
		\includegraphics[width=.48\textwidth, trim=2cm 4cm 3.8cm 4cm, clip=true]{LC/2102_lat+30lon036_v190906_20191127_initial_spinstate_01_fix.pdf} 
		\includegraphics[width=.48\textwidth, trim=2cm 4cm 3.8cm 4cm, clip=true]{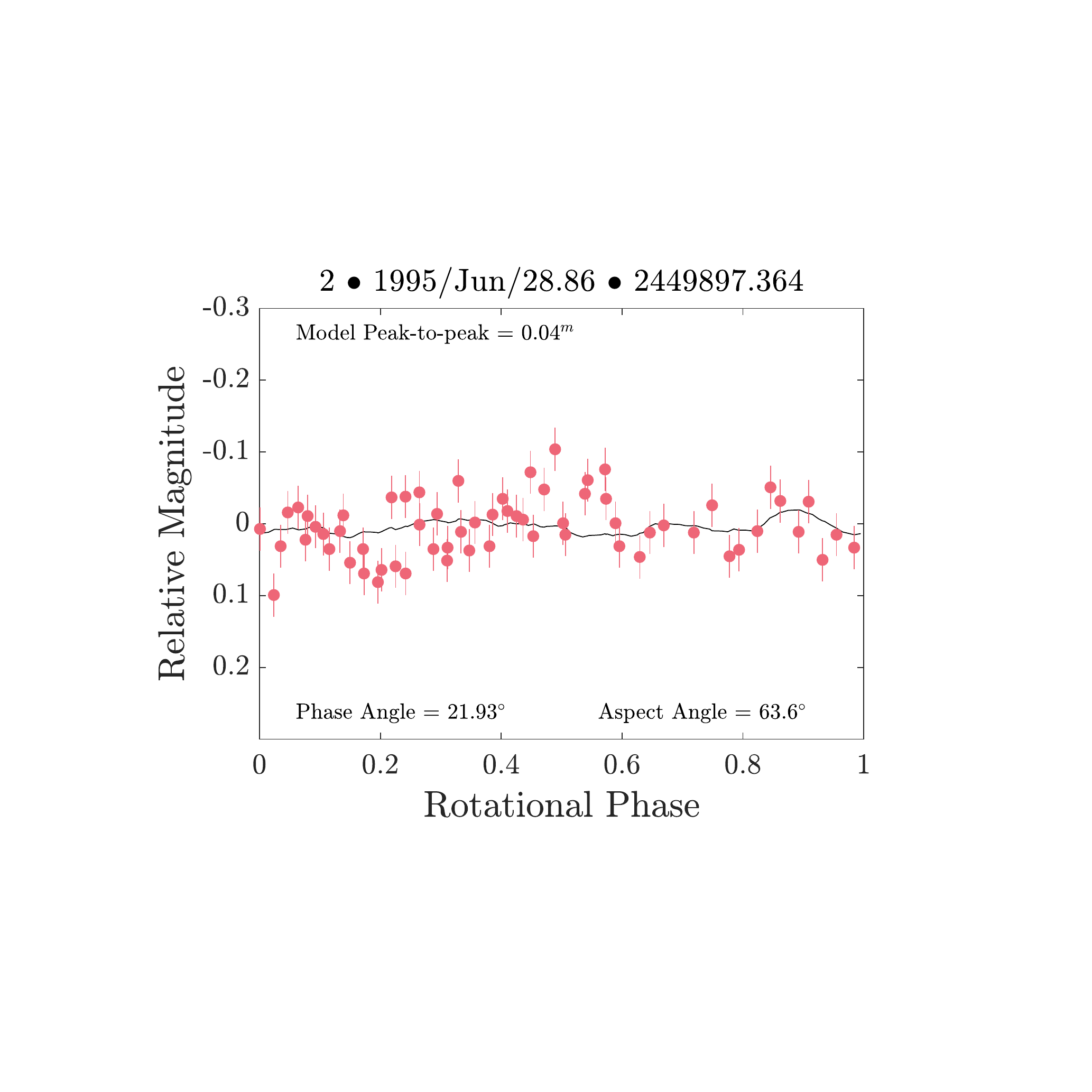} 		
		\includegraphics[width=.48\textwidth, trim=2cm 4cm 3.8cm 4cm, clip=true]{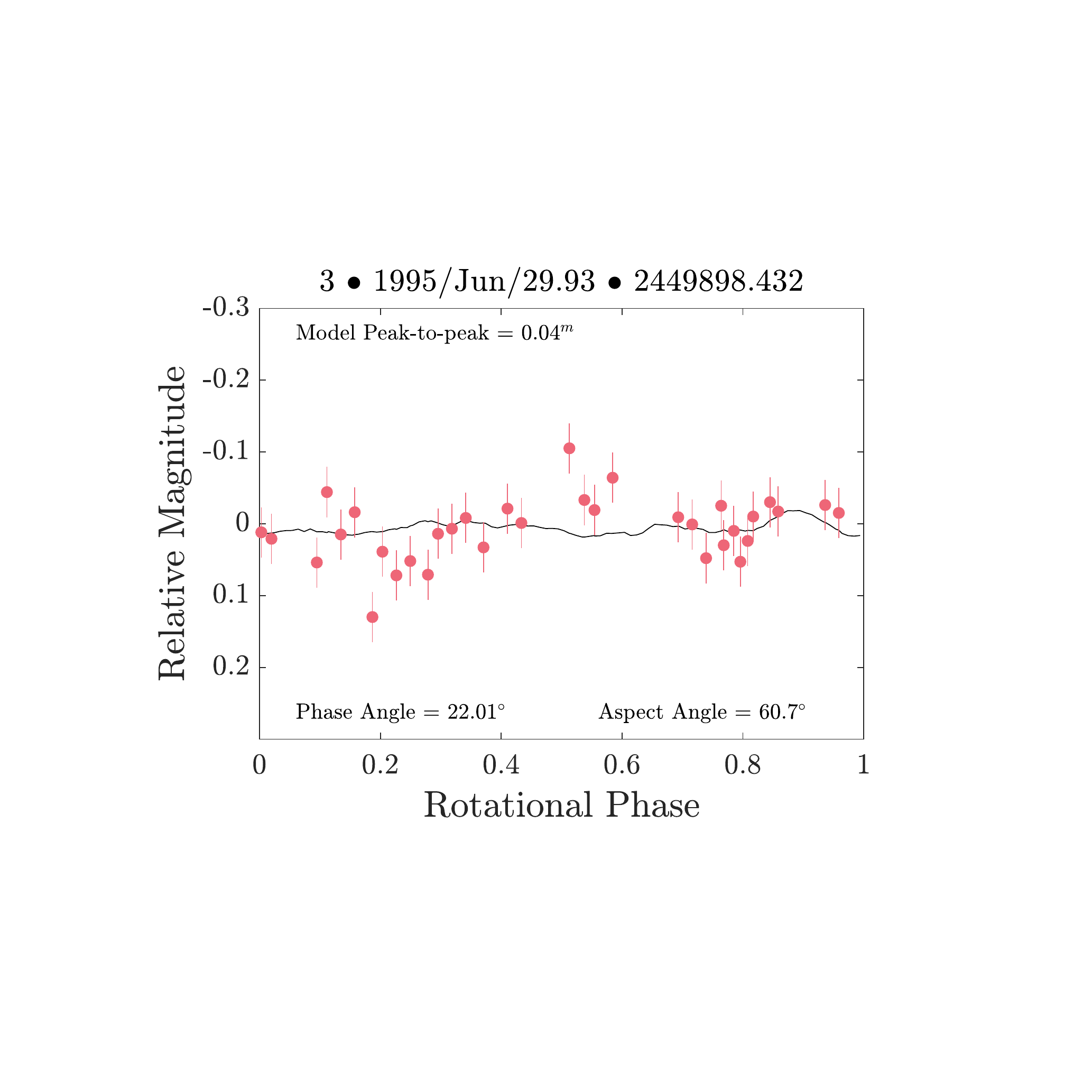} 
	}
	
	\resizebox{\hsize}{!}{
		\includegraphics[width=.48\textwidth, trim=2cm 4cm 3.8cm 4cm, clip=true]{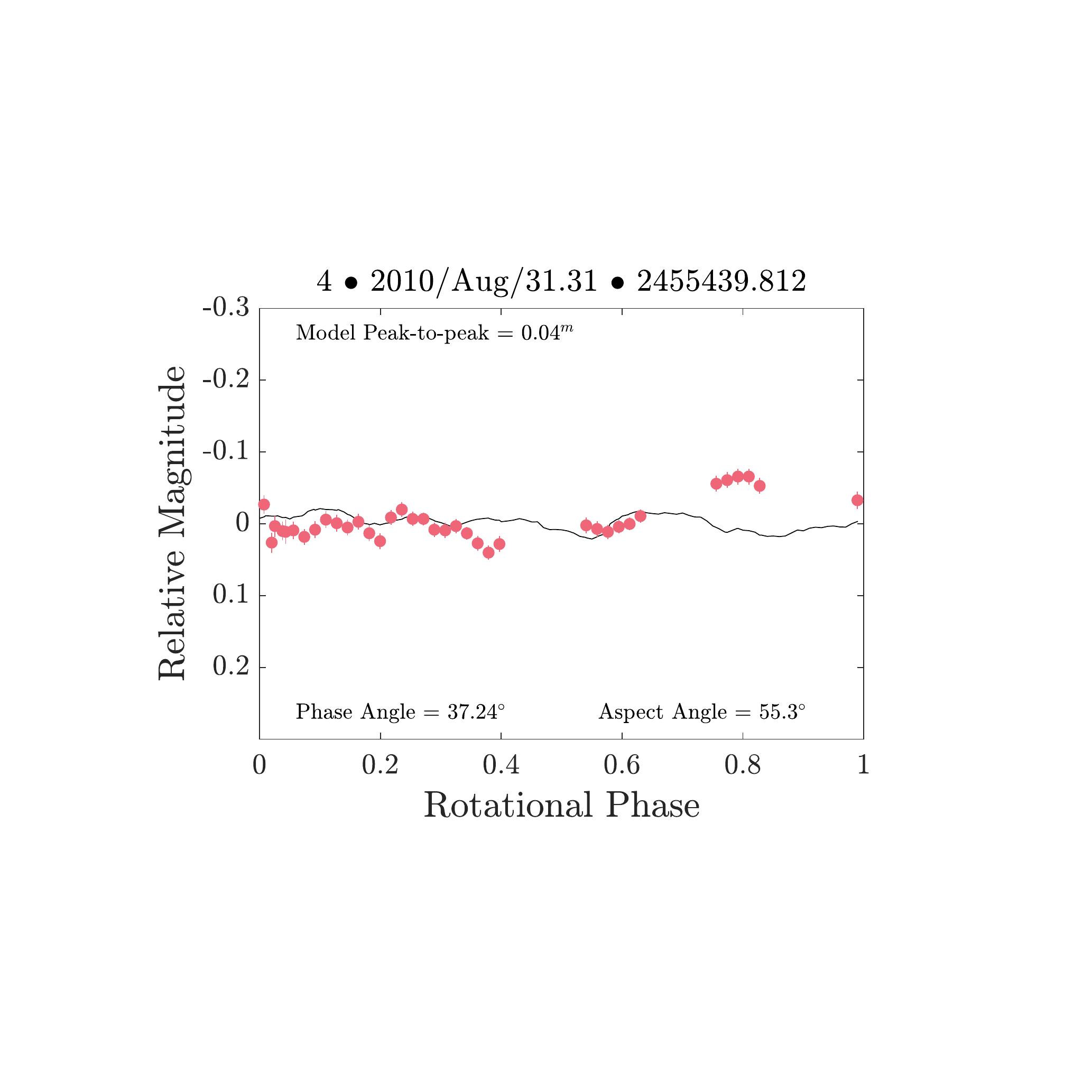} 
		\includegraphics[width=.48\textwidth, trim=2cm 4cm 3.8cm 4cm, clip=true]{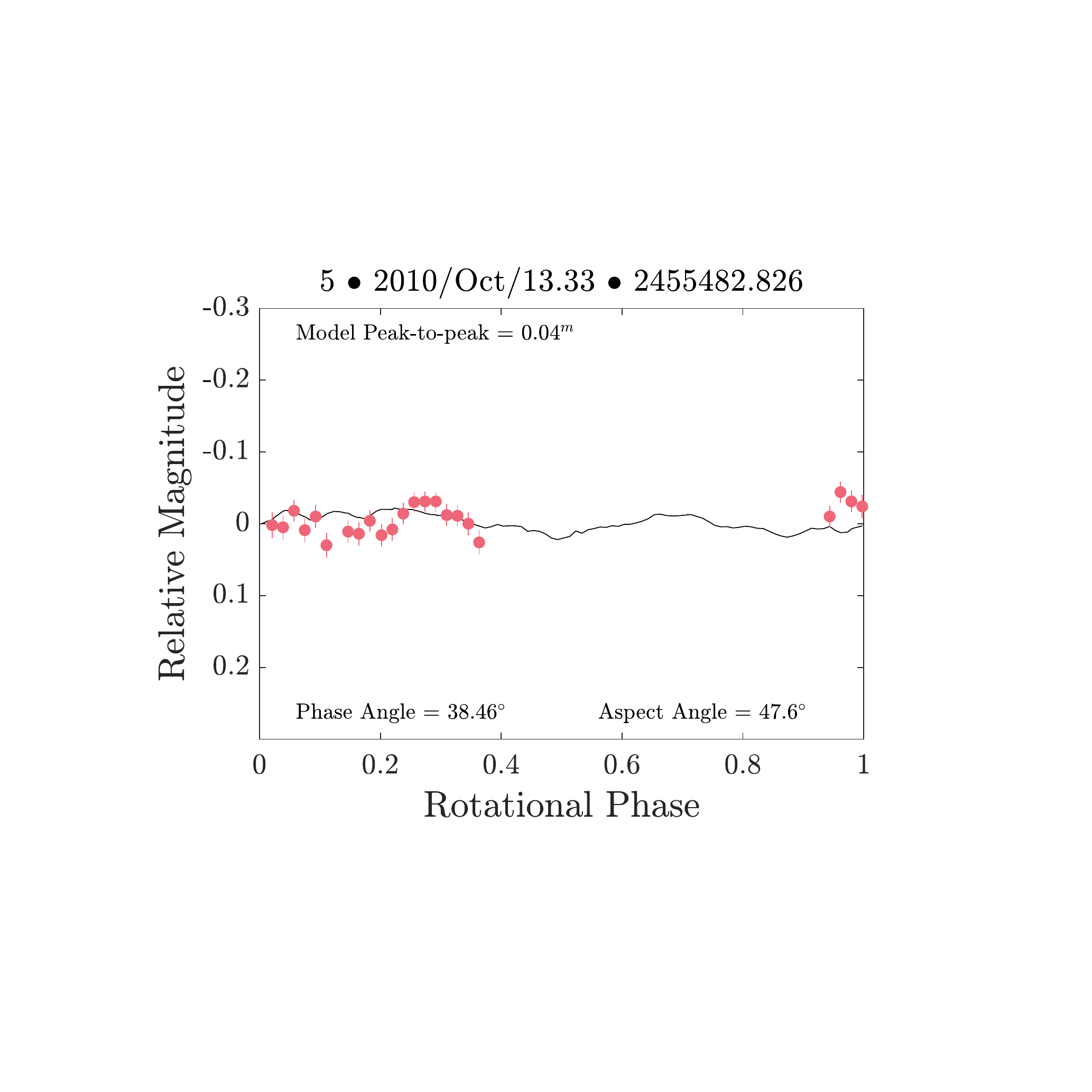} 		
		\includegraphics[width=.48\textwidth, trim=2cm 4cm 3.8cm 4cm, clip=true]{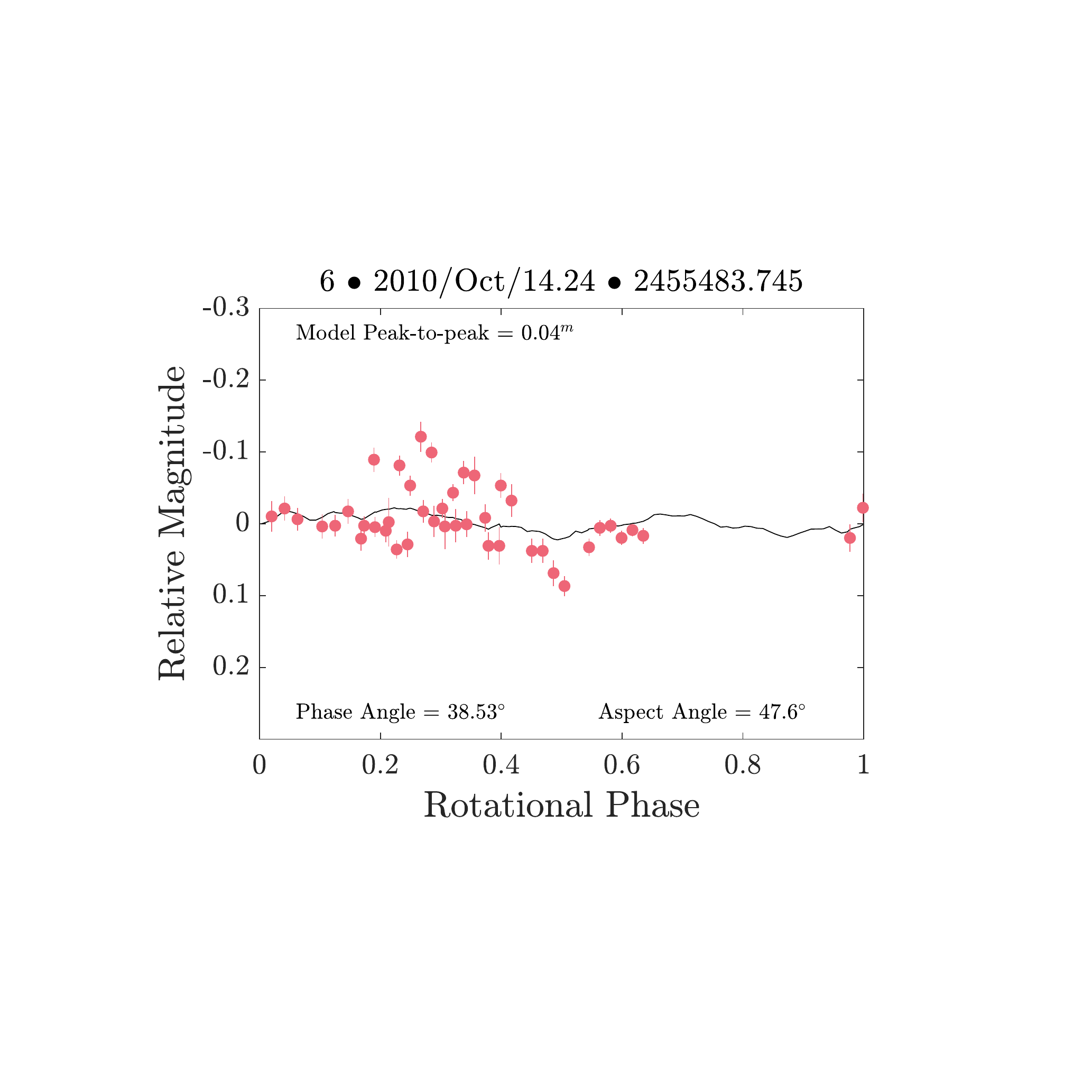} 
	}
	
	\resizebox{\hsize}{!}{
		\includegraphics[width=.48\textwidth, trim=2cm 4cm 3.8cm 4cm, clip=true]{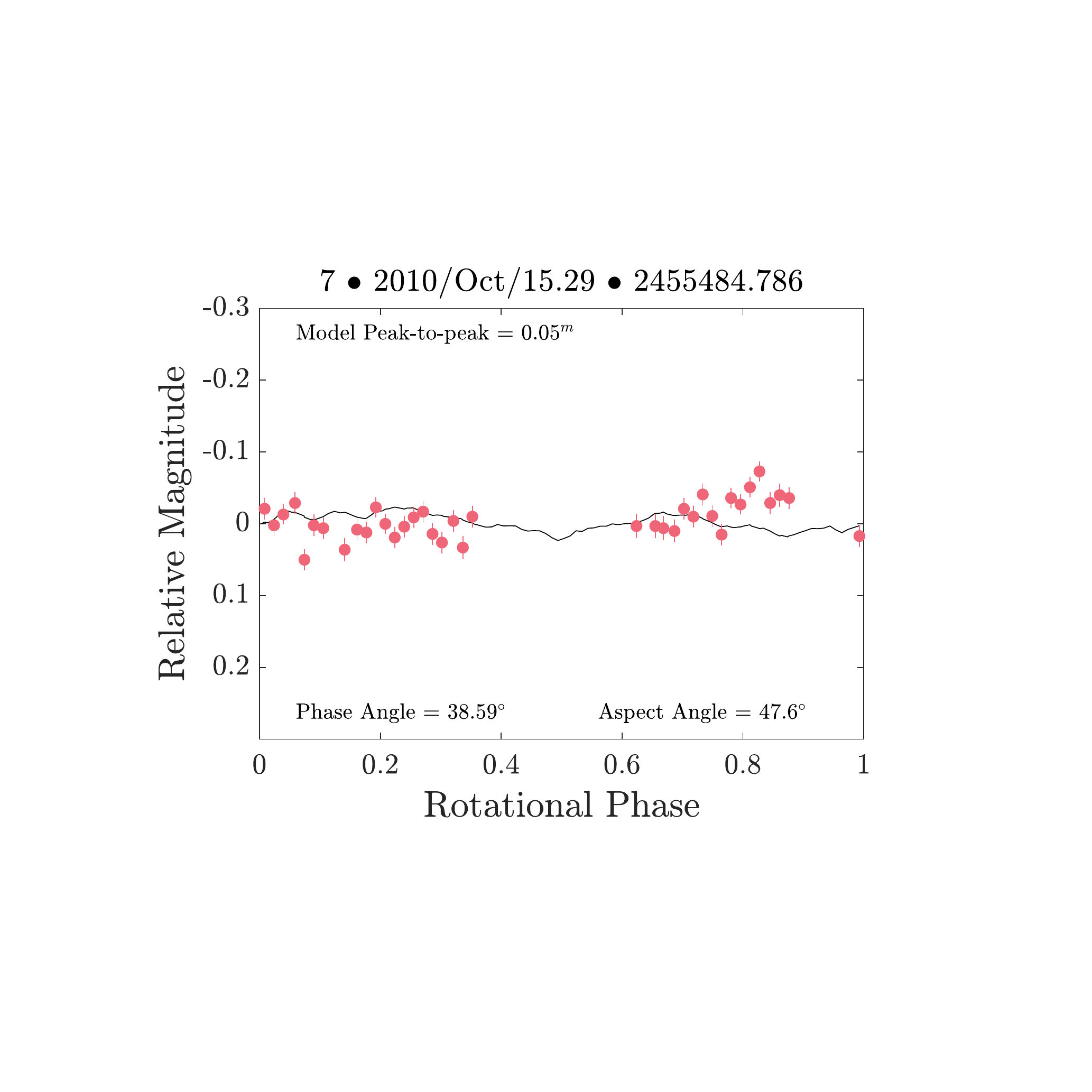} 
		\includegraphics[width=.48\textwidth, trim=2cm 4cm 3.8cm 4cm, clip=true]{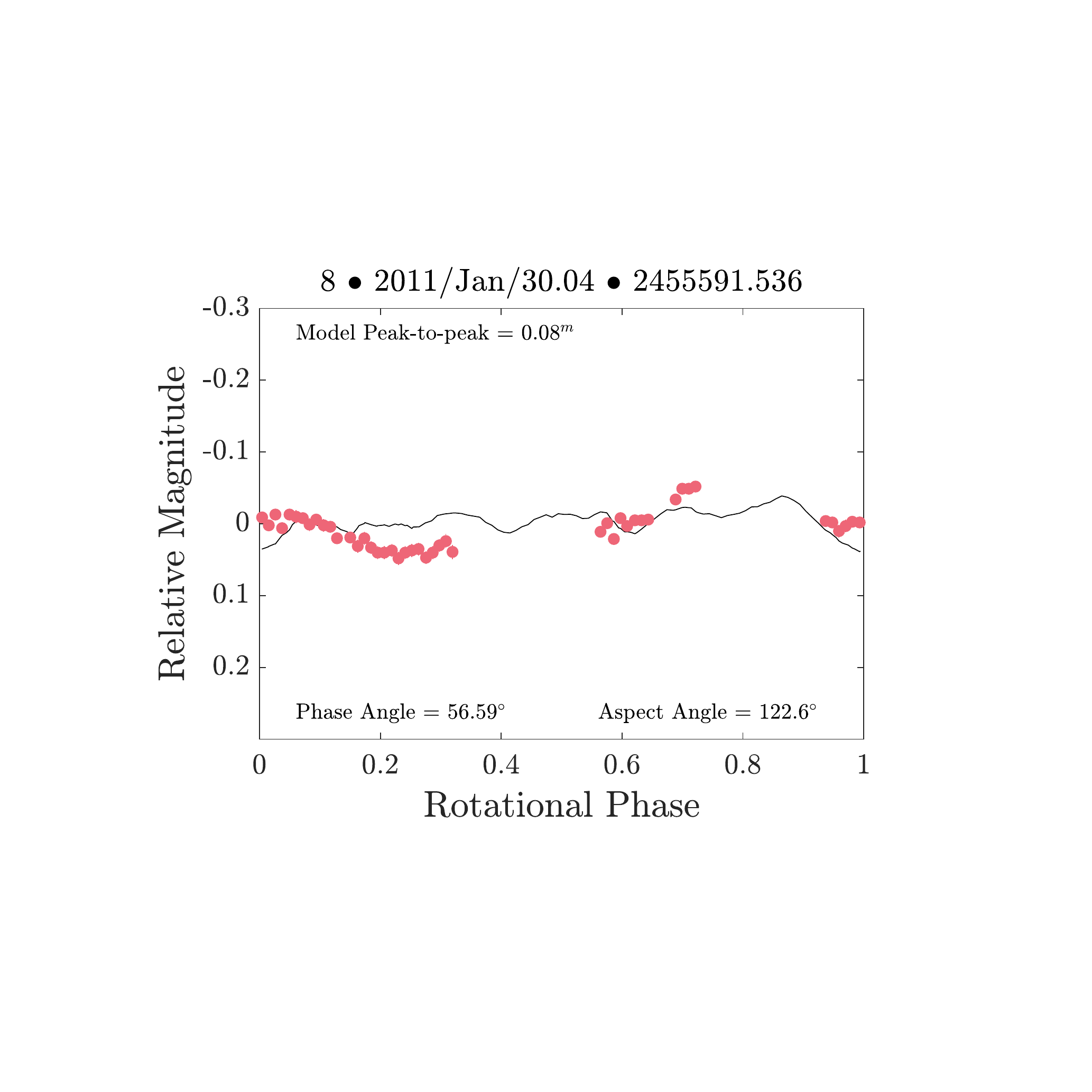} 		
		\includegraphics[width=.48\textwidth, trim=2cm 4cm 3.8cm 4cm, clip=true]{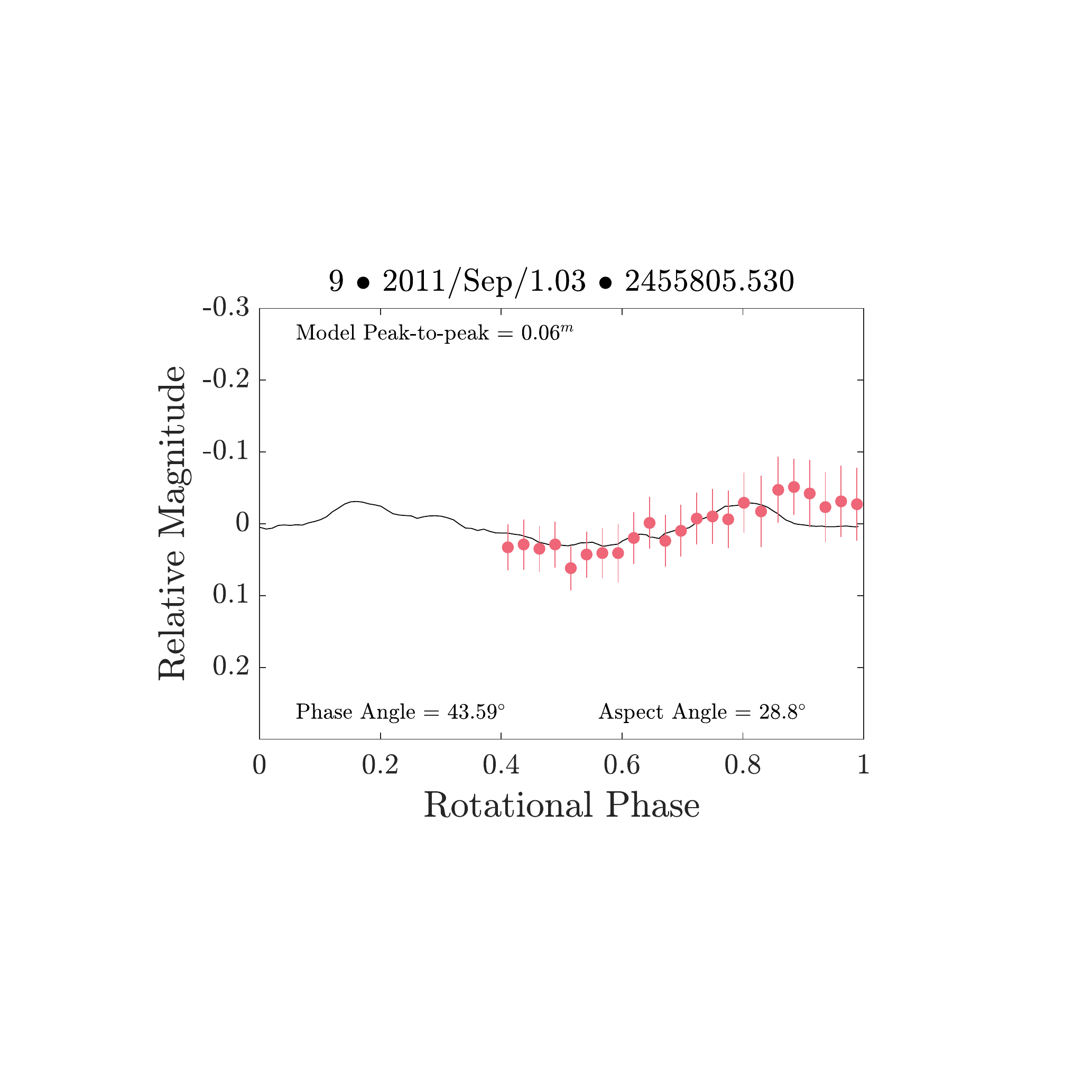} 
	}

	\resizebox{\hsize}{!}{
		\includegraphics[width=.48\textwidth, trim=2cm 4cm 3.8cm 4cm, clip=true]{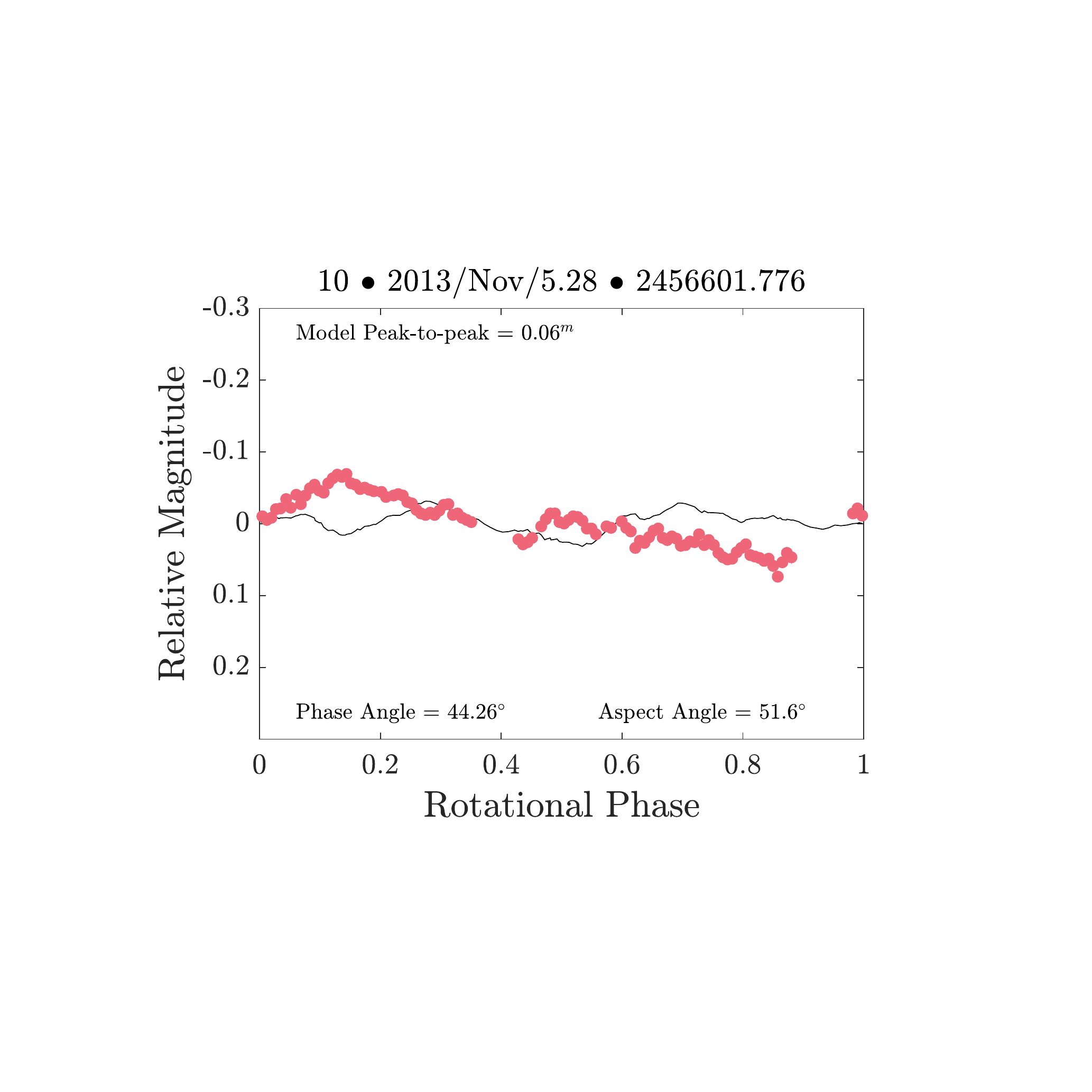} 
		\includegraphics[width=.48\textwidth, trim=2cm 4cm 3.8cm 4cm, clip=true]{LC/2102_lat+30lon036_v190906_20191127_initial_spinstate_11_fix.pdf} 		
		\includegraphics[width=.48\textwidth, trim=2cm 4cm 3.8cm 4cm, clip=true]{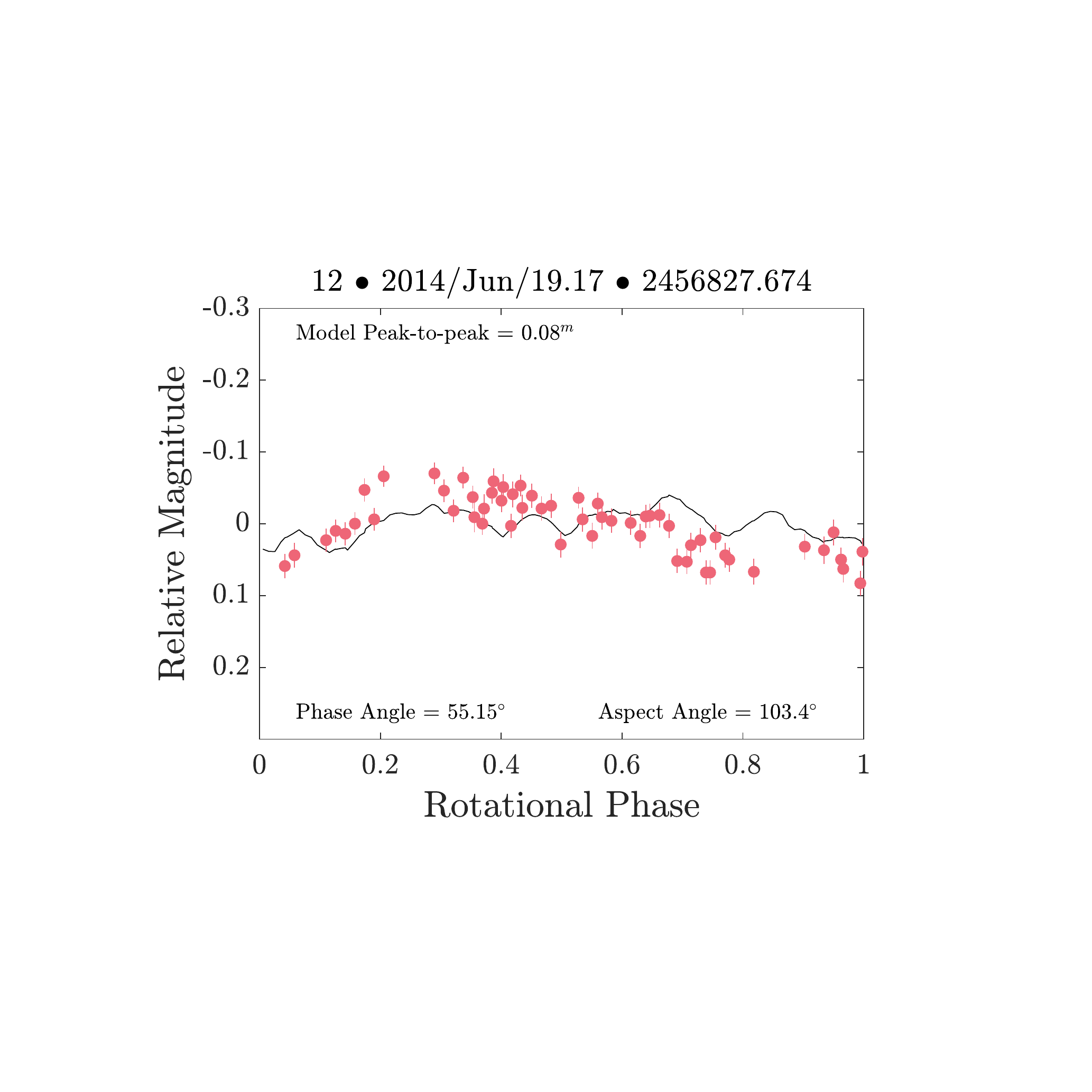} 
	}
	
	\caption{All available data plotted over synthetic light curves generated with the prograde radar shape model of asteroid (2102) Tantalus. This shape model utilises a subset of available optical light-curves and the radar data. 
		\label{fig:radar-pro1}}
\end{figure*}

\addtocounter{figure}{-1}

\begin{figure*}
	\resizebox{\hsize}{!}{
		\includegraphics[width=.48\textwidth, trim=2cm 4cm 3.8cm 4cm, clip=true]{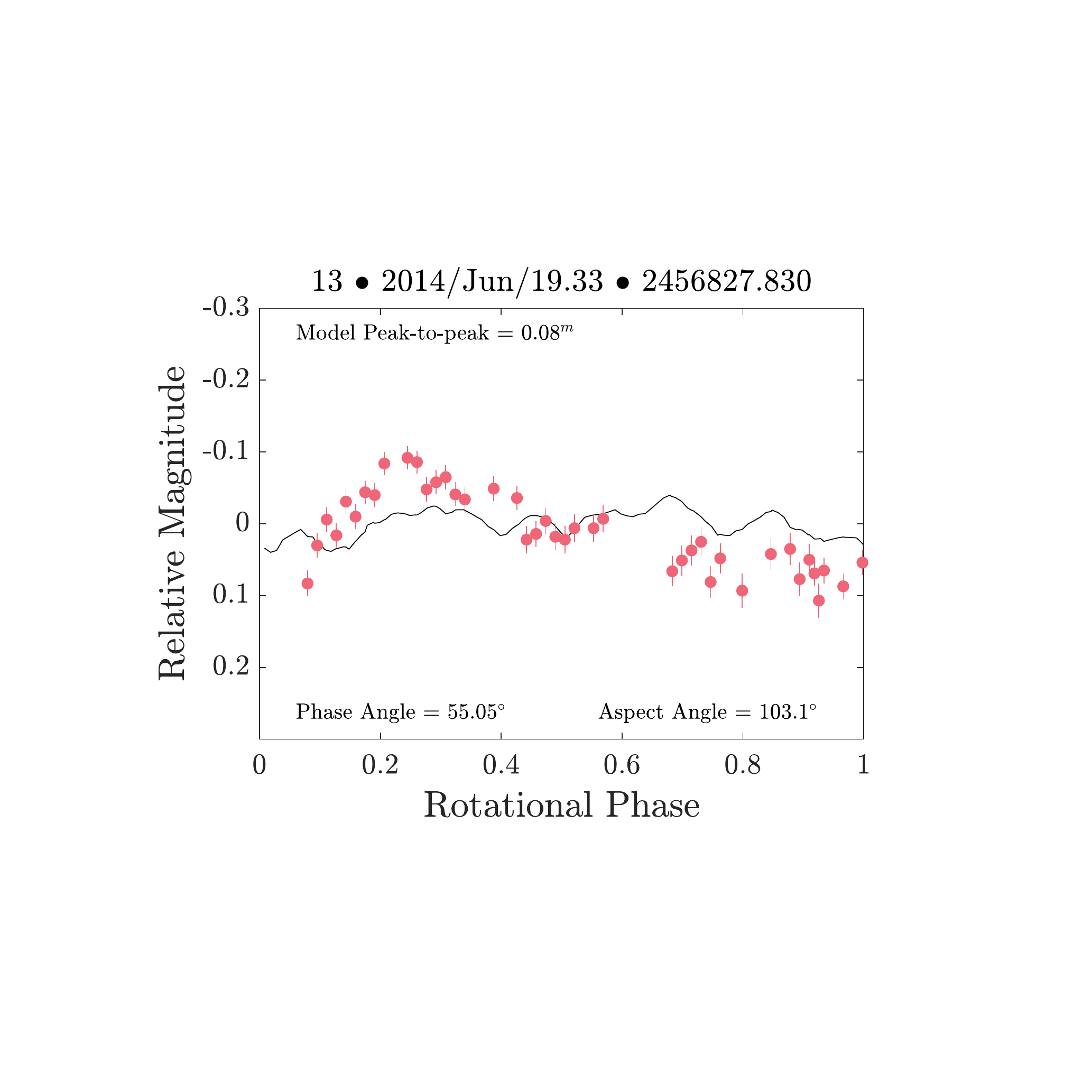} 
		\includegraphics[width=.48\textwidth, trim=2cm 4cm 3.8cm 4cm, clip=true]{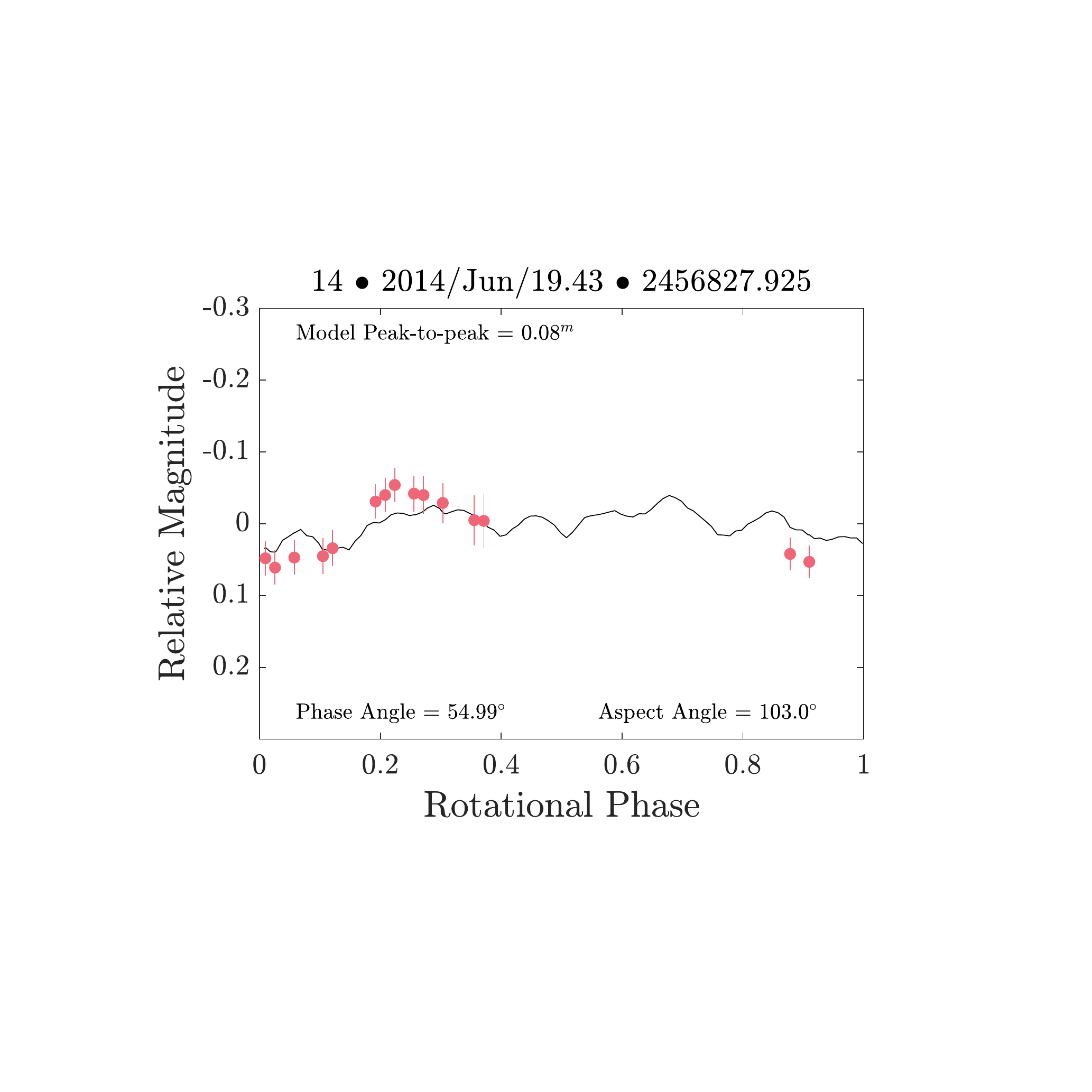} 		
		\includegraphics[width=.48\textwidth, trim=2cm 4cm 3.8cm 4cm, clip=true]{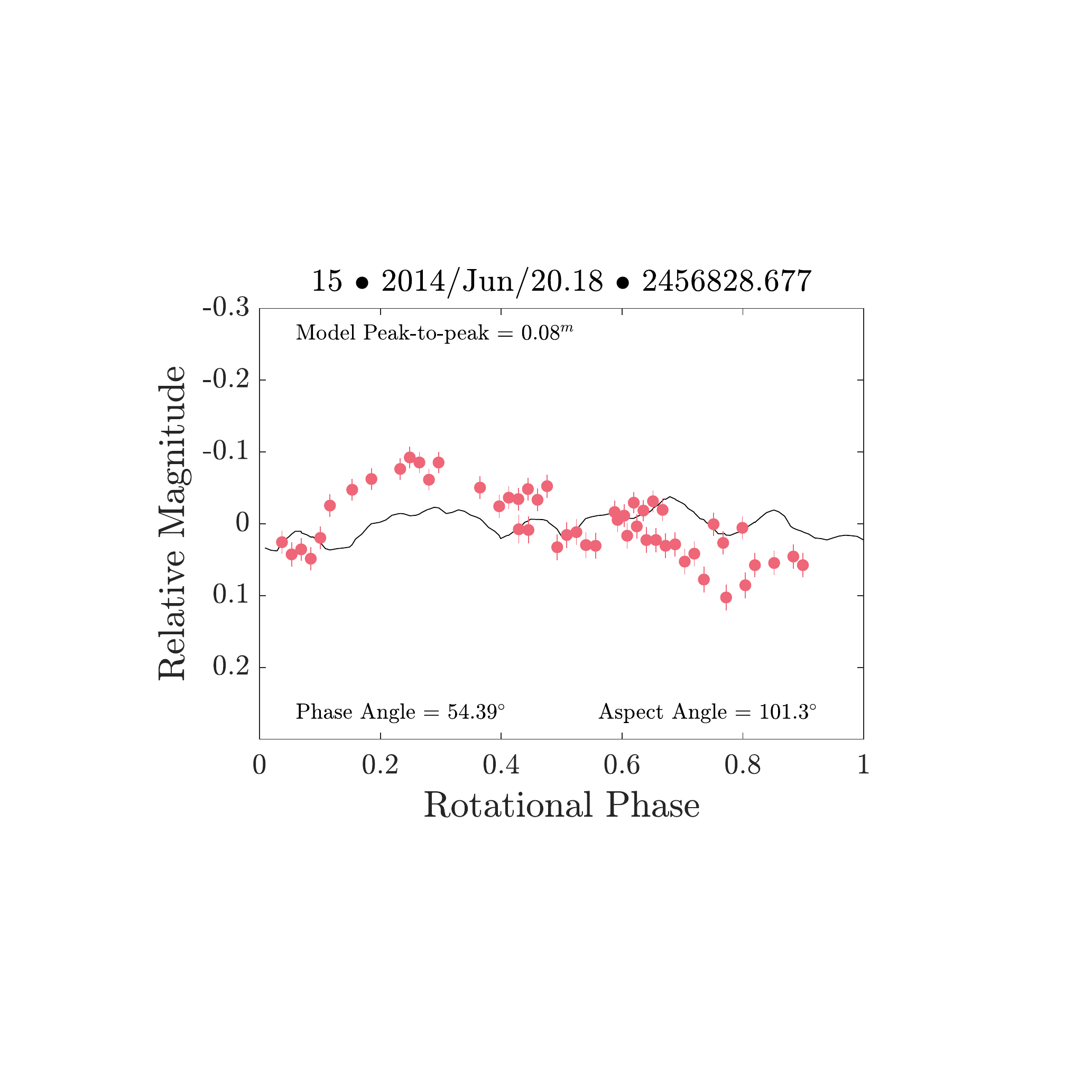} 
	}
	
	\resizebox{\hsize}{!}{
		\includegraphics[width=.48\textwidth, trim=2cm 4cm 3.8cm 4cm, clip=true]{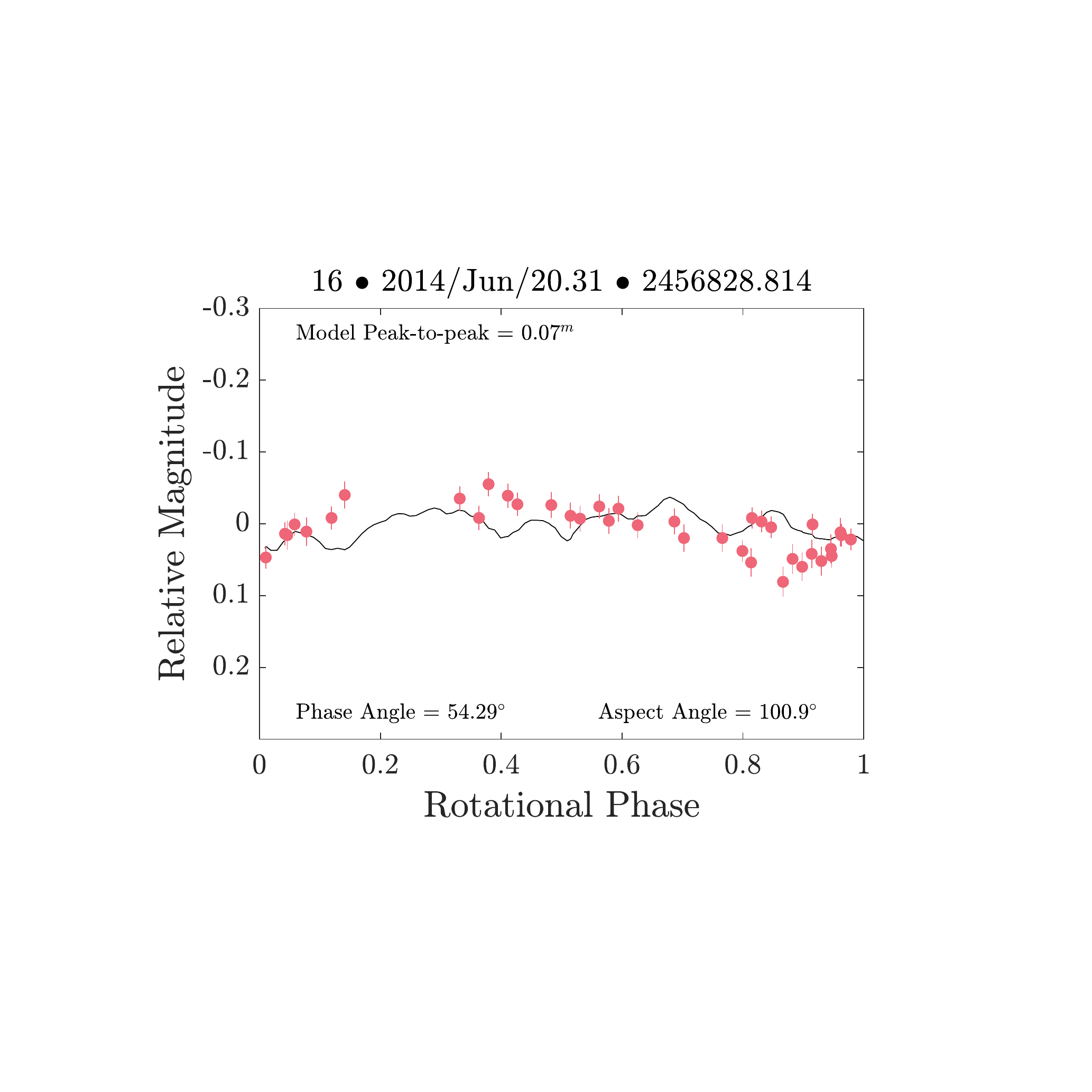} 
		\includegraphics[width=.48\textwidth, trim=2cm 4cm 3.8cm 4cm, clip=true]{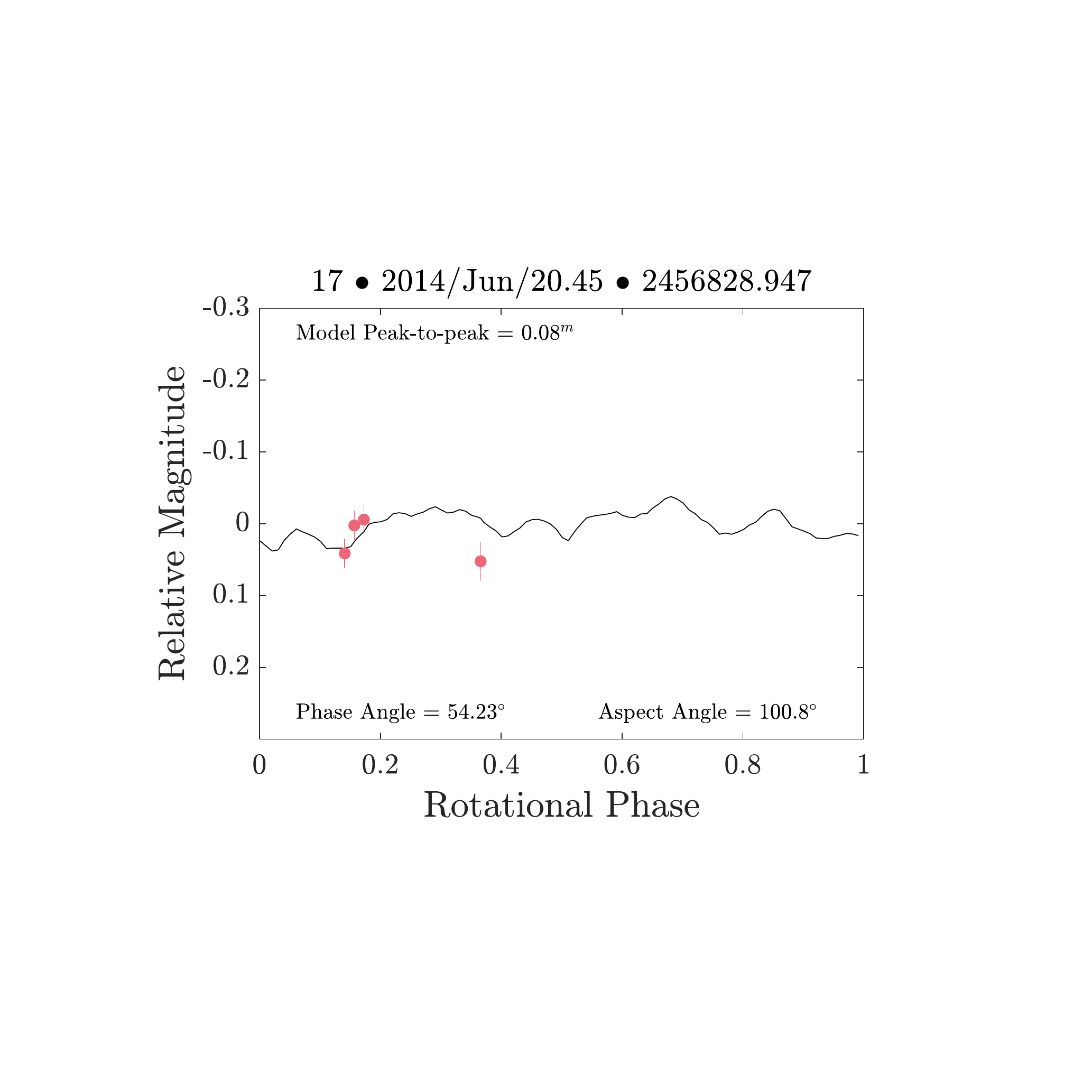} 		
		\includegraphics[width=.48\textwidth, trim=2cm 4cm 3.8cm 4cm, clip=true]{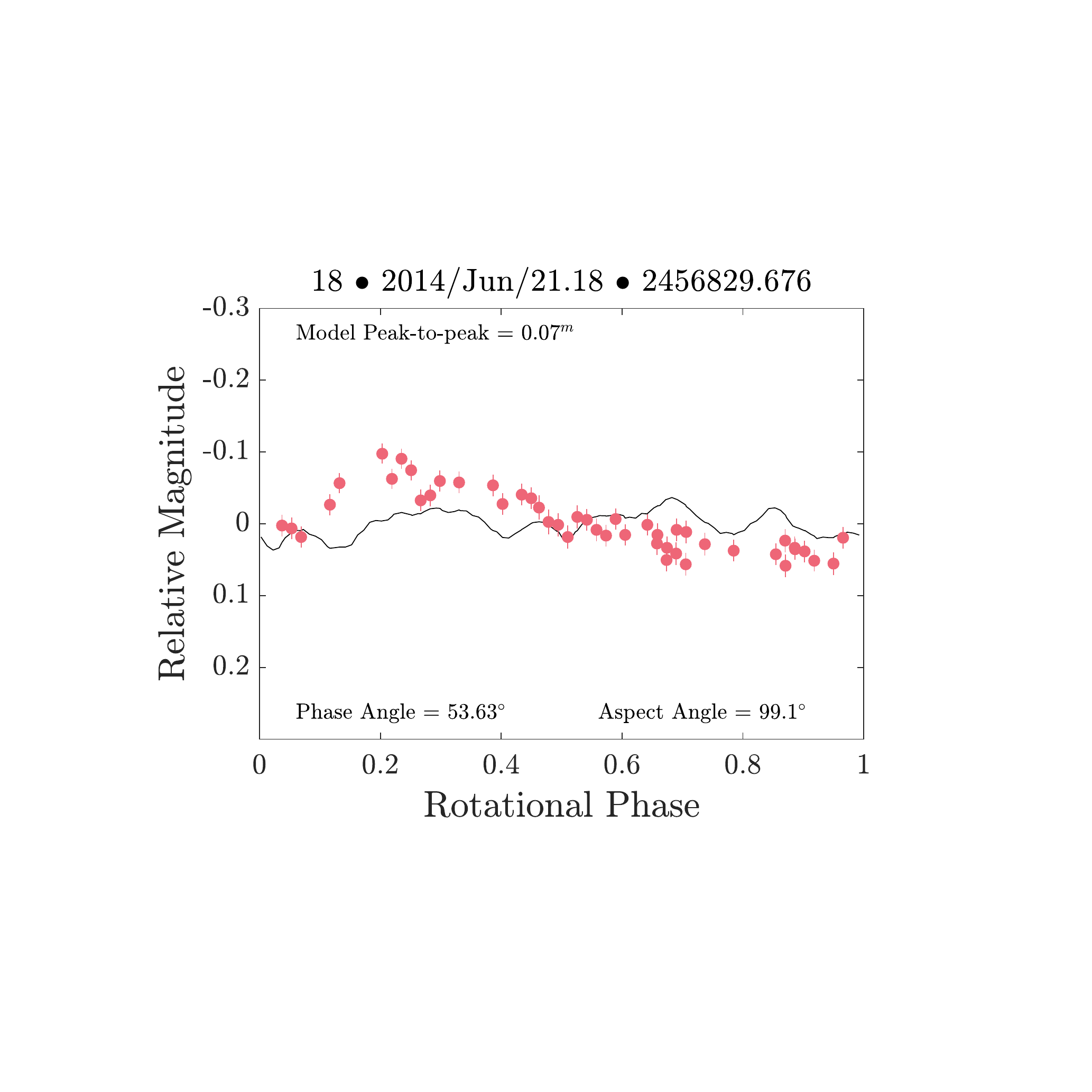} 
	}
	
	\resizebox{\hsize}{!}{
		\includegraphics[width=.48\textwidth, trim=2cm 4cm 3.8cm 4cm, clip=true]{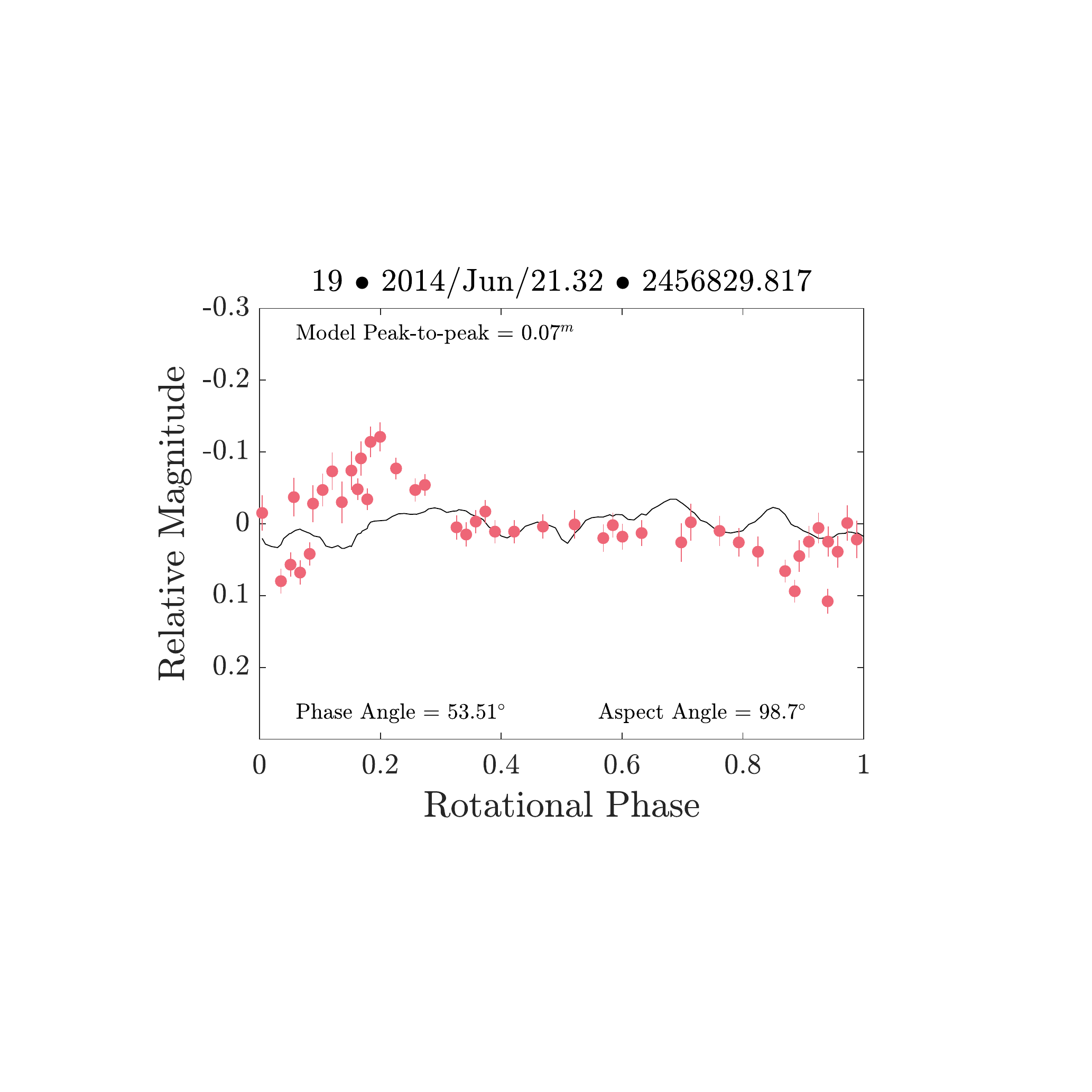} 
		\includegraphics[width=.48\textwidth, trim=2cm 4cm 3.8cm 4cm, clip=true]{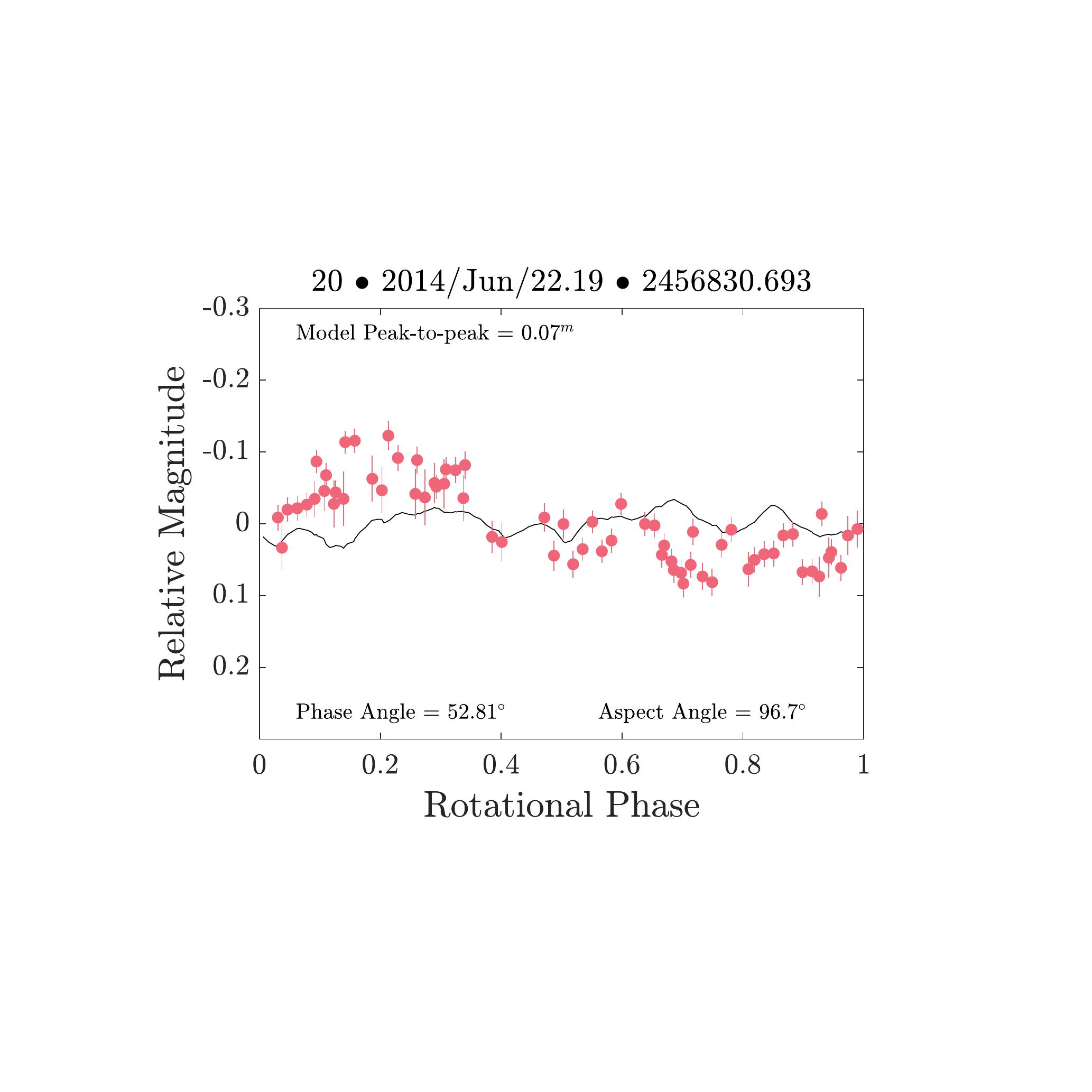} 		
		\includegraphics[width=.48\textwidth, trim=2cm 4cm 3.8cm 4cm, clip=true]{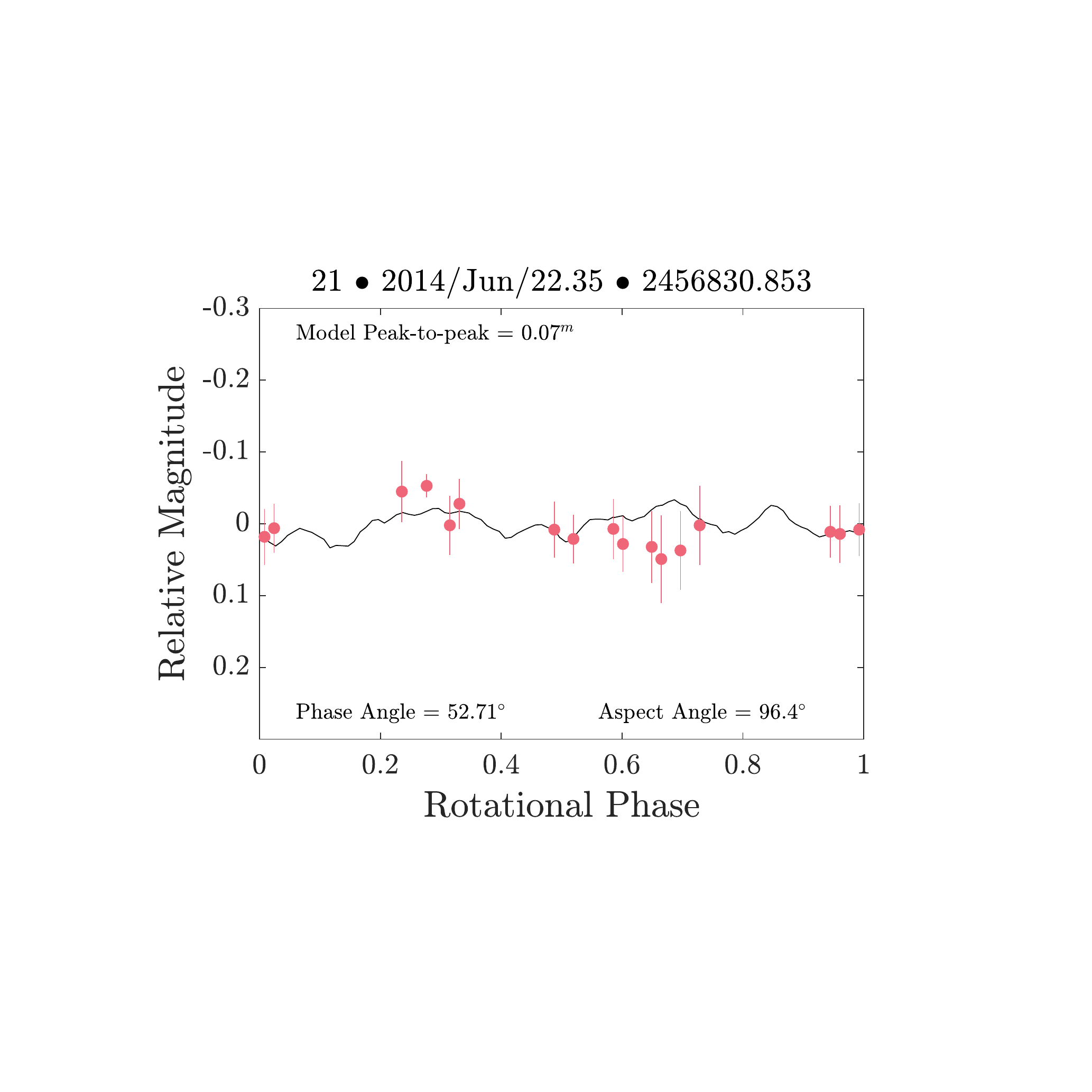} 
	}

	\resizebox{\hsize}{!}{
		\includegraphics[width=.48\textwidth, trim=2cm 4cm 3.8cm 4cm, clip=true]{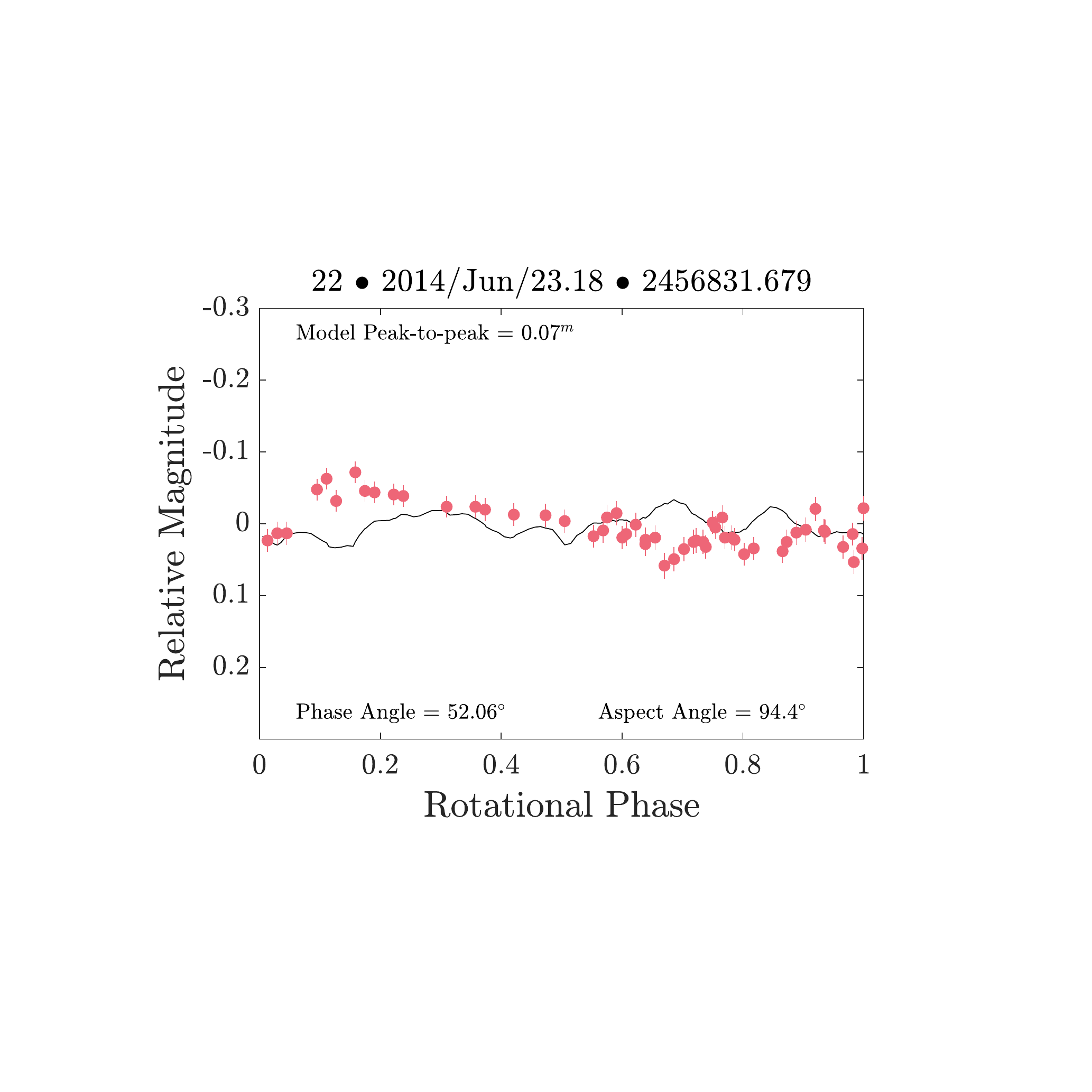} 
		\includegraphics[width=.48\textwidth, trim=2cm 4cm 3.8cm 4cm, clip=true]{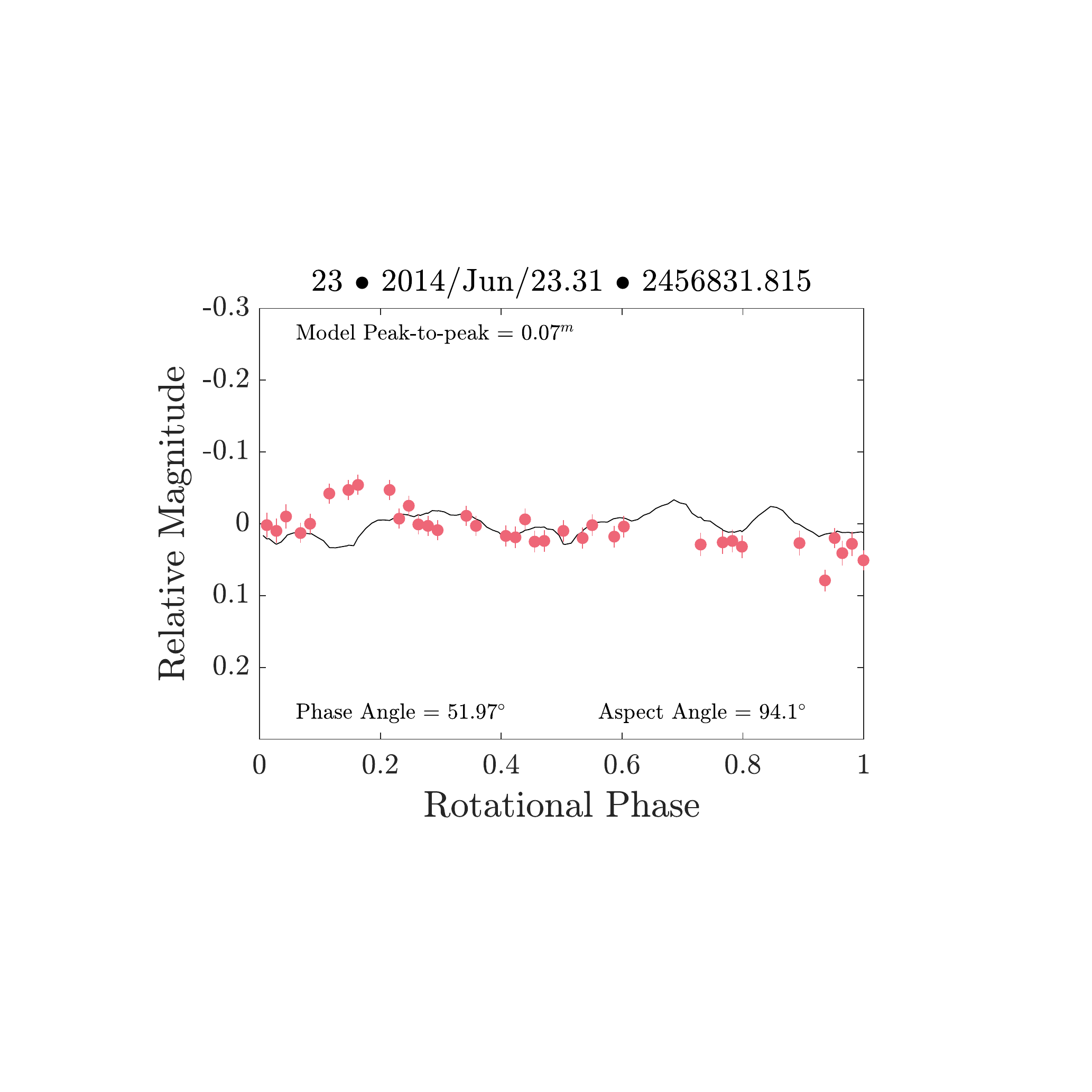} 		
		\includegraphics[width=.48\textwidth, trim=2cm 4cm 3.8cm 4cm, clip=true]{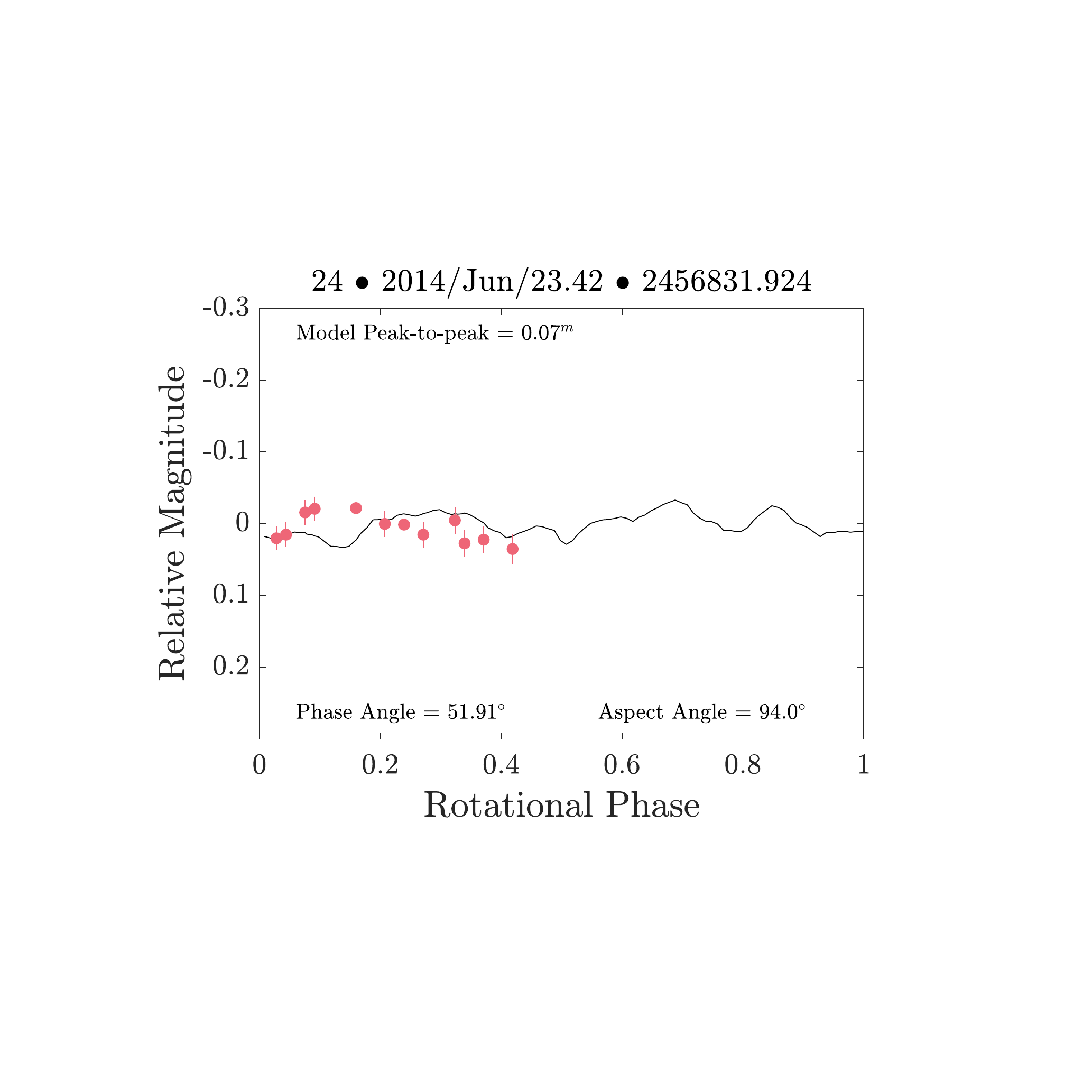} 
	}
	\caption{[Continued]
		\label{fig:radar-pro2}}
\end{figure*}

\addtocounter{figure}{-1}

\begin{figure*}
	\resizebox{\hsize}{!}{
		\includegraphics[width=.48\textwidth, trim=2cm 4cm 3.8cm 4cm, clip=true]{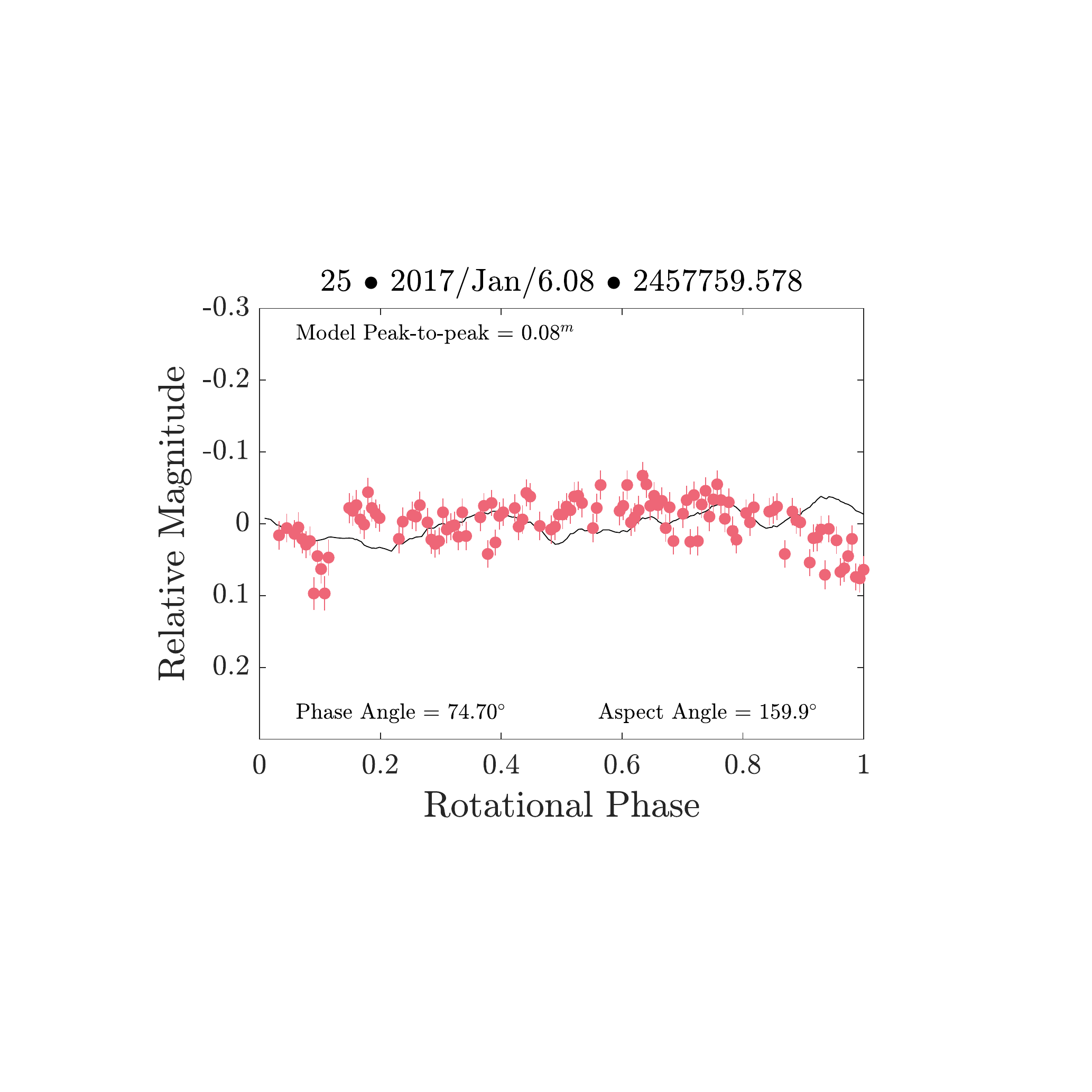} 
		\includegraphics[width=.48\textwidth, trim=2cm 4cm 3.8cm 4cm, clip=true]{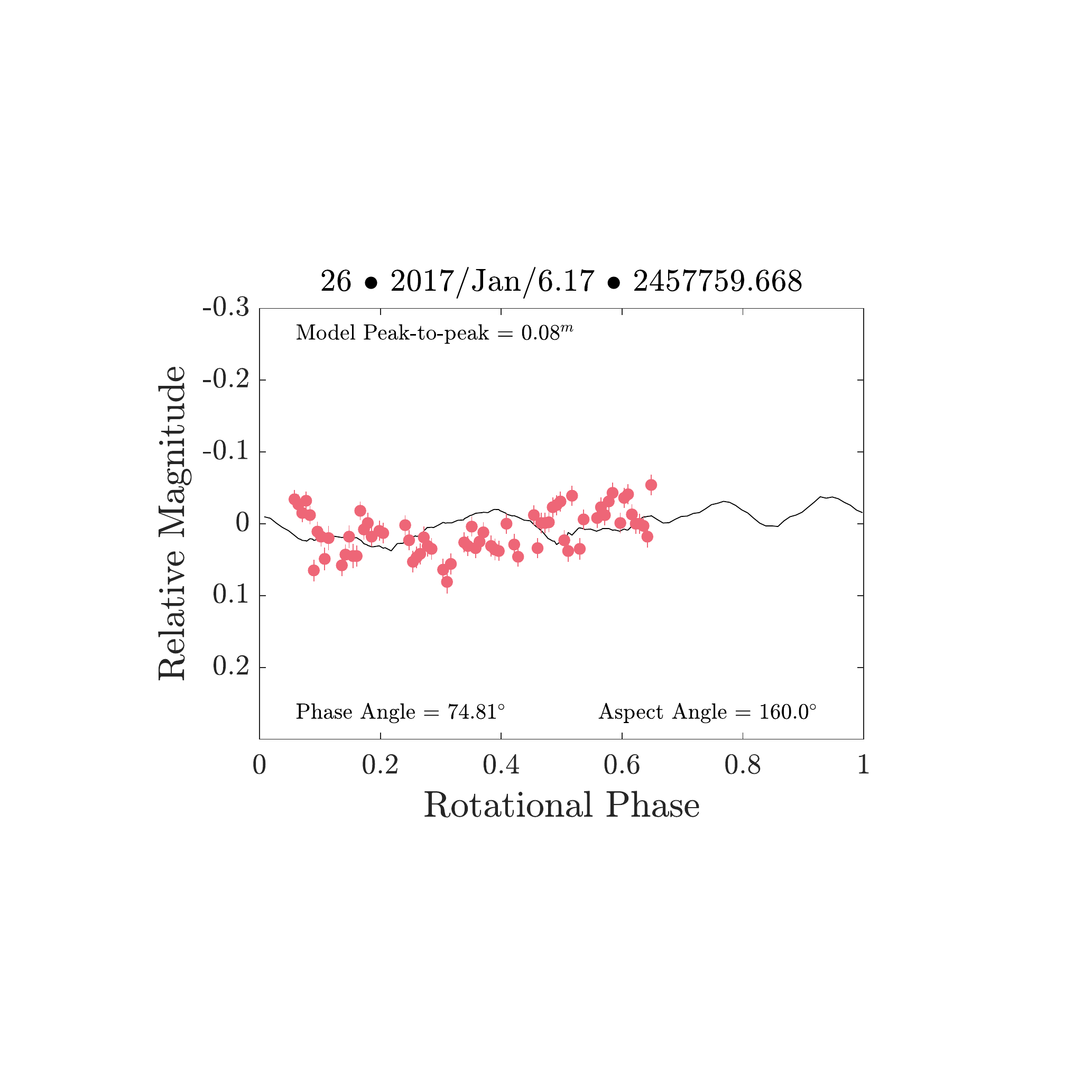} 		
		\includegraphics[width=.48\textwidth, trim=2cm 4cm 3.8cm 4cm, clip=true]{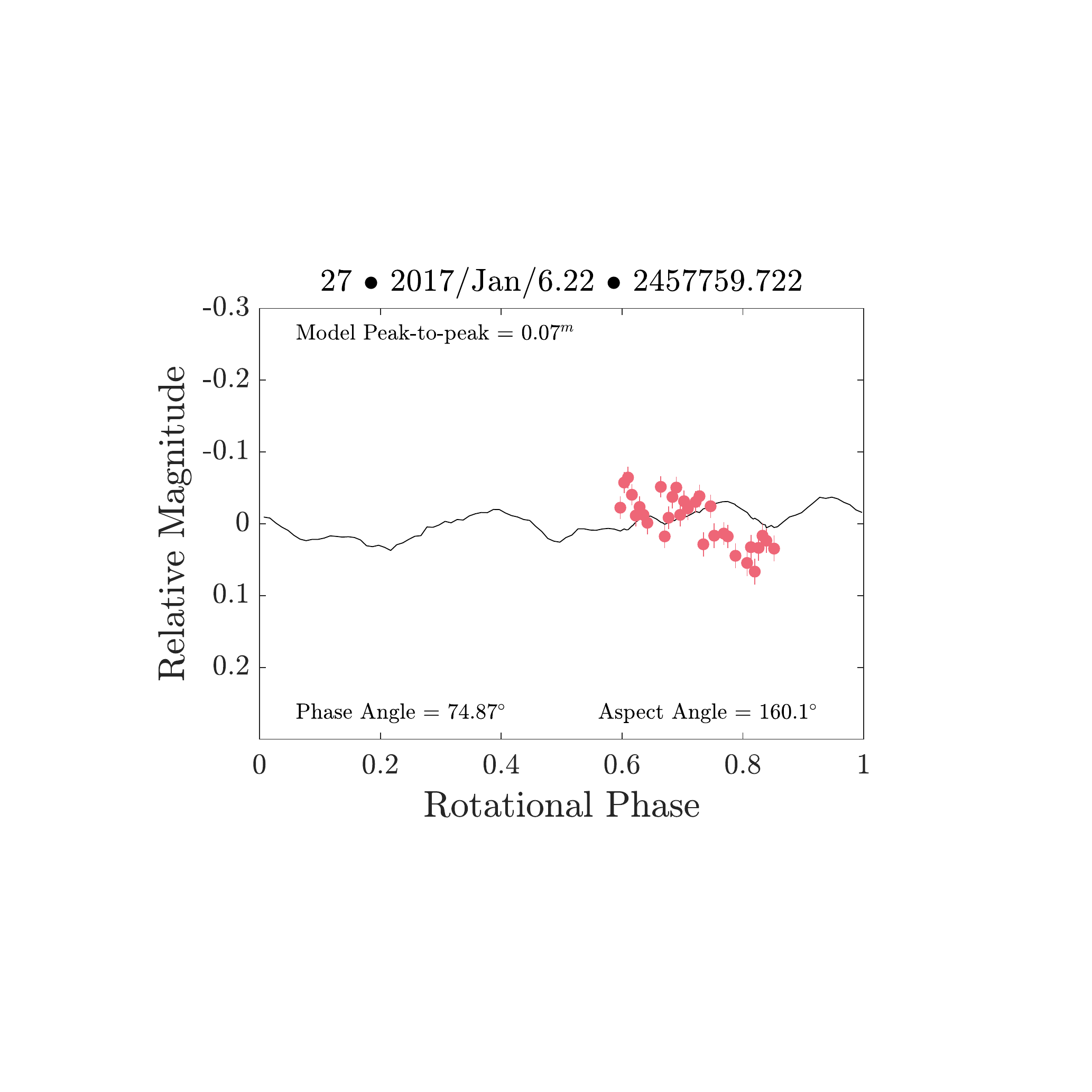} 
	}
	
	\resizebox{\hsize}{!}{
		\includegraphics[width=.48\textwidth, trim=2cm 4cm 3.8cm 4cm, clip=true]{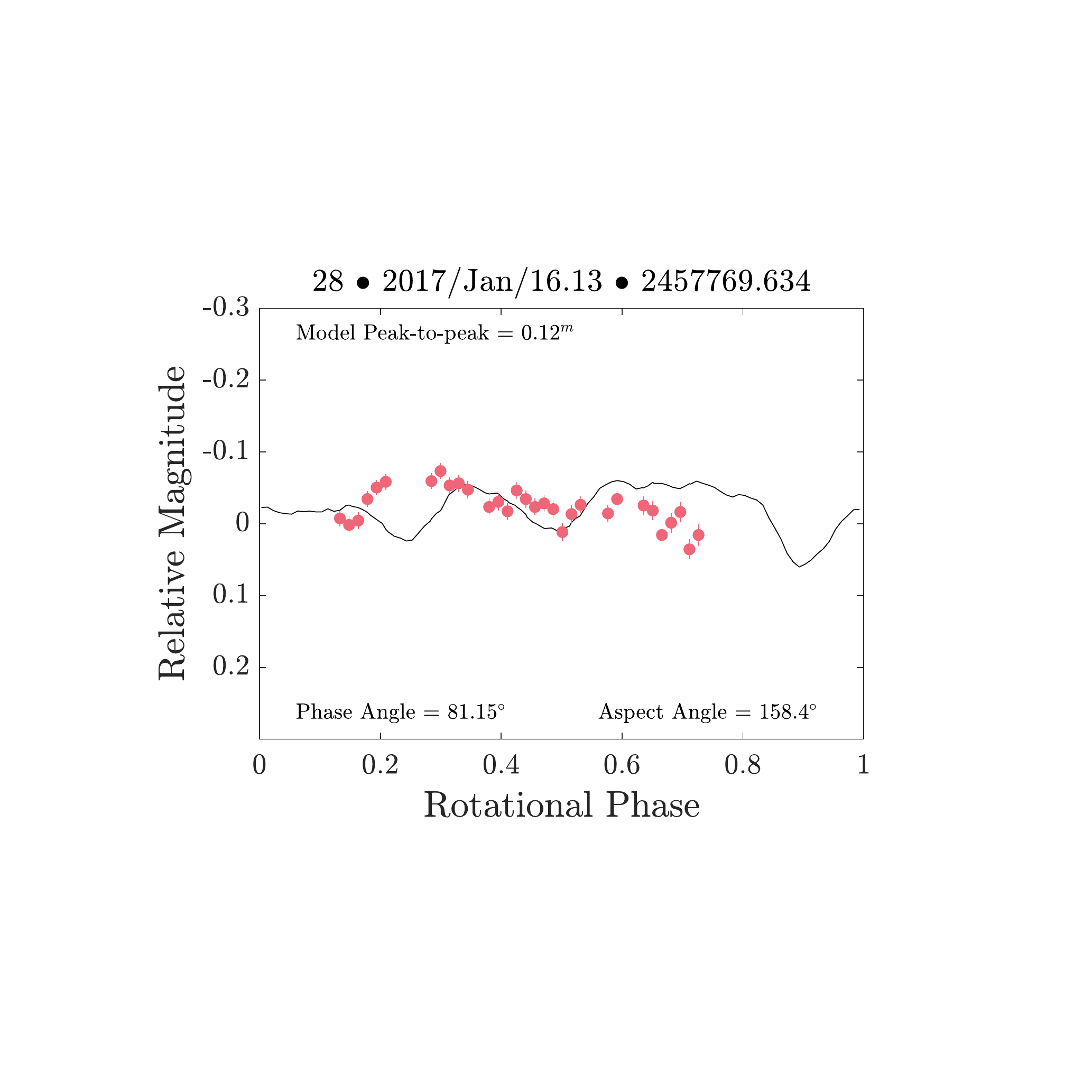} 
		\includegraphics[width=.48\textwidth, trim=2cm 4cm 3.8cm 4cm, clip=true]{LC/2102_lat+30lon036_v190906_20191127_initial_spinstate_29_fix.pdf} 		
		\includegraphics[width=.48\textwidth, trim=2cm 4cm 3.8cm 4cm, clip=true]{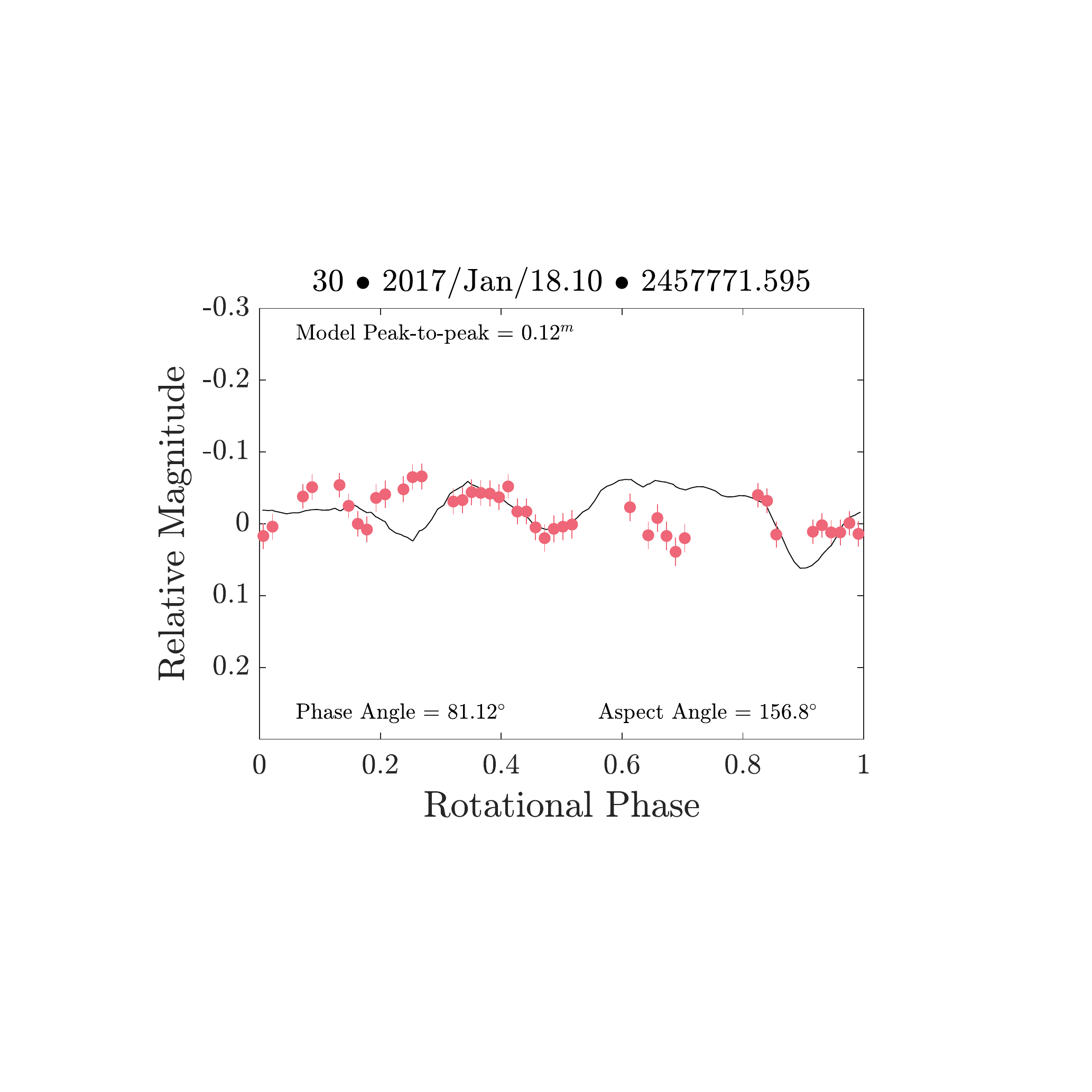} 
	}
	
	\resizebox{.6666\hsize}{!}{
		\includegraphics[width=.48\textwidth, trim=2cm 4cm 3.8cm 4cm, clip=true]{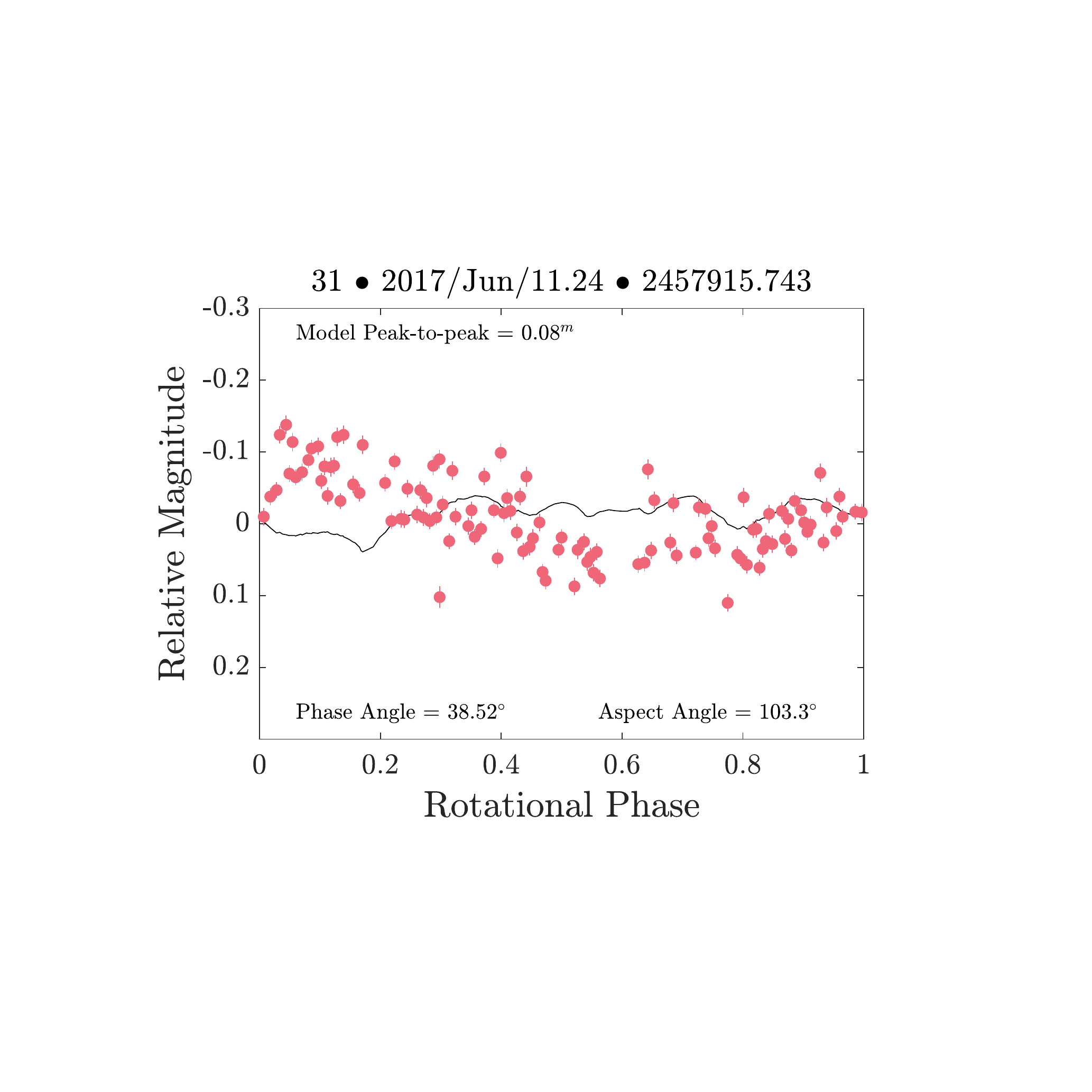} 
		\includegraphics[width=.48\textwidth, trim=2cm 4cm 3.8cm 4cm, clip=true]{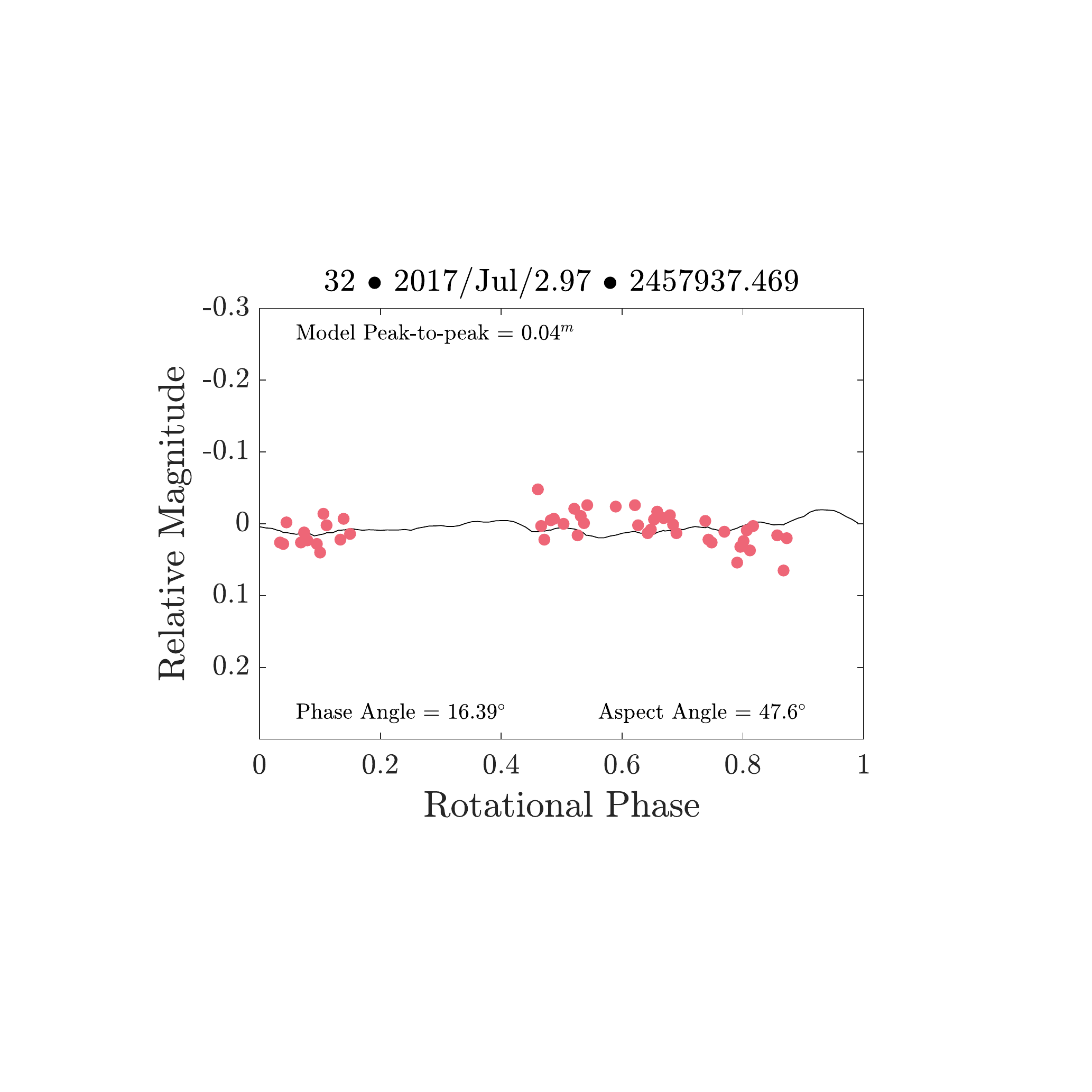} 	
	}
	
	\caption{[Continued]
		\label{fig:radar-pro3}}
\end{figure*}


\bsp	
\label{lastpage}
\end{document}